\documentclass[reprint,twocolumn,bibnotes,amsmath,amssymb,aps,prb,showpacs,longbibliography,floatfix]{revtex4-1}
\usepackage{amsmath,amssymb,bm,mathrsfs,graphicx, braket, times,amsthm,enumerate, scrextend}
\usepackage[colorlinks=true,citecolor=blue,linkcolor=blue]{hyperref}
\usepackage{longtable}
\usepackage{multirow}
\usepackage{array}
\usepackage{bigstrut}
\usepackage[all,cmtip]{xy}
\usepackage[normalem]{ulem}
\usepackage[usenames,dvipsnames]{color}
\usepackage{makecell}

\begin{document}
\title{Symmetry-enforced Band Nodes in 230 Space Groups}
\author{Lin Wu}
\author{Feng Tang}
\email{fengtang@nju.edu.cn}
\author{Xiangang Wan}
\affiliation{National Laboratory of Solid State Microstructures and School of Physics, Nanjing University, Nanjing 210093, China}
\affiliation{Collaborative Innovation Center of Advanced Microstructures, Nanjing University, Nanjing 210093, China}

\begin{abstract}
Crystallographic symmetries enforcing band touchings (BTs) in the Brillouin zone (BZ) have been utilized to classify and predict the topological semimetals. Though the early proposed topological semimetals contain isolated nodal points in the BZ, the proposed nodal line semimetals later could host various structures of several nodal lines/loops: nodal chains, nodal nets or Hopf-links, etc.  In this work, using compatibility relations, we first list all possible high symmetry lines (HSLs) that can be nodal lines itself, high symmetry planes (HSPLs) that can host nodal loops, high symmetry planes (HSPLs) that are nodal surfaces  for all 230 SGs, with spin-orbit coupling and time-reversal symmetry considered or not. We then show  how to diagnose a nodal loop from  the band crossing in an HSL, or nodal line/surface from irreducible representation (irrep) of an high-symmetry point (HSP), while the rest cases  correspond to nodal points.  Among our results, those essential cases, for which the nodal points/lines/loops/surfaces must exist, are highlighted since they are promising for the realizations of (nearly) ideal nodal point/line/loop/surface semimetals, as well as systems with flexible tunability owning fixed structure of topological nodal points/lines/loops/surfaces. Based on our results,  SGs allowing Hopf-link structure with one straight nodal line threading a nodal loop, or  two nesting nodal loops lying in two respective high symmetry planes, are highlighted, with the predicted materials being B$_5$Pb$_2$IO$_9$ in SG 34 and SrAl$_2$Au$_3$ in SG 62, respectively.  Our exhaustive results could serve as a useful guide for efficiently predicting and designing materials or artificial systems owning exotic geometric nodal structures of energy bands simply based on structure symmetries.
\end{abstract}

\date{\today}
\maketitle
\section{Introduction}\label{intro}
 Over the past fifteen years, topological materials in condensed physics have attracted broad interest due to their novel properties related with the nontrivial band topology and potential in device applications with low-energy consumptions \cite{Qi-RMP, Kane-RMP, Bansil-RMP,Weyl-Dirac-RMP,ando-jpsj,Balatsky-AIP}. Symmetry, e.g. time-reversal symmetry (TRS) and space group (SG) symmetry, plays a vital role in the classification, protection or prediction of various topological phases in realistic materials \cite{tenfold-1,tenfold-2,Fu-Kane,RMP-Chiu}. As with TRS that protects $\mathbb{Z}_2$ strong topological insulators,  crystallographic symmetry could protect  topological crystalline insulators (TCIs) with fancy degeneracies on symmetry-respected boundaries \cite{Fu-TCI, Ando-TCI}. The fruitful crystallographic symmetries in 230 SGs for three dimensional (3D) systems thus give rise to various TCIs: mirror Chern insulators \cite{mirrorchernTI}, hourglass insulators \cite{hourglass-Nature}, higher-order topological insulators \cite{Higher-orderTI-1,Higher-orderTI-2,Higher-orderTI-3,Higher-orderTI-4}, TCIs with rotation-anomaly \cite{Fang-Rot} and so on.

Other than topological insulating phases, various topological semimetals (TSMs) were proposed such as: Dirac semimetals \cite{SMYoung,Na3Bi,Cd3As2}, Weyl semimetals \cite{Weyl-Wan,TaAs-th-1,TaAs-th-2}, nodal line/loop semimetals \cite{Balents-NodalLine,ChenFang-NodalLine,NodalChain}, nodal surface semimetals \cite{NS-1,NS-2,NS-3}, nodal-link semimetals \cite{Hopf-1,Hopf-2,Hopf-3,Hopf-4} and so on. The crystallographic symmetries are also very important as in TCIs while different from TIs/TCIs, the nontrivial band closings occur in the bulk Brillouin zone (BZ) for TSMs. No matter for TCIs or TSMs, the band touching (BT) or band node in the boundary or bulk BZ can result in fascinating consequences such as chiral anomaly \cite{chiral-anomaly} and Klein tunneling \cite{klein}. Hence, an exhaustive classification of BTs based on group theory is of great importance. To our best knowledge, the exhaustive studies of BTs based on 230 SGs were mainly focused on nodal points \cite{manes, newfermions, SMYoung, Kane-8, Kramers-Weyl, nagaosa-dsm},  most of which were on nodal points pinned at high-symmetry points (HSPs). Furthermore,  compared with nodal point semimetals, the experimental realizations of nodal line/loop/surface semimetals are relatively scarce \cite{ex-1,ex-2,nodal-line-new-exp} though there have been many first-principles predictions \cite{MTC,Cu3PdN-1,Cu3PdN-2,CaTe,X3SiTe6,SrIrO3,ZrSiS,ReO2,dexi-1,dexi-2}. The reason is as follows: First, the SG symmetry protecting nodal lines/loops/surfaces should be respected in the boundaries to detect surface states related with bulk nodal line/surface structures; Second, the nodal lines/loops/surfaces tend to own finite bandwidth and coexist with other trivial bands in the same energy window in realistic materials. However, the nodal lines/loops/surfaces could cause interesting consequences such as novel long-range Coulomb interaction \cite{long-range} and intriguing magnetotransport properties \cite{cpl}.  Besides,  the surface states of nodal line/loop/surface semimetals are expected to be relatively flat so that electronic interaction could induce instabilities towards superconductivity \cite{NL-SC}.

Hence, developing a useful  guide to find and design nodal line/loop/surface semimetals  could benefit  realizations of ideal nodal line/loop/surface semimetals that are suitable for experiments  as well as providing a  platform  with coexisting nontrivial  band topology and prominent electron correlation.  At the same time, the  nodal line/loop or surface TSMs could also serve as a good starting point for creating nodal point TSMs or TIs/TCIs by introducing nontrivial energy gap through spin-orbit coupling (SOC) \cite{TaAs-th-1,TaAs-th-2} or appropriate external control such as strain and electric and magnetic fields. Very recently, based on crystallographic symmetries,  thousands of topological materials by large-scale database searches \cite{N2,N3,N1,tang-sa,2d-prb,s4} were predicted by  symmetry-indicators  \cite{SI,tang-np} or topological quantum chemistry \cite{TQC}. In the three topological materials databases built in Refs. \cite{N2,N3,N1}, the concrete types of BTs in the category of TSMs are still not clarified clearly, especially for nodal lines/loops/surfaces.  To identify a nodal line/loop or surface in a realistic material, conventionally one may firstly clarify the mechanism of forming nodal lines/loops/surfaces and then calculate the corresponding topological invariants such as Chern number or Berry phase \cite{Qi-RMP, Kane-RMP, Bansil-RMP,Weyl-Dirac-RMP,ando-jpsj,Balatsky-AIP}. Besides, $k\cdot p$ low energy effective models (allowed by symmetries) were usually built to verify the existence of nodal line/loop/surface \cite{Qi-RMP, Kane-RMP, Bansil-RMP,Weyl-Dirac-RMP,ando-jpsj,Balatsky-AIP}. Obviously, these two methods are not very efficient for a large-scale study. However, related with the SG symmetries, the breaking of compatibility relations (CRs) can serve as the mechanism or an indicator of BTs in HSPs, high-symmetry lines or planes (HSLs or HSPLs), which is very convenient.

In this work, we list all possible SGs with concrete positions hosting nodal lines/loops/surfaces based on the irreducible representations (irreps) of little group of HSLs and HSPLs, and their CRs (hence these nodal lines/loops/surfaces are symmetry-enforced). We focus on four settings as (TRS,SOC) where both TRS and SOC are considered, (TRS,NSOC) where TRS is considered while SOC is neglected, (NTRS,SOC) where TRS is not present while SOC is considered and (NTRS,NSOC) where both TRS and SOC are absent.  We also show that from the knowledge of band crossings in the HSL, one can diagnose  a nodal loop in a neighboring HSPL  or even the configuration of several nodal loops. Similarly, the irrep at HSP can imply a nodal line or surface coindicing with the neighboring HSL or HSPL, respectively. The nodal points lying in HSLs or pinned at HSPs are also found concomitantly.  Hence, our work gives a full classification of symmetry-enforced all types of band nodes in 230 SGs.  Based on the results of nodal loops implied by the band crossings in an HSL,  we propose a practical strategy of realizing Hopf-link semimetal with two nesting nodal loops from an hourglass band structure. The essential cases are highlighted since they are guaranteed to exist by the SG symmetry in corresponding setting in the same spirit of filling-enforced band crossings \cite{filling-1, filling-2, filling-3}.

This paper is organized as follows. In Sec. \ref{CRs}, we give a brief overview of CRs in energy bands and sketch the strategy of finding symmetry-enforced nodal points/lines/loops/surfaces using CRs.  The main results are also summarized. In Sec. \ref{HSPL}, we discuss single-valued and double-valued irreps for HSPLs considering the effect of TRS or not, where all possible positions of nodal loops are found which lies in the HSPLs. The HSPLs with degenerate irrep which splits in general direction are nodal surfaces and are further shown to be essential. Then in Sec. \ref{HSL},  we show that many HSLs  themselves are (straight) nodal lines.
Besides, the band crossing in an HSL may indicate a nodal loop in neighboring HSPL and configurations of  several nodal loops in the neighboring HSPLs are discussed. Then Sec. \ref{HSP} deals with the irreps at HSPs. Sec. \ref{essential} is devoted to the essential results mainly related with the hourglass band connectivity. We then take SGs 34 and 62 as SG examples to demonstrate how to apply our results to figure out all possible nodal lines/loops and highlight two kinds of Hopf-link nodal structures in Secs. \ref{hopf-1} and \ref{hopf-2}, respectively. The materials  realizations of these two kinds of Hopf-link nodal structures are predicted by first-principles calculations in Sec. \ref{material}. Finally Sec. \ref{conclu} contains Conclusions and Perspectives.

\section{The strategy and brief summary of the results}\label{CRs}
In this section, we first present a short overview of the CRs for energy bands. Let us assume two associated momenta named by $k_1$ and $k_2$ in the BZ, respectively. With no loss of generality,  assume that $k_1$ has an equal or higher symmetry than $k_2$: For example, $k_1=(0,0,0)$ and $k_2=(0,0,k_z)$ so that $k_2\rightarrow k_1$ if $k_z\rightarrow 0$.  Hence, energy bands originated from one degenerate energy level at $k_1$ may  split in $k_2$ and the corresponding splitting pattern can be found based on the CRs \cite{bradley, Bilbao}. Such degenerate energy level at $k_1$ corresponds to some irrep of little group of $k_1$ ($G(k_1)$),  denoted by $D_{k_1}^i$. $D_{k_1}^i$ can be written in the form of a direct sum of the irreps of little group of $k_2$, $G(k_2)$ with the irreps denoted by $D_{k_2}^j$ as follows:
\begin{equation}\label{CR}
D_{k_1}^i\rightarrow \mathop{\oplus}\limits_{j=1}^{N_{k_2}} d_j D_{k_2}^j, \\
\end{equation}
where $N_{k_2}$ is the total number of different irreps at $k_2$ and $d_j$ in Eq. \ref{CR} is the number of occurrences of $D_{k_2}^j$ in the decomposition which must be non-negative integers.  Note that conventional symmetry analysis using operator algebra for representation of an SG is equivalent with the abstract group method \cite{bradley}. All irreps (single-valued or double-valued) and their CRs of 230 SGs have been listed exhaustively on the Bilbao sever \cite{Bilbao,remark}, based on which we could make an exhaustive study on nodal points/lines/loops/surfaces for all 230 SGs.   For nodal points which can be diagnosed by CRs, it is required that away from the BT (which may be located at HSP, or lying in an HSL), the band should split in any direction. On the other hand, once in some direction away from a BT, the band don't split, higher dimensional BT (nodal lines/loops/surfaces) could occur in this direction. Our strategy for identifying the all kinds of BTs  is summarized as below:\\

\textbf{Case (a)}: $k$ is an HSP. If there exists a degenerate irrep of $G(k)$ and this irrep can decompose into more than one irrep  as Eq. \ref{CR} in all neighboring HSLs, HSPLs and in general points (GPs), $k$ can thus be a  nodal point.  Furthermore, If all the irreps of $G(k)$ satisfy the condition,  the HSP is thus an  essential nodal point. These results are listed in Sec. I A of the Supplementary Material (SM) \cite{SM}. When the point of the HSP contains no achiral operations, the nodal point \cite{Kramers-Weyl, CoSi} would be a chiral point. Besides, the degeneracy of the nodal points (as well as other types of band nodes) can be easily obtained and thus given explicitly in our results;

For the rest degenerate irreps, namely, they would not split in some HSL or HSPL containing the HSP, which just imply a nodal line (if the irreps split in  neighboring HSPLs and GPs) or surface (if the irreps split in GPs) coinciding with the HSL or HSPL, respectively.  We list these results in Secs. I B and I C of the SM \cite{SM}, respectively.

\textbf{Case (b)}: $k$ is an HSL. If there exists a degenerate irrep of $G(k)$ and this irrep can decompose into more than one irrep  as Eq. \ref{CR} in all neighboring HSPLs  and in GPs, $k$ can thus be a  nodal line (in this work, when we mention ``nodal line'', the nodal line is straight coinciding with an HSL).  Furthermore, If all the irreps of $G(k)$ satisfy the condition, this HSL must be straight nodal line, which means that it is essential. These results are listed in Sec. II A of the SM \cite{SM};

For the rest degenerate irreps, namely, they would not split in some HSPL containing the HSL, which just imply a nodal surface coinciding with the HSPL if the irreps split in  GPs.  We list these results in Sec. II B of the SM \cite{SM}.

Furthermore, when there exist at least two different irreps in the HSL, there could exist a band crossing comprised of these two irreps. For this case, the band crossing could be just a nodal point if the two different irreps are  found not to  keep being two different irreps in all the neighboring HSPLs. In this case, the nodal point can be a chiral point when the point group of the HSL contains no achiral operations.  The results are listed in Sec. II C of the SM \cite{SM}. Otherwise, the band crossing would lie in a nodal loop in the neighboring HSPL, with the results listed in Sec. II D of the SM \cite{SM}.

\textbf{Case (c)}: $k$ is an HSPL. The same as cases (a) and (b), we should check whether it can own a degenerate irrep which splits into GPs, and if so, $k$ can be a nodal surface. Such nodal surface just coincides with the HSPL.  These results are listed in Sec. III A of the SM \cite{SM}.

Besides, if there exist two different irreps of $G(k)$, $k$ can thus host a nodal loop of which the Bloch states in the HSPL correspond to these two different irreps of $G(k)$.  We list the corresponding results in Sec. III B of the SM \cite{SM}.

\begin{table}
\begin{tabular}{c|c|c|c}
  \hline
  % after \\: \hline or \cline{col1-col2} \cline{col3-col4} ...
  nodal point & nodal line & nodal loop & nodal surface \\
  \hline\hline
  Secs. I A, II C   & Secs. II A, I B & Secs. III B, II D  & Secs. III A, I C, II B\\
  \hline
\end{tabular}
\caption{The sections in the SM \cite{SM} for the three types of band nodes.}\label{results}
\end{table}

In Table \ref{results}, we show the concrete sections where the results for specific type of band node are listed in the SM \cite{SM}. As applications, we propose  two Hopf-link structures from our results: one is formed by a straight nodal line threading a nodal loop while the other is formed by two nesting nodal loops. For the former case, all possible results are given in Sec. IV of the SM \cite{SM}, from which, we list essential cases in Table \ref{table-hopf} in the main text. For the latter, we first list all possible combinations of two different irreps in HSL in Sec. V of the SM \cite{SM}, for which, two bands in the HSL with two different irreps cross each other and the band crossing point could lie in the nodal loop(s) within neighboring HSPL(s) (there are two inequivalent HSPLs containing the HSL). Then if two or more HSPLs containing the HSL could host  nodal loops originated from one band crossing in the HSL, these nodal loops  would link each other by  the band crossing point. On the contrary, if the two nodal loops in the two inequivalent HSPLs are not linked by a point,   they have a chance to form a nesting Hopf-link structure as shown in Fig. \ref{demon}(c3). For this, the band crossing in the HSL could indicate only one nodal loop in one HSPL while there also exists another band crossing in this HSL indicating another nodal loop in another HSPL. These results are listed in Sec. VI of the SM \cite{SM}. Though such situation may not imply a Hopf-link to exist necessarily, we propose a strategy by which we can tune a Hopf-link to appear easily:  we can require one nodal loop in the Hopf-link to be essential, for example, due to hourglass band connectivity while the other nodal loop is accidental, as demonstrated in Fig. \ref{demon}(c3) by a ``twisted-hourglass''. We list all possible such ``twisted-hourglasses'' in Table \ref{hopf-2loop-1ess} of the main text and also show concrete material example following the strategy.

In the following sections, we first illustrate the irreps in HSPL for which the situation is very simple and the results for nodal surfaces and loops can be obtained.

\section{HSPL: being nodal surface or hosting nodal loop }\label{HSPL}
By definition, the little group of the HSPL must contain two elements: identity $E$ and mirror (or glide) $M$. In general $M^2=T_{\mathbf{R}}E$ where $T_\mathbf{R}$ is a translation operator and the translation $\mathbf{R}$ is the translation part of $M^2$, parallel with the HSPL. Concretely, for $\mathbf{k}$ in the HSPL, $M^2\rightarrow e^{-i\mathbf{k}\cdot\mathbf{R}}$ so that the eigenvalue of $M$ can be $\lambda_M=\pm e^{-i\mathbf{k}\cdot\mathbf{R}/2}$ or $\lambda_M=\pm i e^{-i\mathbf{k}\cdot\mathbf{R}/2}$, for single and double valued representations, respectively (in other words, for negligible and significant SOC, respectively). In fact, the little group $G(\mathbf{k})$ owns two different irreps corresponding to two opposite eigenvalues of $M$. Since in the HSPL, the Bloch Hamiltonian $H_{\mathbf{k}}$ commutes with $M$, each band in the HSPL can be labeled by $\lambda_M$ or equivalently the irrep of $G(\mathbf{k})$. Note that we don't consider TRS in the above discussion and in this setting, HSPL thus could host a nodal loop when two bands with inverse values of $\lambda_M$ cross each other. These results are listed for all 230 SGs in Sec. III B of the SM \cite{SM}, from which we note that the results for negligible and significant SOC are the same which can be easily understood from above.

Furthermore, since both of the two irreps in the HSPL without TRS are non-degenerate,  the HSPL cannot become a nodal surface. However, when considering TRS, one should check whether there exists an SG operator, denoted by $\beta$, whose point part being $p_\beta$,  acts on $\mathbf{k}$ resulting in that $p_\beta \mathbf{k}=-\mathbf{k}+\mathbf{G}$ where $\mathbf{G}$ is an arbitrary reciprocal lattice vector. If this is true, one should check the effect of $T\beta$ ($T$ is the time-reversal operator) on the two irreps of $G(k)$. Then one could be encountered with three possible situations \cite{bradley} listed in the following:

1. each of the irrep can be related with itself preserving the non-degeneracy;

2. each of the irrep can be related with itself doubling the degeneracy resulting in a two-dimensional (2D) (co-)irrep;

3. or two different irreps  are  related with each other leading to a 2D (co-)irrep. \\

Note that in the strict manner we should use ``co-irrep'' when antiunitary symmetry is considered but we still use the term ``irrep'' without any ambiguity in this work. The above three situations have been attributed to any irrep of little group on the Bilbao server \cite{Bilbao} by the ``reality'' of the irrep. In the above first two situations, there are still two different irreps in the HSPL while in the third, there is only one irrep. As discussed in Case (c) in Sec. \ref{CRs}, when two different irreps exist, namely, in the 1st or 2nd situation, two-fold or four-fold degenerate BTs may occur in the HSPL, respectively. Such BTs contain the two different irreps of the HSPL and constitute a nodal loop. On the other hand, when 2D irrep(s) are formed for the 2nd and 3rd situations, the HSPL may become a flat nodal surface if the SG is not centrosymmetric when SOC is included, otherwise the bands in the HSPL would not split in any direction away from the HSPL due to Kramers degeneracy. And when SOC is negligible, all cases in the 2nd and 3rd situations are flat nodal surfaces. The results for all nodal surfaces can be found in Sec. III A of SM \cite{SM}, and it is easy to conclude that they also essentially form nodal surfaces.

\section{HSL: being straight nodal line or hosting band crossing lying in nodal loop or implying a nodal surface}\label{HSL}

In the following, we discuss nodal lines/loops which are related with the irreps and CRs in the HSL. Different from HSLs, the point group of HSL may contain $c_n$ rotation with the rotation axis along the HSL, thus more irreps would be found. The same as the above discussion, the HSL may itself form a nodal line. For this case, we also require that the HSL should allow at least one degenerate irrep which splits in any direction away from the HSL. For the effect of  anti-unitary symmetry $T\beta$ when TRS is considered, it may remove the splitting  and thus may turn the original straight nodal line to lie in a nodal surface, or it may create a straight nodal line when it pairs two irreps in the HSL together. For any degenerate irrep in the HSL, it may imply a nodal surface if the irrep keeps being the sole irrep in neighboring HSPL while this irrep splits in GPs.

\begin{figure*}[!hbtp]
	\includegraphics[width=1. \textwidth]{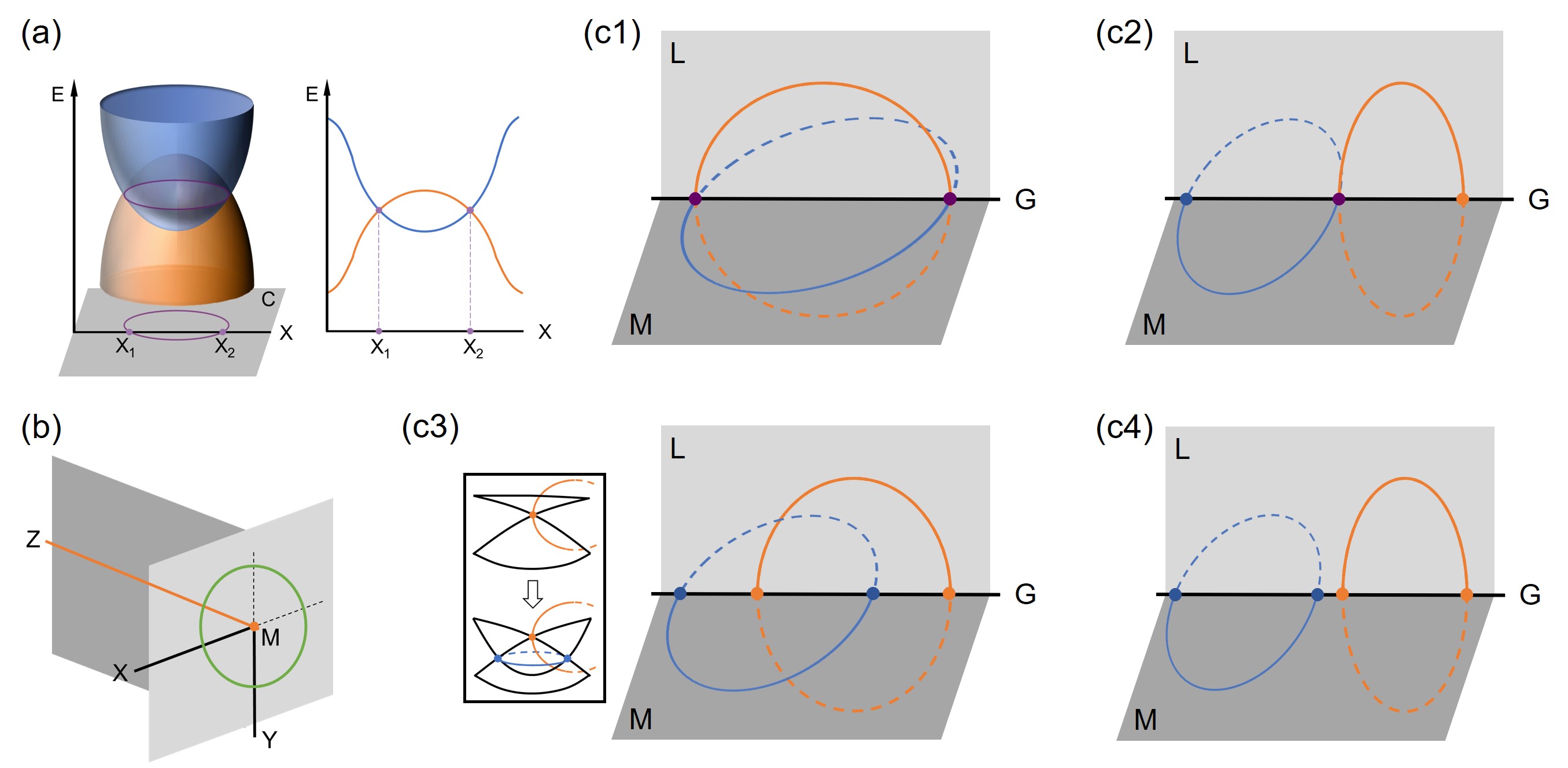}
   \caption{(a) Demonstration of diagnosing nodal loop from band structure along an HSL. Two bands  plotted in different colors in HSPL C cross each other to form a  nodal loop indicated by the purple curve. Only from the HSL X $\in$ C, the band structure shown in the right panel clearly shows the band crossing by X$_1$ and X$_2$ , we can know the existence of a nodal loop lying in C, required by the CRs. (b) The Hopf-link nodal structure consisting of a straight nodal line (itself being an HSL,  and here it's MZ) in orange threading a nodal loop (in green) lying in an HSPL (here, MXY) where M is the intersection point of the HSL and HSPL. (c1-c4) Consider  two HSLs, named by L  and M  and the intersection of them, the HSL G. L and M could host a nodal loop, respectively. The nodal loops can be diagnosed from the band crossing points lying in G. Owing to the CRs, the symmetry-content of the band crossing, uniquely determine where the resulting nodal loop(s). For (c1), all of the band crossings in G lie in two nodal loops within L and M. For (c2), some band crossing in G can lie in two nodal loops in L and M while some can only lie in one nodal loop within L or M. These two cases correspond to nodal chain structure. For (c3) and (c4), all the band crossings in G can only lie in one nodal loop in L or M, providing the possibility of realizing a nesting Hopf-link nodal structure. In the inset of (c3), we display a twisted hourglass where the top is dragged downward to cross the two necks, resulting in two band crossings (in blue) and form a nodal loop. The nodal loop possibly nest with another one formed by the essential nodal point (in orange) corresponding to the crossing of the two necks.  }
\label{demon}
\end{figure*}

 Besides, two bands with different irrep in the HSL may cross each other resulting a band crossing point in the HSL.  Such a band crossing point may simply be a nodal point, namely, the bands split in any direction away from it. Otherwise it must lie in a nodal loop within some HSPL which contains the HSL.  In order to know which case to occur, we should firstly know all possible (inequivalent) HSPLs containing the HSL (denoted by $k$) as shown in Fig. \ref{demon}(c), denoted by $K_a$,$K_b$,\ldots Assume that the band crossing in the HSL contains two different irreps as $D(k)^1$ and $D(k)^2$. Then we should subsequently check whether these two irreps can still maintain to be two different irreps in $K_j$ ($j=a,b,\ldots$) according to the CRs, i.e.:
\begin{equation}\label{hsl-hspl}
    D(k)^1\rightarrow D(K_j)^m, D(k)^2\rightarrow D(K_j)^{m'},
\end{equation}
where $m,m'$ denote two different irreps in $K_j$. If the condition in Eq. \ref{hsl-hspl} is satisfied, there must be a nodal loop in $K_j$. As shown in Fig. \ref{demon}(a), two bands with different irreps indicated by different colors in HSPL, C, cross each other to form a nodal loop. Conventionally one need to calculate band structures along many paths in the HSPL to identify the existence of a nodal loop. This method is not very convenient especially when the nodal loop is very small in size. However, as discussed above,  one can diagnose nodal loop from band crossing in HSL, and such  identification of nodal loop is more operable. As in Fig. \ref{demon}(a), X $\in$ C, is an HSL and could host different irreps, as denoted by X$_1$ and X$_2$, thus allowing robust symmetry-enforced band crossing. This band crossing can be found quickly in the band structure  along X in Fig. \ref{demon}(c). Actually the band crossing containing X$_1$ and X$_2$ must lie in a nodal loop in HSPL C, from the knowledge of CRs.  We list all doublets of irreps in the HSL exhaustively which could form a band crossing in this HSL and preserve to be two different irreps in the neighboring HSPL, namely resulting in a nodal loop in this HSPL, in Sec. II D of the SM \cite{SM}.  In total, there are 466, 466, 428, 194 HSPLs which can host nodal loops in the settings of (TRS,SOC),(TRS,NSOC),(NTRS,SOC),(NTRS,NSOC), respectively. Among these HSPLs, 440, 356, 398, 185 HSPLs could contain at least one HSL of which the band crossing could imply a nodal loop in the HSPL.

 Next we should point out that such simple strategy could be used to find various geometrical structures of different nodal loops. Note that when the HSL has symmetry of $c_n$ when $n=2,4,6$, there would be two inequivalent HSPLs associated with one HSL.  And for $n=3,4,6$, $c_n$ could relate one HSL with the other $2, 1, 2$ equivalent HSPLs, respectively. For the latter case, when the band crossing in the HSL which indicates a nodal loop exist, there  would be naturally a nodal chain structure formed by the other $2, 1, 2$  nodal loops, respectively. In the following, we consider the former situation, where two nodal loops can form a Hopf-link structure.

\subsection{Possible configurations of nodal loops in several HSPLs}\label{HSL-2}
Different from isolated nodal points, nodal loops own more kobs to tune the Fermi surface geometry and its topology. The nodal loops can be isolated, which can be tuned to gapped or contract into nodal points. Interestingly, the nodal loops can also be linked at a point or nested, allowing more fruitful evolutions  under external  perturbations. We here proposed that,  among the two  HSPLs that contain the HSL, if both of the two HSPLs could host a nodal loop from the band crossings in the HSL, we can obtain possible configurations of these nodal loops simply from CRs.  For simplicity, as shown in Fig. \ref{demon}(c1-c4), we show two HSPLs where there could exist a nodal loop in respective HSPL.  Generally, we could classify the band crossing in the HSL into two types: for one type, it lies in both nodal loops in the two HSPLs while for the other type, it only lies in one HSPL. Obviously, the first type requires that the two nodal loops in the two HSPLs link each other at the band crossing point in the HSL as shown in Fig. \ref{demon}(c1), forming a nodal chain. For the second type, if all possible band crossings in the HSL result in  only one nodal loop in the respective HSPL, the nodal loops must not be linked with each other as shown in Figs. \ref{demon}(c3) and (c4): for Fig. \ref{demon}(c3), the two nodal loops constitute a Hopf-link nodal structure \cite{co2mnga}.  Note that  for Fig. \ref{demon}(c2), both of the two types of band crossings in the HSL exist, and it shows another possible configurations forming a nodal chain structure.

Note the inset of Fig. \ref{demon}(c3). There we show a ``twisted hourglass'' where the top of the hourglass is sketched to intersect with the two necking, and these two band crossings (in blue) are accidental (while the band crossing (in orange) in two necks is essential).   When these two accidental band crossings form a nodal loop (in blue) in one HSPL (M), this nodal loop would possibly nest with another nodal loop ( in orange, which is indicated by the hourglass essential band crossing) in another HSPL (G).

\section{HSP: nodal points or implying nodal lines/surfaces}\label{HSP}
Lastly, we consider the HSPs.  Note that different from HSL or HSPL, even though two energy bands at the HSP with two different irreps are tuned to be accidentally degenerate, such BT is easy to be gapped.   In order to own a band node related with the HSP, the HSP must own a degenerate irrep. To guarantee the irrep to constitute a nodal point, it is required that the irrep should split in any direction away from the HSP. And since the HSP owns a higher symmetry than any neighborhood, it is natural to think that the irrep should be split naively.  However, this is not always true. If the irrep cannot split to the neighboring HSL, the irrep can be used as the indicator of a nodal line (simply being the HSL), when it is further found that the irrep in the HSL splits in all neighboring HSPLs. Besides, when the irrep in the HSP is found to become one irrep in a neighboring HSPL, it then indicates the nodal surface (simply being the HSPL), when the irrep could split in GPs. Note that the discussions above are all based on CRs, not replying on any $k\cdot p$ models \cite{manes, SMYoung, nagaosa-dsm, newfermions, Kane-8} as does conventionally,  thus there should be no worry about the cutoff of the $k\cdot p$ expansion.
\section{Essential cases}\label{essential}

So far we are mainly concerned with the possible positions in the BZ where the nodal points/lines/loops/surfaces could appear. The identification for them may need further calculations of irreps. However, some of them  are found to be essential, which means that they must exist as long as the material belongs to the required SGs in suitable setting even without checking the electronic band structures. We have pointed out that the nodal surfaces as listed in Sec. III A of SM \cite{SM} are all essential and thus we only discuss essential nodal points/lines (loops) in the following. The essential cases are of important significance since they are promising to bring about ideal semimetals which host band nodes nearly around the Fermi level.

For essential nodal points pinned at  HSP or straight nodal lines that coincides with  HSL, it is required that all possible irreps in the HSP/HSL should be degenerate and furthermore, they all split in the neighborhood. The results are printed in red as shown in Secs. I A and II A of SM \cite{SM}. For essential nodal points/loops lying in  HSL/HSPL,  we restrict our discussions to those guaranteed by hourglass band connectivity. As discussed in Ref. \cite{wl}, hourglass band structure can occur in R-X-B where X is HSL or HSPL connecting high symmetry momenta R and B. When X is an HSPL, the hourglass band connectivity leads to nodal loop in the HSPL, and if the hourglass band connectivity is essential, the nodal loop is thus essential. These results have been summarized in Table IV of Ref. \cite{wl} for which each point in the nodal loop corresponds to an hourglass structure. Besides, when X is an HSL, essential hourglass band connectivity could guarantee an band crossing in the HSL. Further analysis on splitting patterns based on CRs (see Sec. \ref{HSL}) could also give rise to essential nodal loop in the neighboring HSPL containing the HSL or simply the essential nodal points. These results are printed in red in Secs II C and II D of the SM \cite{SM}.

\begin{table}[b]
	\centering
	\caption{The SGs allowing Hopf-link structure with essential straight nodal line threading one essential nodal loop within a HSPL. The first column lists the corresponding SGs. The second column contains the positions of HSPLs where the essential nodal loops occur. The third column lists HSLs forming essential straight nodal line. Note that there is no result for the setting with no TRS and negligible SOC, that is to say, the setting only allows such Hopf-link structures with nonessential nodal line or loop. Our name convention follows the Bilbao server \cite{Bilbao} and the coordinates can be found there.}
	\begin{tabular}{cp{4em}<{\centering}p{4em}<{\centering}|cp{4em}<{\centering}p{4em}<{\centering}}
		\hline
		\hline
		SG     & HSPL   & HSL    & SG     & HSPL   & HSL \\
		\hline
		\multicolumn{6}{c}{Time-reversal symmetry broken, neglecting SOC} \\
		\hline
		\multirow{3}[4]{*}{52} & L      & E      & 60     & W      & H \\
		\cline{4-6}           & N      & D      & \multirow{3}[4]{*}{61} & K      & A \\
		& V      & H      &        & M      & D \\
		\cline{1-3}    \multirow{2}[4]{*}{54} & M      & D      &        & V      & H \\
		\cline{4-6}           & N      & D      & \multirow{2}[4]{*}{62} & V      & H \\
		\cline{1-3}    \multirow{4}[6]{*}{56} & K      & C      &        & W      & Q \\
		\cline{4-6}           & L      & A      & \multirow{2}[2]{*}{130} & B      & Y \\
		& M      & D      &        & F      & U \\
		\cline{4-6}           & N      & B      & 135    & E      & W \\
		\hline
		\multirow{2}[2]{*}{57} & K      & A      & \multirow{2}[2]{*}{138} & B      & Y \\
		& L      & A      &        & F      & U \\
		\hline
		\multirow{2}[4]{*}{60} & K      & A      & 205    & A      & ZA \\
		\cline{4-6}           & M      & D      &        &        &  \\
		\hline
		\multicolumn{6}{c}{Time-reversal symmetry broken, considering SOC} \\
		\hline
		\multirow{6}[6]{*}{48} & K      & E      & \multirow{3}[2]{*}{61} & L      & C \\
		& L      & SM     &        & N      & B \\
		& M      & P      &        & W      & G \\
		\cline{4-6}           & N      & DT     & \multirow{2}[2]{*}{62} & K      & E \\
		& V      & Q      &        & L      & C \\
		\cline{4-6}           & W      & LD     & \multirow{3}[4]{*}{126} & B      & T \\
		\cline{1-3}    \multirow{4}[6]{*}{50} & L      & A      &        & E      & LD \\
		& N      & B      &        & F      & DT \\
		\cline{4-6}           & V      & Q      & 130    & E      & LD \\
		\cline{4-6}           & W      & Q      & 132    & E      & W \\
		\hline
		\multirow{3}[2]{*}{52} & L      & SM     & \multirow{3}[2]{*}{133} & E      & V \\
		& M      & P      &        & F      & DT \\
		& W      & LD     &        & F      & U \\
		\hline
		\multirow{2}[2]{*}{56} & V      & Q      & \multirow{2}[2]{*}{134} & B      & T \\
		& W      & LD     &        & F      & DT \\
		\hline
		\multirow{2}[4]{*}{59} & V      & Q      & 137    & E      & V \\
		\cline{4-6}           & W      & Q      & \multirow{2}[4]{*}{201} & A      & T \\
		\cline{1-3}    \multirow{3}[6]{*}{60} & L      & E      &        & B      & DT \\
		\cline{4-6}           & N      & DT     & 222    & B      & DT \\
		\cline{4-6}           & W      & G      &        &        &  \\
		\hline
		\multicolumn{6}{c}{Time-reversal symmetric, considering SOC} \\
		\hline
		\multirow{2}[2]{*}{29} & N      & D      & \multirow{3}[4]{*}{61} & W      & G \\
		& M      & D      &        & L      & C \\
		\cline{1-3}    \multirow{2}[4]{*}{30} & L      & C      &        & N      & B \\
		\cline{4-6}           & K      & C      & 62     & L      & C \\
		\hline
		\multirow{2}[2]{*}{31} & N      & D      & \multirow{2}[2]{*}{102} & B      & Y \\
		& M      & D      &        & F      & U \\
		\hline
		\multirow{2}[2]{*}{33} & K      & C      & \multirow{2}[2]{*}{104} & B      & Y \\
		& M      & D      &        & F      & U \\
		\hline
		\multirow{4}[6]{*}{34} & K      & C      & 109    & A      & Y \\
		\cline{4-6}           & N      & B      & 110    & A      & Y \\
		\cline{4-6}           & L      & A      & \multirow{2}[2]{*}{118} & B      & Y \\
		& M      & D      &        & F      & U \\
		\hline
		\multirow{2}[4]{*}{60} & W      & G      & 122    & A      & Y \\
		\cline{4-6}           & L      & E      & 205    & B      & Z \\
		\hline
		\hline
	\end{tabular}%
\label{table-hopf}%
\end{table}%% Table generated by Excel2LaTeX from sheet '1ess'
\begin{table}[b]
	\caption{All the positions of the Hopf-link structure formed by two perpendicular nodal loops, one of which is essential nodal loop. The first column lists the corresponding SGs. The second column lists HSLs hosted more than one kind of irreps doublet. In the third and fourth column it contains the positions of two HSPLs where the nodal loops occur. The HSPLs hosted essential nodal loops are marked in red.}
	\begin{tabular}{p{4em}<{\centering}p{4em}<{\centering}cc}
		%\begin{tabular}{p{4em}<{\centering}p{4em}<{\centering}p{4em}<{\centering}p{4em}<{\centering}}
		\hline
		\hline
		SG     & HSL    & \multicolumn{2}{c}{HSPL} \\
		\hline
		\multicolumn{4}{c}{Time-reversal symmetric, considering SOC}\\
		\hline
		29     & G      & \hspace{2em}{\color{red}L} & M\\
		29     & Q      & \hspace{2em}{\color{red}L} & N \\
		31     & G      & \hspace{2em}L      & {\color{red}M} \\
		31     & Q      & \hspace{2em}L      & {\color{red}N} \\
		33     & G      & \hspace{2em}{\color{red}L} & M \\
		33     & H      & \hspace{2em}{\color{red}K} & N \\
		62     & Q      & \hspace{2em}{\color{red}L} & N\\
		\hline
		\multicolumn{4}{c}{Time-reversal symmetric, neglecting SOC}\\
		\hline
		33     & LD     &\hspace{2em} K      & M\\
		52     & P      &\hspace{2em} {\color{red}L} & W\\
		\hline
		\multicolumn{4}{c}{Time-reversal symmetry broken,  considering SOC}\\
		\hline
		52     & D      &\hspace{2em} {\color{red}L} & V\\
		54     & A      &\hspace{2em} M      & {\color{red}W} \\
		54     & E      &\hspace{2em} N      & {\color{red}W} \\
		56     & A      &\hspace{2em} M      & {\color{red}W} \\
		56     & B      &\hspace{2em} K      & {\color{red}W} \\
		56     & C      &\hspace{2em} N      & {\color{red}V} \\
		56     & D      &\hspace{2em} L      & {\color{red}V} \\
		57     & H      &\hspace{2em} K      & {\color{red}N} \\
		57     & Q      &\hspace{2em} L      & {\color{red}N} \\
		59     & C      &\hspace{2em} N      & {\color{red}V} \\
		59     & D      &\hspace{2em} L      & {\color{red}V} \\
		59     & E      &\hspace{2em} N      & {\color{red}W} \\
		59     & P      &\hspace{2em} L      & {\color{red}W} \\
		60     & A      &\hspace{2em} M      & {\color{red}W} \\
		60     & H      &\hspace{2em} K      & {\color{red}N} \\
		61     & A      &\hspace{2em} M      & {\color{red}W} \\
		61     & D      &\hspace{2em} {\color{red}L} & V \\
		61     & H      &\hspace{2em} K      & {\color{red}N} \\
		62     & B      &\hspace{2em} {\color{red}K} & W \\
		62     & D      &\hspace{2em} {\color{red}L} & V \\
		62     & H      &\hspace{2em} {\color{red}K} & N \\
		62     & Q      &\hspace{2em} {\color{red}L} & N \\
		130    & U      &\hspace{2em} B      & {\color{red}E} \\
		137    & T      &\hspace{2em} {\color{red}E} & F \\
		205    & ZA     &\hspace{2em} A      & {\color{red}B}\\
		\hline
		\multicolumn{4}{c}{Time-reversal symmetry broken,  neglecting SOC}\\
		\hline
		52     & P      &\hspace{2em} {\color{red}L} & W\\
		57     & B      &\hspace{2em} {\color{red}K} & W \\
		57     & P      &\hspace{2em} {\color{red}L} & W \\
		60     & E      &\hspace{2em} N      & {\color{red}W} \\
		60     & G      &\hspace{2em} L      & {\color{red}M} \\
		61     & B      &\hspace{2em} {\color{red}K} & W \\
		61     & C      &\hspace{2em} N      & {\color{red}V} \\
		61     & G      &\hspace{2em} L      & {\color{red}M} \\
		62     & C      &\hspace{2em} N      & {\color{red}V} \\
		62     & E      &\hspace{2em} N      & {\color{red}W} \\
		205    & Z      &\hspace{2em} {\color{red}A} & B\\
		\hline
		\hline
	\end{tabular}%
	\label{hopf-2loop-1ess}%
\end{table}%

\section{Applications}\label{exam}
Hereafter we focus our discussions on the formation of nodal lines and loops.  With the vast body of all possible nodal lines (loops) organized as in Table \ref{results}  in the SM \cite{SM}, one could find materials or systems by fixing the SG symmetry. As long as the required SG symmetry is fulfilled, the system thus has a chance to  host nodal lines/loops. Furthermore, if the SG could essentially host nodal lines/loops, the nodal structures must exist. Given concrete  types of nodal lines/loops and their positions, we can also design target nodal structures. A straight nodal line threads a nodal loop as shown in Fig. \ref{demon}(b) forming a Hopf-link \cite{Volovik-hopf}, for which the essential results are shown in Table \ref{table-hopf} where all the nodal lines and loops are essential. For two nodal loops, as described in Sec. \ref{HSL-2}, when two band crossings in an HSL result in two nodal loops  lying in two different HSPLs containing the HSL, there is a chance for the two nodal loop to nest with each other \cite{co2mnga}. For materials realization, we propose a more feasible strategy: we first require one nodal loop to be essential, for example, formed by an hourglass band connectivity while the other nodal loop is formed by an accidental band crossing in the hourglass structure. This is schematically shown in the inset of Fig. \ref{demon}(c3) by an ``twisted hourglass''. All such hourglasses are listed in Table \ref{hopf-2loop-1ess}. From this table, we find that when the TRS is broken, such as for magnetic materials, there is still or even a better chance of realizing the Hopf-link nodal structure. This affords a guide of finding spin-polarized topological Hopf-link semimetals.  In the following, we show two SGs: SGs 34 and 62 and analyze all possible nodal lines/loops. Interestingly, they could allow the two kinds of Hopf-line nodal structures regarding  nesting structures consisting of one line and one loop or two loops.

\subsection{SG 34: Hopf-link structure of nodal line and loop}\label{hopf-1}
In this section, we take SG 34 as examples to demonstrate detailed analysis on all possible nodal lines/loops considering TRS and SOC.  In the following, we show in detail how to obtain these results.

\begin{figure*}[!hbtp]
	% Requires \usepackage{graphicx}
	\includegraphics[width=0.9 \textwidth]{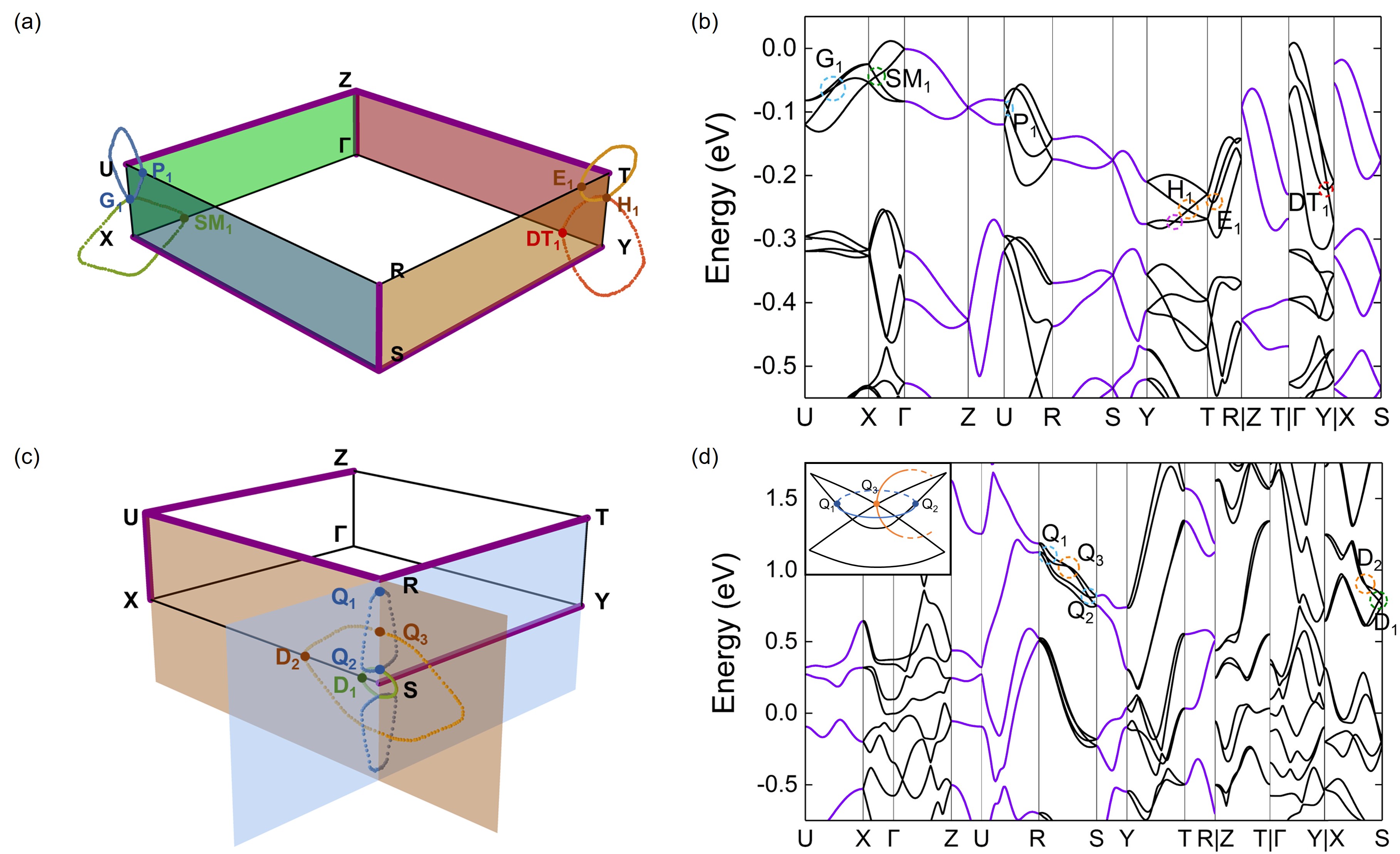}\\
	\caption{
(a) The BZ of SG 34 (primitive orthorhombic lattice). All possible nodal lines (in purple) are shown as $\Gamma$-Z, U-Z, T-Z, S-R, S-Y and S-X. The band crossing in T-Y (denoted by H$_1$) lie in two nodal loops in the HSPLs ZY and ST, respectively. Besides, these nodal loops can also be diagnosed from the band crossings E$_1$ and DT$_1$, within HSLs T-R and $\Gamma$-Y. Here we also demonstrate that the nodal loop around T is threaded by the nodal line Z-T while the nodal loop around Y is threaded by the nodal line S-Y. Similarly, the band crossing in the HSL U-X, denoted by G$_1$, lie in two nodal loops within two HSPLs, ZX and XR, respectively. These two nodal loops can also be diagnosed by the band crossings of P$_1$ and SM$_1$, and they are also threaded by two nodal lines. (b) The first-principles calculated electronic band structure of B$_5$Pb$_2$IO$_9$ in SG 34: All related HSLs for nodal lines/loops are chosen. Note that there is an accidental band crossing in HSL Y-T labeled by dashed purple circle, which is further found to be a nodal point. The band structures in the nodal lines are plotted in purple. (c) The BZ of SG 62 (the primitive orthorhombic lattice). The nodal lines are printed in purple including U-Z, U-X, U-R, T-R and S-Y. Here we show a complex nodal structure comprised of both types of Hopf-links and a nodal chain. The band crossings Q$_3$ and D$_1$ are essential which imply two nodal loops within the HSPL RX. The accidental band crossings Q$_1$ and Q$_2$ imply a nodal loop in HSPL TS which is linked with the nodal loop from D$_1$ and nested with the nodal loop from Q$_3$. Note that the accidental band crossing D$_2$ also implies the nodal loop from Q$_3$. (d) The first-principles calculated electronic band structure of  SrAl$_2$Au$_3$ in SG 62 where the band crossings shown in (c) are denoted. The inset shows an twisted hourglass in the HSL R-S. The band structures in the nodal lines are plotted in purple. }
	\label{34+62}
\end{figure*}

SG 34 is nonsymmorphic with two glide planes and the point group of SG  34 is $C_{2v}$ \cite{bradley}, so SG 34 is noncentrosymmetric. The BZ for SG 34 is shown in Fig. \ref{34+62}(a). As listed on the Bilbao server \cite{Bilbao}, SG 34 owns 12  HSLs as:\\
\begin{widetext}
\centering
G(1/2,0,w), H(0,1/2,w), LD(0,0,w), Q(1/2,1/2,w),
A(u,0,1/2), B(0,v,1/2), \\ C(u,1/2,0), D(1/2,v,0),
DT(0,v,0), E(u,1/2,1/2), P(1/2,v,1/2), SM(u,0,0),
\end{widetext}
and the following parentheses contain the coordinates adopting the convention in the Bilbao server \cite{Bilbao}. Through checking their irreps, we find that LD, Q, A, B, C,D own only one 2D irrep, G and H own four 1D irreps while the rest HSLs own two 1D irreps. For LD, Q, A, B, C and D, all of them are essential nodal lines since their sole irreps split in neighboring HSPLs. Take the HSL C as the example, the 2D irrep is represented by C3,4 (which means that irreps C3 and C4 are paired considering TRS). HSPLs N (u,1/2,w) and V(u,v,0)  can be associated with C, while the irreps of N and V are represented by N3, N4 and V2, respectively, all of which are 1D \cite{Bilbao}. Through  the CRs \cite{Bilbao}:
\begin{equation}\label{205-Z-CRs}
\begin{split}
  &  \mathrm{C}3,4\rightarrow\mathrm{N}3+\mathrm{N}4,\\
  &  \mathrm{C}3,4\rightarrow2\mathrm{V}2,\\
\end{split}
\end{equation}
we know that the bands in C must split in N and V, thus C is an essential straight nodal line. The essential nodal lines in HSLs LD, Q, A, B and D could also be obtained based on similar analysis. The configuration of these nodal lines  are displayed in Fig. \ref{34+62}(a) by thick purple lines including S-R (Q), S-X (D), S-Y (C), Z-U (A), Z-GM (LD) and Z-T (B).

Next, let's consider the HSPLs of SG 34, namely, K(0,v,w), L(1/2,v,w), M(u,0,w), N (u,1/2,w), V(u,v,0), and W(u,v,1/2). All of these HSPLs could only host 1D irreps, and K, L, M, N could host two 1D irreps while V and W could only host one. Hence, all these HSPLs cannot be a nodal surface but for K, L, M, N, they could host nodal loops. Furthermore, all these nodal loops are essential and hourglass nodal loops \cite{wl}. Here we should point out that they can be related with essential band crossings in related HSLs. For example, the band crossings from irreps DT3 and DT4 in HSL DT must lie in the nodal loops in K, due to that:
\begin{equation}\label{DT2K}
\begin{split}
&\mathrm{DT}3\rightarrow\mathrm{K}3,\\
& \mathrm{DT}4\rightarrow\mathrm{K}4.\\
\end{split}
\end{equation}

As displayed in Fig. \ref{34+62}(a), four nodal loops lie in HSPLs XZ (M), XR (L), YZ (K) and YR (N). From Sec. II D1 in the SM \cite{SM}, we find that, the band crossing  in HSL GM-X (SM) indicates an essential nodal loop in the HSPL XZ (M), the band crossings in HSL U-R (P) indicates an essential nodal loop in HSPL XR (L). Note that these band crossings can only indicate one nodal loop. For example, though the HSL U-R (P) lies in HSPL ZR (W), the two irreps in P both decompose to the sole irrep in W, so it is impossible for a  robust nodal loop to exist in the HSPL W. Similarly,  the band crossing  in HSL R-T (E) indicates an essential nodal loop in the HSPL YR (N), the band crossings in HSL GM-Y (DT) indicates an essential nodal loop in HSPL YZ (K) and  these band crossings can only indicate one nodal loop. Interestingly, the essential band crossings in HSL U-X (G) originated by G2+G4 or G3+G5,  both could indicate two essential nodal loops as in Fig. \ref{34+62}(a). Similarly, the essential band crossings in HSL T-Y (H) originated by H2+H4 or H3+H5,  both could indicate two essential nodal loops as in Fig. \ref{34+62}(a). Hence, the essential band crossings in G or H could indicate a nodal chain structure. Here we highlight the Hopf-link structure when including the essential straight nodal lines in S-X,Z-U, S-Y and Z-T which just threading the four essential nodal loops shown in Fig. \ref{34+62}(a).

It is worth mentioning that other than the above essential nodal lines/loops, the accidental band crossings in G formed by G2+G3 or G4+G5 could only indicate a nodal loop in M while the band crossings formed by G2+G5 or G3+G4 could only indicate a nodal loop in L. Similarly,  the accidental band crossings in H formed by H2+H3 or H4+H5 could only indicate a nodal loop in N while the band crossings formed by H2+H5 or H3+H4 could only indicate a nodal loop in K.

\subsection{SG 62: Hopf-link structure of two nodal loops }\label{hopf-2}
In the following we'll demonstrate that SG 62 protects the existence of another type of Hopf-link nodal loops. The analyses are as follows. Since SG 62 is centrosymmetric, no symmetry-enforced nodal surface could exist. With respect to nodal lines coinciding with HSLs, we should consider the HSLs for SG 62, which are the same as those for SG 34.  And HSLs A, C, E, G and P could host only one 4D irrep due to nonsymmorphic symmetries, B, D and H could host 2 2D irreps, DT, LD and SM could host only one 2D irrep while Q could host 4 2D irreps. Thus only A,C,E,G and P could be straight nodal lines since the 2D irreps in other HSLs cannot split in general point due to Kramers degeneracy. They are actually essential straight nodal lines from the CRs on the Bilbao server \cite{Bilbao} and displayed in Fig. \ref{34+62}(c) by thick purple lines where A is Z-U, C is S-Y, E is T-R, G is X-U and P is U-R.

Then investigate the HSPLs. Only HSPLs L (XR), N (YR) and W (ZR) could host nodal loops for they can host two 2D irreps. They are all located in the BZ boundaries.  First consider W, and the band crossing formed by the two irreps in HSL B (ZT) could indicate the nodal loop in W, but the band crossing is not essential. For the rest two HSPLs, namely, L and N, the nodal loop in L is essential while the nodal loop in N is accidental. However, these two nodal loops could form a Hopf-link nodal loop structure as shown in Fig. \ref{34+62}(c).
Consider two HSLs, S-X (D) and S-R (Q) that related with L or N and could own band crossings. The two 2D irreps in D must form a band crossing due to the hourglass band connectivity \cite{wl} and lie in the nodal loop in the HSPL L as shown by the orange ring in Fig. \ref{34+62}(c). As a matter of fact, such orange ring can also be indicated essentially by the band crossing in the HSL Q which have four different 2D irreps Q2,2, Q3,3, Q4,4 and Q5,5. In fact, the band crossing by Q2,2+Q5,5 or Q3,3+Q4,4 are essential due to hourglass connectivity \cite{wl}.  From the CRs as  follows:
\begin{equation}\label{62-hl-CRs}
\begin{split}
&  \mathrm{Q}2,2\rightarrow\mathrm{L}3,3,\quad\mathrm{Q}3,3\rightarrow\mathrm{L}3,3,\\
&  \mathrm{Q}4,4\rightarrow\mathrm{L}4,4,\quad\mathrm{Q}5,5\rightarrow\mathrm{L}4,4,\\
&  \mathrm{Q}2,2\rightarrow\mathrm{N}4,4,\quad\mathrm{Q}3,3\rightarrow\mathrm{N}3,3,\\
&  \mathrm{Q}4,4\rightarrow\mathrm{N}3,3,\quad\mathrm{Q}5,5\rightarrow\mathrm{N}4,4,
\end{split}
\end{equation}
 we know that there is an essential nodal loop within the HSPL L due to band crossing in Q due to  Q2,2+Q5,5 or Q3,3+Q4,4. However, such band crossing cannot result in nodal loop in another HSPL N since  both Q2,2 and  Q5,5 (Q3,3 and Q4,4) decompose to the same irrep N4,4 (N3,3) in N. If the accidental band crossings formed by Q2,2+Q3,3 or Q4,4+Q5,5  as in an ``twisted hourglass'',  the band crossing formed by Q2,2+Q3,3 or Q4,4+Q5,5 can  result in only one nodal loop that lies in the HSPL N from Eq. \ref{62-hl-CRs}. Hence, it is thus possible to realize a Hopf-link structure with the two loops in L and N.  Moreover, the nodal loop in the HSPL L is essential which means the nodal loop in L must exist and we only need to tune material parameters (keeping SG symmetry) to make required accidental band crossing in Q which leads to the other nodal loop in N. It is worth mentioning that for the band crossing by Q2,2+Q4,4 or Q3,3+Q5,5 which is nonessential, it could indicate two nodal loops in the two HSLs L and N, respectively, and result in a nodal chain structure.

\subsection{Nodal line/loop materials}\label{material}
Based on the essential results shown in Secs. II A and III B of the  SM \cite{SM} and the Inorganic Crystal Structure Database \cite{icsd}, we find hundreds of promising materials with nodal lines/loops shown in Sec. VII of the SM \cite{SM} combined with first-principles calculations, where the concrete positions of the nodal lines/loops near the Fermi level are also given.  These materials are all non-magnetic materials, and their number of elements is less than five. More than half of the materials are binary or ternary compounds.  Besides, these materials have relatively band nodes near the Fermi level. It is worth pointing out that Refs. \cite{N2,N3,N1} diagnosed topological semimetal phase by checking breaking of CRs, not caring too much on the concrete types and positions of the nontrivial band crossings, while here we also check the electronic band structures through detailed band plots and then identify detailed information (shape, energetics etc.) of the nodal lines/loops around the Fermi level. In Sec. IX of the SM \cite{SM}, we choose 5 materials to show the concrete electronic band structures and the nodal loops.

In the following we take  B$_5$Pb$_2$IO$_9 $\cite{B5Pb2IO9} crystallizing in SG 34 and SrAl$_2$Au$_3$ \cite{sral2au3} crystallizing in SG 62 as two materials examples in the main text. These two SGs could showcase various essential nodal loops in several HSPLs with coexisting essential straight nodal lines. Here we highlight that they could present two kinds of Hopf-link structures as discussed above.

Firstly we showcase B$_5$Pb$_2$IO$_9$\cite{B5Pb2IO9} crystallizing in SG 34 with orthorhombic lattice, whose first-principles calculated band structure is shown in Fig. \ref{34+62}(b), where the k-paths are chosen to cover the theoretical predicted k$^,s$ which could be essential nodal line or host essential band crossing implying a nodal loop,  as in Sec. \ref{hopf-1}.   From the band structure in HSLs U-X (G), U-R (P), $\Gamma$-X (SM), T-Y (H), T-R (E) and $\Gamma$-Y (DT), we can find essential hourglass band crossings label as G$_1$, P$_1$, SM$_1$, H$_1$, E$_1$ and DT$_1$ respectively. As shown in the above, these band crossings could imply nodal loops in respective HSPLs:
G$_1$ and H$_1$ indicate two nodal loops in two perpendicular HSPLs (thus forming a nodal chain) while the rest band crossings only lie in one nodal loop within corresponding HSPL. These nodal loops from first-principles calculations are demonstrated in Fig. \ref{34+62}(a) by green or orange cicles. Besides, the nodal lines that coincide with HSLs are plotted in purple in the BZ where the electronic band structure is also shown by purple curves in Fig. \ref{34+62}(b). Other than the nodal chain structures, the straight nodal line threading the nodal loop also constitute a Hopf-link structure.

Next we showcase the materials SrAl$_2$Au$_3$ \cite{sral2au3} crystallizing in SG 62 with orthorhombic lattice, whose first-principles calculated electronic band structure and fascinating nodal loop structure are shown in Figs. \ref{34+62}(d) and (c), respectively.  The band crossings in S-X (D) and S-R (Q) that are relevant with the nodal loops we display are denoted by orange or green dashed circle in Fig. \ref{34+62}(d). These band crossings are also named by D$_i$,Q$_j$ ($i=1,2, j=1,2,3$) labeled near the dashed circles.  Based on first-principles results, we find that D$_1$ and D$_2$ contain the irreps D2,3+D4,5 while Q$_1$, Q$_2$, Q$_3$ contain irreps of Q4,4+Q5,5, Q2,2+Q4,4, Q2,2+Q5,5 respectively.  We also show these nodes in Fig. \ref{34+62}(c) in the BZ, exactly corresponding to those in Fig. \ref{34+62}(d). Consistent with the above theoretical analysis, it can be found that along with the band nodes of D$_i$,Q$_j$ ($i=1,2, j=1,2,3$), there exist four nodal loops forming very intriguing structures: The orange nodal loops from D$_2$ and Q$_3$  is essential (Q$_3$ is essential while D$_2$ is not), which is nested with the Q$_{1,2}$ related non-essential nodal loops. Note that Q$_2$ also indicates another nodal loop in green so that the blue nodal loops are linked with the green one. Interestingly, the green nodal loop contains a band crossing D$_1$ which is essential.  Since these nodal loops are not very far from the Fermi level, this material is thus expected to be experimentally studied on the predicted nodal loops and the related consequences in future.

\section{Conclusions and Perspectives}\label{conclu}
We study exhaustively all types of band nodes based on CRs for all 230 SGs. The single-valued and doubled irreps are considered with respect to spin-orbit coupled and free systems, respectively. And the TRS is also considered to affect the irreps and CRs, which could lead to significant effects on the formation of nodal loops and surfaces in HSPLs. The nodal surfaces we found are all enforced by crystallographic symmetries. For these flat nodal surfaces, TRS should be considered otherwise the HSPL only own two nondegenerate irreps. Besides, the SGs allowing flat nodal surfaces should be noncentrosymmetric when SOC is also included. Furthermore, these nodal surfaces are found to be all essential. Furthermore, the HSPLs which own two different irreps whether TRS is considered or not could host a nodal loop. We also study straight nodal lines coinciding with the HSLs, some of which are found to be essential. The band crossing in a HSL and the irrep at HSP can be exploited to diagnose nodal loops in a convenient way. The essential band crossings in HSLs indicating nodal loops in neighboring HSPLs and these essential nodal loops enforced by hourglass band connectivity are also given. When the band crossing in the HSL cannot imply a nodal loop, it is just a nodal point. The irreps at HSP can not only imply nodal lines/surfaces, they can also be a nodal point.  Finally, we take SGs  34 and 62 as the examples to demonstrate comprehensive analysis on all possible nodal lines/loops based on our results in the SM \cite{SM}.  Interestingly, these two SGs could host two kinds of Hopf-link nodal structures formed by a nodal line and a nodal loop, or formed by two nodal loops. B$_5$Pb$_2$IO$_9$ in SG 34 and SrAl$_2$Au$_3$ in SG 62 can host these two kinds of novel Hopf-link structures, respectively. The promising SGs for these two kinds of Hopf-link structure are listed in the main text, respectively. These results are expected to be used in designing artificial systems that can own such Hopf-link structure, simply based on the crystal symmetry. They also throw light on constructing low-energy models with Hopf-links.

It is worth mentioning that our results can be also used to find materials realizations other than nonmagnetic electronic systems. Besides, type-III and IV magnetic SGs are out of our scope of study, but our strategy can be easily generalized to these settings once their CRs are obtained. And 80 layer groups for 2D materials or surfaces of 3D materials, though being the subsets of 230 SGs, could bring about nontrivial results since the high symmetry momenta for layer groups are much simpler than SG cases thus the HSPLs accommodating the nodal loops in SGs may not exist in layer groups. We expect that our tables for all types of symmetry-enforced band nodes can be applied to the design for more new and even ideal semimetals  with targeted nodal structure in near future.

\section{Acknowledgement}
We were supported by the National Key R\&D Program of China (Grants No. 2017YFA0303203 and No. 2018YFA0305704), the National Natural Science Foundation of China (NSFC Grants No. 11525417, No. 11834006, No. 51721001, and No. 11790311) and the excellent program at Nanjing University. X.W. also acknowledges the support from the Tencent Foundation through the XPLORER PRIZE. F.T. was supported by the Fundamental Research Fund for the Central Universities (No. 14380144, 14380157) and thanks Prof. Dingyu Xing and Prof. Baigen Wang for their kind and substantial support on scientific research. F.T. also thanks stimulating discussions with Prof. Wei Chen.

\bibliography{biblio}

\begin{thebibliography}{86}
\bibitem [{rem()}]{remark}%
  \BibitemOpen
\bibfield  {journal} {  }\href@noop {} {}\bibinfo {note} {Using the CRs for 230
  SGs negletcing the effect of TRS, we can obtain the CRs including TRS based
  on the pairing or not of irreps of little group by TRS.}\BibitemShut {Stop}%
\end{thebibliography}
\clearpage

\onecolumngrid

{\centering \bf Supplementary Material}

%\tableofcontents
\setcounter{section}{0}
\setcounter{figure}{0}
\setcounter{equation}{0}

~\\
~\\\indent
In this Supplementary Material, we present the exhaustive results for all symmetry-enforced nodal points/lines/loops/surfaces using the irreducible representations (irreps) and compatibility relations (CRs) listed on the Bilbao server \cite{Bilbao}. In Sec. \ref{hsp}, we listed all band nodes related with the high-symmetry points (HSPs): the single irrep of the HSP implies a nodal point in Sec. \ref{hsp1}, the single irrep of the HSP implies a nodal line coinciding with a neighboring high-symmetry line (HSPL) in Sec. \ref{hsp2}, and the single irrep of the HSP implies a nodal surface coinciding with a neighboring high-symmetry plane (HSPL) in Sec. \ref{hsp3}.

In Sec. \ref{hsl}, we list all band nodes related with the HSLs: the single irrep of the HSL implies a nodal line coinciding with the HSL in Sec. \ref{hsl1}, the single irrep of the HSL implies a nodal surface coinciding with a neighboring HSPL in Sec. \ref{hsl2}, two irreps (forming band crossing accidently) of the HSL implies a nodal point in Sec. \ref{hsl3} and two irreps of the HSL implies a nodal loop within a neighboring high-symmetry plane (HSPL) in Sec. \ref{hsl4}.

In Sec. \ref{hspl}, we listed all band nodes related with the HSPLs: the single irrep of the HSPL implies a nodal surface coinciding with the HSPL in Sec. \ref{hspl1} and two irreps imply a nodal loop within the HSPL  in Sec. \ref{hsl2}.

In these sections, the four settings (with/without time-reversal symmetry (TRS) and spin-orbit coupling (SOC)) are considered. The essential cases are highlighted by red color. The name convention and the coordinates for the momenta $k^,s$ follow those on the Bilbao server \cite{Bilbao}. We also list the dimension of the band nodes. The symmetry-related band nodes can be found from the Bilbao server \cite{Bilbao} from the wave vector star (if TRS exists, $-k$ should be considered when $k$ owns no antiunitary symmetry). There the little group can be found. Furthermore, the $k\cdot p$ model can also be constructed around these band nodes from the irrep matrices.

Then Sec. \ref{hopf1} is devoted to the type-I Hopf-link consisting of one nodal line threading a nodal loop. There we give all positions possibly hosting the type-I Hopf-link structure. In Sec. \ref{twohspls}, we consider two (inequivalent) HSPLs which intersects with each other while the intersection must be an HSL. There we show the doublets of irreps in the HSL which could imply a/two nodal loop/loops in neighboring HSPL/HSPLs. From these results, we list all positions possibly hosting the type-II Hopf-link consisting of two nesting nodal loops in Sec. \ref{hopf2}.

In Sec. \ref{materials}, we list the realistic materials that could host nodal lines/loops near the Fermi level and the first-principles electronic band structures for the selected examples are shown in Sec. \ref{materials-demon}, where the band crossing(BC) in an HSL implying a nodal loop is indicated and the nodal loop is also shown from the principles results.

\clearpage
\section{High-symmetry point}\label{hsp}
\subsection{Nodal point}\label{hsp1}

The results for all HSPs with nodal points are shown in the table below.
The first column contains labels for the HSPs. In the second column, it contains SG numbers for the corresponding HSPs. SGs hosting essential nodal points are printed in red.
\begin{table}[!bhtp]
  \centering
    \begin{tabular}{ll||ll}
    \hline
    \hline
    \multicolumn{1}{c}{HSP} & \multicolumn{1}{c||}{SG} & \multicolumn{1}{c}{HSP} & \multicolumn{1}{c}{SG} \\
    \hline
    \multicolumn{2}{c||}{Time-reversal symmetry broken,  neglecting SOC} & \multicolumn{2}{c}{Time-reversal symmetry broken,  considering SOC} \\
    \hline
    C      & {\color{red}11},{\color{red}14} & C      & {\color{red}11},{\color{red}14} \\
    \hline
    D      & {\color{red}11},{\color{red}13} & D      & {\color{red}11},{\color{red}13} \\
    \hline
    E      & {\color{red}11},{\color{red}13} & E      & {\color{red}11},{\color{red}13} \\
    \hline
    Z      & {\color{red}11},{\color{red}14},{\color{red}17},{\color{red}19},{\color{red}20},{\color{red}84},{\color{red}86},89,90,{\color{red}91},{\color{red}92},93,94,{\color{red}95},{\color{red}96},111-118 & Z      & {\color{red}11},{\color{red}14},{\color{red}16},{\color{red}18},{\color{red}21},{\color{red}22},{\color{red}84},{\color{red}86},{\color{red}89},{\color{red}90},91,92,{\color{red}93},{\color{red}94},95,96,{\color{red}131-138} \\
    \hline
    \multirow{2}[2]{*}{A} & {\color{red}13-15},{\color{red}84},{\color{red}85},89,90,{\color{red}91},{\color{red}92},93,94,{\color{red}95},{\color{red}96},111-118,149-154,{\color{red}163}, & \multirow{2}[2]{*}{A} & {\color{red}13-15},{\color{red}84},{\color{red}85},{\color{red}89},{\color{red}90},91,92,{\color{red}93},{\color{red}94},95,96,{\color{red}125},{\color{red}126},{\color{red}129-132},{\color{red}135}, \\
           & {\color{red}165},{\color{red}176},177,{\color{red}178},{\color{red}179},180,181,{\color{red}182},193,194 &        & {\color{red}136},149-154,{\color{red}163},{\color{red}165},{\color{red}176},{\color{red}177},178,179,{\color{red}180},{\color{red}181},182,193,194 \\
    \hline
    B      & {\color{red}13},{\color{red}14} & B      & {\color{red}13},{\color{red}14} \\
    \hline
    M      & {\color{red}15},{\color{red}85},{\color{red}86},{\color{red}88},89-98,111-122,207,208,212,213,215,218 & \multirow{2}[4]{*}{M} & {\color{red}15},{\color{red}85},{\color{red}86},{\color{red}88-98},{\color{red}125},{\color{red}126},{\color{red}129},{\color{red}130},{\color{red}133},{\color{red}134},{\color{red}137},{\color{red}138},{\color{red}141}, \\
\cline{1-2}    \multirow{2}[4]{*}{R} & {\color{red}17},{\color{red}19},{\color{red}63},{\color{red}67},{\color{red}68},{\color{red}72},{\color{red}73},{\color{red}85},{\color{red}86},{\color{red}90},{\color{red}91},{\color{red}94},{\color{red}95},{\color{red}113},{\color{red}114},195, &        & {\color{red}142},{\color{red}177-182},{\color{red}195},{\color{red}198},{\color{red}207},{\color{red}208},{\color{red}212},{\color{red}213},{\color{red}222},{\color{red}224} \\
\cline{3-4}           & {\color{red}198},200,201,207,208,212,213,215,218,221-224 & GM     & {\color{red}16-24},{\color{red}89-98},149-155,{\color{red}177-182},{\color{red}195-199},207-230 \\
    \hline
    T      & {\color{red}17},{\color{red}18},{\color{red}20},{\color{red}73},{\color{red}74},155,{\color{red}167} & \multirow{2}[4]{*}{R} & {\color{red}16},{\color{red}18},{\color{red}63},{\color{red}67},{\color{red}68},{\color{red}72},{\color{red}73},{\color{red}85},{\color{red}86},{\color{red}89},{\color{red}92},{\color{red}93},{\color{red}96},{\color{red}111},{\color{red}112},{\color{red}195}, \\
\cline{1-2}    U      & {\color{red}17},{\color{red}18} &        & 198,205,207,208,212,213,215,218,221,{\color{red}222},{\color{red}223},224 \\
    \hline
    \multirow{2}[4]{*}{X} & {\color{red}18},{\color{red}19},{\color{red}85},{\color{red}86},{\color{red}88},{\color{red}90},{\color{red}92},{\color{red}94},{\color{red}96},{\color{red}113},{\color{red}114},{\color{red}198},207-210, & S      & {\color{red}16-19},{\color{red}64},{\color{red}67},{\color{red}68},{\color{red}72},{\color{red}73} \\
\cline{3-4}           & {\color{red}212},{\color{red}213},215,216,218,219 & T      & {\color{red}16},{\color{red}19},{\color{red}21},{\color{red}22},{\color{red}73},{\color{red}74},155,{\color{red}167} \\
    \hline
    Y      & {\color{red}18},{\color{red}19} & U      & {\color{red}16},{\color{red}19} \\
    \hline
    W      & {\color{red}24},{\color{red}73},{\color{red}74},{\color{red}210},225,226 & \multirow{2}[4]{*}{X} & {\color{red}16},{\color{red}17},{\color{red}23},{\color{red}24},{\color{red}85},{\color{red}86},{\color{red}88},{\color{red}89},{\color{red}91},{\color{red}93},{\color{red}95},{\color{red}97},{\color{red}98},{\color{red}111},{\color{red}112},{\color{red}119}, \\
\cline{1-2}    WA     & {\color{red}24}   &        & {\color{red}120},{\color{red}195},{\color{red}196},{\color{red}207-210},212,213,{\color{red}223},{\color{red}224},{\color{red}227},{\color{red}228} \\
    \hline
    S      & {\color{red}64},{\color{red}67},{\color{red}68},{\color{red}72},{\color{red}73} & Y      & {\color{red}16},{\color{red}17},{\color{red}20-22} \\
    \hline
    GM     & 89-98,111-122,149-155,177-182,195-230 & W      & {\color{red}23},{\color{red}71},{\color{red}72},{\color{red}209},227,228 \\
    \hline
    P      & {\color{red}98},121,139,140,197,{\color{red}199},204,{\color{red}206},211,{\color{red}214},217,220,229,230 & WA     & {\color{red}23} \\
    \hline
    PA     & 121,197,{\color{red}199},217,220 & P      & {\color{red}97},122,141,142,{\color{red}197},199,{\color{red}204},206,{\color{red}211},214,217,220,229,230 \\
    \hline
    N      & {\color{red}140},{\color{red}142},{\color{red}206} & PA     & 122,{\color{red}197},199,217,220 \\
    \hline
    \multirow{2}[4]{*}{H} & 150,152,154,164,165,177-182,197,199,204,206,211,214, & N      & {\color{red}140},{\color{red}142},{\color{red}206},{\color{red}211},{\color{red}214} \\
\cline{3-4}           & 217,220,229,230 & H      & 150,152,154,164,165,177-182,{\color{red}197},{\color{red}199},211,214,217,220,229,{\color{red}230} \\
    \hline
    HA     & 150,152,154 & HA     & 150,152,154 \\
    \hline
    K      & 150,152,154,164,165,177-182 & K      & 150,152,154,164,165,177-182 \\
    \hline
    KA     & 150,152,154 & KA     & 150,152,154 \\
    \hline
    L      & {\color{red}163},{\color{red}165},{\color{red}167},{\color{red}176},{\color{red}178},{\color{red}179},{\color{red}182},209,210,{\color{red}226},{\color{red}228} & L      & {\color{red}163},{\color{red}165},{\color{red}167},{\color{red}176},{\color{red}177},{\color{red}180},{\color{red}181},209,210,{\color{red}226},{\color{red}228} \\
    \hline
    \hline
    \end{tabular}%
\end{table}%

\clearpage
\begin{table}[!bhtp]
  \centering
(Continued from previous table: the HSP implies a nodal point)\\
    \begin{tabular}{ll||ll}
    \hline
    \hline
    \multicolumn{1}{c}{HSP} & \multicolumn{1}{c||}{SG} & \multicolumn{1}{c}{HSP} & \multicolumn{1}{c}{SG} \\
    \hline
    \multicolumn{2}{c||}{Time-reversal symmetric, neglecting SOC} & \multicolumn{2}{c}{Time-reversal symmetric, considering SOC} \\
    \hline
    \multirow{2}[4]{*}{R} & {\color{red}19},{\color{red}29},{\color{red}54},{\color{red}130},{\color{red}138},195,{\color{red}198},200,201,207,208,{\color{red}212},{\color{red}213},215, & Y      & \multicolumn{1}{l}{{\color{red}3-9},{\color{red}16},{\color{red}17},{\color{red}20-22},{\color{red}48},{\color{red}50},{\color{red}60},{\color{red}70}} \\
\cline{3-4}           & 218,221-224 & \multirow{3}[6]{*}{Z} & \multicolumn{1}{l}{{\color{red}3},{\color{red}6},{\color{red}7},{\color{red}11},{\color{red}14},{\color{red}16},{\color{red}18},{\color{red}21},{\color{red}22},{\color{red}26},{\color{red}27},{\color{red}29-31},{\color{red}33},{\color{red}34},{\color{red}36},{\color{red}37},{\color{red}43},} \\
\cline{1-2}    W      & {\color{red}24},{\color{red}73},209,{\color{red}210},225,{\color{red}228} &        & \multicolumn{1}{l}{{\color{red}48},{\color{red}49},{\color{red}52},{\color{red}54},{\color{red}56},{\color{red}58},{\color{red}66},{\color{red}68},{\color{red}70},{\color{red}75},{\color{red}77},{\color{red}84},{\color{red}86},{\color{red}89},{\color{red}90},{\color{red}93},{\color{red}94},} \\
\cline{1-2}    WA     & {\color{red}24}   &        & \multicolumn{1}{l}{{\color{red}101-106},{\color{red}112},{\color{red}114},{\color{red}116},{\color{red}118},{\color{red}124},{\color{red}126},{\color{red}128},{\color{red}130-138}} \\
    \hline
    U      & {\color{red}29},{\color{red}33},{\color{red}54},{\color{red}56} & \multirow{3}[6]{*}{M} & \multicolumn{1}{l}{{\color{red}5},{\color{red}8},{\color{red}15},{\color{red}75-80},{\color{red}85},{\color{red}86},{\color{red}88-98},{\color{red}100},{\color{red}102},{\color{red}104},{\color{red}106},{\color{red}109},{\color{red}110},} \\
\cline{1-2}    S      & {\color{red}52}   &        & \multicolumn{1}{l}{{\color{red}117},{\color{red}118},{\color{red}122},{\color{red}125},{\color{red}126},{\color{red}133},{\color{red}134},{\color{red}141-145},{\color{red}149-154},{\color{red}156-159},} \\
\cline{1-2}    T      & {\color{red}56},{\color{red}60},146,155,161,167 &        & \multicolumn{1}{l}{{\color{red}168-174},{\color{red}177-182},{\color{red}195},{\color{red}198},{\color{red}201},{\color{red}207},{\color{red}208},{\color{red}212},{\color{red}213},{\color{red}222},{\color{red}224}} \\
    \hline
    \multirow{3}[2]{*}{A} & 75,77,{\color{red}84},{\color{red}85},89,{\color{red}92},93,{\color{red}96},103,106,114,124,128,{\color{red}130},133,{\color{red}135}, & \multirow{4}[4]{*}{R} & \multicolumn{1}{l}{{\color{red}16},{\color{red}18},{\color{red}21},{\color{red}23},{\color{red}24},{\color{red}26},{\color{red}27},{\color{red}32},{\color{red}33},{\color{red}35},{\color{red}37},{\color{red}38},{\color{red}40},{\color{red}44},{\color{red}49},{\color{red}50},{\color{red}63},{\color{red}67},} \\
           & 137,143-145,149-154,158,159,163,165,168,171,172,176, &        & \multicolumn{1}{l}{{\color{red}68},{\color{red}72},{\color{red}73},{\color{red}75},{\color{red}77},{\color{red}81},{\color{red}85},{\color{red}86},{\color{red}89},{\color{red}92},{\color{red}93},{\color{red}96},{\color{red}101},{\color{red}103},{\color{red}111},} \\
           & 177,180,181,184-186,188,190-194 &        & \multicolumn{1}{l}{{\color{red}112},{\color{red}116},{\color{red}124},{\color{red}125},{\color{red}126},{\color{red}132-134},195,198,200,201,205,207,208,} \\
\cline{1-2}    GM     & 75-80,89-98,143-146,149-155,168-173,177-182,195-230 &        & \multicolumn{1}{l}{212,213,215,218,221-224} \\
    \hline
    M      & 75-80,{\color{red}85},{\color{red}86},{\color{red}88},89,91,93,95,97,98,207,208 & S      & \multicolumn{1}{l}{{\color{red}16-21},{\color{red}23},{\color{red}24},{\color{red}32-38},{\color{red}40},{\color{red}44},{\color{red}46},{\color{red}48},{\color{red}50},{\color{red}53},{\color{red}64},{\color{red}67},{\color{red}68},{\color{red}72},{\color{red}73}} \\
    \hline
    Z      & 75,77,{\color{red}84},{\color{red}86},89,90,93,94,103,104,124,126,128,130 & \multirow{2}[4]{*}{T} & \multicolumn{1}{l}{{\color{red}16},{\color{red}19},{\color{red}21-24},{\color{red}26},{\color{red}27},{\color{red}29},{\color{red}31},{\color{red}36},{\color{red}37},{\color{red}44-46},{\color{red}48-50},{\color{red}54},{\color{red}66},} \\
\cline{1-2}    \multirow{2}[4]{*}{P} & 79,{\color{red}80},87,97,{\color{red}98},107,{\color{red}110},121,139,{\color{red}142},197,{\color{red}199},204,{\color{red}206},211, &        & \multicolumn{1}{l}{{\color{red}68},{\color{red}70},{\color{red}73},{\color{red}74},{\color{red}146},{\color{red}155},{\color{red}160},161,{\color{red}167}} \\
\cline{3-4}           & {\color{red}214},217,220,229,{\color{red}230} & U      & \multicolumn{1}{l}{{\color{red}16},{\color{red}19},{\color{red}26},{\color{red}27},{\color{red}30},{\color{red}48-50},{\color{red}52}} \\
    \hline
    PA     & 121,197,{\color{red}199},217,220 & \multirow{3}[4]{*}{X} & \multicolumn{1}{l}{{\color{red}16},{\color{red}17},{\color{red}23},{\color{red}24},{\color{red}48},{\color{red}50},{\color{red}52},{\color{red}53},{\color{red}75-82},{\color{red}85},{\color{red}86},{\color{red}88},{\color{red}89},{\color{red}91},{\color{red}93},} \\
\cline{1-2}    \multirow{2}[2]{*}{H} & 150,152,154,165,168,171,172,177,180,181,184,185,192,193, &        & \multicolumn{1}{l}{{\color{red}95},{\color{red}97},{\color{red}98},{\color{red}111},{\color{red}112},{\color{red}119},{\color{red}120},{\color{red}125},{\color{red}126},{\color{red}133},{\color{red}134},{\color{red}141},{\color{red}142},} \\
           & 197,199,204,206,211,214,217,220,229,230 &        & \multicolumn{1}{l}{{\color{red}195},{\color{red}196},{\color{red}201},{\color{red}203},{\color{red}207-210},{\color{red}218},{\color{red}222-224},{\color{red}227},{\color{red}228}} \\
    \hline
    HA     & 150,152,154 & W      & \multicolumn{1}{l}{{\color{red}23},{\color{red}72},{\color{red}209},210,{\color{red}226},228} \\
    \hline
    K      & 150,152,154,168-173,177-182 & WA     & \multicolumn{1}{l}{{\color{red}23}} \\
    \hline
    KA     & 150,152,154 & \multirow{2}[4]{*}{N} & \multicolumn{1}{l}{{\color{red}79},{\color{red}80},{\color{red}82},{\color{red}97},{\color{red}98},{\color{red}107},{\color{red}109},{\color{red}119},{\color{red}121},{\color{red}122},{\color{red}140},{\color{red}142},{\color{red}197},} \\
\cline{1-2}    L      & 196,209,210,219,226,228 &        & \multicolumn{1}{l}{{\color{red}199},{\color{red}206},{\color{red}211},{\color{red}214},{\color{red}230}} \\
    \hline
    X      & 207-210,222 & \multirow{2}[4]{*}{P} & \multicolumn{1}{l}{{\color{red}79},80,{\color{red}97},98,{\color{red}108},109,122,{\color{red}140},142,{\color{red}197},199,204,206,{\color{red}211},} \\
\cline{1-2}    \multicolumn{2}{c||}{Time-reversal symmetric, considering SOC} &        & \multicolumn{1}{l}{214,217,220,229,230} \\
    \hline
    \multirow{4}[4]{*}{A} & {\color{red}3-6},{\color{red}8},{\color{red}13-15},{\color{red}75},{\color{red}77},{\color{red}84},{\color{red}85},{\color{red}89},{\color{red}90},92,{\color{red}93},{\color{red}94},96,{\color{red}100},{\color{red}101}, & PA     & \multicolumn{1}{l}{122,{\color{red}197},199,217,220} \\
\cline{3-4}           & {\color{red}103-106},{\color{red}112},{\color{red}116},{\color{red}117},{\color{red}124-126},{\color{red}130-133},{\color{red}135},{\color{red}143-145}, & \multirow{3}[2]{*}{L} & \multicolumn{1}{l}{{\color{red}143-146},{\color{red}149-157},{\color{red}160},{\color{red}163},{\color{red}165},{\color{red}167},{\color{red}168},{\color{red}171},{\color{red}172},{\color{red}174},} \\
           & {\color{red}149-154},{\color{red}156},{\color{red}157},158,159,{\color{red}163},{\color{red}165},{\color{red}168},{\color{red}171},{\color{red}172},{\color{red}174}, &        & \multicolumn{1}{l}{{\color{red}176},{\color{red}177},{\color{red}180},{\color{red}181},{\color{red}184-186},{\color{red}192},{\color{red}196},{\color{red}209},{\color{red}210},{\color{red}216},219,} \\
           & {\color{red}176},{\color{red}177},{\color{red}180},{\color{red}181},{\color{red}184},{\color{red}185},{\color{red}186},188,190,{\color{red}192} &        & \multicolumn{1}{l}{{\color{red}226},{\color{red}228}} \\
    \hline
    B      & {\color{red}3},{\color{red}4},{\color{red}6},{\color{red}13},{\color{red}14} & F      & \multicolumn{1}{l}{{\color{red}146},{\color{red}155},{\color{red}160},{\color{red}161}} \\
    \hline
    C      & {\color{red}3},{\color{red}6},{\color{red}7},{\color{red}11},{\color{red}14} & \multirow{2}[4]{*}{H} & \multicolumn{1}{l}{150,152,154,165,168,171,172,177,180,181,184,185,192,} \\
\cline{1-2}    D      & {\color{red}3},{\color{red}6},{\color{red}11},{\color{red}13} &        & \multicolumn{1}{l}{197,199,204,206,211,214,217,220,229,230} \\
    \hline
    E      & {\color{red}3},{\color{red}6},{\color{red}11},{\color{red}13} & HA     & \multicolumn{1}{l}{150,152,154} \\
    \hline
    \multirow{2}[4]{*}{GM} & {\color{red}3-9},{\color{red}16-24},{\color{red}75-80},{\color{red}89-98},{\color{red}143-146},{\color{red}149-161},{\color{red}168-174}, & K      & \multicolumn{1}{l}{150,152,154,168-173,177-182} \\
\cline{3-4}           & {\color{red}177-182},195-230 & KA     & 150,152,154 \\
    \hline
    \hline
    \end{tabular}%
\end{table}%

\subsection{Nodal line}\label{hsp2}
In this section we list all the positions of HSPs which host degenerate energy levels implying  neighboring HSL to be a nodal line for four settings. Take U in SG 25 in the setting with TRS and significant SOC as the example, namely, the first item in the table of Sec.~\ref{hsp2trssoc}. U is an HSP in SG 25 and its coordinate in the conventional basis is $(\frac{1}{2}, 0, \frac{1}{2})$ (following convention in the Bilbao server \cite{Bilbao}), could host a degenerate energy level which corresponds to the 2D double-valued irrep U5 in the little group and the straight nodal line structure is implied to occur in the associated HSL G whose coordinate is $(\frac{1}{2}, 0, w)$. We further find that such a nodal line is essential so G$(\frac{1}{2}, 0, w)$ is printed in red. For the other three settings, the corresponding tables can also be read following the same way as described here.
\subsubsection{Results for materials with TRS and significant SOC}\label{hsp2trssoc}
\newpage
\begin{center}\begin{table*}[!h]
(Continued from previous table: Time-reversal symmetric, considering SOC)\\
\begin{tabular}{c|c|c|c|c||c|c|c|c|c}\hline\hline
\multicolumn{1}{l|}{SG} & HSP & Irrep & HSL & Irrep & \multicolumn{1}{l|}{SG} & HSP & Irrep & HSL & Irrep \bigstrut\\
\hline
25&U$(\frac{1}{2},0,\frac{1}{2})$&U5(2)&{\color{red}G$(\frac{1}{2},0,w)$}&G5(2)
&25&X$(\frac{1}{2},0,0)$&X5(2)&{\color{red}G$(\frac{1}{2},0,w)$}&G5(2)
\bigstrut\\
\hline
25&T$(0,\frac{1}{2},\frac{1}{2})$&T5(2)&{\color{red}H$(0,\frac{1}{2},w)$}&H5(2)
&25&Y$(0,\frac{1}{2},0)$&Y5(2)&{\color{red}H$(0,\frac{1}{2},w)$}&H5(2)
\bigstrut\\
\hline
25&GM$(0,0,0)$&GM5(2)&{\color{red}LD$(0,0,w)$}&LD5(2)
&25&Z$(0,0,\frac{1}{2})$&Z5(2)&{\color{red}LD$(0,0,w)$}&LD5(2)
\bigstrut\\
\hline
25&R$(\frac{1}{2},\frac{1}{2},\frac{1}{2})$&R5(2)&{\color{red}Q$(\frac{1}{2},\frac{1}{2},w)$}&Q5(2)
&25&S$(\frac{1}{2},\frac{1}{2},0)$&S5(2)&{\color{red}Q$(\frac{1}{2},\frac{1}{2},w)$}&Q5(2)
\bigstrut\\
\hline
26&U$(\frac{1}{2},0,\frac{1}{2})$&U5,5(4)&{\color{red}G$(\frac{1}{2},0,w)$}&G5(2)
&26&X$(\frac{1}{2},0,0)$&X5(2)&{\color{red}G$(\frac{1}{2},0,w)$}&G5(2)
\bigstrut\\
\hline
26&T$(0,\frac{1}{2},\frac{1}{2})$&T5,5(4)&{\color{red}H$(0,\frac{1}{2},w)$}&H5(2)
&26&Y$(0,\frac{1}{2},0)$&Y5(2)&{\color{red}H$(0,\frac{1}{2},w)$}&H5(2)
\bigstrut\\
\hline
26&GM$(0,0,0)$&GM5(2)&{\color{red}LD$(0,0,w)$}&LD5(2)
&26&Z$(0,0,\frac{1}{2})$&Z5,5(4)&{\color{red}LD$(0,0,w)$}&LD5(2)
\bigstrut\\
\hline
\hline
\end{tabular}
\end{table*}\end{center}

\newpage
\begin{center}\begin{table*}[!h]
(Continued from previous table: Time-reversal symmetric, considering SOC)\\
\begin{tabular}{c|c|c|c|c||c|c|c|c|c}\hline\hline
\multicolumn{1}{l|}{SG} & HSP & Irrep & HSL & Irrep & \multicolumn{1}{l|}{SG} & HSP & Irrep & HSL & Irrep \bigstrut\\
\hline
32&X$(\frac{1}{2},0,0)$&X2,5(2)&{\color{red}D$(\frac{1}{2},v,0)$}&D3,4(2)
&32&X$(\frac{1}{2},0,0)$&X3,4(2)&{\color{red}D$(\frac{1}{2},v,0)$}&D3,4(2)
\bigstrut\\
\hline
32&R$(\frac{1}{2},\frac{1}{2},\frac{1}{2})$&R5,5(4)&{\color{red}E$(u,\frac{1}{2},\frac{1}{2})$}&E3,4(2)
&32&T$(0,\frac{1}{2},\frac{1}{2})$&T2,3(2)&{\color{red}E$(u,\frac{1}{2},\frac{1}{2})$}&E3,4(2)
\bigstrut\\
\hline
32&T$(0,\frac{1}{2},\frac{1}{2})$&T4,5(2)&{\color{red}E$(u,\frac{1}{2},\frac{1}{2})$}&E3,4(2)
&32&GM$(0,0,0)$&GM5(2)&{\color{red}LD$(0,0,w)$}&LD5(2)
\bigstrut\\
\hline
32&Z$(0,0,\frac{1}{2})$&Z5(2)&{\color{red}LD$(0,0,w)$}&LD5(2)
&32&R$(\frac{1}{2},\frac{1}{2},\frac{1}{2})$&R5,5(4)&{\color{red}P$(\frac{1}{2},v,\frac{1}{2})$}&P3,4(2)
\bigstrut\\
\hline
32&U$(\frac{1}{2},0,\frac{1}{2})$&U2,5(2)&{\color{red}P$(\frac{1}{2},v,\frac{1}{2})$}&P3,4(2)
&32&U$(\frac{1}{2},0,\frac{1}{2})$&U3,4(2)&{\color{red}P$(\frac{1}{2},v,\frac{1}{2})$}&P3,4(2)
\bigstrut\\
\hline
32&R$(\frac{1}{2},\frac{1}{2},\frac{1}{2})$&R5,5(4)&{\color{red}Q$(\frac{1}{2},\frac{1}{2},w)$}&Q5(2)
&32&S$(\frac{1}{2},\frac{1}{2},0)$&S5,5(4)&{\color{red}Q$(\frac{1}{2},\frac{1}{2},w)$}&Q5(2)
\bigstrut\\
\hline
33&S$(\frac{1}{2},\frac{1}{2},0)$&S5,5(4)&{\color{red}C$(u,\frac{1}{2},0)$}&C3,4(2)
&33&Y$(0,\frac{1}{2},0)$&Y2,3(2)&{\color{red}C$(u,\frac{1}{2},0)$}&C3,4(2)
\bigstrut\\
\hline
\hline
\end{tabular}
\end{table*}\end{center}

\newpage
\begin{center}\begin{table*}[!h]
(Continued from previous table: Time-reversal symmetric, considering SOC)\\
\begin{tabular}{c|c|c|c|c||c|c|c|c|c}\hline\hline
\multicolumn{1}{l|}{SG} & HSP & Irrep & HSL & Irrep & \multicolumn{1}{l|}{SG} & HSP & Irrep & HSL & Irrep \bigstrut\\
\hline
43&Z$(0,0,1)$&Z5,5(4)&{\color{red}A$(u,0,1)$}&A3,4(2)
&43&T$(1,0,0)$&T2,5(2)&{\color{red}B$(0,v,1)$}&B3,4(2)
\bigstrut\\
\hline
43&T$(1,0,0)$&T3,4(2)&{\color{red}B$(0,v,1)$}&B3,4(2)
&43&Z$(0,0,1)$&Z5,5(4)&{\color{red}B$(0,v,1)$}&B3,4(2)
\bigstrut\\
\hline
43&GM$(0,0,0)$&GM5(2)&{\color{red}LD$(0,0,w)$}&LD5(2)
&43&Z$(0,0,1)$&Z5,5(4)&{\color{red}LD$(0,0,w)$}&LD5(2)
\bigstrut\\
\hline
44&GM$(0,0,0)$&GM5(2)&{\color{red}LD$(0,0,w)$}&LD5(2)
&44&X$(1,1,1)$&X5(2)&{\color{red}LD$(0,0,w)$}&LD5(2)
\bigstrut\\
\hline
45&GM$(0,0,0)$&GM5(2)&{\color{red}LD$(0,0,w)$}&LD5(2)
&45&X$(1,1,1)$&X5(2)&{\color{red}LD$(0,0,w)$}&LD5(2)
\bigstrut\\
\hline
46&GM$(0,0,0)$&GM5(2)&{\color{red}LD$(0,0,w)$}&LD5(2)
&46&X$(1,1,1)$&X5(2)&{\color{red}LD$(0,0,w)$}&LD5(2)
\bigstrut\\
\hline
51&U$(\frac{1}{2},0,\frac{1}{2})$&U3,4(4)&{\color{red}G$(\frac{1}{2},0,w)$}&G5,5(4)
&51&X$(\frac{1}{2},0,0)$&X3,4(4)&{\color{red}G$(\frac{1}{2},0,w)$}&G5,5(4)
\bigstrut\\
\hline
\hline
\end{tabular}
\end{table*}\end{center}

\newpage
\begin{center}\begin{table*}[!h]
(Continued from previous table: Time-reversal symmetric, considering SOC)\\
\begin{tabular}{c|c|c|c|c||c|c|c|c|c}\hline\hline
\multicolumn{1}{l|}{SG} & HSP & Irrep & HSL & Irrep & \multicolumn{1}{l|}{SG} & HSP & Irrep & HSL & Irrep \bigstrut\\
\hline
58&X$(\frac{1}{2},0,0)$&X3,4(4)&{\color{red}D$(\frac{1}{2},v,0)$}&D5,5(4)
&58&R$(\frac{1}{2},\frac{1}{2},\frac{1}{2})$&R3,4(4)&{\color{red}Q$(\frac{1}{2},\frac{1}{2},w)$}&Q5,5(4)
\bigstrut\\
\hline
58&S$(\frac{1}{2},\frac{1}{2},0)$&S5,5(4)&{\color{red}Q$(\frac{1}{2},\frac{1}{2},w)$}&Q5,5(4)
&58&S$(\frac{1}{2},\frac{1}{2},0)$&S6,6(4)&{\color{red}Q$(\frac{1}{2},\frac{1}{2},w)$}&Q5,5(4)
\bigstrut\\
\hline
59&U$(\frac{1}{2},0,\frac{1}{2})$&U3,4(4)&{\color{red}G$(\frac{1}{2},0,w)$}&G5,5(4)
&59&X$(\frac{1}{2},0,0)$&X3,4(4)&{\color{red}G$(\frac{1}{2},0,w)$}&G5,5(4)
\bigstrut\\
\hline
59&T$(0,\frac{1}{2},\frac{1}{2})$&T3,4(4)&{\color{red}H$(0,\frac{1}{2},w)$}&H5,5(4)
&59&Y$(0,\frac{1}{2},0)$&Y3,4(4)&{\color{red}H$(0,\frac{1}{2},w)$}&H5,5(4)
\bigstrut\\
\hline
59&R$(\frac{1}{2},\frac{1}{2},\frac{1}{2})$&R3,4(4)&{\color{red}Q$(\frac{1}{2},\frac{1}{2},w)$}&Q5,5(4)
&59&S$(\frac{1}{2},\frac{1}{2},0)$&S3,4(4)&{\color{red}Q$(\frac{1}{2},\frac{1}{2},w)$}&Q5,5(4)
\bigstrut\\
\hline
60&T$(0,\frac{1}{2},\frac{1}{2})$&T3,3(4)&{\color{red}B$(0,v,\frac{1}{2})$}&B5,5(4)
&60&T$(0,\frac{1}{2},\frac{1}{2})$&T4,4(4)&{\color{red}B$(0,v,\frac{1}{2})$}&B5,5(4)
\bigstrut\\
\hline
60&Z$(0,0,\frac{1}{2})$&Z3,4(4)&{\color{red}B$(0,v,\frac{1}{2})$}&B5,5(4)
&60&R$(\frac{1}{2},\frac{1}{2},\frac{1}{2})$&R3,3(4)&{\color{red}E$(u,\frac{1}{2},\frac{1}{2})$}&E5,5(4)
\bigstrut\\
\hline
\hline
\end{tabular}
\end{table*}\end{center}

\newpage
\begin{center}\begin{table*}[!h]
(Continued from previous table: Time-reversal symmetric, considering SOC)\\
\begin{tabular}{c|c|c|c|c||c|c|c|c|c}\hline\hline
\multicolumn{1}{l|}{SG} & HSP & Irrep & HSL & Irrep & \multicolumn{1}{l|}{SG} & HSP & Irrep & HSL & Irrep \bigstrut\\
\hline
100&M$(\frac{1}{2},\frac{1}{2},0)$&M6,7(4)&{\color{red}Y$(u,\frac{1}{2},0)$}&Y3,4(2)
&100&X$(0,\frac{1}{2},0)$&X2,3(2)&{\color{red}Y$(u,\frac{1}{2},0)$}&Y3,4(2)
\bigstrut\\
\hline
100&X$(0,\frac{1}{2},0)$&X4,5(2)&{\color{red}Y$(u,\frac{1}{2},0)$}&Y3,4(2)
&101&GM$(0,0,0)$&GM6(2)&{\color{red}LD$(0,0,w)$}&LD6(2)
\bigstrut\\
\hline
101&Z$(0,0,\frac{1}{2})$&Z6,7(4)&{\color{red}LD$(0,0,w)$}&LD6(2)
&101&GM$(0,0,0)$&GM7(2)&{\color{red}LD$(0,0,w)$}&LD7(2)
\bigstrut\\
\hline
101&A$(\frac{1}{2},\frac{1}{2},\frac{1}{2})$&A6,7(4)&{\color{red}T$(u,\frac{1}{2},\frac{1}{2})$}&T3,4(2)
&101&R$(0,\frac{1}{2},\frac{1}{2})$&R5,5(4)&{\color{red}T$(u,\frac{1}{2},\frac{1}{2})$}&T3,4(2)
\bigstrut\\
\hline
101&R$(0,\frac{1}{2},\frac{1}{2})$&R5,5(4)&{\color{red}U$(0,v,\frac{1}{2})$}&U3,4(2)
&101&Z$(0,0,\frac{1}{2})$&Z6,7(4)&{\color{red}U$(0,v,\frac{1}{2})$}&U3,4(2)
\bigstrut\\
\hline
101&A$(\frac{1}{2},\frac{1}{2},\frac{1}{2})$&A6,7(4)&{\color{red}V$(\frac{1}{2},\frac{1}{2},w)$}&V6(2)
&101&M$(\frac{1}{2},\frac{1}{2},0)$&M6(2)&{\color{red}V$(\frac{1}{2},\frac{1}{2},w)$}&V6(2)
\bigstrut\\
\hline
101&M$(\frac{1}{2},\frac{1}{2},0)$&M7(2)&{\color{red}V$(\frac{1}{2},\frac{1}{2},w)$}&V7(2)
&101&R$(0,\frac{1}{2},\frac{1}{2})$&R5,5(4)&{\color{red}W$(0,\frac{1}{2},w)$}&W5(2)
\bigstrut\\
\hline
\hline
\end{tabular}
\end{table*}\end{center}

\newpage
\begin{center}\begin{table*}[!h]
(Continued from previous table: Time-reversal symmetric, considering SOC)\\
\begin{tabular}{c|c|c|c|c||c|c|c|c|c}\hline\hline
\multicolumn{1}{l|}{SG} & HSP & Irrep & HSL & Irrep & \multicolumn{1}{l|}{SG} & HSP & Irrep & HSL & Irrep \bigstrut\\
\hline
106&A$(\frac{1}{2},\frac{1}{2},\frac{1}{2})$&A7,7(4)&{\color{red}T$(u,\frac{1}{2},\frac{1}{2})$}&T3,4(2)
&106&R$(0,\frac{1}{2},\frac{1}{2})$&R2,3(2)&{\color{red}T$(u,\frac{1}{2},\frac{1}{2})$}&T3,4(2)
\bigstrut\\
\hline
106&R$(0,\frac{1}{2},\frac{1}{2})$&R4,5(2)&{\color{red}T$(u,\frac{1}{2},\frac{1}{2})$}&T3,4(2)
&106&A$(\frac{1}{2},\frac{1}{2},\frac{1}{2})$&A6,6(4)&{\color{red}V$(\frac{1}{2},\frac{1}{2},w)$}&V6(2)
\bigstrut\\
\hline
106&M$(\frac{1}{2},\frac{1}{2},0)$&M6,7(4)&{\color{red}V$(\frac{1}{2},\frac{1}{2},w)$}&V6(2)
&106&A$(\frac{1}{2},\frac{1}{2},\frac{1}{2})$&A7,7(4)&{\color{red}V$(\frac{1}{2},\frac{1}{2},w)$}&V7(2)
\bigstrut\\
\hline
106&M$(\frac{1}{2},\frac{1}{2},0)$&M6,7(4)&{\color{red}Y$(u,\frac{1}{2},0)$}&Y3,4(2)
&106&X$(0,\frac{1}{2},0)$&X2,3(2)&{\color{red}Y$(u,\frac{1}{2},0)$}&Y3,4(2)
\bigstrut\\
\hline
106&X$(0,\frac{1}{2},0)$&X4,5(2)&{\color{red}Y$(u,\frac{1}{2},0)$}&Y3,4(2)
&107&GM$(0,0,0)$&GM6(2)&{\color{red}LD$(0,0,w)$}&LD6(2)
\bigstrut\\
\hline
107&M$(1,1,1)$&M6(2)&{\color{red}LD$(0,0,w)$}&LD6(2)
&107&GM$(0,0,0)$&GM7(2)&{\color{red}LD$(0,0,w)$}&LD7(2)
\bigstrut\\
\hline
107&M$(1,1,1)$&M7(2)&{\color{red}LD$(0,0,w)$}&LD7(2)
&107&P$(\frac{1}{2},\frac{1}{2},\frac{1}{2})$&P5(2)&{\color{red}W$(\frac{1}{2},\frac{1}{2},w)$}&W5(2)
\bigstrut\\
\hline
\hline
\end{tabular}
\end{table*}\end{center}

\newpage
\begin{center}\begin{table*}[!h]
(Continued from previous table: Time-reversal symmetric, considering SOC)\\
\begin{tabular}{c|c|c|c|c||c|c|c|c|c}\hline\hline
\multicolumn{1}{l|}{SG} & HSP & Irrep & HSL & Irrep & \multicolumn{1}{l|}{SG} & HSP & Irrep & HSL & Irrep \bigstrut\\
\hline
116&Z$(0,0,\frac{1}{2})$&Z6,7(4)&{\color{red}U$(0,v,\frac{1}{2})$}&U3,4(2)
&116&A$(\frac{1}{2},\frac{1}{2},\frac{1}{2})$&A6,7(4)&{\color{red}V$(\frac{1}{2},\frac{1}{2},w)$}&V5(2)
\bigstrut\\
\hline
116&M$(\frac{1}{2},\frac{1}{2},0)$&M6(2)&{\color{red}V$(\frac{1}{2},\frac{1}{2},w)$}&V5(2)
&116&M$(\frac{1}{2},\frac{1}{2},0)$&M7(2)&{\color{red}V$(\frac{1}{2},\frac{1}{2},w)$}&V5(2)
\bigstrut\\
\hline
116&R$(0,\frac{1}{2},\frac{1}{2})$&R5,5(4)&{\color{red}W$(0,\frac{1}{2},w)$}&W5(2)
&116&X$(0,\frac{1}{2},0)$&X5(2)&{\color{red}W$(0,\frac{1}{2},w)$}&W5(2)
\bigstrut\\
\hline
117&GM$(0,0,0)$&GM6(2)&{\color{red}LD$(0,0,w)$}&LD5(2)
&117&GM$(0,0,0)$&GM7(2)&{\color{red}LD$(0,0,w)$}&LD5(2)
\bigstrut\\
\hline
117&Z$(0,0,\frac{1}{2})$&Z6(2)&{\color{red}LD$(0,0,w)$}&LD5(2)
&117&Z$(0,0,\frac{1}{2})$&Z7(2)&{\color{red}LD$(0,0,w)$}&LD5(2)
\bigstrut\\
\hline
117&A$(\frac{1}{2},\frac{1}{2},\frac{1}{2})$&A6,7(4)&{\color{red}T$(u,\frac{1}{2},\frac{1}{2})$}&T3,4(2)
&117&R$(0,\frac{1}{2},\frac{1}{2})$&R2,3(2)&{\color{red}T$(u,\frac{1}{2},\frac{1}{2})$}&T3,4(2)
\bigstrut\\
\hline
117&R$(0,\frac{1}{2},\frac{1}{2})$&R4,5(2)&{\color{red}T$(u,\frac{1}{2},\frac{1}{2})$}&T3,4(2)
&117&A$(\frac{1}{2},\frac{1}{2},\frac{1}{2})$&A6,7(4)&{\color{red}V$(\frac{1}{2},\frac{1}{2},w)$}&V5(2)
\bigstrut\\
\hline
\hline
\end{tabular}
\end{table*}\end{center}

\newpage
\begin{center}\begin{table*}[!h]
(Continued from previous table: Time-reversal symmetric, considering SOC)\\
\begin{tabular}{c|c|c|c|c||c|c|c|c|c}\hline\hline
\multicolumn{1}{l|}{SG} & HSP & Irrep & HSL & Irrep & \multicolumn{1}{l|}{SG} & HSP & Irrep & HSL & Irrep \bigstrut\\
\hline
130&R$(0,\frac{1}{2},\frac{1}{2})$&R4,4(4)&{\color{red}W$(0,\frac{1}{2},w)$}&W5,5(4)
&130&X$(0,\frac{1}{2},0)$&X3,4(4)&{\color{red}W$(0,\frac{1}{2},w)$}&W5,5(4)
\bigstrut\\
\hline
135&A$(\frac{1}{2},\frac{1}{2},\frac{1}{2})$&A5,5(8)&{\color{red}T$(u,\frac{1}{2},\frac{1}{2})$}&T5,5(4)
&135&R$(0,\frac{1}{2},\frac{1}{2})$&R3,4(4)&{\color{red}T$(u,\frac{1}{2},\frac{1}{2})$}&T5,5(4)
\bigstrut\\
\hline
135&A$(\frac{1}{2},\frac{1}{2},\frac{1}{2})$&A5,5(8)&{\color{red}V$(\frac{1}{2},\frac{1}{2},w)$}&V6,7(4)
&135&M$(\frac{1}{2},\frac{1}{2},0)$&M6,7(4)&{\color{red}V$(\frac{1}{2},\frac{1}{2},w)$}&V6,7(4)
\bigstrut\\
\hline
135&M$(\frac{1}{2},\frac{1}{2},0)$&M8,9(4)&{\color{red}V$(\frac{1}{2},\frac{1}{2},w)$}&V6,7(4)
&135&M$(\frac{1}{2},\frac{1}{2},0)$&M6,7(4)&{\color{red}Y$(u,\frac{1}{2},0)$}&Y5,5(4)
\bigstrut\\
\hline
135&M$(\frac{1}{2},\frac{1}{2},0)$&M8,9(4)&{\color{red}Y$(u,\frac{1}{2},0)$}&Y5,5(4)
&135&X$(0,\frac{1}{2},0)$&X3,4(4)&{\color{red}Y$(u,\frac{1}{2},0)$}&Y5,5(4)
\bigstrut\\
\hline
136&A$(\frac{1}{2},\frac{1}{2},\frac{1}{2})$&A5(4)&{\color{red}V$(\frac{1}{2},\frac{1}{2},w)$}&V6,7(4)
&136&M$(\frac{1}{2},\frac{1}{2},0)$&M6,7(4)&{\color{red}V$(\frac{1}{2},\frac{1}{2},w)$}&V6,7(4)
\bigstrut\\
\hline
136&M$(\frac{1}{2},\frac{1}{2},0)$&M8,9(4)&{\color{red}V$(\frac{1}{2},\frac{1}{2},w)$}&V6,7(4)
&136&M$(\frac{1}{2},\frac{1}{2},0)$&M6,7(4)&{\color{red}Y$(u,\frac{1}{2},0)$}&Y5,5(4)
\bigstrut\\
\hline
\hline
\end{tabular}
\end{table*}\end{center}

\newpage
\begin{center}\begin{table*}[!h]
(Continued from previous table: Time-reversal symmetric, considering SOC)\\
\begin{tabular}{c|c|c|c|c||c|c|c|c|c}\hline\hline
\multicolumn{1}{l|}{SG} & HSP & Irrep & HSL & Irrep & \multicolumn{1}{l|}{SG} & HSP & Irrep & HSL & Irrep \bigstrut\\
\hline
184&A$(0,0,\frac{1}{2})$&A8,8(4)&{\color{red}DT$(0,0,w)$}&DT8(2)
&184&GM$(0,0,0)$&GM8(2)&{\color{red}DT$(0,0,w)$}&DT8(2)
\bigstrut\\
\hline
184&A$(0,0,\frac{1}{2})$&A9,9(4)&{\color{red}DT$(0,0,w)$}&DT9(2)
&184&GM$(0,0,0)$&GM9(2)&{\color{red}DT$(0,0,w)$}&DT9(2)
\bigstrut\\
\hline
184&H$(\frac{1}{3},\frac{1}{3},\frac{1}{2})$&H6,6(4)&P$(\frac{1}{3},\frac{1}{3},w)$&P6(2)
&184&K$(\frac{1}{3},\frac{1}{3},0)$&K6(2)&P$(\frac{1}{3},\frac{1}{3},w)$&P6(2)
\bigstrut\\
\hline
184&A$(0,0,\frac{1}{2})$&A7,7(4)&{\color{red}Q$(u,u,\frac{1}{2})$}&Q3,4(2)
&184&A$(0,0,\frac{1}{2})$&A8,8(4)&{\color{red}Q$(u,u,\frac{1}{2})$}&Q3,4(2)
\bigstrut\\
\hline
184&A$(0,0,\frac{1}{2})$&A9,9(4)&{\color{red}Q$(u,u,\frac{1}{2})$}&Q3,4(2)
&184&H$(\frac{1}{3},\frac{1}{3},\frac{1}{2})$&H4,5(2)&{\color{red}Q$(u,u,\frac{1}{2})$}&Q3,4(2)
\bigstrut\\
\hline
184&H$(\frac{1}{3},\frac{1}{3},\frac{1}{2})$&H6,6(4)&{\color{red}Q$(u,u,\frac{1}{2})$}&Q3,4(2)
&184&L$(\frac{1}{2},0,\frac{1}{2})$&L5,5(4)&{\color{red}Q$(u,u,\frac{1}{2})$}&Q3,4(2)
\bigstrut\\
\hline
184&A$(0,0,\frac{1}{2})$&A7,7(4)&{\color{red}R$(u,0,\frac{1}{2})$}&R3,4(2)
&184&A$(0,0,\frac{1}{2})$&A8,8(4)&{\color{red}R$(u,0,\frac{1}{2})$}&R3,4(2)
\bigstrut\\
\hline
\hline
\end{tabular}
\end{table*}\end{center}

\newpage
\begin{center}\begin{table*}[!h]
(Continued from previous table: Time-reversal symmetric, considering SOC)\\
\begin{tabular}{c|c|c|c|c||c|c|c|c|c}\hline\hline
\multicolumn{1}{l|}{SG} & HSP & Irrep & HSL & Irrep & \multicolumn{1}{l|}{SG} & HSP & Irrep & HSL & Irrep \bigstrut\\
\hline
189&K$(\frac{1}{3},\frac{1}{3},0)$&K9(2)&P$(\frac{1}{3},\frac{1}{3},w)$&P6(2)
&189&HA$(-\frac{1}{3},-\frac{1}{3},-\frac{1}{2})$&HA8(2)&PA$(-\frac{1}{3},-\frac{1}{3},-w)$&PA6(2)
\bigstrut\\
\hline
189&HA$(-\frac{1}{3},-\frac{1}{3},-\frac{1}{2})$&HA9(2)&PA$(-\frac{1}{3},-\frac{1}{3},-w)$&PA6(2)
&189&KA$(-\frac{1}{3},-\frac{1}{3},0)$&KA8(2)&PA$(-\frac{1}{3},-\frac{1}{3},-w)$&PA6(2)
\bigstrut\\
\hline
189&KA$(-\frac{1}{3},-\frac{1}{3},0)$&KA9(2)&PA$(-\frac{1}{3},-\frac{1}{3},-w)$&PA6(2)
&189&A$(0,0,\frac{1}{2})$&A7(2)&{\color{red}Q$(u,u,\frac{1}{2})$}&Q5(2)
\bigstrut\\
\hline
189&A$(0,0,\frac{1}{2})$&A8(2)&{\color{red}Q$(u,u,\frac{1}{2})$}&Q5(2)
&189&A$(0,0,\frac{1}{2})$&A9(2)&{\color{red}Q$(u,u,\frac{1}{2})$}&Q5(2)
\bigstrut\\
\hline
189&H$(\frac{1}{3},\frac{1}{3},\frac{1}{2})$&H7(2)&{\color{red}Q$(u,u,\frac{1}{2})$}&Q5(2)
&189&H$(\frac{1}{3},\frac{1}{3},\frac{1}{2})$&H8(2)&{\color{red}Q$(u,u,\frac{1}{2})$}&Q5(2)
\bigstrut\\
\hline
189&H$(\frac{1}{3},\frac{1}{3},\frac{1}{2})$&H9(2)&{\color{red}Q$(u,u,\frac{1}{2})$}&Q5(2)
&189&HA$(-\frac{1}{3},-\frac{1}{3},-\frac{1}{2})$&HA7(2)&{\color{red}Q$(u,u,\frac{1}{2})$}&Q5(2)
\bigstrut\\
\hline
189&HA$(-\frac{1}{3},-\frac{1}{3},-\frac{1}{2})$&HA8(2)&{\color{red}Q$(u,u,\frac{1}{2})$}&Q5(2)
&189&HA$(-\frac{1}{3},-\frac{1}{3},-\frac{1}{2})$&HA9(2)&{\color{red}Q$(u,u,\frac{1}{2})$}&Q5(2)
\bigstrut\\
\hline
\hline
\end{tabular}
\end{table*}\end{center}

\newpage
\begin{center}\begin{table*}[!h]
(Continued from previous table: Time-reversal symmetric, considering SOC)\\
\begin{tabular}{c|c|c|c|c||c|c|c|c|c}\hline\hline
\multicolumn{1}{l|}{SG} & HSP & Irrep & HSL & Irrep & \multicolumn{1}{l|}{SG} & HSP & Irrep & HSL & Irrep \bigstrut\\
\hline
217&N$(\frac{1}{2},\frac{1}{2},0)$&N5(2)&{\color{red}D$(\frac{1}{2},\frac{1}{2},w)$}&D5(2)
&217&GM$(0,0,0)$&GM6(2)&{\color{red}DT$(0,v,0)$}&DT5(2)
\bigstrut\\
\hline
217&GM$(0,0,0)$&GM7(2)&{\color{red}DT$(0,v,0)$}&DT5(2)
&217&GM$(0,0,0)$&GM8(4)&{\color{red}DT$(0,v,0)$}&DT5(2)
\bigstrut\\
\hline
217&H$(1,1,1)$&H6(2)&{\color{red}DT$(0,v,0)$}&DT5(2)
&217&H$(1,1,1)$&H7(2)&{\color{red}DT$(0,v,0)$}&DT5(2)
\bigstrut\\
\hline
217&H$(1,1,1)$&H8(4)&{\color{red}DT$(0,v,0)$}&DT5(2)
&217&GM$(0,0,0)$&GM6(2)&LD$(u,u,u)$&LD6(2)
\bigstrut\\
\hline
217&GM$(0,0,0)$&GM7(2)&LD$(u,u,u)$&LD6(2)
&217&H$(1,1,1)$&H6(2)&LD$(u,u,u)$&LD6(2)
\bigstrut\\
\hline
217&H$(1,1,1)$&H7(2)&LD$(u,u,u)$&LD6(2)
&217&P$(\frac{1}{2},\frac{1}{2},\frac{1}{2})$&P6(2)&LD$(u,u,u)$&LD6(2)
\bigstrut\\
\hline
217&P$(\frac{1}{2},\frac{1}{2},\frac{1}{2})$&P7(2)&LD$(u,u,u)$&LD6(2)
&217&PA$(-\frac{1}{2},-\frac{1}{2},-\frac{1}{2})$&PA6(2)&LD$(u,u,u)$&LD6(2)
\bigstrut\\
\hline\hline
\end{tabular}
\end{table*}\end{center}

\subsubsection{Results for materials with TRS and negligible SOC}\label{hsp2trsnosoc}
\newpage
\begin{center}\begin{table*}[!h]
(Continued from previous table: Time-reversal symmetric, neglecting SOC)\\
\begin{tabular}{c|c|c|c|c||c|c|c|c|c}\hline\hline
\multicolumn{1}{l|}{SG} & HSP & Irrep & HSL & Irrep & \multicolumn{1}{l|}{SG} & HSP & Irrep & HSL & Irrep \bigstrut\\
\hline
27&U$(\frac{1}{2},0,\frac{1}{2})$&U1,2(2)&{\color{red}A$(u,0,\frac{1}{2})$}&A1,2(2)
&27&U$(\frac{1}{2},0,\frac{1}{2})$&U3,4(2)&{\color{red}A$(u,0,\frac{1}{2})$}&A1,2(2)
\bigstrut\\
\hline
27&Z$(0,0,\frac{1}{2})$&Z1,2(2)&{\color{red}A$(u,0,\frac{1}{2})$}&A1,2(2)
&27&Z$(0,0,\frac{1}{2})$&Z3,4(2)&{\color{red}A$(u,0,\frac{1}{2})$}&A1,2(2)
\bigstrut\\
\hline
27&T$(0,\frac{1}{2},\frac{1}{2})$&T1,2(2)&{\color{red}B$(0,v,\frac{1}{2})$}&B1,2(2)
&27&T$(0,\frac{1}{2},\frac{1}{2})$&T3,4(2)&{\color{red}B$(0,v,\frac{1}{2})$}&B1,2(2)
\bigstrut\\
\hline
27&Z$(0,0,\frac{1}{2})$&Z1,2(2)&{\color{red}B$(0,v,\frac{1}{2})$}&B1,2(2)
&27&Z$(0,0,\frac{1}{2})$&Z3,4(2)&{\color{red}B$(0,v,\frac{1}{2})$}&B1,2(2)
\bigstrut\\
\hline
27&R$(\frac{1}{2},\frac{1}{2},\frac{1}{2})$&R1,2(2)&{\color{red}E$(u,\frac{1}{2},\frac{1}{2})$}&E1,2(2)
&27&R$(\frac{1}{2},\frac{1}{2},\frac{1}{2})$&R3,4(2)&{\color{red}E$(u,\frac{1}{2},\frac{1}{2})$}&E1,2(2)
\bigstrut\\
\hline
27&T$(0,\frac{1}{2},\frac{1}{2})$&T1,2(2)&{\color{red}E$(u,\frac{1}{2},\frac{1}{2})$}&E1,2(2)
&27&T$(0,\frac{1}{2},\frac{1}{2})$&T3,4(2)&{\color{red}E$(u,\frac{1}{2},\frac{1}{2})$}&E1,2(2)
\bigstrut\\
\hline
27&R$(\frac{1}{2},\frac{1}{2},\frac{1}{2})$&R1,2(2)&{\color{red}P$(\frac{1}{2},v,\frac{1}{2})$}&P1,2(2)
&27&R$(\frac{1}{2},\frac{1}{2},\frac{1}{2})$&R3,4(2)&{\color{red}P$(\frac{1}{2},v,\frac{1}{2})$}&P1,2(2)
\bigstrut\\
\hline
\hline
\end{tabular}
\end{table*}\end{center}

\newpage
\begin{center}\begin{table*}[!h]
(Continued from previous table: Time-reversal symmetric, neglecting SOC)\\
\begin{tabular}{c|c|c|c|c||c|c|c|c|c}\hline\hline
\multicolumn{1}{l|}{SG} & HSP & Irrep & HSL & Irrep & \multicolumn{1}{l|}{SG} & HSP & Irrep & HSL & Irrep \bigstrut\\
\hline
34&Z$(0,0,\frac{1}{2})$&Z3,4(2)&{\color{red}A$(u,0,\frac{1}{2})$}&A1,2(2)
&34&T$(0,\frac{1}{2},\frac{1}{2})$&T1(2)&{\color{red}B$(0,v,\frac{1}{2})$}&B1,2(2)
\bigstrut\\
\hline
34&Z$(0,0,\frac{1}{2})$&Z1,2(2)&{\color{red}B$(0,v,\frac{1}{2})$}&B1,2(2)
&34&Z$(0,0,\frac{1}{2})$&Z3,4(2)&{\color{red}B$(0,v,\frac{1}{2})$}&B1,2(2)
\bigstrut\\
\hline
34&S$(\frac{1}{2},\frac{1}{2},0)$&S1,2(2)&{\color{red}C$(u,\frac{1}{2},0)$}&C1,2(2)
&34&S$(\frac{1}{2},\frac{1}{2},0)$&S3,4(2)&{\color{red}C$(u,\frac{1}{2},0)$}&C1,2(2)
\bigstrut\\
\hline
34&Y$(0,\frac{1}{2},0)$&Y1(2)&{\color{red}C$(u,\frac{1}{2},0)$}&C1,2(2)
&34&S$(\frac{1}{2},\frac{1}{2},0)$&S1,2(2)&{\color{red}D$(\frac{1}{2},v,0)$}&D1,2(2)
\bigstrut\\
\hline
34&S$(\frac{1}{2},\frac{1}{2},0)$&S3,4(2)&{\color{red}D$(\frac{1}{2},v,0)$}&D1,2(2)
&34&X$(\frac{1}{2},0,0)$&X1(2)&{\color{red}D$(\frac{1}{2},v,0)$}&D1,2(2)
\bigstrut\\
\hline
34&U$(\frac{1}{2},0,\frac{1}{2})$&U1(2)&{\color{red}G$(\frac{1}{2},0,w)$}&G1(2)
&34&X$(\frac{1}{2},0,0)$&X1(2)&{\color{red}G$(\frac{1}{2},0,w)$}&G1(2)
\bigstrut\\
\hline
34&T$(0,\frac{1}{2},\frac{1}{2})$&T1(2)&{\color{red}H$(0,\frac{1}{2},w)$}&H1(2)
&34&Y$(0,\frac{1}{2},0)$&Y1(2)&{\color{red}H$(0,\frac{1}{2},w)$}&H1(2)
\bigstrut\\
\hline
\hline
\end{tabular}
\end{table*}\end{center}

\newpage
\begin{center}\begin{table*}[!h]
(Continued from previous table: Time-reversal symmetric, neglecting SOC)\\
\begin{tabular}{c|c|c|c|c||c|c|c|c|c}\hline\hline
\multicolumn{1}{l|}{SG} & HSP & Irrep & HSL & Irrep & \multicolumn{1}{l|}{SG} & HSP & Irrep & HSL & Irrep \bigstrut\\
\hline
50&Y$(0,\frac{1}{2},0)$&Y1(2)&{\color{red}C$(u,\frac{1}{2},0)$}&C1(2)
&50&Y$(0,\frac{1}{2},0)$&Y2(2)&{\color{red}C$(u,\frac{1}{2},0)$}&C1(2)
\bigstrut\\
\hline
50&S$(\frac{1}{2},\frac{1}{2},0)$&S1(2)&{\color{red}D$(\frac{1}{2},v,0)$}&D1(2)
&50&S$(\frac{1}{2},\frac{1}{2},0)$&S2(2)&{\color{red}D$(\frac{1}{2},v,0)$}&D1(2)
\bigstrut\\
\hline
50&X$(\frac{1}{2},0,0)$&X1(2)&{\color{red}D$(\frac{1}{2},v,0)$}&D1(2)
&50&X$(\frac{1}{2},0,0)$&X2(2)&{\color{red}D$(\frac{1}{2},v,0)$}&D1(2)
\bigstrut\\
\hline
50&R$(\frac{1}{2},\frac{1}{2},\frac{1}{2})$&R1(2)&{\color{red}E$(u,\frac{1}{2},\frac{1}{2})$}&E1(2)
&50&R$(\frac{1}{2},\frac{1}{2},\frac{1}{2})$&R2(2)&{\color{red}E$(u,\frac{1}{2},\frac{1}{2})$}&E1(2)
\bigstrut\\
\hline
50&T$(0,\frac{1}{2},\frac{1}{2})$&T1(2)&{\color{red}E$(u,\frac{1}{2},\frac{1}{2})$}&E1(2)
&50&T$(0,\frac{1}{2},\frac{1}{2})$&T2(2)&{\color{red}E$(u,\frac{1}{2},\frac{1}{2})$}&E1(2)
\bigstrut\\
\hline
50&U$(\frac{1}{2},0,\frac{1}{2})$&U1(2)&{\color{red}G$(\frac{1}{2},0,w)$}&G1(2)
&50&U$(\frac{1}{2},0,\frac{1}{2})$&U2(2)&{\color{red}G$(\frac{1}{2},0,w)$}&G1(2)
\bigstrut\\
\hline
50&X$(\frac{1}{2},0,0)$&X1(2)&{\color{red}G$(\frac{1}{2},0,w)$}&G1(2)
&50&X$(\frac{1}{2},0,0)$&X2(2)&{\color{red}G$(\frac{1}{2},0,w)$}&G1(2)
\bigstrut\\
\hline
\hline
\end{tabular}
\end{table*}\end{center}

\newpage
\begin{center}\begin{table*}[!h]
(Continued from previous table: Time-reversal symmetric, neglecting SOC)\\
\begin{tabular}{c|c|c|c|c||c|c|c|c|c}\hline\hline
\multicolumn{1}{l|}{SG} & HSP & Irrep & HSL & Irrep & \multicolumn{1}{l|}{SG} & HSP & Irrep & HSL & Irrep \bigstrut\\
\hline
60&Y$(0,\frac{1}{2},0)$&Y1(2)&{\color{red}H$(0,\frac{1}{2},w)$}&H1(2)
&60&Y$(0,\frac{1}{2},0)$&Y2(2)&{\color{red}H$(0,\frac{1}{2},w)$}&H1(2)
\bigstrut\\
\hline
60&R$(\frac{1}{2},\frac{1}{2},\frac{1}{2})$&R1,2(4)&{\color{red}P$(\frac{1}{2},v,\frac{1}{2})$}&P1,1(4)
&60&U$(\frac{1}{2},0,\frac{1}{2})$&U1,2(4)&{\color{red}P$(\frac{1}{2},v,\frac{1}{2})$}&P1,1(4)
\bigstrut\\
\hline
61&R$(\frac{1}{2},\frac{1}{2},\frac{1}{2})$&R1+,1+(4)&{\color{red}E$(u,\frac{1}{2},\frac{1}{2})$}&E1,1(4)
&61&R$(\frac{1}{2},\frac{1}{2},\frac{1}{2})$&R1-,1-(4)&{\color{red}E$(u,\frac{1}{2},\frac{1}{2})$}&E1,1(4)
\bigstrut\\
\hline
61&T$(0,\frac{1}{2},\frac{1}{2})$&T1,2(4)&{\color{red}E$(u,\frac{1}{2},\frac{1}{2})$}&E1,1(4)
&61&R$(\frac{1}{2},\frac{1}{2},\frac{1}{2})$&R1+,1+(4)&{\color{red}P$(\frac{1}{2},v,\frac{1}{2})$}&P1,1(4)
\bigstrut\\
\hline
61&R$(\frac{1}{2},\frac{1}{2},\frac{1}{2})$&R1-,1-(4)&{\color{red}P$(\frac{1}{2},v,\frac{1}{2})$}&P1,1(4)
&61&U$(\frac{1}{2},0,\frac{1}{2})$&U1,2(4)&{\color{red}P$(\frac{1}{2},v,\frac{1}{2})$}&P1,1(4)
\bigstrut\\
\hline
61&R$(\frac{1}{2},\frac{1}{2},\frac{1}{2})$&R1+,1+(4)&{\color{red}Q$(\frac{1}{2},\frac{1}{2},w)$}&Q1,1(4)
&61&R$(\frac{1}{2},\frac{1}{2},\frac{1}{2})$&R1-,1-(4)&{\color{red}Q$(\frac{1}{2},\frac{1}{2},w)$}&Q1,1(4)
\bigstrut\\
\hline
61&S$(\frac{1}{2},\frac{1}{2},0)$&S1,2(4)&{\color{red}Q$(\frac{1}{2},\frac{1}{2},w)$}&Q1,1(4)
&62&R$(\frac{1}{2},\frac{1}{2},\frac{1}{2})$&R1,2(4)&{\color{red}Q$(\frac{1}{2},\frac{1}{2},w)$}&Q1,1(4)
\bigstrut\\
\hline
\hline
\end{tabular}
\end{table*}\end{center}

\newpage
\begin{center}\begin{table*}[!h]
(Continued from previous table: Time-reversal symmetric, neglecting SOC)\\
\begin{tabular}{c|c|c|c|c||c|c|c|c|c}\hline\hline
\multicolumn{1}{l|}{SG} & HSP & Irrep & HSL & Irrep & \multicolumn{1}{l|}{SG} & HSP & Irrep & HSL & Irrep \bigstrut\\
\hline
87&GM$(0,0,0)$&GM3-,4-(2)&LD$(0,0,w)$&LD3,4(2)
&87&M$(1,1,1)$&M3+,4+(2)&LD$(0,0,w)$&LD3,4(2)
\bigstrut\\
\hline
87&M$(1,1,1)$&M3-,4-(2)&LD$(0,0,w)$&LD3,4(2)
&88&GM$(0,0,0)$&GM3+,4+(2)&LD$(0,0,w)$&LD3,4(2)
\bigstrut\\
\hline
88&GM$(0,0,0)$&GM3-,4-(2)&LD$(0,0,w)$&LD3,4(2)
&88&M$(1,1,1)$&M2(2)&LD$(0,0,w)$&LD3,4(2)
\bigstrut\\
\hline
88&P$(\frac{1}{2},\frac{1}{2},\frac{1}{2})$&P1,4(2)&{\color{red}W$(\frac{1}{2},\frac{1}{2},w)$}&W1,2(2)
&88&P$(\frac{1}{2},\frac{1}{2},\frac{1}{2})$&P2,3(2)&{\color{red}W$(\frac{1}{2},\frac{1}{2},w)$}&W1,2(2)
\bigstrut\\
\hline
88&X$(\frac{1}{2},\frac{1}{2},0)$&X1(2)&{\color{red}W$(\frac{1}{2},\frac{1}{2},w)$}&W1,2(2)
&99&GM$(0,0,0)$&GM5(2)&LD$(0,0,w)$&LD5(2)
\bigstrut\\
\hline
99&Z$(0,0,\frac{1}{2})$&Z5(2)&LD$(0,0,w)$&LD5(2)
&99&A$(\frac{1}{2},\frac{1}{2},\frac{1}{2})$&A5(2)&V$(\frac{1}{2},\frac{1}{2},w)$&V5(2)
\bigstrut\\
\hline
99&M$(\frac{1}{2},\frac{1}{2},0)$&M5(2)&V$(\frac{1}{2},\frac{1}{2},w)$&V5(2)
&100&GM$(0,0,0)$&GM5(2)&LD$(0,0,w)$&LD5(2)
\bigstrut\\
\hline
\hline
\end{tabular}
\end{table*}\end{center}

\newpage
\begin{center}\begin{table*}[!h]
(Continued from previous table: Time-reversal symmetric, neglecting SOC)\\
\begin{tabular}{c|c|c|c|c||c|c|c|c|c}\hline\hline
\multicolumn{1}{l|}{SG} & HSP & Irrep & HSL & Irrep & \multicolumn{1}{l|}{SG} & HSP & Irrep & HSL & Irrep \bigstrut\\
\hline
104&Z$(0,0,\frac{1}{2})$&Z1,4(2)&{\color{red}S$(u,u,\frac{1}{2})$}&S1,2(2)
&104&Z$(0,0,\frac{1}{2})$&Z2,3(2)&{\color{red}S$(u,u,\frac{1}{2})$}&S1,2(2)
\bigstrut\\
\hline
104&Z$(0,0,\frac{1}{2})$&Z5,5(4)&{\color{red}S$(u,u,\frac{1}{2})$}&S1,2(2)
&104&R$(0,\frac{1}{2},\frac{1}{2})$&R1(2)&{\color{red}U$(0,v,\frac{1}{2})$}&U1,2(2)
\bigstrut\\
\hline
104&Z$(0,0,\frac{1}{2})$&Z1,4(2)&{\color{red}U$(0,v,\frac{1}{2})$}&U1,2(2)
&104&Z$(0,0,\frac{1}{2})$&Z2,3(2)&{\color{red}U$(0,v,\frac{1}{2})$}&U1,2(2)
\bigstrut\\
\hline
104&Z$(0,0,\frac{1}{2})$&Z5,5(4)&{\color{red}U$(0,v,\frac{1}{2})$}&U1,2(2)
&104&A$(\frac{1}{2},\frac{1}{2},\frac{1}{2})$&A5(2)&V$(\frac{1}{2},\frac{1}{2},w)$&V5(2)
\bigstrut\\
\hline
104&M$(\frac{1}{2},\frac{1}{2},0)$&M5(2)&V$(\frac{1}{2},\frac{1}{2},w)$&V5(2)
&104&R$(0,\frac{1}{2},\frac{1}{2})$&R1(2)&{\color{red}W$(0,\frac{1}{2},w)$}&W1(2)
\bigstrut\\
\hline
104&X$(0,\frac{1}{2},0)$&X1(2)&{\color{red}W$(0,\frac{1}{2},w)$}&W1(2)
&104&M$(\frac{1}{2},\frac{1}{2},0)$&M1,3(2)&{\color{red}Y$(u,\frac{1}{2},0)$}&Y1,2(2)
\bigstrut\\
\hline
104&M$(\frac{1}{2},\frac{1}{2},0)$&M2,4(2)&{\color{red}Y$(u,\frac{1}{2},0)$}&Y1,2(2)
&104&M$(\frac{1}{2},\frac{1}{2},0)$&M5(2)&{\color{red}Y$(u,\frac{1}{2},0)$}&Y1,2(2)
\bigstrut\\
\hline
\hline
\end{tabular}
\end{table*}\end{center}

\newpage
\begin{center}\begin{table*}[!h]
(Continued from previous table: Time-reversal symmetric, neglecting SOC)\\
\begin{tabular}{c|c|c|c|c||c|c|c|c|c}\hline\hline
\multicolumn{1}{l|}{SG} & HSP & Irrep & HSL & Irrep & \multicolumn{1}{l|}{SG} & HSP & Irrep & HSL & Irrep \bigstrut\\
\hline
112&M$(\frac{1}{2},\frac{1}{2},0)$&M5(2)&V$(\frac{1}{2},\frac{1}{2},w)$&V3,4(2)
&113&GM$(0,0,0)$&GM5(2)&LD$(0,0,w)$&LD3,4(2)
\bigstrut\\
\hline
113&Z$(0,0,\frac{1}{2})$&Z5(2)&LD$(0,0,w)$&LD3,4(2)
&114&GM$(0,0,0)$&GM5(2)&LD$(0,0,w)$&LD3,4(2)
\bigstrut\\
\hline
114&Z$(0,0,\frac{1}{2})$&Z1,2(2)&LD$(0,0,w)$&LD3,4(2)
&114&Z$(0,0,\frac{1}{2})$&Z3,4(2)&LD$(0,0,w)$&LD3,4(2)
\bigstrut\\
\hline
114&A$(\frac{1}{2},\frac{1}{2},\frac{1}{2})$&A1,4(2)&{\color{red}S$(u,u,\frac{1}{2})$}&S1,2(2)
&114&A$(\frac{1}{2},\frac{1}{2},\frac{1}{2})$&A2,3(2)&{\color{red}S$(u,u,\frac{1}{2})$}&S1,2(2)
\bigstrut\\
\hline
114&A$(\frac{1}{2},\frac{1}{2},\frac{1}{2})$&A5,5(4)&{\color{red}S$(u,u,\frac{1}{2})$}&S1,2(2)
&114&Z$(0,0,\frac{1}{2})$&Z1,2(2)&{\color{red}S$(u,u,\frac{1}{2})$}&S1,2(2)
\bigstrut\\
\hline
114&Z$(0,0,\frac{1}{2})$&Z3,4(2)&{\color{red}S$(u,u,\frac{1}{2})$}&S1,2(2)
&114&Z$(0,0,\frac{1}{2})$&Z5(2)&{\color{red}S$(u,u,\frac{1}{2})$}&S1,2(2)
\bigstrut\\
\hline
115&GM$(0,0,0)$&GM5(2)&LD$(0,0,w)$&LD3,4(2)
&115&Z$(0,0,\frac{1}{2})$&Z5(2)&LD$(0,0,w)$&LD3,4(2)
\bigstrut\\
\hline
\hline
\end{tabular}
\end{table*}\end{center}

\newpage
\begin{center}\begin{table*}[!h]
(Continued from previous table: Time-reversal symmetric, neglecting SOC)\\
\begin{tabular}{c|c|c|c|c||c|c|c|c|c}\hline\hline
\multicolumn{1}{l|}{SG} & HSP & Irrep & HSL & Irrep & \multicolumn{1}{l|}{SG} & HSP & Irrep & HSL & Irrep \bigstrut\\
\hline
122&M$(1,1,1)$&M5(2)&{\color{red}Y$(u,1-u,0)$}&Y1,2(2)
&122&X$(\frac{1}{2},\frac{1}{2},0)$&X1(2)&{\color{red}Y$(u,1-u,0)$}&Y1,2(2)
\bigstrut\\
\hline
123&GM$(0,0,0)$&GM5+(2)&LD$(0,0,w)$&LD5(2)
&123&GM$(0,0,0)$&GM5-(2)&LD$(0,0,w)$&LD5(2)
\bigstrut\\
\hline
123&Z$(0,0,\frac{1}{2})$&Z5+(2)&LD$(0,0,w)$&LD5(2)
&123&Z$(0,0,\frac{1}{2})$&Z5-(2)&LD$(0,0,w)$&LD5(2)
\bigstrut\\
\hline
123&A$(\frac{1}{2},\frac{1}{2},\frac{1}{2})$&A5+(2)&V$(\frac{1}{2},\frac{1}{2},w)$&V5(2)
&123&A$(\frac{1}{2},\frac{1}{2},\frac{1}{2})$&A5-(2)&V$(\frac{1}{2},\frac{1}{2},w)$&V5(2)
\bigstrut\\
\hline
123&M$(\frac{1}{2},\frac{1}{2},0)$&M5+(2)&V$(\frac{1}{2},\frac{1}{2},w)$&V5(2)
&123&M$(\frac{1}{2},\frac{1}{2},0)$&M5-(2)&V$(\frac{1}{2},\frac{1}{2},w)$&V5(2)
\bigstrut\\
\hline
124&GM$(0,0,0)$&GM5+(2)&LD$(0,0,w)$&LD5(2)
&124&GM$(0,0,0)$&GM5-(2)&LD$(0,0,w)$&LD5(2)
\bigstrut\\
\hline
124&Z$(0,0,\frac{1}{2})$&Z3,4(4)&LD$(0,0,w)$&LD5(2)
&124&A$(\frac{1}{2},\frac{1}{2},\frac{1}{2})$&A1(2)&{\color{red}S$(u,u,\frac{1}{2})$}&S1(2)
\bigstrut\\
\hline
\hline
\end{tabular}
\end{table*}\end{center}

\newpage
\begin{center}\begin{table*}[!h]
(Continued from previous table: Time-reversal symmetric, neglecting SOC)\\
\begin{tabular}{c|c|c|c|c||c|c|c|c|c}\hline\hline
\multicolumn{1}{l|}{SG} & HSP & Irrep & HSL & Irrep & \multicolumn{1}{l|}{SG} & HSP & Irrep & HSL & Irrep \bigstrut\\
\hline
126&M$(\frac{1}{2},\frac{1}{2},0)$&M2(2)&{\color{red}Y$(u,\frac{1}{2},0)$}&Y1(2)
&126&M$(\frac{1}{2},\frac{1}{2},0)$&M3(2)&{\color{red}Y$(u,\frac{1}{2},0)$}&Y1(2)
\bigstrut\\
\hline
126&M$(\frac{1}{2},\frac{1}{2},0)$&M4(2)&{\color{red}Y$(u,\frac{1}{2},0)$}&Y1(2)
&126&X$(0,\frac{1}{2},0)$&X1(2)&{\color{red}Y$(u,\frac{1}{2},0)$}&Y1(2)
\bigstrut\\
\hline
126&X$(0,\frac{1}{2},0)$&X2(2)&{\color{red}Y$(u,\frac{1}{2},0)$}&Y1(2)
&127&GM$(0,0,0)$&GM5+(2)&LD$(0,0,w)$&LD5(2)
\bigstrut\\
\hline
127&GM$(0,0,0)$&GM5-(2)&LD$(0,0,w)$&LD5(2)
&127&Z$(0,0,\frac{1}{2})$&Z5+(2)&LD$(0,0,w)$&LD5(2)
\bigstrut\\
\hline
127&Z$(0,0,\frac{1}{2})$&Z5-(2)&LD$(0,0,w)$&LD5(2)
&128&GM$(0,0,0)$&GM5+(2)&LD$(0,0,w)$&LD5(2)
\bigstrut\\
\hline
128&GM$(0,0,0)$&GM5-(2)&LD$(0,0,w)$&LD5(2)
&128&Z$(0,0,\frac{1}{2})$&Z3,4(4)&LD$(0,0,w)$&LD5(2)
\bigstrut\\
\hline
128&A$(\frac{1}{2},\frac{1}{2},\frac{1}{2})$&A1(2)&{\color{red}S$(u,u,\frac{1}{2})$}&S1(2)
&128&A$(\frac{1}{2},\frac{1}{2},\frac{1}{2})$&A2(2)&{\color{red}S$(u,u,\frac{1}{2})$}&S1(2)
\bigstrut\\
\hline
\hline
\end{tabular}
\end{table*}\end{center}

\newpage
\begin{center}\begin{table*}[!h]
(Continued from previous table: Time-reversal symmetric, neglecting SOC)\\
\begin{tabular}{c|c|c|c|c||c|c|c|c|c}\hline\hline
\multicolumn{1}{l|}{SG} & HSP & Irrep & HSL & Irrep & \multicolumn{1}{l|}{SG} & HSP & Irrep & HSL & Irrep \bigstrut\\
\hline
133&A$(\frac{1}{2},\frac{1}{2},\frac{1}{2})$&A2(2)&{\color{red}S$(u,u,\frac{1}{2})$}&S1(2)
&133&A$(\frac{1}{2},\frac{1}{2},\frac{1}{2})$&A3,4(4)&{\color{red}S$(u,u,\frac{1}{2})$}&S1(2)
\bigstrut\\
\hline
133&Z$(0,0,\frac{1}{2})$&Z1(2)&{\color{red}S$(u,u,\frac{1}{2})$}&S1(2)
&133&Z$(0,0,\frac{1}{2})$&Z2(2)&{\color{red}S$(u,u,\frac{1}{2})$}&S1(2)
\bigstrut\\
\hline
133&Z$(0,0,\frac{1}{2})$&Z3(2)&{\color{red}S$(u,u,\frac{1}{2})$}&S1(2)
&133&Z$(0,0,\frac{1}{2})$&Z4(2)&{\color{red}S$(u,u,\frac{1}{2})$}&S1(2)
\bigstrut\\
\hline
133&A$(\frac{1}{2},\frac{1}{2},\frac{1}{2})$&A1(2)&{\color{red}T$(u,\frac{1}{2},\frac{1}{2})$}&T1(2)
&133&A$(\frac{1}{2},\frac{1}{2},\frac{1}{2})$&A2(2)&{\color{red}T$(u,\frac{1}{2},\frac{1}{2})$}&T1(2)
\bigstrut\\
\hline
133&A$(\frac{1}{2},\frac{1}{2},\frac{1}{2})$&A3,4(4)&{\color{red}T$(u,\frac{1}{2},\frac{1}{2})$}&T1(2)
&133&R$(0,\frac{1}{2},\frac{1}{2})$&R1(2)&{\color{red}T$(u,\frac{1}{2},\frac{1}{2})$}&T1(2)
\bigstrut\\
\hline
133&R$(0,\frac{1}{2},\frac{1}{2})$&R2(2)&{\color{red}T$(u,\frac{1}{2},\frac{1}{2})$}&T1(2)
&133&A$(\frac{1}{2},\frac{1}{2},\frac{1}{2})$&A3,4(4)&V$(\frac{1}{2},\frac{1}{2},w)$&V5(2)
\bigstrut\\
\hline
133&M$(\frac{1}{2},\frac{1}{2},0)$&M3(2)&V$(\frac{1}{2},\frac{1}{2},w)$&V5(2)
&133&M$(\frac{1}{2},\frac{1}{2},0)$&M4(2)&V$(\frac{1}{2},\frac{1}{2},w)$&V5(2)
\bigstrut\\
\hline
\hline
\end{tabular}
\end{table*}\end{center}

\newpage
\begin{center}\begin{table*}[!h]
(Continued from previous table: Time-reversal symmetric, neglecting SOC)\\
\begin{tabular}{c|c|c|c|c||c|c|c|c|c}\hline\hline
\multicolumn{1}{l|}{SG} & HSP & Irrep & HSL & Irrep & \multicolumn{1}{l|}{SG} & HSP & Irrep & HSL & Irrep \bigstrut\\
\hline
138&GM$(0,0,0)$&GM5-(2)&LD$(0,0,w)$&LD5(2)
&138&Z$(0,0,\frac{1}{2})$&Z3(2)&LD$(0,0,w)$&LD5(2)
\bigstrut\\
\hline
138&Z$(0,0,\frac{1}{2})$&Z4(2)&LD$(0,0,w)$&LD5(2)
&138&R$(0,\frac{1}{2},\frac{1}{2})$&R1,2(4)&{\color{red}U$(0,v,\frac{1}{2})$}&U1(2)
\bigstrut\\
\hline
138&Z$(0,0,\frac{1}{2})$&Z1(2)&{\color{red}U$(0,v,\frac{1}{2})$}&U1(2)
&138&Z$(0,0,\frac{1}{2})$&Z2(2)&{\color{red}U$(0,v,\frac{1}{2})$}&U1(2)
\bigstrut\\
\hline
138&Z$(0,0,\frac{1}{2})$&Z3(2)&{\color{red}U$(0,v,\frac{1}{2})$}&U1(2)
&138&Z$(0,0,\frac{1}{2})$&Z4(2)&{\color{red}U$(0,v,\frac{1}{2})$}&U1(2)
\bigstrut\\
\hline
139&GM$(0,0,0)$&GM5+(2)&LD$(0,0,w)$&LD5(2)
&139&GM$(0,0,0)$&GM5-(2)&LD$(0,0,w)$&LD5(2)
\bigstrut\\
\hline
139&M$(1,1,1)$&M5+(2)&LD$(0,0,w)$&LD5(2)
&139&M$(1,1,1)$&M5-(2)&LD$(0,0,w)$&LD5(2)
\bigstrut\\
\hline
140&GM$(0,0,0)$&GM5+(2)&LD$(0,0,w)$&LD5(2)
&140&GM$(0,0,0)$&GM5-(2)&LD$(0,0,w)$&LD5(2)
\bigstrut\\
\hline
\hline
\end{tabular}
\end{table*}\end{center}

\newpage
\begin{center}\begin{table*}[!h]
(Continued from previous table: Time-reversal symmetric, neglecting SOC)\\
\begin{tabular}{c|c|c|c|c||c|c|c|c|c}\hline\hline
\multicolumn{1}{l|}{SG} & HSP & Irrep & HSL & Irrep & \multicolumn{1}{l|}{SG} & HSP & Irrep & HSL & Irrep \bigstrut\\
\hline
161&GM$(0,0,0)$&GM3(2)&LD$(0,0,w)$&LD3(2)
&161&T$(0,0,3/2)$&T3,3(4)&LD$(0,0,w)$&LD3(2)
\bigstrut\\
\hline
162&A$(0,0,\frac{1}{2})$&A3+(2)&DT$(0,0,w)$&DT3(2)
&162&A$(0,0,\frac{1}{2})$&A3-(2)&DT$(0,0,w)$&DT3(2)
\bigstrut\\
\hline
162&GM$(0,0,0)$&GM3+(2)&DT$(0,0,w)$&DT3(2)
&162&GM$(0,0,0)$&GM3-(2)&DT$(0,0,w)$&DT3(2)
\bigstrut\\
\hline
162&H$(\frac{1}{3},\frac{1}{3},\frac{1}{2})$&H3(2)&P$(\frac{1}{3},\frac{1}{3},w)$&P3(2)
&162&K$(\frac{1}{3},\frac{1}{3},0)$&K3(2)&P$(\frac{1}{3},\frac{1}{3},w)$&P3(2)
\bigstrut\\
\hline
163&A$(0,0,\frac{1}{2})$&A1,2(4)&DT$(0,0,w)$&DT3(2)
&163&GM$(0,0,0)$&GM3+(2)&DT$(0,0,w)$&DT3(2)
\bigstrut\\
\hline
163&GM$(0,0,0)$&GM3-(2)&DT$(0,0,w)$&DT3(2)
&163&H$(\frac{1}{3},\frac{1}{3},\frac{1}{2})$&H3(2)&P$(\frac{1}{3},\frac{1}{3},w)$&P3(2)
\bigstrut\\
\hline
163&K$(\frac{1}{3},\frac{1}{3},0)$&K3(2)&P$(\frac{1}{3},\frac{1}{3},w)$&P3(2)
&163&A$(0,0,\frac{1}{2})$&A1,2(4)&{\color{red}R$(u,0,\frac{1}{2})$}&R1,2(2)
\bigstrut\\
\hline
\hline
\end{tabular}
\end{table*}\end{center}

\newpage
\begin{center}\begin{table*}[!h]
(Continued from previous table: Time-reversal symmetric, neglecting SOC)\\
\begin{tabular}{c|c|c|c|c||c|c|c|c|c}\hline\hline
\multicolumn{1}{l|}{SG} & HSP & Irrep & HSL & Irrep & \multicolumn{1}{l|}{SG} & HSP & Irrep & HSL & Irrep \bigstrut\\
\hline
184&H$(\frac{1}{3},\frac{1}{3},\frac{1}{2})$&H1,2(2)&{\color{red}Q$(u,u,\frac{1}{2})$}&Q1,2(2)
&184&H$(\frac{1}{3},\frac{1}{3},\frac{1}{2})$&H3,3(4)&{\color{red}Q$(u,u,\frac{1}{2})$}&Q1,2(2)
\bigstrut\\
\hline
184&L$(\frac{1}{2},0,\frac{1}{2})$&L1,2(2)&{\color{red}Q$(u,u,\frac{1}{2})$}&Q1,2(2)
&184&L$(\frac{1}{2},0,\frac{1}{2})$&L3,4(2)&{\color{red}Q$(u,u,\frac{1}{2})$}&Q1,2(2)
\bigstrut\\
\hline
184&A$(0,0,\frac{1}{2})$&A1,2(2)&{\color{red}R$(u,0,\frac{1}{2})$}&R1,2(2)
&184&A$(0,0,\frac{1}{2})$&A3,4(2)&{\color{red}R$(u,0,\frac{1}{2})$}&R1,2(2)
\bigstrut\\
\hline
184&A$(0,0,\frac{1}{2})$&A5,5(4)&{\color{red}R$(u,0,\frac{1}{2})$}&R1,2(2)
&184&A$(0,0,\frac{1}{2})$&A6,6(4)&{\color{red}R$(u,0,\frac{1}{2})$}&R1,2(2)
\bigstrut\\
\hline
184&L$(\frac{1}{2},0,\frac{1}{2})$&L1,2(2)&{\color{red}R$(u,0,\frac{1}{2})$}&R1,2(2)
&184&L$(\frac{1}{2},0,\frac{1}{2})$&L3,4(2)&{\color{red}R$(u,0,\frac{1}{2})$}&R1,2(2)
\bigstrut\\
\hline
185&A$(0,0,\frac{1}{2})$&A5,6(4)&DT$(0,0,w)$&DT5(2)
&185&GM$(0,0,0)$&GM5(2)&DT$(0,0,w)$&DT5(2)
\bigstrut\\
\hline
185&GM$(0,0,0)$&GM6(2)&DT$(0,0,w)$&DT6(2)
&185&H$(\frac{1}{3},\frac{1}{3},\frac{1}{2})$&H3,3(4)&P$(\frac{1}{3},\frac{1}{3},w)$&P3(2)
\bigstrut\\
\hline
\hline
\end{tabular}
\end{table*}\end{center}

\newpage
\begin{center}\begin{table*}[!h]
(Continued from previous table: Time-reversal symmetric, neglecting SOC)\\
\begin{tabular}{c|c|c|c|c||c|c|c|c|c}\hline\hline
\multicolumn{1}{l|}{SG} & HSP & Irrep & HSL & Irrep & \multicolumn{1}{l|}{SG} & HSP & Irrep & HSL & Irrep \bigstrut\\
\hline
191&H$(\frac{1}{3},\frac{1}{3},\frac{1}{2})$&H6(2)&P$(\frac{1}{3},\frac{1}{3},w)$&P3(2)
&191&K$(\frac{1}{3},\frac{1}{3},0)$&K5(2)&P$(\frac{1}{3},\frac{1}{3},w)$&P3(2)
\bigstrut\\
\hline
191&K$(\frac{1}{3},\frac{1}{3},0)$&K6(2)&P$(\frac{1}{3},\frac{1}{3},w)$&P3(2)
&192&A$(0,0,\frac{1}{2})$&A1,2(4)&DT$(0,0,w)$&DT5(2)
\bigstrut\\
\hline
192&GM$(0,0,0)$&GM5+(2)&DT$(0,0,w)$&DT5(2)
&192&GM$(0,0,0)$&GM5-(2)&DT$(0,0,w)$&DT5(2)
\bigstrut\\
\hline
192&A$(0,0,\frac{1}{2})$&A3,4(4)&DT$(0,0,w)$&DT6(2)
&192&GM$(0,0,0)$&GM6+(2)&DT$(0,0,w)$&DT6(2)
\bigstrut\\
\hline
192&GM$(0,0,0)$&GM6-(2)&DT$(0,0,w)$&DT6(2)
&192&H$(\frac{1}{3},\frac{1}{3},\frac{1}{2})$&H1,2(4)&P$(\frac{1}{3},\frac{1}{3},w)$&P3(2)
\bigstrut\\
\hline
192&K$(\frac{1}{3},\frac{1}{3},0)$&K5(2)&P$(\frac{1}{3},\frac{1}{3},w)$&P3(2)
&192&K$(\frac{1}{3},\frac{1}{3},0)$&K6(2)&P$(\frac{1}{3},\frac{1}{3},w)$&P3(2)
\bigstrut\\
\hline
192&A$(0,0,\frac{1}{2})$&A1,2(4)&{\color{red}Q$(u,u,\frac{1}{2})$}&Q1(2)
&192&A$(0,0,\frac{1}{2})$&A3,4(4)&{\color{red}Q$(u,u,\frac{1}{2})$}&Q1(2)
\bigstrut\\
\hline
\hline
\end{tabular}
\end{table*}\end{center}

\newpage
\begin{center}\begin{table*}[!h]
(Continued from previous table: Time-reversal symmetric, neglecting SOC)\\
\begin{tabular}{c|c|c|c|c||c|c|c|c|c}\hline\hline
\multicolumn{1}{l|}{SG} & HSP & Irrep & HSL & Irrep & \multicolumn{1}{l|}{SG} & HSP & Irrep & HSL & Irrep \bigstrut\\
\hline
205&R$(\frac{1}{2},\frac{1}{2},\frac{1}{2})$&R1+,3+(4)&{\color{red}T$(\frac{1}{2},\frac{1}{2},w)$}&T1,1(4)
&205&R$(\frac{1}{2},\frac{1}{2},\frac{1}{2})$&R1-,3-(4)&{\color{red}T$(\frac{1}{2},\frac{1}{2},w)$}&T1,1(4)
\bigstrut\\
\hline
205&R$(\frac{1}{2},\frac{1}{2},\frac{1}{2})$&R2+,2+(4)&{\color{red}T$(\frac{1}{2},\frac{1}{2},w)$}&T1,1(4)
&205&R$(\frac{1}{2},\frac{1}{2},\frac{1}{2})$&R2-,2-(4)&{\color{red}T$(\frac{1}{2},\frac{1}{2},w)$}&T1,1(4)
\bigstrut\\
\hline
206&P$(\frac{1}{2},\frac{1}{2},\frac{1}{2})$&P1,3(4)&{\color{red}D$(\frac{1}{2},\frac{1}{2},w)$}&D1,2(2)
&206&P$(\frac{1}{2},\frac{1}{2},\frac{1}{2})$&P2,2(4)&{\color{red}D$(\frac{1}{2},\frac{1}{2},w)$}&D1,2(2)
\bigstrut\\
\hline
206&N$(\frac{1}{2},\frac{1}{2},0)$&N1(2)&{\color{red}D$(\frac{1}{2},\frac{1}{2},w)$}&D1,2(2)
&206&GM$(0,0,0)$&GM2+,3+(2)&LD$(u,u,u)$&LD2,3(2)
\bigstrut\\
\hline
206&GM$(0,0,0)$&GM2-,3-(2)&LD$(u,u,u)$&LD2,3(2)
&206&H$(1,1,1)$&H2+,3+(2)&LD$(u,u,u)$&LD2,3(2)
\bigstrut\\
\hline
206&H$(1,1,1)$&H2-,3-(2)&LD$(u,u,u)$&LD2,3(2)
&206&P$(\frac{1}{2},\frac{1}{2},\frac{1}{2})$&P2,2(4)&LD$(u,u,u)$&LD2,3(2)
\bigstrut\\
\hline
215&X$(0,\frac{1}{2},0)$&X5(2)&DT$(0,v,0)$&DT3,4(2)
&215&GM$(0,0,0)$&GM3(2)&LD$(u,u,u)$&LD3(2)
\bigstrut\\
\hline
\hline
\end{tabular}
\end{table*}\end{center}

\newpage
\begin{center}\begin{table*}[!h]
(Continued from previous table: Time-reversal symmetric, neglecting SOC)\\
\begin{tabular}{c|c|c|c|c||c|c|c|c|c}\hline\hline
\multicolumn{1}{l|}{SG} & HSP & Irrep & HSL & Irrep & \multicolumn{1}{l|}{SG} & HSP & Irrep & HSL & Irrep \bigstrut\\
\hline
222&R$(\frac{1}{2},\frac{1}{2},\frac{1}{2})$&R2,3(4)&T$(\frac{1}{2},\frac{1}{2},w)$&T5(2)
&222&M$(\frac{1}{2},\frac{1}{2},0)$&M1(2)&{\color{red}Z$(u,\frac{1}{2},0)$}&Z1(2)
\bigstrut\\
\hline
222&M$(\frac{1}{2},\frac{1}{2},0)$&M2(2)&{\color{red}Z$(u,\frac{1}{2},0)$}&Z1(2)
&222&M$(\frac{1}{2},\frac{1}{2},0)$&M3(2)&{\color{red}Z$(u,\frac{1}{2},0)$}&Z1(2)
\bigstrut\\
\hline
222&M$(\frac{1}{2},\frac{1}{2},0)$&M4(2)&{\color{red}Z$(u,\frac{1}{2},0)$}&Z1(2)
&222&X$(0,\frac{1}{2},0)$&X1(2)&{\color{red}Z$(u,\frac{1}{2},0)$}&Z1(2)
\bigstrut\\
\hline
222&X$(0,\frac{1}{2},0)$&X2(2)&{\color{red}Z$(u,\frac{1}{2},0)$}&Z1(2)
&222&X$(0,\frac{1}{2},0)$&X3,4(4)&{\color{red}Z$(u,\frac{1}{2},0)$}&Z1(2)
\bigstrut\\
\hline
223&X$(0,\frac{1}{2},0)$&X3(2)&DT$(0,v,0)$&DT5(2)
&223&X$(0,\frac{1}{2},0)$&X4(2)&DT$(0,v,0)$&DT5(2)
\bigstrut\\
\hline
223&GM$(0,0,0)$&GM3+(2)&LD$(u,u,u)$&LD3(2)
&223&GM$(0,0,0)$&GM3-(2)&LD$(u,u,u)$&LD3(2)
\bigstrut\\
\hline
223&R$(\frac{1}{2},\frac{1}{2},\frac{1}{2})$&R2,3(4)&LD$(u,u,u)$&LD3(2)
&223&R$(\frac{1}{2},\frac{1}{2},\frac{1}{2})$&R1(2)&{\color{red}S$(u,\frac{1}{2},u)$}&S1(2)
\bigstrut\\
\hline
\hline
\end{tabular}
\end{table*}\end{center}

\newpage
\begin{center}\begin{table*}[!h]
(Continued from previous table: Time-reversal symmetric, neglecting SOC)\\
\begin{tabular}{c|c|c|c|c||c|c|c|c|c}\hline\hline
\multicolumn{1}{l|}{SG} & HSP & Irrep & HSL & Irrep & \multicolumn{1}{l|}{SG} & HSP & Irrep & HSL & Irrep \bigstrut\\
\hline
230&N$(\frac{1}{2},\frac{1}{2},0)$&N1(2)&{\color{red}D$(\frac{1}{2},\frac{1}{2},w)$}&D1(2)
&230&N$(\frac{1}{2},\frac{1}{2},0)$&N2(2)&{\color{red}D$(\frac{1}{2},\frac{1}{2},w)$}&D1(2)
\bigstrut\\
\hline
230&H$(1,1,1)$&H1(2)&DT$(0,v,0)$&DT5(2)
&230&H$(1,1,1)$&H2,3(4)&DT$(0,v,0)$&DT5(2)
\bigstrut\\
\hline
230&H$(1,1,1)$&H1(2)&{\color{red}G$(1+u,1-u,1)$}&G1(2)
&230&H$(1,1,1)$&H2,3(4)&{\color{red}G$(1+u,1-u,1)$}&G1(2)
\bigstrut\\
\hline
230&H$(1,1,1)$&H4(6)&{\color{red}G$(1+u,1-u,1)$}&G1(2)
&230&N$(\frac{1}{2},\frac{1}{2},0)$&N1(2)&{\color{red}G$(1+u,1-u,1)$}&G1(2)
\bigstrut\\
\hline
230&N$(\frac{1}{2},\frac{1}{2},0)$&N2(2)&{\color{red}G$(1+u,1-u,1)$}&G1(2)
&230&GM$(0,0,0)$&GM3+(2)&LD$(u,u,u)$&LD3(2)
\bigstrut\\
\hline
230&GM$(0,0,0)$&GM3-(2)&LD$(u,u,u)$&LD3(2)
&230&H$(1,1,1)$&H2,3(4)&LD$(u,u,u)$&LD3(2)
\bigstrut\\
\hline
230&P$(\frac{1}{2},\frac{1}{2},\frac{1}{2})$&P1,2(4)&LD$(u,u,u)$&LD3(2)
& & & & \bigstrut\\
\hline\hline
\end{tabular}
\end{table*}\end{center}

\subsubsection{Results for materials without TRS and with significant SOC}\label{hsp2notrssoc}
\newpage
\begin{center}\begin{table*}[!h]
(Continued from previous table: Time-reversal symmetry broken, considering SOC)\\
\begin{tabular}{c|c|c|c|c||c|c|c|c|c}\hline\hline
\multicolumn{1}{l|}{SG} & HSP & Irrep & HSL & Irrep & \multicolumn{1}{l|}{SG} & HSP & Irrep & HSL & Irrep \bigstrut\\
\hline
25&U$(\frac{1}{2},0,\frac{1}{2})$&U5(2)&{\color{red}G$(\frac{1}{2},0,w)$}&G5(2)
&25&X$(\frac{1}{2},0,0)$&X5(2)&{\color{red}G$(\frac{1}{2},0,w)$}&G5(2)
\bigstrut\\
\hline
25&T$(0,\frac{1}{2},\frac{1}{2})$&T5(2)&{\color{red}H$(0,\frac{1}{2},w)$}&H5(2)
&25&Y$(0,\frac{1}{2},0)$&Y5(2)&{\color{red}H$(0,\frac{1}{2},w)$}&H5(2)
\bigstrut\\
\hline
25&GM$(0,0,0)$&GM5(2)&{\color{red}LD$(0,0,w)$}&LD5(2)
&25&Z$(0,0,\frac{1}{2})$&Z5(2)&{\color{red}LD$(0,0,w)$}&LD5(2)
\bigstrut\\
\hline
25&R$(\frac{1}{2},\frac{1}{2},\frac{1}{2})$&R5(2)&{\color{red}Q$(\frac{1}{2},\frac{1}{2},w)$}&Q5(2)
&25&S$(\frac{1}{2},\frac{1}{2},0)$&S5(2)&{\color{red}Q$(\frac{1}{2},\frac{1}{2},w)$}&Q5(2)
\bigstrut\\
\hline
26&U$(\frac{1}{2},0,\frac{1}{2})$&U5(2)&{\color{red}G$(\frac{1}{2},0,w)$}&G5(2)
&26&X$(\frac{1}{2},0,0)$&X5(2)&{\color{red}G$(\frac{1}{2},0,w)$}&G5(2)
\bigstrut\\
\hline
26&T$(0,\frac{1}{2},\frac{1}{2})$&T5(2)&{\color{red}H$(0,\frac{1}{2},w)$}&H5(2)
&26&Y$(0,\frac{1}{2},0)$&Y5(2)&{\color{red}H$(0,\frac{1}{2},w)$}&H5(2)
\bigstrut\\
\hline
26&GM$(0,0,0)$&GM5(2)&{\color{red}LD$(0,0,w)$}&LD5(2)
&26&Z$(0,0,\frac{1}{2})$&Z5(2)&{\color{red}LD$(0,0,w)$}&LD5(2)
\bigstrut\\
\hline
\hline
\end{tabular}
\end{table*}\end{center}

\newpage
\begin{center}\begin{table*}[!h]
(Continued from previous table: Time-reversal symmetry broken, considering SOC)\\
\begin{tabular}{c|c|c|c|c||c|c|c|c|c}\hline\hline
\multicolumn{1}{l|}{SG} & HSP & Irrep & HSL & Irrep & \multicolumn{1}{l|}{SG} & HSP & Irrep & HSL & Irrep \bigstrut\\
\hline
43&GM$(0,0,0)$&GM5(2)&{\color{red}LD$(0,0,w)$}&LD5(2)
&43&Z$(0,0,1)$&Z5(2)&{\color{red}LD$(0,0,w)$}&LD5(2)
\bigstrut\\
\hline
44&GM$(0,0,0)$&GM5(2)&{\color{red}LD$(0,0,w)$}&LD5(2)
&44&X$(1,1,1)$&X5(2)&{\color{red}LD$(0,0,w)$}&LD5(2)
\bigstrut\\
\hline
45&GM$(0,0,0)$&GM5(2)&{\color{red}LD$(0,0,w)$}&LD5(2)
&45&X$(1,1,1)$&X5(2)&{\color{red}LD$(0,0,w)$}&LD5(2)
\bigstrut\\
\hline
46&GM$(0,0,0)$&GM5(2)&{\color{red}LD$(0,0,w)$}&LD5(2)
&46&X$(1,1,1)$&X5(2)&{\color{red}LD$(0,0,w)$}&LD5(2)
\bigstrut\\
\hline
47&U$(\frac{1}{2},0,\frac{1}{2})$&U5(2)&{\color{red}A$(u,0,\frac{1}{2})$}&A5(2)
&47&U$(\frac{1}{2},0,\frac{1}{2})$&U6(2)&{\color{red}A$(u,0,\frac{1}{2})$}&A5(2)
\bigstrut\\
\hline
47&Z$(0,0,\frac{1}{2})$&Z5(2)&{\color{red}A$(u,0,\frac{1}{2})$}&A5(2)
&47&Z$(0,0,\frac{1}{2})$&Z6(2)&{\color{red}A$(u,0,\frac{1}{2})$}&A5(2)
\bigstrut\\
\hline
47&T$(0,\frac{1}{2},\frac{1}{2})$&T5(2)&{\color{red}B$(0,v,\frac{1}{2})$}&B5(2)
&47&T$(0,\frac{1}{2},\frac{1}{2})$&T6(2)&{\color{red}B$(0,v,\frac{1}{2})$}&B5(2)
\bigstrut\\
\hline
\hline
\end{tabular}
\end{table*}\end{center}

\newpage
\begin{center}\begin{table*}[!h]
(Continued from previous table: Time-reversal symmetry broken, considering SOC)\\
\begin{tabular}{c|c|c|c|c||c|c|c|c|c}\hline\hline
\multicolumn{1}{l|}{SG} & HSP & Irrep & HSL & Irrep & \multicolumn{1}{l|}{SG} & HSP & Irrep & HSL & Irrep \bigstrut\\
\hline
49&S$(\frac{1}{2},\frac{1}{2},0)$&S5(2)&{\color{red}C$(u,\frac{1}{2},0)$}&C5(2)
&49&S$(\frac{1}{2},\frac{1}{2},0)$&S6(2)&{\color{red}C$(u,\frac{1}{2},0)$}&C5(2)
\bigstrut\\
\hline
49&Y$(0,\frac{1}{2},0)$&Y5(2)&{\color{red}C$(u,\frac{1}{2},0)$}&C5(2)
&49&Y$(0,\frac{1}{2},0)$&Y6(2)&{\color{red}C$(u,\frac{1}{2},0)$}&C5(2)
\bigstrut\\
\hline
49&S$(\frac{1}{2},\frac{1}{2},0)$&S5(2)&{\color{red}D$(\frac{1}{2},v,0)$}&D5(2)
&49&S$(\frac{1}{2},\frac{1}{2},0)$&S6(2)&{\color{red}D$(\frac{1}{2},v,0)$}&D5(2)
\bigstrut\\
\hline
49&X$(\frac{1}{2},0,0)$&X5(2)&{\color{red}D$(\frac{1}{2},v,0)$}&D5(2)
&49&X$(\frac{1}{2},0,0)$&X6(2)&{\color{red}D$(\frac{1}{2},v,0)$}&D5(2)
\bigstrut\\
\hline
49&GM$(0,0,0)$&GM5(2)&{\color{red}DT$(0,v,0)$}&DT5(2)
&49&GM$(0,0,0)$&GM6(2)&{\color{red}DT$(0,v,0)$}&DT5(2)
\bigstrut\\
\hline
49&Y$(0,\frac{1}{2},0)$&Y5(2)&{\color{red}DT$(0,v,0)$}&DT5(2)
&49&Y$(0,\frac{1}{2},0)$&Y6(2)&{\color{red}DT$(0,v,0)$}&DT5(2)
\bigstrut\\
\hline
49&U$(\frac{1}{2},0,\frac{1}{2})$&U3(2)&{\color{red}G$(\frac{1}{2},0,w)$}&G5(2)
&49&U$(\frac{1}{2},0,\frac{1}{2})$&U4(2)&{\color{red}G$(\frac{1}{2},0,w)$}&G5(2)
\bigstrut\\
\hline
\hline
\end{tabular}
\end{table*}\end{center}

\newpage
\begin{center}\begin{table*}[!h]
(Continued from previous table: Time-reversal symmetry broken, considering SOC)\\
\begin{tabular}{c|c|c|c|c||c|c|c|c|c}\hline\hline
\multicolumn{1}{l|}{SG} & HSP & Irrep & HSL & Irrep & \multicolumn{1}{l|}{SG} & HSP & Irrep & HSL & Irrep \bigstrut\\
\hline
51&T$(0,\frac{1}{2},\frac{1}{2})$&T5(2)&{\color{red}H$(0,\frac{1}{2},w)$}&H5(2)
&51&T$(0,\frac{1}{2},\frac{1}{2})$&T6(2)&{\color{red}H$(0,\frac{1}{2},w)$}&H5(2)
\bigstrut\\
\hline
51&Y$(0,\frac{1}{2},0)$&Y5(2)&{\color{red}H$(0,\frac{1}{2},w)$}&H5(2)
&51&Y$(0,\frac{1}{2},0)$&Y6(2)&{\color{red}H$(0,\frac{1}{2},w)$}&H5(2)
\bigstrut\\
\hline
51&GM$(0,0,0)$&GM5(2)&{\color{red}LD$(0,0,w)$}&LD5(2)
&51&GM$(0,0,0)$&GM6(2)&{\color{red}LD$(0,0,w)$}&LD5(2)
\bigstrut\\
\hline
51&Z$(0,0,\frac{1}{2})$&Z5(2)&{\color{red}LD$(0,0,w)$}&LD5(2)
&51&Z$(0,0,\frac{1}{2})$&Z6(2)&{\color{red}LD$(0,0,w)$}&LD5(2)
\bigstrut\\
\hline
51&R$(\frac{1}{2},\frac{1}{2},\frac{1}{2})$&R3(2)&{\color{red}Q$(\frac{1}{2},\frac{1}{2},w)$}&Q5(2)
&51&R$(\frac{1}{2},\frac{1}{2},\frac{1}{2})$&R4(2)&{\color{red}Q$(\frac{1}{2},\frac{1}{2},w)$}&Q5(2)
\bigstrut\\
\hline
51&S$(\frac{1}{2},\frac{1}{2},0)$&S3(2)&{\color{red}Q$(\frac{1}{2},\frac{1}{2},w)$}&Q5(2)
&51&S$(\frac{1}{2},\frac{1}{2},0)$&S4(2)&{\color{red}Q$(\frac{1}{2},\frac{1}{2},w)$}&Q5(2)
\bigstrut\\
\hline
51&GM$(0,0,0)$&GM5(2)&{\color{red}SM$(u,0,0)$}&SM5(2)
&51&GM$(0,0,0)$&GM6(2)&{\color{red}SM$(u,0,0)$}&SM5(2)
\bigstrut\\
\hline
\hline
\end{tabular}
\end{table*}\end{center}

\newpage
\begin{center}\begin{table*}[!h]
(Continued from previous table: Time-reversal symmetry broken, considering SOC)\\
\begin{tabular}{c|c|c|c|c||c|c|c|c|c}\hline\hline
\multicolumn{1}{l|}{SG} & HSP & Irrep & HSL & Irrep & \multicolumn{1}{l|}{SG} & HSP & Irrep & HSL & Irrep \bigstrut\\
\hline
54&GM$(0,0,0)$&GM5(2)&{\color{red}LD$(0,0,w)$}&LD5(2)
&54&GM$(0,0,0)$&GM6(2)&{\color{red}LD$(0,0,w)$}&LD5(2)
\bigstrut\\
\hline
54&Z$(0,0,\frac{1}{2})$&Z3(2)&{\color{red}LD$(0,0,w)$}&LD5(2)
&54&Z$(0,0,\frac{1}{2})$&Z4(2)&{\color{red}LD$(0,0,w)$}&LD5(2)
\bigstrut\\
\hline
54&R$(\frac{1}{2},\frac{1}{2},\frac{1}{2})$&R3(2)&{\color{red}P$(\frac{1}{2},v,\frac{1}{2})$}&P5(2)
&54&R$(\frac{1}{2},\frac{1}{2},\frac{1}{2})$&R4(2)&{\color{red}P$(\frac{1}{2},v,\frac{1}{2})$}&P5(2)
\bigstrut\\
\hline
54&U$(\frac{1}{2},0,\frac{1}{2})$&U3(2)&{\color{red}P$(\frac{1}{2},v,\frac{1}{2})$}&P5(2)
&54&U$(\frac{1}{2},0,\frac{1}{2})$&U4(2)&{\color{red}P$(\frac{1}{2},v,\frac{1}{2})$}&P5(2)
\bigstrut\\
\hline
54&R$(\frac{1}{2},\frac{1}{2},\frac{1}{2})$&R3(2)&{\color{red}Q$(\frac{1}{2},\frac{1}{2},w)$}&Q5(2)
&54&R$(\frac{1}{2},\frac{1}{2},\frac{1}{2})$&R4(2)&{\color{red}Q$(\frac{1}{2},\frac{1}{2},w)$}&Q5(2)
\bigstrut\\
\hline
54&S$(\frac{1}{2},\frac{1}{2},0)$&S3(2)&{\color{red}Q$(\frac{1}{2},\frac{1}{2},w)$}&Q5(2)
&54&S$(\frac{1}{2},\frac{1}{2},0)$&S4(2)&{\color{red}Q$(\frac{1}{2},\frac{1}{2},w)$}&Q5(2)
\bigstrut\\
\hline
54&GM$(0,0,0)$&GM5(2)&{\color{red}SM$(u,0,0)$}&SM5(2)
&54&GM$(0,0,0)$&GM6(2)&{\color{red}SM$(u,0,0)$}&SM5(2)
\bigstrut\\
\hline
\hline
\end{tabular}
\end{table*}\end{center}

\newpage
\begin{center}\begin{table*}[!h]
(Continued from previous table: Time-reversal symmetry broken, considering SOC)\\
\begin{tabular}{c|c|c|c|c||c|c|c|c|c}\hline\hline
\multicolumn{1}{l|}{SG} & HSP & Irrep & HSL & Irrep & \multicolumn{1}{l|}{SG} & HSP & Irrep & HSL & Irrep \bigstrut\\
\hline
56&R$(\frac{1}{2},\frac{1}{2},\frac{1}{2})$&R5(2)&{\color{red}Q$(\frac{1}{2},\frac{1}{2},w)$}&Q5(2)
&56&R$(\frac{1}{2},\frac{1}{2},\frac{1}{2})$&R6(2)&{\color{red}Q$(\frac{1}{2},\frac{1}{2},w)$}&Q5(2)
\bigstrut\\
\hline
56&S$(\frac{1}{2},\frac{1}{2},0)$&S3(2)&{\color{red}Q$(\frac{1}{2},\frac{1}{2},w)$}&Q5(2)
&56&S$(\frac{1}{2},\frac{1}{2},0)$&S4(2)&{\color{red}Q$(\frac{1}{2},\frac{1}{2},w)$}&Q5(2)
\bigstrut\\
\hline
56&GM$(0,0,0)$&GM5(2)&{\color{red}SM$(u,0,0)$}&SM5(2)
&56&GM$(0,0,0)$&GM6(2)&{\color{red}SM$(u,0,0)$}&SM5(2)
\bigstrut\\
\hline
56&X$(\frac{1}{2},0,0)$&X3(2)&{\color{red}SM$(u,0,0)$}&SM5(2)
&56&X$(\frac{1}{2},0,0)$&X4(2)&{\color{red}SM$(u,0,0)$}&SM5(2)
\bigstrut\\
\hline
57&T$(0,\frac{1}{2},\frac{1}{2})$&T3(2)&{\color{red}B$(0,v,\frac{1}{2})$}&B5(2)
&57&T$(0,\frac{1}{2},\frac{1}{2})$&T4(2)&{\color{red}B$(0,v,\frac{1}{2})$}&B5(2)
\bigstrut\\
\hline
57&Z$(0,0,\frac{1}{2})$&Z3(2)&{\color{red}B$(0,v,\frac{1}{2})$}&B5(2)
&57&Z$(0,0,\frac{1}{2})$&Z4(2)&{\color{red}B$(0,v,\frac{1}{2})$}&B5(2)
\bigstrut\\
\hline
57&S$(\frac{1}{2},\frac{1}{2},0)$&S3(2)&{\color{red}C$(u,\frac{1}{2},0)$}&C5(2)
&57&S$(\frac{1}{2},\frac{1}{2},0)$&S4(2)&{\color{red}C$(u,\frac{1}{2},0)$}&C5(2)
\bigstrut\\
\hline
\hline
\end{tabular}
\end{table*}\end{center}

\newpage
\begin{center}\begin{table*}[!h]
(Continued from previous table: Time-reversal symmetry broken, considering SOC)\\
\begin{tabular}{c|c|c|c|c||c|c|c|c|c}\hline\hline
\multicolumn{1}{l|}{SG} & HSP & Irrep & HSL & Irrep & \multicolumn{1}{l|}{SG} & HSP & Irrep & HSL & Irrep \bigstrut\\
\hline
59&T$(0,\frac{1}{2},\frac{1}{2})$&T3(2)&{\color{red}H$(0,\frac{1}{2},w)$}&H5(2)
&59&T$(0,\frac{1}{2},\frac{1}{2})$&T4(2)&{\color{red}H$(0,\frac{1}{2},w)$}&H5(2)
\bigstrut\\
\hline
59&Y$(0,\frac{1}{2},0)$&Y3(2)&{\color{red}H$(0,\frac{1}{2},w)$}&H5(2)
&59&Y$(0,\frac{1}{2},0)$&Y4(2)&{\color{red}H$(0,\frac{1}{2},w)$}&H5(2)
\bigstrut\\
\hline
59&GM$(0,0,0)$&GM5(2)&{\color{red}LD$(0,0,w)$}&LD5(2)
&59&GM$(0,0,0)$&GM6(2)&{\color{red}LD$(0,0,w)$}&LD5(2)
\bigstrut\\
\hline
59&Z$(0,0,\frac{1}{2})$&Z5(2)&{\color{red}LD$(0,0,w)$}&LD5(2)
&59&Z$(0,0,\frac{1}{2})$&Z6(2)&{\color{red}LD$(0,0,w)$}&LD5(2)
\bigstrut\\
\hline
59&R$(\frac{1}{2},\frac{1}{2},\frac{1}{2})$&R3(2)&{\color{red}Q$(\frac{1}{2},\frac{1}{2},w)$}&Q5(2)
&59&R$(\frac{1}{2},\frac{1}{2},\frac{1}{2})$&R4(2)&{\color{red}Q$(\frac{1}{2},\frac{1}{2},w)$}&Q5(2)
\bigstrut\\
\hline
59&S$(\frac{1}{2},\frac{1}{2},0)$&S3(2)&{\color{red}Q$(\frac{1}{2},\frac{1}{2},w)$}&Q5(2)
&59&S$(\frac{1}{2},\frac{1}{2},0)$&S4(2)&{\color{red}Q$(\frac{1}{2},\frac{1}{2},w)$}&Q5(2)
\bigstrut\\
\hline
59&GM$(0,0,0)$&GM5(2)&{\color{red}SM$(u,0,0)$}&SM5(2)
&59&GM$(0,0,0)$&GM6(2)&{\color{red}SM$(u,0,0)$}&SM5(2)
\bigstrut\\
\hline
\hline
\end{tabular}
\end{table*}\end{center}

\newpage
\begin{center}\begin{table*}[!h]
(Continued from previous table: Time-reversal symmetry broken, considering SOC)\\
\begin{tabular}{c|c|c|c|c||c|c|c|c|c}\hline\hline
\multicolumn{1}{l|}{SG} & HSP & Irrep & HSL & Irrep & \multicolumn{1}{l|}{SG} & HSP & Irrep & HSL & Irrep \bigstrut\\
\hline
62&U$(\frac{1}{2},0,\frac{1}{2})$&U5(2)&{\color{red}G$(\frac{1}{2},0,w)$}&G5(2)
&62&U$(\frac{1}{2},0,\frac{1}{2})$&U6(2)&{\color{red}G$(\frac{1}{2},0,w)$}&G5(2)
\bigstrut\\
\hline
62&X$(\frac{1}{2},0,0)$&X3(2)&{\color{red}G$(\frac{1}{2},0,w)$}&G5(2)
&62&X$(\frac{1}{2},0,0)$&X4(2)&{\color{red}G$(\frac{1}{2},0,w)$}&G5(2)
\bigstrut\\
\hline
62&GM$(0,0,0)$&GM5(2)&{\color{red}LD$(0,0,w)$}&LD5(2)
&62&GM$(0,0,0)$&GM6(2)&{\color{red}LD$(0,0,w)$}&LD5(2)
\bigstrut\\
\hline
62&Z$(0,0,\frac{1}{2})$&Z3(2)&{\color{red}LD$(0,0,w)$}&LD5(2)
&62&Z$(0,0,\frac{1}{2})$&Z4(2)&{\color{red}LD$(0,0,w)$}&LD5(2)
\bigstrut\\
\hline
62&R$(\frac{1}{2},\frac{1}{2},\frac{1}{2})$&R3(2)&{\color{red}P$(\frac{1}{2},v,\frac{1}{2})$}&P5(2)
&62&R$(\frac{1}{2},\frac{1}{2},\frac{1}{2})$&R4(2)&{\color{red}P$(\frac{1}{2},v,\frac{1}{2})$}&P5(2)
\bigstrut\\
\hline
62&U$(\frac{1}{2},0,\frac{1}{2})$&U5(2)&{\color{red}P$(\frac{1}{2},v,\frac{1}{2})$}&P5(2)
&62&U$(\frac{1}{2},0,\frac{1}{2})$&U6(2)&{\color{red}P$(\frac{1}{2},v,\frac{1}{2})$}&P5(2)
\bigstrut\\
\hline
62&GM$(0,0,0)$&GM5(2)&{\color{red}SM$(u,0,0)$}&SM5(2)
&62&GM$(0,0,0)$&GM6(2)&{\color{red}SM$(u,0,0)$}&SM5(2)
\bigstrut\\
\hline
\hline
\end{tabular}
\end{table*}\end{center}

\newpage
\begin{center}\begin{table*}[!h]
(Continued from previous table: Time-reversal symmetry broken, considering SOC)\\
\begin{tabular}{c|c|c|c|c||c|c|c|c|c}\hline\hline
\multicolumn{1}{l|}{SG} & HSP & Irrep & HSL & Irrep & \multicolumn{1}{l|}{SG} & HSP & Irrep & HSL & Irrep \bigstrut\\
\hline
66&GM$(0,0,0)$&GM5(2)&{\color{red}DT$(0,v,0)$}&DT5(2)
&66&GM$(0,0,0)$&GM6(2)&{\color{red}DT$(0,v,0)$}&DT5(2)
\bigstrut\\
\hline
66&Y$(1,0,0)$&Y5(2)&{\color{red}DT$(0,v,0)$}&DT5(2)
&66&Y$(1,0,0)$&Y6(2)&{\color{red}DT$(0,v,0)$}&DT5(2)
\bigstrut\\
\hline
66&T$(1,0,\frac{1}{2})$&T3(2)&{\color{red}H$(1,0,w)$}&H5(2)
&66&T$(1,0,\frac{1}{2})$&T4(2)&{\color{red}H$(1,0,w)$}&H5(2)
\bigstrut\\
\hline
66&Y$(1,0,0)$&Y5(2)&{\color{red}H$(1,0,w)$}&H5(2)
&66&Y$(1,0,0)$&Y6(2)&{\color{red}H$(1,0,w)$}&H5(2)
\bigstrut\\
\hline
66&GM$(0,0,0)$&GM5(2)&{\color{red}LD$(0,0,w)$}&LD5(2)
&66&GM$(0,0,0)$&GM6(2)&{\color{red}LD$(0,0,w)$}&LD5(2)
\bigstrut\\
\hline
66&Z$(0,0,\frac{1}{2})$&Z3(2)&{\color{red}LD$(0,0,w)$}&LD5(2)
&66&Z$(0,0,\frac{1}{2})$&Z4(2)&{\color{red}LD$(0,0,w)$}&LD5(2)
\bigstrut\\
\hline
66&GM$(0,0,0)$&GM5(2)&{\color{red}SM$(u,0,0)$}&SM5(2)
&66&GM$(0,0,0)$&GM6(2)&{\color{red}SM$(u,0,0)$}&SM5(2)
\bigstrut\\
\hline
\hline
\end{tabular}
\end{table*}\end{center}

\newpage
\begin{center}\begin{table*}[!h]
(Continued from previous table: Time-reversal symmetry broken, considering SOC)\\
\begin{tabular}{c|c|c|c|c||c|c|c|c|c}\hline\hline
\multicolumn{1}{l|}{SG} & HSP & Irrep & HSL & Irrep & \multicolumn{1}{l|}{SG} & HSP & Irrep & HSL & Irrep \bigstrut\\
\hline
70&GM$(0,0,0)$&GM5(2)&{\color{red}DT$(0,v,0)$}&DT5(2)
&70&GM$(0,0,0)$&GM6(2)&{\color{red}DT$(0,v,0)$}&DT5(2)
\bigstrut\\
\hline
70&Y$(0,1,0)$&Y3(2)&{\color{red}DT$(0,v,0)$}&DT5(2)
&70&Y$(0,1,0)$&Y4(2)&{\color{red}DT$(0,v,0)$}&DT5(2)
\bigstrut\\
\hline
70&GM$(0,0,0)$&GM5(2)&{\color{red}LD$(0,0,w)$}&LD5(2)
&70&GM$(0,0,0)$&GM6(2)&{\color{red}LD$(0,0,w)$}&LD5(2)
\bigstrut\\
\hline
70&Z$(0,0,1)$&Z3(2)&{\color{red}LD$(0,0,w)$}&LD5(2)
&70&Z$(0,0,1)$&Z4(2)&{\color{red}LD$(0,0,w)$}&LD5(2)
\bigstrut\\
\hline
70&GM$(0,0,0)$&GM5(2)&{\color{red}SM$(u,0,0)$}&SM5(2)
&70&GM$(0,0,0)$&GM6(2)&{\color{red}SM$(u,0,0)$}&SM5(2)
\bigstrut\\
\hline
70&T$(1,0,0)$&T3(2)&{\color{red}SM$(u,0,0)$}&SM5(2)
&70&T$(1,0,0)$&T4(2)&{\color{red}SM$(u,0,0)$}&SM5(2)
\bigstrut\\
\hline
71&GM$(0,0,0)$&GM5(2)&{\color{red}DT$(0,v,0)$}&DT5(2)
&71&GM$(0,0,0)$&GM6(2)&{\color{red}DT$(0,v,0)$}&DT5(2)
\bigstrut\\
\hline
\hline
\end{tabular}
\end{table*}\end{center}

\newpage
\begin{center}\begin{table*}[!h]
(Continued from previous table: Time-reversal symmetry broken, considering SOC)\\
\begin{tabular}{c|c|c|c|c||c|c|c|c|c}\hline\hline
\multicolumn{1}{l|}{SG} & HSP & Irrep & HSL & Irrep & \multicolumn{1}{l|}{SG} & HSP & Irrep & HSL & Irrep \bigstrut\\
\hline
101&GM$(0,0,0)$&GM7(2)&{\color{red}LD$(0,0,w)$}&LD7(2)
&101&Z$(0,0,\frac{1}{2})$&Z7(2)&{\color{red}LD$(0,0,w)$}&LD7(2)
\bigstrut\\
\hline
101&A$(\frac{1}{2},\frac{1}{2},\frac{1}{2})$&A6(2)&{\color{red}V$(\frac{1}{2},\frac{1}{2},w)$}&V6(2)
&101&M$(\frac{1}{2},\frac{1}{2},0)$&M6(2)&{\color{red}V$(\frac{1}{2},\frac{1}{2},w)$}&V6(2)
\bigstrut\\
\hline
101&A$(\frac{1}{2},\frac{1}{2},\frac{1}{2})$&A7(2)&{\color{red}V$(\frac{1}{2},\frac{1}{2},w)$}&V7(2)
&101&M$(\frac{1}{2},\frac{1}{2},0)$&M7(2)&{\color{red}V$(\frac{1}{2},\frac{1}{2},w)$}&V7(2)
\bigstrut\\
\hline
101&R$(0,\frac{1}{2},\frac{1}{2})$&R5(2)&{\color{red}W$(0,\frac{1}{2},w)$}&W5(2)
&101&X$(0,\frac{1}{2},0)$&X5(2)&{\color{red}W$(0,\frac{1}{2},w)$}&W5(2)
\bigstrut\\
\hline
102&GM$(0,0,0)$&GM6(2)&{\color{red}LD$(0,0,w)$}&LD6(2)
&102&Z$(0,0,\frac{1}{2})$&Z6(2)&{\color{red}LD$(0,0,w)$}&LD6(2)
\bigstrut\\
\hline
102&GM$(0,0,0)$&GM7(2)&{\color{red}LD$(0,0,w)$}&LD7(2)
&102&Z$(0,0,\frac{1}{2})$&Z7(2)&{\color{red}LD$(0,0,w)$}&LD7(2)
\bigstrut\\
\hline
102&A$(\frac{1}{2},\frac{1}{2},\frac{1}{2})$&A7(2)&{\color{red}V$(\frac{1}{2},\frac{1}{2},w)$}&V6(2)
&102&M$(\frac{1}{2},\frac{1}{2},0)$&M6(2)&{\color{red}V$(\frac{1}{2},\frac{1}{2},w)$}&V6(2)
\bigstrut\\
\hline
\hline
\end{tabular}
\end{table*}\end{center}

\newpage
\begin{center}\begin{table*}[!h]
(Continued from previous table: Time-reversal symmetry broken, considering SOC)\\
\begin{tabular}{c|c|c|c|c||c|c|c|c|c}\hline\hline
\multicolumn{1}{l|}{SG} & HSP & Irrep & HSL & Irrep & \multicolumn{1}{l|}{SG} & HSP & Irrep & HSL & Irrep \bigstrut\\
\hline
112&GM$(0,0,0)$&GM6(2)&{\color{red}LD$(0,0,w)$}&LD5(2)
&112&GM$(0,0,0)$&GM7(2)&{\color{red}LD$(0,0,w)$}&LD5(2)
\bigstrut\\
\hline
112&Z$(0,0,\frac{1}{2})$&Z6(2)&{\color{red}LD$(0,0,w)$}&LD5(2)
&112&Z$(0,0,\frac{1}{2})$&Z7(2)&{\color{red}LD$(0,0,w)$}&LD5(2)
\bigstrut\\
\hline
112&A$(\frac{1}{2},\frac{1}{2},\frac{1}{2})$&A6(2)&{\color{red}V$(\frac{1}{2},\frac{1}{2},w)$}&V5(2)
&112&A$(\frac{1}{2},\frac{1}{2},\frac{1}{2})$&A7(2)&{\color{red}V$(\frac{1}{2},\frac{1}{2},w)$}&V5(2)
\bigstrut\\
\hline
112&M$(\frac{1}{2},\frac{1}{2},0)$&M6(2)&{\color{red}V$(\frac{1}{2},\frac{1}{2},w)$}&V5(2)
&112&M$(\frac{1}{2},\frac{1}{2},0)$&M7(2)&{\color{red}V$(\frac{1}{2},\frac{1}{2},w)$}&V5(2)
\bigstrut\\
\hline
113&GM$(0,0,0)$&GM6(2)&{\color{red}LD$(0,0,w)$}&LD5(2)
&113&GM$(0,0,0)$&GM7(2)&{\color{red}LD$(0,0,w)$}&LD5(2)
\bigstrut\\
\hline
113&Z$(0,0,\frac{1}{2})$&Z6(2)&{\color{red}LD$(0,0,w)$}&LD5(2)
&113&Z$(0,0,\frac{1}{2})$&Z7(2)&{\color{red}LD$(0,0,w)$}&LD5(2)
\bigstrut\\
\hline
113&A$(\frac{1}{2},\frac{1}{2},\frac{1}{2})$&A6(2)&{\color{red}V$(\frac{1}{2},\frac{1}{2},w)$}&V5(2)
&113&A$(\frac{1}{2},\frac{1}{2},\frac{1}{2})$&A7(2)&{\color{red}V$(\frac{1}{2},\frac{1}{2},w)$}&V5(2)
\bigstrut\\
\hline
\hline
\end{tabular}
\end{table*}\end{center}

\newpage
\begin{center}\begin{table*}[!h]
(Continued from previous table: Time-reversal symmetry broken, considering SOC)\\
\begin{tabular}{c|c|c|c|c||c|c|c|c|c}\hline\hline
\multicolumn{1}{l|}{SG} & HSP & Irrep & HSL & Irrep & \multicolumn{1}{l|}{SG} & HSP & Irrep & HSL & Irrep \bigstrut\\
\hline
122&M$(1,1,1)$&M7(2)&{\color{red}LD$(0,0,w)$}&LD5(2)
&123&GM$(0,0,0)$&GM6(2)&{\color{red}DT$(0,v,0)$}&DT5(2)
\bigstrut\\
\hline
123&GM$(0,0,0)$&GM7(2)&{\color{red}DT$(0,v,0)$}&DT5(2)
&123&GM$(0,0,0)$&GM8(2)&{\color{red}DT$(0,v,0)$}&DT5(2)
\bigstrut\\
\hline
123&GM$(0,0,0)$&GM9(2)&{\color{red}DT$(0,v,0)$}&DT5(2)
&123&X$(0,\frac{1}{2},0)$&X5(2)&{\color{red}DT$(0,v,0)$}&DT5(2)
\bigstrut\\
\hline
123&X$(0,\frac{1}{2},0)$&X6(2)&{\color{red}DT$(0,v,0)$}&DT5(2)
&123&GM$(0,0,0)$&GM6(2)&{\color{red}LD$(0,0,w)$}&LD6(2)
\bigstrut\\
\hline
123&GM$(0,0,0)$&GM8(2)&{\color{red}LD$(0,0,w)$}&LD6(2)
&123&Z$(0,0,\frac{1}{2})$&Z6(2)&{\color{red}LD$(0,0,w)$}&LD6(2)
\bigstrut\\
\hline
123&Z$(0,0,\frac{1}{2})$&Z8(2)&{\color{red}LD$(0,0,w)$}&LD6(2)
&123&GM$(0,0,0)$&GM7(2)&{\color{red}LD$(0,0,w)$}&LD7(2)
\bigstrut\\
\hline
123&GM$(0,0,0)$&GM9(2)&{\color{red}LD$(0,0,w)$}&LD7(2)
&123&Z$(0,0,\frac{1}{2})$&Z7(2)&{\color{red}LD$(0,0,w)$}&LD7(2)
\bigstrut\\
\hline
\hline
\end{tabular}
\end{table*}\end{center}

\newpage
\begin{center}\begin{table*}[!h]
(Continued from previous table: Time-reversal symmetry broken, considering SOC)\\
\begin{tabular}{c|c|c|c|c||c|c|c|c|c}\hline\hline
\multicolumn{1}{l|}{SG} & HSP & Irrep & HSL & Irrep & \multicolumn{1}{l|}{SG} & HSP & Irrep & HSL & Irrep \bigstrut\\
\hline
124&M$(\frac{1}{2},\frac{1}{2},0)$&M7(2)&{\color{red}SM$(u,u,0)$}&SM5(2)
&124&M$(\frac{1}{2},\frac{1}{2},0)$&M8(2)&{\color{red}SM$(u,u,0)$}&SM5(2)
\bigstrut\\
\hline
124&M$(\frac{1}{2},\frac{1}{2},0)$&M9(2)&{\color{red}SM$(u,u,0)$}&SM5(2)
&124&A$(\frac{1}{2},\frac{1}{2},\frac{1}{2})$&A5(2)&{\color{red}V$(\frac{1}{2},\frac{1}{2},w)$}&V6(2)
\bigstrut\\
\hline
124&A$(\frac{1}{2},\frac{1}{2},\frac{1}{2})$&A7(2)&{\color{red}V$(\frac{1}{2},\frac{1}{2},w)$}&V6(2)
&124&M$(\frac{1}{2},\frac{1}{2},0)$&M6(2)&{\color{red}V$(\frac{1}{2},\frac{1}{2},w)$}&V6(2)
\bigstrut\\
\hline
124&M$(\frac{1}{2},\frac{1}{2},0)$&M8(2)&{\color{red}V$(\frac{1}{2},\frac{1}{2},w)$}&V6(2)
&124&A$(\frac{1}{2},\frac{1}{2},\frac{1}{2})$&A6(2)&{\color{red}V$(\frac{1}{2},\frac{1}{2},w)$}&V7(2)
\bigstrut\\
\hline
124&A$(\frac{1}{2},\frac{1}{2},\frac{1}{2})$&A8(2)&{\color{red}V$(\frac{1}{2},\frac{1}{2},w)$}&V7(2)
&124&M$(\frac{1}{2},\frac{1}{2},0)$&M7(2)&{\color{red}V$(\frac{1}{2},\frac{1}{2},w)$}&V7(2)
\bigstrut\\
\hline
124&M$(\frac{1}{2},\frac{1}{2},0)$&M9(2)&{\color{red}V$(\frac{1}{2},\frac{1}{2},w)$}&V7(2)
&124&R$(0,\frac{1}{2},\frac{1}{2})$&R3(2)&{\color{red}W$(0,\frac{1}{2},w)$}&W5(2)
\bigstrut\\
\hline
124&R$(0,\frac{1}{2},\frac{1}{2})$&R4(2)&{\color{red}W$(0,\frac{1}{2},w)$}&W5(2)
&124&X$(0,\frac{1}{2},0)$&X5(2)&{\color{red}W$(0,\frac{1}{2},w)$}&W5(2)
\bigstrut\\
\hline
\hline
\end{tabular}
\end{table*}\end{center}

\newpage
\begin{center}\begin{table*}[!h]
(Continued from previous table: Time-reversal symmetry broken, considering SOC)\\
\begin{tabular}{c|c|c|c|c||c|c|c|c|c}\hline\hline
\multicolumn{1}{l|}{SG} & HSP & Irrep & HSL & Irrep & \multicolumn{1}{l|}{SG} & HSP & Irrep & HSL & Irrep \bigstrut\\
\hline
127&GM$(0,0,0)$&GM9(2)&{\color{red}DT$(0,v,0)$}&DT5(2)
&127&X$(0,\frac{1}{2},0)$&X3(2)&{\color{red}DT$(0,v,0)$}&DT5(2)
\bigstrut\\
\hline
127&X$(0,\frac{1}{2},0)$&X4(2)&{\color{red}DT$(0,v,0)$}&DT5(2)
&127&GM$(0,0,0)$&GM6(2)&{\color{red}LD$(0,0,w)$}&LD6(2)
\bigstrut\\
\hline
127&GM$(0,0,0)$&GM8(2)&{\color{red}LD$(0,0,w)$}&LD6(2)
&127&Z$(0,0,\frac{1}{2})$&Z6(2)&{\color{red}LD$(0,0,w)$}&LD6(2)
\bigstrut\\
\hline
127&Z$(0,0,\frac{1}{2})$&Z8(2)&{\color{red}LD$(0,0,w)$}&LD6(2)
&127&GM$(0,0,0)$&GM7(2)&{\color{red}LD$(0,0,w)$}&LD7(2)
\bigstrut\\
\hline
127&GM$(0,0,0)$&GM9(2)&{\color{red}LD$(0,0,w)$}&LD7(2)
&127&Z$(0,0,\frac{1}{2})$&Z7(2)&{\color{red}LD$(0,0,w)$}&LD7(2)
\bigstrut\\
\hline
127&Z$(0,0,\frac{1}{2})$&Z9(2)&{\color{red}LD$(0,0,w)$}&LD7(2)
&127&A$(\frac{1}{2},\frac{1}{2},\frac{1}{2})$&A6(2)&{\color{red}S$(u,u,\frac{1}{2})$}&S5(2)
\bigstrut\\
\hline
127&A$(\frac{1}{2},\frac{1}{2},\frac{1}{2})$&A7(2)&{\color{red}S$(u,u,\frac{1}{2})$}&S5(2)
&127&A$(\frac{1}{2},\frac{1}{2},\frac{1}{2})$&A8(2)&{\color{red}S$(u,u,\frac{1}{2})$}&S5(2)
\bigstrut\\
\hline
\hline
\end{tabular}
\end{table*}\end{center}

\newpage
\begin{center}\begin{table*}[!h]
(Continued from previous table: Time-reversal symmetry broken, considering SOC)\\
\begin{tabular}{c|c|c|c|c||c|c|c|c|c}\hline\hline
\multicolumn{1}{l|}{SG} & HSP & Irrep & HSL & Irrep & \multicolumn{1}{l|}{SG} & HSP & Irrep & HSL & Irrep \bigstrut\\
\hline
128&A$(\frac{1}{2},\frac{1}{2},\frac{1}{2})$&A7(2)&{\color{red}V$(\frac{1}{2},\frac{1}{2},w)$}&V7(2)
&128&M$(\frac{1}{2},\frac{1}{2},0)$&M7(2)&{\color{red}V$(\frac{1}{2},\frac{1}{2},w)$}&V7(2)
\bigstrut\\
\hline
128&M$(\frac{1}{2},\frac{1}{2},0)$&M9(2)&{\color{red}V$(\frac{1}{2},\frac{1}{2},w)$}&V7(2)
&128&M$(\frac{1}{2},\frac{1}{2},0)$&M6(2)&{\color{red}Y$(u,\frac{1}{2},0)$}&Y5(2)
\bigstrut\\
\hline
128&M$(\frac{1}{2},\frac{1}{2},0)$&M7(2)&{\color{red}Y$(u,\frac{1}{2},0)$}&Y5(2)
&128&M$(\frac{1}{2},\frac{1}{2},0)$&M8(2)&{\color{red}Y$(u,\frac{1}{2},0)$}&Y5(2)
\bigstrut\\
\hline
128&M$(\frac{1}{2},\frac{1}{2},0)$&M9(2)&{\color{red}Y$(u,\frac{1}{2},0)$}&Y5(2)
&128&X$(0,\frac{1}{2},0)$&X3(2)&{\color{red}Y$(u,\frac{1}{2},0)$}&Y5(2)
\bigstrut\\
\hline
128&X$(0,\frac{1}{2},0)$&X4(2)&{\color{red}Y$(u,\frac{1}{2},0)$}&Y5(2)
&129&GM$(0,0,0)$&GM6(2)&{\color{red}DT$(0,v,0)$}&DT5(2)
\bigstrut\\
\hline
129&GM$(0,0,0)$&GM7(2)&{\color{red}DT$(0,v,0)$}&DT5(2)
&129&GM$(0,0,0)$&GM8(2)&{\color{red}DT$(0,v,0)$}&DT5(2)
\bigstrut\\
\hline
129&GM$(0,0,0)$&GM9(2)&{\color{red}DT$(0,v,0)$}&DT5(2)
&129&X$(0,\frac{1}{2},0)$&X3(2)&{\color{red}DT$(0,v,0)$}&DT5(2)
\bigstrut\\
\hline
\hline
\end{tabular}
\end{table*}\end{center}

\newpage
\begin{center}\begin{table*}[!h]
(Continued from previous table: Time-reversal symmetry broken, considering SOC)\\
\begin{tabular}{c|c|c|c|c||c|c|c|c|c}\hline\hline
\multicolumn{1}{l|}{SG} & HSP & Irrep & HSL & Irrep & \multicolumn{1}{l|}{SG} & HSP & Irrep & HSL & Irrep \bigstrut\\
\hline
131&GM$(0,0,0)$&GM8(2)&{\color{red}LD$(0,0,w)$}&LD6(2)
&131&Z$(0,0,\frac{1}{2})$&Z5(4)&{\color{red}LD$(0,0,w)$}&LD6(2)
\bigstrut\\
\hline
131&GM$(0,0,0)$&GM7(2)&{\color{red}LD$(0,0,w)$}&LD7(2)
&131&GM$(0,0,0)$&GM9(2)&{\color{red}LD$(0,0,w)$}&LD7(2)
\bigstrut\\
\hline
131&GM$(0,0,0)$&GM6(2)&{\color{red}SM$(u,u,0)$}&SM5(2)
&131&GM$(0,0,0)$&GM7(2)&{\color{red}SM$(u,u,0)$}&SM5(2)
\bigstrut\\
\hline
131&GM$(0,0,0)$&GM8(2)&{\color{red}SM$(u,u,0)$}&SM5(2)
&131&GM$(0,0,0)$&GM9(2)&{\color{red}SM$(u,u,0)$}&SM5(2)
\bigstrut\\
\hline
131&M$(\frac{1}{2},\frac{1}{2},0)$&M6(2)&{\color{red}SM$(u,u,0)$}&SM5(2)
&131&M$(\frac{1}{2},\frac{1}{2},0)$&M7(2)&{\color{red}SM$(u,u,0)$}&SM5(2)
\bigstrut\\
\hline
131&M$(\frac{1}{2},\frac{1}{2},0)$&M8(2)&{\color{red}SM$(u,u,0)$}&SM5(2)
&131&M$(\frac{1}{2},\frac{1}{2},0)$&M9(2)&{\color{red}SM$(u,u,0)$}&SM5(2)
\bigstrut\\
\hline
131&A$(\frac{1}{2},\frac{1}{2},\frac{1}{2})$&A5(4)&{\color{red}T$(u,\frac{1}{2},\frac{1}{2})$}&T5(2)
&131&R$(0,\frac{1}{2},\frac{1}{2})$&R5(2)&{\color{red}T$(u,\frac{1}{2},\frac{1}{2})$}&T5(2)
\bigstrut\\
\hline
\hline
\end{tabular}
\end{table*}\end{center}

\newpage
\begin{center}\begin{table*}[!h]
(Continued from previous table: Time-reversal symmetry broken, considering SOC)\\
\begin{tabular}{c|c|c|c|c||c|c|c|c|c}\hline\hline
\multicolumn{1}{l|}{SG} & HSP & Irrep & HSL & Irrep & \multicolumn{1}{l|}{SG} & HSP & Irrep & HSL & Irrep \bigstrut\\
\hline
133&GM$(0,0,0)$&GM6(2)&{\color{red}SM$(u,u,0)$}&SM5(2)
&133&GM$(0,0,0)$&GM7(2)&{\color{red}SM$(u,u,0)$}&SM5(2)
\bigstrut\\
\hline
133&GM$(0,0,0)$&GM8(2)&{\color{red}SM$(u,u,0)$}&SM5(2)
&133&GM$(0,0,0)$&GM9(2)&{\color{red}SM$(u,u,0)$}&SM5(2)
\bigstrut\\
\hline
133&M$(\frac{1}{2},\frac{1}{2},0)$&M5(4)&{\color{red}SM$(u,u,0)$}&SM5(2)
&133&R$(0,\frac{1}{2},\frac{1}{2})$&R3(2)&{\color{red}U$(0,v,\frac{1}{2})$}&U5(2)
\bigstrut\\
\hline
133&R$(0,\frac{1}{2},\frac{1}{2})$&R4(2)&{\color{red}U$(0,v,\frac{1}{2})$}&U5(2)
&133&Z$(0,0,\frac{1}{2})$&Z5(4)&{\color{red}U$(0,v,\frac{1}{2})$}&U5(2)
\bigstrut\\
\hline
133&A$(\frac{1}{2},\frac{1}{2},\frac{1}{2})$&A5(2)&{\color{red}V$(\frac{1}{2},\frac{1}{2},w)$}&V6(2)
&133&A$(\frac{1}{2},\frac{1}{2},\frac{1}{2})$&A7(2)&{\color{red}V$(\frac{1}{2},\frac{1}{2},w)$}&V6(2)
\bigstrut\\
\hline
133&M$(\frac{1}{2},\frac{1}{2},0)$&M5(4)&{\color{red}V$(\frac{1}{2},\frac{1}{2},w)$}&V6(2)
&133&A$(\frac{1}{2},\frac{1}{2},\frac{1}{2})$&A6(2)&{\color{red}V$(\frac{1}{2},\frac{1}{2},w)$}&V7(2)
\bigstrut\\
\hline
133&A$(\frac{1}{2},\frac{1}{2},\frac{1}{2})$&A8(2)&{\color{red}V$(\frac{1}{2},\frac{1}{2},w)$}&V7(2)
&134&GM$(0,0,0)$&GM6(2)&{\color{red}DT$(0,v,0)$}&DT5(2)
\bigstrut\\
\hline
\hline
\end{tabular}
\end{table*}\end{center}

\newpage
\begin{center}\begin{table*}[!h]
(Continued from previous table: Time-reversal symmetry broken, considering SOC)\\
\begin{tabular}{c|c|c|c|c||c|c|c|c|c}\hline\hline
\multicolumn{1}{l|}{SG} & HSP & Irrep & HSL & Irrep & \multicolumn{1}{l|}{SG} & HSP & Irrep & HSL & Irrep \bigstrut\\
\hline
135&X$(0,\frac{1}{2},0)$&X4(2)&{\color{red}Y$(u,\frac{1}{2},0)$}&Y5(2)
&136&GM$(0,0,0)$&GM6(2)&{\color{red}DT$(0,v,0)$}&DT5(2)
\bigstrut\\
\hline
136&GM$(0,0,0)$&GM7(2)&{\color{red}DT$(0,v,0)$}&DT5(2)
&136&GM$(0,0,0)$&GM8(2)&{\color{red}DT$(0,v,0)$}&DT5(2)
\bigstrut\\
\hline
136&GM$(0,0,0)$&GM9(2)&{\color{red}DT$(0,v,0)$}&DT5(2)
&136&X$(0,\frac{1}{2},0)$&X3(2)&{\color{red}DT$(0,v,0)$}&DT5(2)
\bigstrut\\
\hline
136&X$(0,\frac{1}{2},0)$&X4(2)&{\color{red}DT$(0,v,0)$}&DT5(2)
&136&GM$(0,0,0)$&GM6(2)&{\color{red}LD$(0,0,w)$}&LD6(2)
\bigstrut\\
\hline
136&GM$(0,0,0)$&GM8(2)&{\color{red}LD$(0,0,w)$}&LD6(2)
&136&Z$(0,0,\frac{1}{2})$&Z5(4)&{\color{red}LD$(0,0,w)$}&LD6(2)
\bigstrut\\
\hline
136&GM$(0,0,0)$&GM7(2)&{\color{red}LD$(0,0,w)$}&LD7(2)
&136&GM$(0,0,0)$&GM9(2)&{\color{red}LD$(0,0,w)$}&LD7(2)
\bigstrut\\
\hline
136&A$(\frac{1}{2},\frac{1}{2},\frac{1}{2})$&A5(4)&{\color{red}S$(u,u,\frac{1}{2})$}&S5(2)
&136&Z$(0,0,\frac{1}{2})$&Z5(4)&{\color{red}S$(u,u,\frac{1}{2})$}&S5(2)
\bigstrut\\
\hline
\hline
\end{tabular}
\end{table*}\end{center}

\newpage
\begin{center}\begin{table*}[!h]
(Continued from previous table: Time-reversal symmetry broken, considering SOC)\\
\begin{tabular}{c|c|c|c|c||c|c|c|c|c}\hline\hline
\multicolumn{1}{l|}{SG} & HSP & Irrep & HSL & Irrep & \multicolumn{1}{l|}{SG} & HSP & Irrep & HSL & Irrep \bigstrut\\
\hline
138&GM$(0,0,0)$&GM9(2)&{\color{red}SM$(u,u,0)$}&SM5(2)
&138&M$(\frac{1}{2},\frac{1}{2},0)$&M5(4)&{\color{red}SM$(u,u,0)$}&SM5(2)
\bigstrut\\
\hline
138&A$(\frac{1}{2},\frac{1}{2},\frac{1}{2})$&A6(2)&{\color{red}T$(u,\frac{1}{2},\frac{1}{2})$}&T5(2)
&138&A$(\frac{1}{2},\frac{1}{2},\frac{1}{2})$&A7(2)&{\color{red}T$(u,\frac{1}{2},\frac{1}{2})$}&T5(2)
\bigstrut\\
\hline
138&A$(\frac{1}{2},\frac{1}{2},\frac{1}{2})$&A8(2)&{\color{red}T$(u,\frac{1}{2},\frac{1}{2})$}&T5(2)
&138&A$(\frac{1}{2},\frac{1}{2},\frac{1}{2})$&A9(2)&{\color{red}T$(u,\frac{1}{2},\frac{1}{2})$}&T5(2)
\bigstrut\\
\hline
138&R$(0,\frac{1}{2},\frac{1}{2})$&R3(2)&{\color{red}T$(u,\frac{1}{2},\frac{1}{2})$}&T5(2)
&138&R$(0,\frac{1}{2},\frac{1}{2})$&R4(2)&{\color{red}T$(u,\frac{1}{2},\frac{1}{2})$}&T5(2)
\bigstrut\\
\hline
138&A$(\frac{1}{2},\frac{1}{2},\frac{1}{2})$&A6(2)&{\color{red}V$(\frac{1}{2},\frac{1}{2},w)$}&V6(2)
&138&A$(\frac{1}{2},\frac{1}{2},\frac{1}{2})$&A8(2)&{\color{red}V$(\frac{1}{2},\frac{1}{2},w)$}&V6(2)
\bigstrut\\
\hline
138&M$(\frac{1}{2},\frac{1}{2},0)$&M5(4)&{\color{red}V$(\frac{1}{2},\frac{1}{2},w)$}&V6(2)
&138&A$(\frac{1}{2},\frac{1}{2},\frac{1}{2})$&A7(2)&{\color{red}V$(\frac{1}{2},\frac{1}{2},w)$}&V7(2)
\bigstrut\\
\hline
138&A$(\frac{1}{2},\frac{1}{2},\frac{1}{2})$&A9(2)&{\color{red}V$(\frac{1}{2},\frac{1}{2},w)$}&V7(2)
&138&R$(0,\frac{1}{2},\frac{1}{2})$&R3(2)&{\color{red}W$(0,\frac{1}{2},w)$}&W5(2)
\bigstrut\\
\hline
\hline
\end{tabular}
\end{table*}\end{center}

\newpage
\begin{center}\begin{table*}[!h]
(Continued from previous table: Time-reversal symmetry broken, considering SOC)\\
\begin{tabular}{c|c|c|c|c||c|c|c|c|c}\hline\hline
\multicolumn{1}{l|}{SG} & HSP & Irrep & HSL & Irrep & \multicolumn{1}{l|}{SG} & HSP & Irrep & HSL & Irrep \bigstrut\\
\hline
140&X$(\frac{1}{2},\frac{1}{2},0)$&X6(2)&{\color{red}Y$(u,1-u,0)$}&Y5(2)
&141&GM$(0,0,0)$&GM6(2)&{\color{red}DT$(u,u,0)$}&DT5(2)
\bigstrut\\
\hline
141&GM$(0,0,0)$&GM7(2)&{\color{red}DT$(u,u,0)$}&DT5(2)
&141&GM$(0,0,0)$&GM8(2)&{\color{red}DT$(u,u,0)$}&DT5(2)
\bigstrut\\
\hline
141&GM$(0,0,0)$&GM9(2)&{\color{red}DT$(u,u,0)$}&DT5(2)
&141&X$(\frac{1}{2},\frac{1}{2},0)$&X3(2)&{\color{red}DT$(u,u,0)$}&DT5(2)
\bigstrut\\
\hline
141&X$(\frac{1}{2},\frac{1}{2},0)$&X4(2)&{\color{red}DT$(u,u,0)$}&DT5(2)
&141&GM$(0,0,0)$&GM6(2)&{\color{red}LD$(0,0,w)$}&LD6(2)
\bigstrut\\
\hline
141&GM$(0,0,0)$&GM8(2)&{\color{red}LD$(0,0,w)$}&LD6(2)
&141&M$(1,1,1)$&M5(4)&{\color{red}LD$(0,0,w)$}&LD6(2)
\bigstrut\\
\hline
141&GM$(0,0,0)$&GM7(2)&{\color{red}LD$(0,0,w)$}&LD7(2)
&141&GM$(0,0,0)$&GM9(2)&{\color{red}LD$(0,0,w)$}&LD7(2)
\bigstrut\\
\hline
141&GM$(0,0,0)$&GM6(2)&{\color{red}SM$(u,0,0)$}&SM5(2)
&141&GM$(0,0,0)$&GM7(2)&{\color{red}SM$(u,0,0)$}&SM5(2)
\bigstrut\\
\hline
\hline
\end{tabular}
\end{table*}\end{center}

\newpage
\begin{center}\begin{table*}[!h]
(Continued from previous table: Time-reversal symmetry broken, considering SOC)\\
\begin{tabular}{c|c|c|c|c||c|c|c|c|c}\hline\hline
\multicolumn{1}{l|}{SG} & HSP & Irrep & HSL & Irrep & \multicolumn{1}{l|}{SG} & HSP & Irrep & HSL & Irrep \bigstrut\\
\hline
167&T$(0,0,3/2)$&T6(2)&LD$(0,0,w)$&LD6(2)
&183&A$(0,0,\frac{1}{2})$&A7(2)&{\color{red}DT$(0,0,w)$}&DT7(2)
\bigstrut\\
\hline
183&GM$(0,0,0)$&GM7(2)&{\color{red}DT$(0,0,w)$}&DT7(2)
&183&A$(0,0,\frac{1}{2})$&A8(2)&{\color{red}DT$(0,0,w)$}&DT8(2)
\bigstrut\\
\hline
183&GM$(0,0,0)$&GM8(2)&{\color{red}DT$(0,0,w)$}&DT8(2)
&183&A$(0,0,\frac{1}{2})$&A9(2)&{\color{red}DT$(0,0,w)$}&DT9(2)
\bigstrut\\
\hline
183&GM$(0,0,0)$&GM9(2)&{\color{red}DT$(0,0,w)$}&DT9(2)
&183&H$(\frac{1}{3},\frac{1}{3},\frac{1}{2})$&H6(2)&P$(\frac{1}{3},\frac{1}{3},w)$&P6(2)
\bigstrut\\
\hline
183&K$(\frac{1}{3},\frac{1}{3},0)$&K6(2)&P$(\frac{1}{3},\frac{1}{3},w)$&P6(2)
&183&L$(\frac{1}{2},0,\frac{1}{2})$&L5(2)&{\color{red}U$(\frac{1}{2},0,w)$}&U5(2)
\bigstrut\\
\hline
183&M$(\frac{1}{2},0,0)$&M5(2)&{\color{red}U$(\frac{1}{2},0,w)$}&U5(2)
&184&A$(0,0,\frac{1}{2})$&A7(2)&{\color{red}DT$(0,0,w)$}&DT7(2)
\bigstrut\\
\hline
184&GM$(0,0,0)$&GM7(2)&{\color{red}DT$(0,0,w)$}&DT7(2)
&184&A$(0,0,\frac{1}{2})$&A8(2)&{\color{red}DT$(0,0,w)$}&DT8(2)
\bigstrut\\
\hline
\hline
\end{tabular}
\end{table*}\end{center}

\newpage
\begin{center}\begin{table*}[!h]
(Continued from previous table: Time-reversal symmetry broken, considering SOC)\\
\begin{tabular}{c|c|c|c|c||c|c|c|c|c}\hline\hline
\multicolumn{1}{l|}{SG} & HSP & Irrep & HSL & Irrep & \multicolumn{1}{l|}{SG} & HSP & Irrep & HSL & Irrep \bigstrut\\
\hline
189&HA$(-\frac{1}{3},-\frac{1}{3},-\frac{1}{2})$&HA9(2)&PA$(-\frac{1}{3},-\frac{1}{3},-w)$&PA6(2)
&189&KA$(-\frac{1}{3},-\frac{1}{3},0)$&KA8(2)&PA$(-\frac{1}{3},-\frac{1}{3},-w)$&PA6(2)
\bigstrut\\
\hline
189&KA$(-\frac{1}{3},-\frac{1}{3},0)$&KA9(2)&PA$(-\frac{1}{3},-\frac{1}{3},-w)$&PA6(2)
&189&A$(0,0,\frac{1}{2})$&A7(2)&{\color{red}Q$(u,u,\frac{1}{2})$}&Q5(2)
\bigstrut\\
\hline
189&A$(0,0,\frac{1}{2})$&A8(2)&{\color{red}Q$(u,u,\frac{1}{2})$}&Q5(2)
&189&A$(0,0,\frac{1}{2})$&A9(2)&{\color{red}Q$(u,u,\frac{1}{2})$}&Q5(2)
\bigstrut\\
\hline
189&H$(\frac{1}{3},\frac{1}{3},\frac{1}{2})$&H7(2)&{\color{red}Q$(u,u,\frac{1}{2})$}&Q5(2)
&189&H$(\frac{1}{3},\frac{1}{3},\frac{1}{2})$&H8(2)&{\color{red}Q$(u,u,\frac{1}{2})$}&Q5(2)
\bigstrut\\
\hline
189&H$(\frac{1}{3},\frac{1}{3},\frac{1}{2})$&H9(2)&{\color{red}Q$(u,u,\frac{1}{2})$}&Q5(2)
&189&HA$(-\frac{1}{3},-\frac{1}{3},-\frac{1}{2})$&HA7(2)&{\color{red}Q$(u,u,\frac{1}{2})$}&Q5(2)
\bigstrut\\
\hline
189&HA$(-\frac{1}{3},-\frac{1}{3},-\frac{1}{2})$&HA8(2)&{\color{red}Q$(u,u,\frac{1}{2})$}&Q5(2)
&189&HA$(-\frac{1}{3},-\frac{1}{3},-\frac{1}{2})$&HA9(2)&{\color{red}Q$(u,u,\frac{1}{2})$}&Q5(2)
\bigstrut\\
\hline
189&L$(\frac{1}{2},0,\frac{1}{2})$&L5(2)&{\color{red}Q$(u,u,\frac{1}{2})$}&Q5(2)
&190&A$(0,0,\frac{1}{2})$&A8(2)&DT$(0,0,w)$&DT6(2)
\bigstrut\\
\hline
\hline
\end{tabular}
\end{table*}\end{center}

\newpage
\begin{center}\begin{table*}[!h]
(Continued from previous table: Time-reversal symmetry broken, considering SOC)\\
\begin{tabular}{c|c|c|c|c||c|c|c|c|c}\hline\hline
\multicolumn{1}{l|}{SG} & HSP & Irrep & HSL & Irrep & \multicolumn{1}{l|}{SG} & HSP & Irrep & HSL & Irrep \bigstrut\\
\hline
191&L$(\frac{1}{2},0,\frac{1}{2})$&L6(2)&{\color{red}R$(u,0,\frac{1}{2})$}&R5(2)
&191&GM$(0,0,0)$&GM7(2)&{\color{red}SM$(u,0,0)$}&SM5(2)
\bigstrut\\
\hline
191&GM$(0,0,0)$&GM8(2)&{\color{red}SM$(u,0,0)$}&SM5(2)
&191&GM$(0,0,0)$&GM9(2)&{\color{red}SM$(u,0,0)$}&SM5(2)
\bigstrut\\
\hline
191&GM$(0,0,0)$&GM10(2)&{\color{red}SM$(u,0,0)$}&SM5(2)
&191&GM$(0,0,0)$&GM11(2)&{\color{red}SM$(u,0,0)$}&SM5(2)
\bigstrut\\
\hline
191&GM$(0,0,0)$&GM12(2)&{\color{red}SM$(u,0,0)$}&SM5(2)
&191&M$(\frac{1}{2},0,0)$&M5(2)&{\color{red}SM$(u,0,0)$}&SM5(2)
\bigstrut\\
\hline
191&M$(\frac{1}{2},0,0)$&M6(2)&{\color{red}SM$(u,0,0)$}&SM5(2)
&191&L$(\frac{1}{2},0,\frac{1}{2})$&L5(2)&{\color{red}U$(\frac{1}{2},0,w)$}&U5(2)
\bigstrut\\
\hline
191&L$(\frac{1}{2},0,\frac{1}{2})$&L6(2)&{\color{red}U$(\frac{1}{2},0,w)$}&U5(2)
&191&M$(\frac{1}{2},0,0)$&M5(2)&{\color{red}U$(\frac{1}{2},0,w)$}&U5(2)
\bigstrut\\
\hline
191&M$(\frac{1}{2},0,0)$&M6(2)&{\color{red}U$(\frac{1}{2},0,w)$}&U5(2)
&192&A$(0,0,\frac{1}{2})$&A7(2)&{\color{red}DT$(0,0,w)$}&DT7(2)
\bigstrut\\
\hline
\hline
\end{tabular}
\end{table*}\end{center}

\newpage
\begin{center}\begin{table*}[!h]
(Continued from previous table: Time-reversal symmetry broken, considering SOC)\\
\begin{tabular}{c|c|c|c|c||c|c|c|c|c}\hline\hline
\multicolumn{1}{l|}{SG} & HSP & Irrep & HSL & Irrep & \multicolumn{1}{l|}{SG} & HSP & Irrep & HSL & Irrep \bigstrut\\
\hline
193&H$(\frac{1}{3},\frac{1}{3},\frac{1}{2})$&H8(2)&{\color{red}Q$(u,u,\frac{1}{2})$}&Q5(2)
&193&H$(\frac{1}{3},\frac{1}{3},\frac{1}{2})$&H9(2)&{\color{red}Q$(u,u,\frac{1}{2})$}&Q5(2)
\bigstrut\\
\hline
193&L$(\frac{1}{2},0,\frac{1}{2})$&L3(2)&{\color{red}Q$(u,u,\frac{1}{2})$}&Q5(2)
&193&L$(\frac{1}{2},0,\frac{1}{2})$&L4(2)&{\color{red}Q$(u,u,\frac{1}{2})$}&Q5(2)
\bigstrut\\
\hline
193&GM$(0,0,0)$&GM7(2)&{\color{red}SM$(u,0,0)$}&SM5(2)
&193&GM$(0,0,0)$&GM8(2)&{\color{red}SM$(u,0,0)$}&SM5(2)
\bigstrut\\
\hline
193&GM$(0,0,0)$&GM9(2)&{\color{red}SM$(u,0,0)$}&SM5(2)
&193&GM$(0,0,0)$&GM10(2)&{\color{red}SM$(u,0,0)$}&SM5(2)
\bigstrut\\
\hline
193&GM$(0,0,0)$&GM11(2)&{\color{red}SM$(u,0,0)$}&SM5(2)
&193&GM$(0,0,0)$&GM12(2)&{\color{red}SM$(u,0,0)$}&SM5(2)
\bigstrut\\
\hline
193&M$(\frac{1}{2},0,0)$&M5(2)&{\color{red}SM$(u,0,0)$}&SM5(2)
&193&M$(\frac{1}{2},0,0)$&M6(2)&{\color{red}SM$(u,0,0)$}&SM5(2)
\bigstrut\\
\hline
193&L$(\frac{1}{2},0,\frac{1}{2})$&L3(2)&{\color{red}U$(\frac{1}{2},0,w)$}&U5(2)
&193&L$(\frac{1}{2},0,\frac{1}{2})$&L4(2)&{\color{red}U$(\frac{1}{2},0,w)$}&U5(2)
\bigstrut\\
\hline
\hline
\end{tabular}
\end{table*}\end{center}

\newpage
\begin{center}\begin{table*}[!h]
(Continued from previous table: Time-reversal symmetry broken, considering SOC)\\
\begin{tabular}{c|c|c|c|c||c|c|c|c|c}\hline\hline
\multicolumn{1}{l|}{SG} & HSP & Irrep & HSL & Irrep & \multicolumn{1}{l|}{SG} & HSP & Irrep & HSL & Irrep \bigstrut\\
\hline
200&X$(0,\frac{1}{2},0)$&X6(2)&{\color{red}ZA$(\frac{1}{2},u,0)$}&ZA5(2)
&201&GM$(0,0,0)$&GM5(2)&{\color{red}DT$(0,v,0)$}&DT5(2)
\bigstrut\\
\hline
201&GM$(0,0,0)$&GM6(2)&{\color{red}DT$(0,v,0)$}&DT5(2)
&201&GM$(0,0,0)$&GM7(2)&{\color{red}DT$(0,v,0)$}&DT5(2)
\bigstrut\\
\hline
201&GM$(0,0,0)$&GM8(2)&{\color{red}DT$(0,v,0)$}&DT5(2)
&201&GM$(0,0,0)$&GM9(2)&{\color{red}DT$(0,v,0)$}&DT5(2)
\bigstrut\\
\hline
201&GM$(0,0,0)$&GM10(2)&{\color{red}DT$(0,v,0)$}&DT5(2)
&201&X$(0,\frac{1}{2},0)$&X3(2)&{\color{red}DT$(0,v,0)$}&DT5(2)
\bigstrut\\
\hline
201&X$(0,\frac{1}{2},0)$&X4(2)&{\color{red}DT$(0,v,0)$}&DT5(2)
&201&M$(\frac{1}{2},\frac{1}{2},0)$&M3(2)&{\color{red}T$(\frac{1}{2},\frac{1}{2},w)$}&T5(2)
\bigstrut\\
\hline
201&M$(\frac{1}{2},\frac{1}{2},0)$&M4(2)&{\color{red}T$(\frac{1}{2},\frac{1}{2},w)$}&T5(2)
&201&R$(\frac{1}{2},\frac{1}{2},\frac{1}{2})$&R5(2)&{\color{red}T$(\frac{1}{2},\frac{1}{2},w)$}&T5(2)
\bigstrut\\
\hline
201&R$(\frac{1}{2},\frac{1}{2},\frac{1}{2})$&R6(2)&{\color{red}T$(\frac{1}{2},\frac{1}{2},w)$}&T5(2)
&201&R$(\frac{1}{2},\frac{1}{2},\frac{1}{2})$&R7(2)&{\color{red}T$(\frac{1}{2},\frac{1}{2},w)$}&T5(2)
\bigstrut\\
\hline
\hline
\end{tabular}
\end{table*}\end{center}

\newpage
\begin{center}\begin{table*}[!h]
(Continued from previous table: Time-reversal symmetry broken, considering SOC)\\
\begin{tabular}{c|c|c|c|c||c|c|c|c|c}\hline\hline
\multicolumn{1}{l|}{SG} & HSP & Irrep & HSL & Irrep & \multicolumn{1}{l|}{SG} & HSP & Irrep & HSL & Irrep \bigstrut\\
\hline
215&R$(\frac{1}{2},\frac{1}{2},\frac{1}{2})$&R7(2)&LD$(u,u,u)$&LD6(2)
&215&M$(\frac{1}{2},\frac{1}{2},0)$&M6(2)&{\color{red}T$(\frac{1}{2},\frac{1}{2},w)$}&T5(2)
\bigstrut\\
\hline
215&M$(\frac{1}{2},\frac{1}{2},0)$&M7(2)&{\color{red}T$(\frac{1}{2},\frac{1}{2},w)$}&T5(2)
&215&R$(\frac{1}{2},\frac{1}{2},\frac{1}{2})$&R6(2)&{\color{red}T$(\frac{1}{2},\frac{1}{2},w)$}&T5(2)
\bigstrut\\
\hline
215&R$(\frac{1}{2},\frac{1}{2},\frac{1}{2})$&R7(2)&{\color{red}T$(\frac{1}{2},\frac{1}{2},w)$}&T5(2)
&215&R$(\frac{1}{2},\frac{1}{2},\frac{1}{2})$&R8(4)&{\color{red}T$(\frac{1}{2},\frac{1}{2},w)$}&T5(2)
\bigstrut\\
\hline
216&GM$(0,0,0)$&GM6(2)&{\color{red}DT$(0,v,0)$}&DT5(2)
&216&GM$(0,0,0)$&GM7(2)&{\color{red}DT$(0,v,0)$}&DT5(2)
\bigstrut\\
\hline
216&GM$(0,0,0)$&GM8(4)&{\color{red}DT$(0,v,0)$}&DT5(2)
&216&X$(0,1,0)$&X6(2)&{\color{red}DT$(0,v,0)$}&DT5(2)
\bigstrut\\
\hline
216&X$(0,1,0)$&X7(2)&{\color{red}DT$(0,v,0)$}&DT5(2)
&216&GM$(0,0,0)$&GM6(2)&LD$(u,u,u)$&LD6(2)
\bigstrut\\
\hline
216&GM$(0,0,0)$&GM7(2)&LD$(u,u,u)$&LD6(2)
&216&L$(\frac{1}{2},\frac{1}{2},\frac{1}{2})$&L6(2)&LD$(u,u,u)$&LD6(2)
\bigstrut\\
\hline
\hline
\end{tabular}
\end{table*}\end{center}

\newpage
\begin{center}\begin{table*}[!h]
(Continued from previous table: Time-reversal symmetry broken, considering SOC)\\
\begin{tabular}{c|c|c|c|c||c|c|c|c|c}\hline\hline
\multicolumn{1}{l|}{SG} & HSP & Irrep & HSL & Irrep & \multicolumn{1}{l|}{SG} & HSP & Irrep & HSL & Irrep \bigstrut\\
\hline
221&GM$(0,0,0)$&GM7(2)&LD$(u,u,u)$&LD6(2)
&221&GM$(0,0,0)$&GM8(2)&LD$(u,u,u)$&LD6(2)
\bigstrut\\
\hline
221&GM$(0,0,0)$&GM9(2)&LD$(u,u,u)$&LD6(2)
&221&R$(\frac{1}{2},\frac{1}{2},\frac{1}{2})$&R6(2)&LD$(u,u,u)$&LD6(2)
\bigstrut\\
\hline
221&R$(\frac{1}{2},\frac{1}{2},\frac{1}{2})$&R7(2)&LD$(u,u,u)$&LD6(2)
&221&R$(\frac{1}{2},\frac{1}{2},\frac{1}{2})$&R8(2)&LD$(u,u,u)$&LD6(2)
\bigstrut\\
\hline
221&R$(\frac{1}{2},\frac{1}{2},\frac{1}{2})$&R9(2)&LD$(u,u,u)$&LD6(2)
&221&R$(\frac{1}{2},\frac{1}{2},\frac{1}{2})$&R6(2)&{\color{red}S$(u,\frac{1}{2},u)$}&S5(2)
\bigstrut\\
\hline
221&R$(\frac{1}{2},\frac{1}{2},\frac{1}{2})$&R7(2)&{\color{red}S$(u,\frac{1}{2},u)$}&S5(2)
&221&R$(\frac{1}{2},\frac{1}{2},\frac{1}{2})$&R8(2)&{\color{red}S$(u,\frac{1}{2},u)$}&S5(2)
\bigstrut\\
\hline
221&R$(\frac{1}{2},\frac{1}{2},\frac{1}{2})$&R9(2)&{\color{red}S$(u,\frac{1}{2},u)$}&S5(2)
&221&R$(\frac{1}{2},\frac{1}{2},\frac{1}{2})$&R10(4)&{\color{red}S$(u,\frac{1}{2},u)$}&S5(2)
\bigstrut\\
\hline
221&R$(\frac{1}{2},\frac{1}{2},\frac{1}{2})$&R11(4)&{\color{red}S$(u,\frac{1}{2},u)$}&S5(2)
&221&X$(0,\frac{1}{2},0)$&X6(2)&{\color{red}S$(u,\frac{1}{2},u)$}&S5(2)
\bigstrut\\
\hline
\hline
\end{tabular}
\end{table*}\end{center}

\newpage
\begin{center}\begin{table*}[!h]
(Continued from previous table: Time-reversal symmetry broken, considering SOC)\\
\begin{tabular}{c|c|c|c|c||c|c|c|c|c}\hline\hline
\multicolumn{1}{l|}{SG} & HSP & Irrep & HSL & Irrep & \multicolumn{1}{l|}{SG} & HSP & Irrep & HSL & Irrep \bigstrut\\
\hline
223&GM$(0,0,0)$&GM8(2)&LD$(u,u,u)$&LD6(2)
&223&GM$(0,0,0)$&GM9(2)&LD$(u,u,u)$&LD6(2)
\bigstrut\\
\hline
223&R$(\frac{1}{2},\frac{1}{2},\frac{1}{2})$&R5(4)&LD$(u,u,u)$&LD6(2)
&223&GM$(0,0,0)$&GM6(2)&{\color{red}SM$(u,u,0)$}&SM5(2)
\bigstrut\\
\hline
223&GM$(0,0,0)$&GM7(2)&{\color{red}SM$(u,u,0)$}&SM5(2)
&223&GM$(0,0,0)$&GM8(2)&{\color{red}SM$(u,u,0)$}&SM5(2)
\bigstrut\\
\hline
223&GM$(0,0,0)$&GM9(2)&{\color{red}SM$(u,u,0)$}&SM5(2)
&223&GM$(0,0,0)$&GM10(4)&{\color{red}SM$(u,u,0)$}&SM5(2)
\bigstrut\\
\hline
223&GM$(0,0,0)$&GM11(4)&{\color{red}SM$(u,u,0)$}&SM5(2)
&223&M$(\frac{1}{2},\frac{1}{2},0)$&M6(2)&{\color{red}SM$(u,u,0)$}&SM5(2)
\bigstrut\\
\hline
223&M$(\frac{1}{2},\frac{1}{2},0)$&M7(2)&{\color{red}SM$(u,u,0)$}&SM5(2)
&223&M$(\frac{1}{2},\frac{1}{2},0)$&M8(2)&{\color{red}SM$(u,u,0)$}&SM5(2)
\bigstrut\\
\hline
223&M$(\frac{1}{2},\frac{1}{2},0)$&M9(2)&{\color{red}SM$(u,u,0)$}&SM5(2)
&223&M$(\frac{1}{2},\frac{1}{2},0)$&M6(2)&{\color{red}T$(\frac{1}{2},\frac{1}{2},w)$}&T6(2)
\bigstrut\\
\hline
\hline
\end{tabular}
\end{table*}\end{center}

\newpage
\begin{center}\begin{table*}[!h]
(Continued from previous table: Time-reversal symmetry broken, considering SOC)\\
\begin{tabular}{c|c|c|c|c||c|c|c|c|c}\hline\hline
\multicolumn{1}{l|}{SG} & HSP & Irrep & HSL & Irrep & \multicolumn{1}{l|}{SG} & HSP & Irrep & HSL & Irrep \bigstrut\\
\hline
225&GM$(0,0,0)$&GM9(2)&{\color{red}SM$(u,u,0)$}&SM5(2)
&225&GM$(0,0,0)$&GM10(4)&{\color{red}SM$(u,u,0)$}&SM5(2)
\bigstrut\\
\hline
225&GM$(0,0,0)$&GM11(4)&{\color{red}SM$(u,u,0)$}&SM5(2)
&225&X$(0,1,0)$&X6(2)&{\color{red}SM$(u,u,0)$}&SM5(2)
\bigstrut\\
\hline
225&X$(0,1,0)$&X7(2)&{\color{red}SM$(u,u,0)$}&SM5(2)
&225&X$(0,1,0)$&X8(2)&{\color{red}SM$(u,u,0)$}&SM5(2)
\bigstrut\\
\hline
225&X$(0,1,0)$&X9(2)&{\color{red}SM$(u,u,0)$}&SM5(2)
&225&X$(0,1,0)$&X6(2)&{\color{red}V$(u,1,0)$}&V5(2)
\bigstrut\\
\hline
225&X$(0,1,0)$&X7(2)&{\color{red}V$(u,1,0)$}&V5(2)
&225&X$(0,1,0)$&X8(2)&{\color{red}V$(u,1,0)$}&V5(2)
\bigstrut\\
\hline
225&X$(0,1,0)$&X9(2)&{\color{red}V$(u,1,0)$}&V5(2)
&225&W$(\frac{1}{2},1,0)$&W6(2)&{\color{red}V$(u,1,0)$}&V5(2)
\bigstrut\\
\hline
225&W$(\frac{1}{2},1,0)$&W7(2)&{\color{red}V$(u,1,0)$}&V5(2)
&226&GM$(0,0,0)$&GM7(2)&{\color{red}DT$(0,v,0)$}&DT6(2)
\bigstrut\\
\hline
\hline
\end{tabular}
\end{table*}\end{center}

\newpage
\begin{center}\begin{table*}[!h]
(Continued from previous table: Time-reversal symmetry broken, considering SOC)\\
\begin{tabular}{c|c|c|c|c||c|c|c|c|c}\hline\hline
\multicolumn{1}{l|}{SG} & HSP & Irrep & HSL & Irrep & \multicolumn{1}{l|}{SG} & HSP & Irrep & HSL & Irrep \bigstrut\\
\hline
228&GM$(0,0,0)$&GM8(2)&{\color{red}SM$(u,u,0)$}&SM5(2)
&228&GM$(0,0,0)$&GM9(2)&{\color{red}SM$(u,u,0)$}&SM5(2)
\bigstrut\\
\hline
228&GM$(0,0,0)$&GM10(4)&{\color{red}SM$(u,u,0)$}&SM5(2)
&228&GM$(0,0,0)$&GM11(4)&{\color{red}SM$(u,u,0)$}&SM5(2)
\bigstrut\\
\hline
228&X$(0,1,0)$&X5(4)&{\color{red}SM$(u,u,0)$}&SM5(2)
&229&P$(\frac{1}{2},\frac{1}{2},\frac{1}{2})$&P6(2)&{\color{red}D$(\frac{1}{2},\frac{1}{2},w)$}&D5(2)
\bigstrut\\
\hline
229&P$(\frac{1}{2},\frac{1}{2},\frac{1}{2})$&P7(2)&{\color{red}D$(\frac{1}{2},\frac{1}{2},w)$}&D5(2)
&229&P$(\frac{1}{2},\frac{1}{2},\frac{1}{2})$&P8(4)&{\color{red}D$(\frac{1}{2},\frac{1}{2},w)$}&D5(2)
\bigstrut\\
\hline
229&N$(\frac{1}{2},\frac{1}{2},0)$&N5(2)&{\color{red}D$(\frac{1}{2},\frac{1}{2},w)$}&D5(2)
&229&N$(\frac{1}{2},\frac{1}{2},0)$&N6(2)&{\color{red}D$(\frac{1}{2},\frac{1}{2},w)$}&D5(2)
\bigstrut\\
\hline
229&GM$(0,0,0)$&GM7(2)&{\color{red}DT$(0,v,0)$}&DT6(2)
&229&GM$(0,0,0)$&GM9(2)&{\color{red}DT$(0,v,0)$}&DT6(2)
\bigstrut\\
\hline
229&GM$(0,0,0)$&GM10(4)&{\color{red}DT$(0,v,0)$}&DT6(2)
&229&GM$(0,0,0)$&GM11(4)&{\color{red}DT$(0,v,0)$}&DT6(2)
\bigstrut\\
\hline\hline
\end{tabular}
\end{table*}\end{center}
\subsubsection{Results for materials without TRS and with negligible SOC}\label{hsp2notrsnosoc}
\newpage
\begin{center}\begin{table*}[!h]
(Continued from previous table: Time-reversal symmetry broken, neglecting SOC)\\
\begin{tabular}{c|c|c|c|c||c|c|c|c|c}\hline\hline
\multicolumn{1}{l|}{SG} & HSP & Irrep & HSL & Irrep & \multicolumn{1}{l|}{SG} & HSP & Irrep & HSL & Irrep \bigstrut\\
\hline
28&U$(\frac{1}{2},0,\frac{1}{2})$&U1(2)&{\color{red}G$(\frac{1}{2},0,w)$}&G1(2)
&28&X$(\frac{1}{2},0,0)$&X1(2)&{\color{red}G$(\frac{1}{2},0,w)$}&G1(2)
\bigstrut\\
\hline
28&R$(\frac{1}{2},\frac{1}{2},\frac{1}{2})$&R1(2)&{\color{red}Q$(\frac{1}{2},\frac{1}{2},w)$}&Q1(2)
&28&S$(\frac{1}{2},\frac{1}{2},0)$&S1(2)&{\color{red}Q$(\frac{1}{2},\frac{1}{2},w)$}&Q1(2)
\bigstrut\\
\hline
29&U$(\frac{1}{2},0,\frac{1}{2})$&U1(2)&{\color{red}G$(\frac{1}{2},0,w)$}&G1(2)
&29&X$(\frac{1}{2},0,0)$&X1(2)&{\color{red}G$(\frac{1}{2},0,w)$}&G1(2)
\bigstrut\\
\hline
29&R$(\frac{1}{2},\frac{1}{2},\frac{1}{2})$&R1(2)&{\color{red}Q$(\frac{1}{2},\frac{1}{2},w)$}&Q1(2)
&29&S$(\frac{1}{2},\frac{1}{2},0)$&S1(2)&{\color{red}Q$(\frac{1}{2},\frac{1}{2},w)$}&Q1(2)
\bigstrut\\
\hline
30&T$(0,\frac{1}{2},\frac{1}{2})$&T1(2)&{\color{red}H$(0,\frac{1}{2},w)$}&H1(2)
&30&Y$(0,\frac{1}{2},0)$&Y1(2)&{\color{red}H$(0,\frac{1}{2},w)$}&H1(2)
\bigstrut\\
\hline
30&R$(\frac{1}{2},\frac{1}{2},\frac{1}{2})$&R1(2)&{\color{red}Q$(\frac{1}{2},\frac{1}{2},w)$}&Q1(2)
&30&S$(\frac{1}{2},\frac{1}{2},0)$&S1(2)&{\color{red}Q$(\frac{1}{2},\frac{1}{2},w)$}&Q1(2)
\bigstrut\\
\hline
31&U$(\frac{1}{2},0,\frac{1}{2})$&U1(2)&{\color{red}G$(\frac{1}{2},0,w)$}&G1(2)
&31&X$(\frac{1}{2},0,0)$&X1(2)&{\color{red}G$(\frac{1}{2},0,w)$}&G1(2)
\bigstrut\\
\hline
\hline
\end{tabular}
\end{table*}\end{center}

\newpage
\begin{center}\begin{table*}[!h]
(Continued from previous table: Time-reversal symmetry broken, neglecting SOC)\\
\begin{tabular}{c|c|c|c|c||c|c|c|c|c}\hline\hline
\multicolumn{1}{l|}{SG} & HSP & Irrep & HSL & Irrep & \multicolumn{1}{l|}{SG} & HSP & Irrep & HSL & Irrep \bigstrut\\
\hline
50&X$(\frac{1}{2},0,0)$&X1(2)&{\color{red}D$(\frac{1}{2},v,0)$}&D1(2)
&50&X$(\frac{1}{2},0,0)$&X2(2)&{\color{red}D$(\frac{1}{2},v,0)$}&D1(2)
\bigstrut\\
\hline
50&R$(\frac{1}{2},\frac{1}{2},\frac{1}{2})$&R1(2)&{\color{red}E$(u,\frac{1}{2},\frac{1}{2})$}&E1(2)
&50&R$(\frac{1}{2},\frac{1}{2},\frac{1}{2})$&R2(2)&{\color{red}E$(u,\frac{1}{2},\frac{1}{2})$}&E1(2)
\bigstrut\\
\hline
50&T$(0,\frac{1}{2},\frac{1}{2})$&T1(2)&{\color{red}E$(u,\frac{1}{2},\frac{1}{2})$}&E1(2)
&50&T$(0,\frac{1}{2},\frac{1}{2})$&T2(2)&{\color{red}E$(u,\frac{1}{2},\frac{1}{2})$}&E1(2)
\bigstrut\\
\hline
50&U$(\frac{1}{2},0,\frac{1}{2})$&U1(2)&{\color{red}G$(\frac{1}{2},0,w)$}&G1(2)
&50&U$(\frac{1}{2},0,\frac{1}{2})$&U2(2)&{\color{red}G$(\frac{1}{2},0,w)$}&G1(2)
\bigstrut\\
\hline
50&X$(\frac{1}{2},0,0)$&X1(2)&{\color{red}G$(\frac{1}{2},0,w)$}&G1(2)
&50&X$(\frac{1}{2},0,0)$&X2(2)&{\color{red}G$(\frac{1}{2},0,w)$}&G1(2)
\bigstrut\\
\hline
50&T$(0,\frac{1}{2},\frac{1}{2})$&T1(2)&{\color{red}H$(0,\frac{1}{2},w)$}&H1(2)
&50&T$(0,\frac{1}{2},\frac{1}{2})$&T2(2)&{\color{red}H$(0,\frac{1}{2},w)$}&H1(2)
\bigstrut\\
\hline
50&Y$(0,\frac{1}{2},0)$&Y1(2)&{\color{red}H$(0,\frac{1}{2},w)$}&H1(2)
&50&Y$(0,\frac{1}{2},0)$&Y2(2)&{\color{red}H$(0,\frac{1}{2},w)$}&H1(2)
\bigstrut\\
\hline
\hline
\end{tabular}
\end{table*}\end{center}

\newpage
\begin{center}\begin{table*}[!h]
(Continued from previous table: Time-reversal symmetry broken, neglecting SOC)\\
\begin{tabular}{c|c|c|c|c||c|c|c|c|c}\hline\hline
\multicolumn{1}{l|}{SG} & HSP & Irrep & HSL & Irrep & \multicolumn{1}{l|}{SG} & HSP & Irrep & HSL & Irrep \bigstrut\\
\hline
54&Z$(0,0,\frac{1}{2})$&Z1(2)&{\color{red}B$(0,v,\frac{1}{2})$}&B1(2)
&54&Z$(0,0,\frac{1}{2})$&Z2(2)&{\color{red}B$(0,v,\frac{1}{2})$}&B1(2)
\bigstrut\\
\hline
54&S$(\frac{1}{2},\frac{1}{2},0)$&S1(2)&{\color{red}D$(\frac{1}{2},v,0)$}&D1(2)
&54&S$(\frac{1}{2},\frac{1}{2},0)$&S2(2)&{\color{red}D$(\frac{1}{2},v,0)$}&D1(2)
\bigstrut\\
\hline
54&X$(\frac{1}{2},0,0)$&X1(2)&{\color{red}D$(\frac{1}{2},v,0)$}&D1(2)
&54&X$(\frac{1}{2},0,0)$&X2(2)&{\color{red}D$(\frac{1}{2},v,0)$}&D1(2)
\bigstrut\\
\hline
54&R$(\frac{1}{2},\frac{1}{2},\frac{1}{2})$&R1(2)&{\color{red}E$(u,\frac{1}{2},\frac{1}{2})$}&E1(2)
&54&R$(\frac{1}{2},\frac{1}{2},\frac{1}{2})$&R2(2)&{\color{red}E$(u,\frac{1}{2},\frac{1}{2})$}&E1(2)
\bigstrut\\
\hline
54&T$(0,\frac{1}{2},\frac{1}{2})$&T1(2)&{\color{red}E$(u,\frac{1}{2},\frac{1}{2})$}&E1(2)
&54&T$(0,\frac{1}{2},\frac{1}{2})$&T2(2)&{\color{red}E$(u,\frac{1}{2},\frac{1}{2})$}&E1(2)
\bigstrut\\
\hline
55&U$(\frac{1}{2},0,\frac{1}{2})$&U1(2)&{\color{red}G$(\frac{1}{2},0,w)$}&G1(2)
&55&U$(\frac{1}{2},0,\frac{1}{2})$&U2(2)&{\color{red}G$(\frac{1}{2},0,w)$}&G1(2)
\bigstrut\\
\hline
55&X$(\frac{1}{2},0,0)$&X1(2)&{\color{red}G$(\frac{1}{2},0,w)$}&G1(2)
&55&X$(\frac{1}{2},0,0)$&X2(2)&{\color{red}G$(\frac{1}{2},0,w)$}&G1(2)
\bigstrut\\
\hline
\hline
\end{tabular}
\end{table*}\end{center}

\newpage
\begin{center}\begin{table*}[!h]
(Continued from previous table: Time-reversal symmetry broken, neglecting SOC)\\
\begin{tabular}{c|c|c|c|c||c|c|c|c|c}\hline\hline
\multicolumn{1}{l|}{SG} & HSP & Irrep & HSL & Irrep & \multicolumn{1}{l|}{SG} & HSP & Irrep & HSL & Irrep \bigstrut\\
\hline
59&X$(\frac{1}{2},0,0)$&X1(2)&{\color{red}D$(\frac{1}{2},v,0)$}&D1(2)
&59&X$(\frac{1}{2},0,0)$&X2(2)&{\color{red}D$(\frac{1}{2},v,0)$}&D1(2)
\bigstrut\\
\hline
59&R$(\frac{1}{2},\frac{1}{2},\frac{1}{2})$&R1(2)&{\color{red}E$(u,\frac{1}{2},\frac{1}{2})$}&E1(2)
&59&R$(\frac{1}{2},\frac{1}{2},\frac{1}{2})$&R2(2)&{\color{red}E$(u,\frac{1}{2},\frac{1}{2})$}&E1(2)
\bigstrut\\
\hline
59&T$(0,\frac{1}{2},\frac{1}{2})$&T1(2)&{\color{red}E$(u,\frac{1}{2},\frac{1}{2})$}&E1(2)
&59&T$(0,\frac{1}{2},\frac{1}{2})$&T2(2)&{\color{red}E$(u,\frac{1}{2},\frac{1}{2})$}&E1(2)
\bigstrut\\
\hline
59&R$(\frac{1}{2},\frac{1}{2},\frac{1}{2})$&R1(2)&{\color{red}P$(\frac{1}{2},v,\frac{1}{2})$}&P1(2)
&59&R$(\frac{1}{2},\frac{1}{2},\frac{1}{2})$&R2(2)&{\color{red}P$(\frac{1}{2},v,\frac{1}{2})$}&P1(2)
\bigstrut\\
\hline
59&U$(\frac{1}{2},0,\frac{1}{2})$&U1(2)&{\color{red}P$(\frac{1}{2},v,\frac{1}{2})$}&P1(2)
&59&U$(\frac{1}{2},0,\frac{1}{2})$&U2(2)&{\color{red}P$(\frac{1}{2},v,\frac{1}{2})$}&P1(2)
\bigstrut\\
\hline
60&U$(\frac{1}{2},0,\frac{1}{2})$&U1(2)&{\color{red}A$(u,0,\frac{1}{2})$}&A1(2)
&60&U$(\frac{1}{2},0,\frac{1}{2})$&U2(2)&{\color{red}A$(u,0,\frac{1}{2})$}&A1(2)
\bigstrut\\
\hline
60&Z$(0,0,\frac{1}{2})$&Z1(2)&{\color{red}A$(u,0,\frac{1}{2})$}&A1(2)
&60&Z$(0,0,\frac{1}{2})$&Z2(2)&{\color{red}A$(u,0,\frac{1}{2})$}&A1(2)
\bigstrut\\
\hline
\hline
\end{tabular}
\end{table*}\end{center}

\newpage
\begin{center}\begin{table*}[!h]
(Continued from previous table: Time-reversal symmetry broken, neglecting SOC)\\
\begin{tabular}{c|c|c|c|c||c|c|c|c|c}\hline\hline
\multicolumn{1}{l|}{SG} & HSP & Irrep & HSL & Irrep & \multicolumn{1}{l|}{SG} & HSP & Irrep & HSL & Irrep \bigstrut\\
\hline
64&Z$(0,0,\frac{1}{2})$&Z1(2)&{\color{red}A$(u,0,\frac{1}{2})$}&A1(2)
&64&Z$(0,0,\frac{1}{2})$&Z2(2)&{\color{red}A$(u,0,\frac{1}{2})$}&A1(2)
\bigstrut\\
\hline
66&T$(1,0,\frac{1}{2})$&T1(2)&{\color{red}A$(u,0,\frac{1}{2})$}&A1(2)
&66&T$(1,0,\frac{1}{2})$&T2(2)&{\color{red}A$(u,0,\frac{1}{2})$}&A1(2)
\bigstrut\\
\hline
66&Z$(0,0,\frac{1}{2})$&Z1(2)&{\color{red}A$(u,0,\frac{1}{2})$}&A1(2)
&66&Z$(0,0,\frac{1}{2})$&Z2(2)&{\color{red}A$(u,0,\frac{1}{2})$}&A1(2)
\bigstrut\\
\hline
66&T$(1,0,\frac{1}{2})$&T1(2)&{\color{red}B$(0,v,\frac{1}{2})$}&B1(2)
&66&T$(1,0,\frac{1}{2})$&T2(2)&{\color{red}B$(0,v,\frac{1}{2})$}&B1(2)
\bigstrut\\
\hline
66&Z$(0,0,\frac{1}{2})$&Z1(2)&{\color{red}B$(0,v,\frac{1}{2})$}&B1(2)
&66&Z$(0,0,\frac{1}{2})$&Z2(2)&{\color{red}B$(0,v,\frac{1}{2})$}&B1(2)
\bigstrut\\
\hline
68&T$(1,0,\frac{1}{2})$&T1(2)&{\color{red}A$(u,0,\frac{1}{2})$}&A1(2)
&68&T$(1,0,\frac{1}{2})$&T2(2)&{\color{red}A$(u,0,\frac{1}{2})$}&A1(2)
\bigstrut\\
\hline
68&Z$(0,0,\frac{1}{2})$&Z1(2)&{\color{red}A$(u,0,\frac{1}{2})$}&A1(2)
&68&Z$(0,0,\frac{1}{2})$&Z2(2)&{\color{red}A$(u,0,\frac{1}{2})$}&A1(2)
\bigstrut\\
\hline
\hline
\end{tabular}%
\end{table*}\end{center}

\subsection{Nodal surface}\label{hsp3}
In this section we list all the positions of HSPs which could host degenerate energy levels implying nodal surfaces coinciding with neighboring HSPLs. Take C in SG 4 in the setting with TRS and significant SOC as the example, namely, the first item in the table of Sec.~\ref{hsp3trssoc}. C is an HSP in SG 4 and its coordinate in the conventional basis is $(\frac{1}{2}, \frac{1}{2}, 0)$. The little group of C has two 2D double-valued irreps C3,3 and C4,4 (here C3,3 means that C3 is paired with C3 by antiunitary symmetry).  And the two irreps would not decompose and furthermore form a nodal surface on the corresponding HSPL G with coordinate $(u, \frac{1}{2}, w)$. According to Sec. III of the main text, nodal surface protected by space group (SG) symmetry must appear in the presence of TRS and all nodal surfaces are 2D degenerate. The results for all HSPs which imply nodal surfaces coinciding with a neighboring HSPL are shown in the table below.
\subsubsection{Results for materials with TRS and significant SOC}\label{hsp3trssoc}

\begin{center}\begin{table*}[!h]
\begin{tabular}{c|c|c|c||c|c|c|c}\hline\hline
\multicolumn{1}{l|}{SG} & HSP & Irrep & HSPL & \multicolumn{1}{l|}{SG} & HSP & Irrep & HSPL \bigstrut\\
\hline
4&C$(\frac{1}{2},\frac{1}{2},0)$&C3,3,C4,4&G$(u,\frac{1}{2},w)$&4&D$(0,\frac{1}{2},\frac{1}{2})$&D3,3,D4,4&G$(u,\frac{1}{2},w)$\bigstrut\\
\hline
4&E$(\frac{1}{2},\frac{1}{2},\frac{1}{2})$&E3,3,E4,4&G$(u,\frac{1}{2},w)$&4&Z$(0,\frac{1}{2},0)$&Z3,3,Z4,4&G$(u,\frac{1}{2},w)$\bigstrut\\
\hline
17&R$(\frac{1}{2},\frac{1}{2},\frac{1}{2})$&R2,4,R3,5&W$(u,v,\frac{1}{2})$&17&T$(0,\frac{1}{2},\frac{1}{2})$&T2,4,T3,5&W$(u,v,\frac{1}{2})$\bigstrut\\
\hline
17&U$(\frac{1}{2},0,\frac{1}{2})$&U2,4,U3,5&W$(u,v,\frac{1}{2})$&17&Z$(0,0,\frac{1}{2})$&Z2,4,Z3,5&W$(u,v,\frac{1}{2})$\bigstrut\\
\hline
18&T$(0,\frac{1}{2},\frac{1}{2})$&T2,5,T3,4&N$(u,\frac{1}{2},w)$&18&U$(\frac{1}{2},0,\frac{1}{2})$&U2,3,U4,5&L$(\frac{1}{2},v,w)$\bigstrut\\
\hline
18&X$(\frac{1}{2},0,0)$&X2,3,X4,5&L$(\frac{1}{2},v,w)$&18&Y$(0,\frac{1}{2},0)$&Y2,5,Y3,4&N$(u,\frac{1}{2},w)$\bigstrut\\
\hline
19&R$(\frac{1}{2},\frac{1}{2},\frac{1}{2})$&R2,2,R3,3,R4,4,R5,5&L$(\frac{1}{2},v,w)$&19&R$(\frac{1}{2},\frac{1}{2},\frac{1}{2})$&R2,2,R3,3,R4,4,R5,5&N$(u,\frac{1}{2},w)$\bigstrut\\
\hline
19&R$(\frac{1}{2},\frac{1}{2},\frac{1}{2})$&R2,2,R3,3,R4,4,R5,5&W$(u,v,\frac{1}{2})$&19&X$(\frac{1}{2},0,0)$&X2,3,X4,5&L$(\frac{1}{2},v,w)$\bigstrut\\
\hline
19&Y$(0,\frac{1}{2},0)$&Y2,5,Y3,4&N$(u,\frac{1}{2},w)$&19&Z$(0,0,\frac{1}{2})$&Z2,4,Z3,5&W$(u,v,\frac{1}{2})$\bigstrut\\
\hline
20&R$(\frac{1}{2},\frac{1}{2},\frac{1}{2})$&R3,3,R4,4&Q$(u,v,\frac{1}{2})$&20&T$(1,0,\frac{1}{2})$&T2,4,T3,5&Q$(u,v,\frac{1}{2})$\bigstrut\\
\hline
20&Z$(0,0,\frac{1}{2})$&Z2,4,Z3,5&Q$(u,v,\frac{1}{2})$&29&R$(\frac{1}{2},\frac{1}{2},\frac{1}{2})$&R2,2,R3,3,R4,4,R5,5&W$(u,v,\frac{1}{2})$\bigstrut\\
\hline
29&U$(\frac{1}{2},0,\frac{1}{2})$&U2,2,U3,3,U4,4,U5,5&W$(u,v,\frac{1}{2})$&31&R$(\frac{1}{2},\frac{1}{2},\frac{1}{2})$&R2,4,R3,5&W$(u,v,\frac{1}{2})$\bigstrut\\
\hline
31&U$(\frac{1}{2},0,\frac{1}{2})$&U2,4,U3,5&W$(u,v,\frac{1}{2})$&33&T$(0,\frac{1}{2},\frac{1}{2})$&T2,4,T3,5&W$(u,v,\frac{1}{2})$\bigstrut\\
\hline
33&U$(\frac{1}{2},0,\frac{1}{2})$&U2,2,U3,3,U4,4,U5,5&W$(u,v,\frac{1}{2})$&36&R$(\frac{1}{2},\frac{1}{2},\frac{1}{2})$&R3,3,R4,4&Q$(u,v,\frac{1}{2})$\bigstrut\\
\hline
76&A$(\frac{1}{2},\frac{1}{2},\frac{1}{2})$&A5,5,A6,6,A7,8&E$(u,v,\frac{1}{2})$&76&R$(0,\frac{1}{2},\frac{1}{2})$&R3,3,R4,4&E$(u,v,\frac{1}{2})$\bigstrut\\
\hline
76&Z$(0,0,\frac{1}{2})$&Z5,5,Z6,6,Z7,8&E$(u,v,\frac{1}{2})$&78&A$(\frac{1}{2},\frac{1}{2},\frac{1}{2})$&A5,6,A7,7,A8,8&E$(u,v,\frac{1}{2})$\bigstrut\\
\hline
78&R$(0,\frac{1}{2},\frac{1}{2})$&R3,3,R4,4&E$(u,v,\frac{1}{2})$&78&Z$(0,0,\frac{1}{2})$&Z5,6,Z7,7,Z8,8&E$(u,v,\frac{1}{2})$\bigstrut\\
\hline
90&R$(0,\frac{1}{2},\frac{1}{2})$&R2,5,R3,4&F$(u,\frac{1}{2},w)$&90&X$(0,\frac{1}{2},0)$&X2,5,X3,4&F$(u,\frac{1}{2},w)$\bigstrut\\
\hline
91&A$(\frac{1}{2},\frac{1}{2},\frac{1}{2})$&A3,5,A4,6,A7&E$(u,v,\frac{1}{2})$&91&R$(0,\frac{1}{2},\frac{1}{2})$&R2,4,R3,5&E$(u,v,\frac{1}{2})$\bigstrut\\
\hline
91&Z$(0,0,\frac{1}{2})$&Z3,5,Z4,6,Z7&E$(u,v,\frac{1}{2})$&92&A$(\frac{1}{2},\frac{1}{2},\frac{1}{2})$&A3,6,A4,5&E$(u,v,\frac{1}{2})$\bigstrut\\
\hline
92&A$(\frac{1}{2},\frac{1}{2},\frac{1}{2})$&A3,6,A4,5&F$(u,\frac{1}{2},w)$&92&X$(0,\frac{1}{2},0)$&X2,5,X3,4&F$(u,\frac{1}{2},w)$\bigstrut\\
\hline
92&Z$(0,0,\frac{1}{2})$&Z3,5,Z4,6,Z7&E$(u,v,\frac{1}{2})$&94&R$(0,\frac{1}{2},\frac{1}{2})$&R2,5,R3,4&F$(u,\frac{1}{2},w)$\bigstrut\\
\hline
94&X$(0,\frac{1}{2},0)$&X2,5,X3,4&F$(u,\frac{1}{2},w)$&95&A$(\frac{1}{2},\frac{1}{2},\frac{1}{2})$&A3,5,A4,6,A7&E$(u,v,\frac{1}{2})$\bigstrut\\
\hline
95&R$(0,\frac{1}{2},\frac{1}{2})$&R2,4,R3,5&E$(u,v,\frac{1}{2})$&95&Z$(0,0,\frac{1}{2})$&Z3,5,Z4,6,Z7&E$(u,v,\frac{1}{2})$\bigstrut\\
\hline
96&A$(\frac{1}{2},\frac{1}{2},\frac{1}{2})$&A3,6,A4,5&E$(u,v,\frac{1}{2})$&96&A$(\frac{1}{2},\frac{1}{2},\frac{1}{2})$&A3,6,A4,5&F$(u,\frac{1}{2},w)$\bigstrut\\
\hline
96&X$(0,\frac{1}{2},0)$&X2,5,X3,4&F$(u,\frac{1}{2},w)$&96&Z$(0,0,\frac{1}{2})$&Z3,5,Z4,6,Z7&E$(u,v,\frac{1}{2})$\bigstrut\\
\hline
113&R$(0,\frac{1}{2},\frac{1}{2})$&R2,5,R3,4&F$(u,\frac{1}{2},w)$&113&X$(0,\frac{1}{2},0)$&X2,5,X3,4&F$(u,\frac{1}{2},w)$\bigstrut\\
\hline
114&R$(0,\frac{1}{2},\frac{1}{2})$&R2,5,R3,4&F$(u,\frac{1}{2},w)$&114&X$(0,\frac{1}{2},0)$&X2,5,X3,4&F$(u,\frac{1}{2},w)$\bigstrut\\
\hline
169&A$(0,0,\frac{1}{2})$&A7,11,A8,12,A9,9,A10,10&E$(u,v,\frac{1}{2})$&169&H$(\frac{1}{3},\frac{1}{3},\frac{1}{2})$&H4,6,H5,5&E$(u,v,\frac{1}{2})$\bigstrut\\
\hline
169&L$(\frac{1}{2},0,\frac{1}{2})$&L3,3,L4,4&E$(u,v,\frac{1}{2})$&170&A$(0,0,\frac{1}{2})$&A7,9,A8,10,A11,11,A12,12&E$(u,v,\frac{1}{2})$\bigstrut\\
\hline
170&H$(\frac{1}{3},\frac{1}{3},\frac{1}{2})$&H4,4,H5,6&E$(u,v,\frac{1}{2})$&170&L$(\frac{1}{2},0,\frac{1}{2})$&L3,3,L4,4&E$(u,v,\frac{1}{2})$\bigstrut\\
\hline
173&A$(0,0,\frac{1}{2})$&A7,7,A8,8,A9,11,A10,12&E$(u,v,\frac{1}{2})$&173&H$(\frac{1}{3},\frac{1}{3},\frac{1}{2})$&H4,4,H5,6&E$(u,v,\frac{1}{2})$\bigstrut\\
\hline
173&L$(\frac{1}{2},0,\frac{1}{2})$&L3,3,L4,4&E$(u,v,\frac{1}{2})$&178&A$(0,0,\frac{1}{2})$&A4,5,A6,7,A8,A9&E$(u,v,\frac{1}{2})$\bigstrut\\
\hline
178&H$(\frac{1}{3},\frac{1}{3},\frac{1}{2})$&H4,5,H6&E$(u,v,\frac{1}{2})$&178&L$(\frac{1}{2},0,\frac{1}{2})$&L2,4,L3,5&E$(u,v,\frac{1}{2})$\bigstrut\\
\hline
179&A$(0,0,\frac{1}{2})$&A4,5,A6,7,A8,A9&E$(u,v,\frac{1}{2})$&179&H$(\frac{1}{3},\frac{1}{3},\frac{1}{2})$&H4,5,H6&E$(u,v,\frac{1}{2})$\bigstrut\\
\hline
179&L$(\frac{1}{2},0,\frac{1}{2})$&L2,4,L3,5&E$(u,v,\frac{1}{2})$&182&A$(0,0,\frac{1}{2})$&A4,5,A6,7,A8,A9&E$(u,v,\frac{1}{2})$\bigstrut\\
\hline
182&H$(\frac{1}{3},\frac{1}{3},\frac{1}{2})$&H4,5,H6&E$(u,v,\frac{1}{2})$&182&L$(\frac{1}{2},0,\frac{1}{2})$&L2,4,L3,5&E$(u,v,\frac{1}{2})$\bigstrut\\
\hline
185&H$(\frac{1}{3},\frac{1}{3},\frac{1}{2})$&H4,4,H5,5&E$(u,v,\frac{1}{2})$&186&H$(\frac{1}{3},\frac{1}{3},\frac{1}{2})$&H4,5,H6&E$(u,v,\frac{1}{2})$\bigstrut\\
\hline
198&R$(\frac{1}{2},\frac{1}{2},\frac{1}{2})$&R4,4,R5,6&B$(u,\frac{1}{2},w)$&198&X$(0,\frac{1}{2},0)$&X2,5,X3,4&B$(u,\frac{1}{2},w)$\bigstrut\\
\hline
212&R$(\frac{1}{2},\frac{1}{2},\frac{1}{2})$&R4,5,R6&B$(u,\frac{1}{2},w)$&212&X$(0,\frac{1}{2},0)$&X3,6,X4,5,X7&B$(u,\frac{1}{2},w)$\bigstrut\\
\hline
213&R$(\frac{1}{2},\frac{1}{2},\frac{1}{2})$&R4,5,R6&B$(u,\frac{1}{2},w)$&213&X$(0,\frac{1}{2},0)$&X3,6,X4,5,X7&B$(u,\frac{1}{2},w)$\bigstrut\\
\hline
\hline
\end{tabular}
\end{table*}\end{center}
\subsubsection{Results for materials with TRS and negligible SOC}\label{hsp3trsnosoc}

\begin{center}\begin{table*}[!h]
\begin{tabular}{c|c|c|c||c|c|c|c}\hline\hline
\multicolumn{1}{l|}{SG} & HSP & Irrep & HSPL & \multicolumn{1}{l|}{SG} & HSP & Irrep & HSPL \bigstrut\\
\hline
4&C$(\frac{1}{2},\frac{1}{2},0)$&C1,2&G$(u,\frac{1}{2},w)$&4&D$(0,\frac{1}{2},\frac{1}{2})$&D1,2&G$(u,\frac{1}{2},w)$\bigstrut\\
\hline
4&E$(\frac{1}{2},\frac{1}{2},\frac{1}{2})$&E1,2&G$(u,\frac{1}{2},w)$&4&Z$(0,\frac{1}{2},0)$&Z1,2&G$(u,\frac{1}{2},w)$\bigstrut\\
\hline
11&C$(\frac{1}{2},\frac{1}{2},0)$&C1&G$(u,\frac{1}{2},w)$&11&D$(0,\frac{1}{2},\frac{1}{2})$&D1&G$(u,\frac{1}{2},w)$\bigstrut\\
\hline
11&E$(\frac{1}{2},\frac{1}{2},\frac{1}{2})$&E1&G$(u,\frac{1}{2},w)$&11&Z$(0,\frac{1}{2},0)$&Z1&G$(u,\frac{1}{2},w)$\bigstrut\\
\hline
14&C$(\frac{1}{2},\frac{1}{2},0)$&C1&G$(u,\frac{1}{2},w)$&14&D$(0,\frac{1}{2},\frac{1}{2})$&D1+,2+,D1-,2-&G$(u,\frac{1}{2},w)$\bigstrut\\
\hline
14&E$(\frac{1}{2},\frac{1}{2},\frac{1}{2})$&E1+,2+,E1-,2-&G$(u,\frac{1}{2},w)$&14&Z$(0,\frac{1}{2},0)$&Z1&G$(u,\frac{1}{2},w)$\bigstrut\\
\hline
17&R$(\frac{1}{2},\frac{1}{2},\frac{1}{2})$&R1&W$(u,v,\frac{1}{2})$&17&T$(0,\frac{1}{2},\frac{1}{2})$&T1&W$(u,v,\frac{1}{2})$\bigstrut\\
\hline
17&U$(\frac{1}{2},0,\frac{1}{2})$&U1&W$(u,v,\frac{1}{2})$&17&Z$(0,0,\frac{1}{2})$&Z1&W$(u,v,\frac{1}{2})$\bigstrut\\
\hline
18&R$(\frac{1}{2},\frac{1}{2},\frac{1}{2})$&R1,2,R3,4&L$(\frac{1}{2},v,w)$&18&R$(\frac{1}{2},\frac{1}{2},\frac{1}{2})$&R1,2,R3,4&N$(u,\frac{1}{2},w)$\bigstrut\\
\hline
18&S$(\frac{1}{2},\frac{1}{2},0)$&S1,2,S3,4&L$(\frac{1}{2},v,w)$&18&S$(\frac{1}{2},\frac{1}{2},0)$&S1,2,S3,4&N$(u,\frac{1}{2},w)$\bigstrut\\
\hline
18&T$(0,\frac{1}{2},\frac{1}{2})$&T1&N$(u,\frac{1}{2},w)$&18&U$(\frac{1}{2},0,\frac{1}{2})$&U1&L$(\frac{1}{2},v,w)$\bigstrut\\
\hline
18&X$(\frac{1}{2},0,0)$&X1&L$(\frac{1}{2},v,w)$&18&Y$(0,\frac{1}{2},0)$&Y1&N$(u,\frac{1}{2},w)$\bigstrut\\
\hline
19&S$(\frac{1}{2},\frac{1}{2},0)$&S1,2,S3,4&L$(\frac{1}{2},v,w)$&19&S$(\frac{1}{2},\frac{1}{2},0)$&S1,2,S3,4&N$(u,\frac{1}{2},w)$\bigstrut\\
\hline
19&T$(0,\frac{1}{2},\frac{1}{2})$&T1,3,T2,4&N$(u,\frac{1}{2},w)$&19&T$(0,\frac{1}{2},\frac{1}{2})$&T1,3,T2,4&W$(u,v,\frac{1}{2})$\bigstrut\\
\hline
19&U$(\frac{1}{2},0,\frac{1}{2})$&U1,4,U2,3&L$(\frac{1}{2},v,w)$&19&U$(\frac{1}{2},0,\frac{1}{2})$&U1,4,U2,3&W$(u,v,\frac{1}{2})$\bigstrut\\
\hline
19&X$(\frac{1}{2},0,0)$&X1&L$(\frac{1}{2},v,w)$&19&Y$(0,\frac{1}{2},0)$&Y1&N$(u,\frac{1}{2},w)$\bigstrut\\
\hline
19&Z$(0,0,\frac{1}{2})$&Z1&W$(u,v,\frac{1}{2})$&20&R$(\frac{1}{2},\frac{1}{2},\frac{1}{2})$&R1,2&Q$(u,v,\frac{1}{2})$\bigstrut\\
\hline
20&T$(1,0,\frac{1}{2})$&T1&Q$(u,v,\frac{1}{2})$&20&Z$(0,0,\frac{1}{2})$&Z1&Q$(u,v,\frac{1}{2})$\bigstrut\\
\hline
26&R$(\frac{1}{2},\frac{1}{2},\frac{1}{2})$&R1,3,R2,4&W$(u,v,\frac{1}{2})$&26&T$(0,\frac{1}{2},\frac{1}{2})$&T1,3,T2,4&W$(u,v,\frac{1}{2})$\bigstrut\\
\hline
26&U$(\frac{1}{2},0,\frac{1}{2})$&U1,3,U2,4&W$(u,v,\frac{1}{2})$&26&Z$(0,0,\frac{1}{2})$&Z1,3,Z2,4&W$(u,v,\frac{1}{2})$\bigstrut\\
\hline
29&T$(0,\frac{1}{2},\frac{1}{2})$&T1,4,T2,3&W$(u,v,\frac{1}{2})$&29&Z$(0,0,\frac{1}{2})$&Z1,4,Z2,3&W$(u,v,\frac{1}{2})$\bigstrut\\
\hline
31&R$(\frac{1}{2},\frac{1}{2},\frac{1}{2})$&R1&W$(u,v,\frac{1}{2})$&31&T$(0,\frac{1}{2},\frac{1}{2})$&T1,3,T2,4&W$(u,v,\frac{1}{2})$\bigstrut\\
\hline
31&U$(\frac{1}{2},0,\frac{1}{2})$&U1&W$(u,v,\frac{1}{2})$&31&Z$(0,0,\frac{1}{2})$&Z1,3,Z2,4&W$(u,v,\frac{1}{2})$\bigstrut\\
\hline
33&R$(\frac{1}{2},\frac{1}{2},\frac{1}{2})$&R1,3,R2,4&W$(u,v,\frac{1}{2})$&33&T$(0,\frac{1}{2},\frac{1}{2})$&T1&W$(u,v,\frac{1}{2})$\bigstrut\\
\hline
33&Z$(0,0,\frac{1}{2})$&Z1,4,Z2,3&W$(u,v,\frac{1}{2})$&36&R$(\frac{1}{2},\frac{1}{2},\frac{1}{2})$&R1,2&Q$(u,v,\frac{1}{2})$\bigstrut\\
\hline
36&T$(1,0,\frac{1}{2})$&T1,3,T2,4&Q$(u,v,\frac{1}{2})$&36&Z$(0,0,\frac{1}{2})$&Z1,3,Z2,4&Q$(u,v,\frac{1}{2})$\bigstrut\\
\hline
51&R$(\frac{1}{2},\frac{1}{2},\frac{1}{2})$&R1,R2&L$(\frac{1}{2},v,w)$&51&S$(\frac{1}{2},\frac{1}{2},0)$&S1,S2&L$(\frac{1}{2},v,w)$\bigstrut\\
\hline
51&U$(\frac{1}{2},0,\frac{1}{2})$&U1,U2&L$(\frac{1}{2},v,w)$&51&X$(\frac{1}{2},0,0)$&X1,X2&L$(\frac{1}{2},v,w)$\bigstrut\\
\hline
52&R$(\frac{1}{2},\frac{1}{2},\frac{1}{2})$&R1,R2&N$(u,\frac{1}{2},w)$&52&T$(0,\frac{1}{2},\frac{1}{2})$&T1+,T1-&N$(u,\frac{1}{2},w)$\bigstrut\\
\hline
52&Y$(0,\frac{1}{2},0)$&Y1,Y2&N$(u,\frac{1}{2},w)$&53&R$(\frac{1}{2},\frac{1}{2},\frac{1}{2})$&R1+,R1-&W$(u,v,\frac{1}{2})$\bigstrut\\
\hline
53&T$(0,\frac{1}{2},\frac{1}{2})$&T1,T2&W$(u,v,\frac{1}{2})$&53&U$(\frac{1}{2},0,\frac{1}{2})$&U1+,U1-&W$(u,v,\frac{1}{2})$\bigstrut\\
\hline
53&Z$(0,0,\frac{1}{2})$&Z1,Z2&W$(u,v,\frac{1}{2})$&54&S$(\frac{1}{2},\frac{1}{2},0)$&S1,S2&L$(\frac{1}{2},v,w)$\bigstrut\\
\hline
54&X$(\frac{1}{2},0,0)$&X1,X2&L$(\frac{1}{2},v,w)$&55&R$(\frac{1}{2},\frac{1}{2},\frac{1}{2})$&R1+,2+,R1-,2-,R3+,4+,R3-,4-&L$(\frac{1}{2},v,w)$\bigstrut\\
\hline
55&R$(\frac{1}{2},\frac{1}{2},\frac{1}{2})$&R1+,2+,R1-,2-,R3+,4+,R3-,4-&N$(u,\frac{1}{2},w)$&55&S$(\frac{1}{2},\frac{1}{2},0)$&S1+,2+,S1-,2-,S3+,4+,S3-,4-&L$(\frac{1}{2},v,w)$\bigstrut\\
\hline
55&S$(\frac{1}{2},\frac{1}{2},0)$&S1+,2+,S1-,2-,S3+,4+,S3-,4-&N$(u,\frac{1}{2},w)$&55&T$(0,\frac{1}{2},\frac{1}{2})$&T1,T2&N$(u,\frac{1}{2},w)$\bigstrut\\
\hline
55&U$(\frac{1}{2},0,\frac{1}{2})$&U1,U2&L$(\frac{1}{2},v,w)$&55&X$(\frac{1}{2},0,0)$&X1,X2&L$(\frac{1}{2},v,w)$\bigstrut\\
\hline
55&Y$(0,\frac{1}{2},0)$&Y1,Y2&N$(u,\frac{1}{2},w)$&56&R$(\frac{1}{2},\frac{1}{2},\frac{1}{2})$&R1+,2+,R1-,2-,R3+,4+,R3-,4-&L$(\frac{1}{2},v,w)$\bigstrut\\
\hline
\hline
\end{tabular}
\end{table*}\end{center}

\newpage
\begin{center}\begin{table*}[!h]
(Continued from previous table: Time-reversal symmetric, neglecting SOC)\\
\begin{tabular}{c|c|c|c||c|c|c|c}\hline\hline
\multicolumn{1}{l|}{SG} & HSP & Irrep & HSPL & \multicolumn{1}{l|}{SG} & HSP & Irrep & HSPL \bigstrut\\
\hline
56&R$(\frac{1}{2},\frac{1}{2},\frac{1}{2})$&R1+,2+,R1-,2-,R3+,4+,R3-,4-&N$(u,\frac{1}{2},w)$&56&S$(\frac{1}{2},\frac{1}{2},0)$&S1,S2&L$(\frac{1}{2},v,w)$\bigstrut\\
\hline
56&S$(\frac{1}{2},\frac{1}{2},0)$&S1,S2&N$(u,\frac{1}{2},w)$&56&X$(\frac{1}{2},0,0)$&X1,X2&L$(\frac{1}{2},v,w)$\bigstrut\\
\hline
56&Y$(0,\frac{1}{2},0)$&Y1,Y2&N$(u,\frac{1}{2},w)$&57&S$(\frac{1}{2},\frac{1}{2},0)$&S1,S2&N$(u,\frac{1}{2},w)$\bigstrut\\
\hline
57&U$(\frac{1}{2},0,\frac{1}{2})$&U1,U2&W$(u,v,\frac{1}{2})$&57&Y$(0,\frac{1}{2},0)$&Y1,Y2&N$(u,\frac{1}{2},w)$\bigstrut\\
\hline
57&Z$(0,0,\frac{1}{2})$&Z1,Z2&W$(u,v,\frac{1}{2})$&58&R$(\frac{1}{2},\frac{1}{2},\frac{1}{2})$&R1,R2&L$(\frac{1}{2},v,w)$\bigstrut\\
\hline
58&S$(\frac{1}{2},\frac{1}{2},0)$&S1+,2+,S1-,2-,S3+,4+,S3-,4-&L$(\frac{1}{2},v,w)$&58&R$(\frac{1}{2},\frac{1}{2},\frac{1}{2})$&R1,R2&N$(u,\frac{1}{2},w)$\bigstrut\\
\hline
58&S$(\frac{1}{2},\frac{1}{2},0)$&S1+,2+,S1-,2-,S3+,4+,S3-,4-&N$(u,\frac{1}{2},w)$&58&T$(0,\frac{1}{2},\frac{1}{2})$&T1+,T1-&N$(u,\frac{1}{2},w)$\bigstrut\\
\hline
58&U$(\frac{1}{2},0,\frac{1}{2})$&U1+,U1-&L$(\frac{1}{2},v,w)$&58&X$(\frac{1}{2},0,0)$&X1,X2&L$(\frac{1}{2},v,w)$\bigstrut\\
\hline
58&Y$(0,\frac{1}{2},0)$&Y1,Y2&N$(u,\frac{1}{2},w)$&59&R$(\frac{1}{2},\frac{1}{2},\frac{1}{2})$&R1,R2&L$(\frac{1}{2},v,w)$\bigstrut\\
\hline
59&R$(\frac{1}{2},\frac{1}{2},\frac{1}{2})$&R1,R2&N$(u,\frac{1}{2},w)$&59&S$(\frac{1}{2},\frac{1}{2},0)$&S1,S2&L$(\frac{1}{2},v,w)$\bigstrut\\
\hline
59&S$(\frac{1}{2},\frac{1}{2},0)$&S1,S2&N$(u,\frac{1}{2},w)$&59&T$(0,\frac{1}{2},\frac{1}{2})$&T1,T2&N$(u,\frac{1}{2},w)$\bigstrut\\
\hline
59&U$(\frac{1}{2},0,\frac{1}{2})$&U1,U2&L$(\frac{1}{2},v,w)$&59&X$(\frac{1}{2},0,0)$&X1,X2&L$(\frac{1}{2},v,w)$\bigstrut\\
\hline
59&Y$(0,\frac{1}{2},0)$&Y1,Y2&N$(u,\frac{1}{2},w)$&60&S$(\frac{1}{2},\frac{1}{2},0)$&S1+,S1-&L$(\frac{1}{2},v,w)$\bigstrut\\
\hline
60&X$(\frac{1}{2},0,0)$&X1,X2&L$(\frac{1}{2},v,w)$&60&Z$(0,0,\frac{1}{2})$&Z1,Z2&W$(u,v,\frac{1}{2})$\bigstrut\\
\hline
61&X$(\frac{1}{2},0,0)$&X1,X2&L$(\frac{1}{2},v,w)$&61&Y$(0,\frac{1}{2},0)$&Y1,Y2&N$(u,\frac{1}{2},w)$\bigstrut\\
\hline
61&Z$(0,0,\frac{1}{2})$&Z1,Z2&W$(u,v,\frac{1}{2})$&62&T$(0,\frac{1}{2},\frac{1}{2})$&T1,T2&N$(u,\frac{1}{2},w)$\bigstrut\\
\hline
62&U$(\frac{1}{2},0,\frac{1}{2})$&U1+,4+,U1-,4-,U2+,3+,U2-,3-&L$(\frac{1}{2},v,w)$&62&T$(0,\frac{1}{2},\frac{1}{2})$&T1,T2&W$(u,v,\frac{1}{2})$\bigstrut\\
\hline
62&U$(\frac{1}{2},0,\frac{1}{2})$&U1+,4+,U1-,4-,U2+,3+,U2-,3-&W$(u,v,\frac{1}{2})$&62&X$(\frac{1}{2},0,0)$&X1,X2&L$(\frac{1}{2},v,w)$\bigstrut\\
\hline
62&Y$(0,\frac{1}{2},0)$&Y1,Y2&N$(u,\frac{1}{2},w)$&62&Z$(0,0,\frac{1}{2})$&Z1,Z2&W$(u,v,\frac{1}{2})$\bigstrut\\
\hline
63&R$(\frac{1}{2},\frac{1}{2},\frac{1}{2})$&R1&Q$(u,v,\frac{1}{2})$&63&T$(1,0,\frac{1}{2})$&T1,T2&Q$(u,v,\frac{1}{2})$\bigstrut\\
\hline
63&Z$(0,0,\frac{1}{2})$&Z1,Z2&Q$(u,v,\frac{1}{2})$&64&R$(\frac{1}{2},\frac{1}{2},\frac{1}{2})$&R1+,2+,R1-,2-&Q$(u,v,\frac{1}{2})$\bigstrut\\
\hline
64&T$(1,0,\frac{1}{2})$&T1,T2&Q$(u,v,\frac{1}{2})$&64&Z$(0,0,\frac{1}{2})$&Z1,Z2&Q$(u,v,\frac{1}{2})$\bigstrut\\
\hline
76&A$(\frac{1}{2},\frac{1}{2},\frac{1}{2})$&A1,3,A2,4&E$(u,v,\frac{1}{2})$&76&R$(0,\frac{1}{2},\frac{1}{2})$&R1,2&E$(u,v,\frac{1}{2})$\bigstrut\\
\hline
76&Z$(0,0,\frac{1}{2})$&Z1,3,Z2,4&E$(u,v,\frac{1}{2})$&78&A$(\frac{1}{2},\frac{1}{2},\frac{1}{2})$&A1,3,A2,4&E$(u,v,\frac{1}{2})$\bigstrut\\
\hline
78&R$(0,\frac{1}{2},\frac{1}{2})$&R1,2&E$(u,v,\frac{1}{2})$&78&Z$(0,0,\frac{1}{2})$&Z1,3,Z2,4&E$(u,v,\frac{1}{2})$\bigstrut\\
\hline
90&A$(\frac{1}{2},\frac{1}{2},\frac{1}{2})$&A1,4,A2,3,A5&F$(u,\frac{1}{2},w)$&90&M$(\frac{1}{2},\frac{1}{2},0)$&M1,4,M2,3,M5&F$(u,\frac{1}{2},w)$\bigstrut\\
\hline
90&R$(0,\frac{1}{2},\frac{1}{2})$&R1&F$(u,\frac{1}{2},w)$&90&X$(0,\frac{1}{2},0)$&X1&F$(u,\frac{1}{2},w)$\bigstrut\\
\hline
91&A$(\frac{1}{2},\frac{1}{2},\frac{1}{2})$&A1,A2&E$(u,v,\frac{1}{2})$&91&R$(0,\frac{1}{2},\frac{1}{2})$&R1&E$(u,v,\frac{1}{2})$\bigstrut\\
\hline
91&Z$(0,0,\frac{1}{2})$&Z1,Z2&E$(u,v,\frac{1}{2})$&92&M$(\frac{1}{2},\frac{1}{2},0)$&M1,4,M2,3,M5&F$(u,\frac{1}{2},w)$\bigstrut\\
\hline
92&R$(0,\frac{1}{2},\frac{1}{2})$&R1,3,R2,4&E$(u,v,\frac{1}{2})$&92&R$(0,\frac{1}{2},\frac{1}{2})$&R1,3,R2,4&F$(u,\frac{1}{2},w)$\bigstrut\\
\hline
92&X$(0,\frac{1}{2},0)$&X1&F$(u,\frac{1}{2},w)$&92&Z$(0,0,\frac{1}{2})$&Z1,Z2&E$(u,v,\frac{1}{2})$\bigstrut\\
\hline
94&A$(\frac{1}{2},\frac{1}{2},\frac{1}{2})$&A1,4,A2,3,A5&F$(u,\frac{1}{2},w)$&94&M$(\frac{1}{2},\frac{1}{2},0)$&M1,4,M2,3,M5&F$(u,\frac{1}{2},w)$\bigstrut\\
\hline
94&R$(0,\frac{1}{2},\frac{1}{2})$&R1&F$(u,\frac{1}{2},w)$&94&X$(0,\frac{1}{2},0)$&X1&F$(u,\frac{1}{2},w)$\bigstrut\\
\hline
95&A$(\frac{1}{2},\frac{1}{2},\frac{1}{2})$&A1,A2&E$(u,v,\frac{1}{2})$&95&R$(0,\frac{1}{2},\frac{1}{2})$&R1&E$(u,v,\frac{1}{2})$\bigstrut\\
\hline
95&Z$(0,0,\frac{1}{2})$&Z1,Z2&E$(u,v,\frac{1}{2})$&96&M$(\frac{1}{2},\frac{1}{2},0)$&M1,4,M2,3,M5&F$(u,\frac{1}{2},w)$\bigstrut\\
\hline
96&R$(0,\frac{1}{2},\frac{1}{2})$&R1,3,R2,4&E$(u,v,\frac{1}{2})$&96&R$(0,\frac{1}{2},\frac{1}{2})$&R1,3,R2,4&F$(u,\frac{1}{2},w)$\bigstrut\\
\hline
96&X$(0,\frac{1}{2},0)$&X1&F$(u,\frac{1}{2},w)$&96&Z$(0,0,\frac{1}{2})$&Z1,Z2&E$(u,v,\frac{1}{2})$\bigstrut\\
\hline
113&A$(\frac{1}{2},\frac{1}{2},\frac{1}{2})$&A1,3,A2,4,A5&F$(u,\frac{1}{2},w)$&113&M$(\frac{1}{2},\frac{1}{2},0)$&M1,3,M2,4,M5&F$(u,\frac{1}{2},w)$\bigstrut\\
\hline
113&R$(0,\frac{1}{2},\frac{1}{2})$&R1&F$(u,\frac{1}{2},w)$&113&X$(0,\frac{1}{2},0)$&X1&F$(u,\frac{1}{2},w)$\bigstrut\\
\hline
114&A$(\frac{1}{2},\frac{1}{2},\frac{1}{2})$&A1,4,A2,3&F$(u,\frac{1}{2},w)$&114&M$(\frac{1}{2},\frac{1}{2},0)$&M1,3,M2,4,M5&F$(u,\frac{1}{2},w)$\bigstrut\\
\hline
114&R$(0,\frac{1}{2},\frac{1}{2})$&R1&F$(u,\frac{1}{2},w)$&114&X$(0,\frac{1}{2},0)$&X1&F$(u,\frac{1}{2},w)$\bigstrut\\
\hline
\hline
\end{tabular}
\end{table*}\end{center}

\newpage
\begin{center}\begin{table*}[!h]
(Continued from previous table: Time-reversal symmetric, neglecting SOC)\\
\begin{tabular}{c|c|c|c||c|c|c|c}\hline\hline
\multicolumn{1}{l|}{SG} & HSP & Irrep & HSPL & \multicolumn{1}{l|}{SG} & HSP & Irrep & HSPL \bigstrut\\
\hline
127&A$(\frac{1}{2},\frac{1}{2},\frac{1}{2})$&A1+,4+,A1-,4-,A2+,3+,A2-,3-,A5+,A5-&F$(u,\frac{1}{2},w)$&127&R$(0,\frac{1}{2},\frac{1}{2})$&R1,R2&F$(u,\frac{1}{2},w)$\bigstrut\\
\hline
127&M$(\frac{1}{2},\frac{1}{2},0)$&M1+,4+,M1-,4-,M2+,3+,M2-,3-,M5+,M5-&F$(u,\frac{1}{2},w)$&127&X$(0,\frac{1}{2},0)$&X1,X2&F$(u,\frac{1}{2},w)$\bigstrut\\
\hline
128&A$(\frac{1}{2},\frac{1}{2},\frac{1}{2})$&A1,A2&F$(u,\frac{1}{2},w)$&128&M$(\frac{1}{2},\frac{1}{2},0)$&M1+,4+,M1-,4-,M2+,3+,M2-,3-,M5+,M5-&F$(u,\frac{1}{2},w)$\bigstrut\\
\hline
128&R$(0,\frac{1}{2},\frac{1}{2})$&R1+,R1-&F$(u,\frac{1}{2},w)$&128&X$(0,\frac{1}{2},0)$&X1,X2&F$(u,\frac{1}{2},w)$\bigstrut\\
\hline
129&A$(\frac{1}{2},\frac{1}{2},\frac{1}{2})$&A1,A2,A3,A4&F$(u,\frac{1}{2},w)$&129&M$(\frac{1}{2},\frac{1}{2},0)$&M1,M2,M3,M4&F$(u,\frac{1}{2},w)$\bigstrut\\
\hline
129&R$(0,\frac{1}{2},\frac{1}{2})$&R1,R2&F$(u,\frac{1}{2},w)$&129&X$(0,\frac{1}{2},0)$&X1,X2&F$(u,\frac{1}{2},w)$\bigstrut\\
\hline
130&M$(\frac{1}{2},\frac{1}{2},0)$&M1,M2,M3,M4&F$(u,\frac{1}{2},w)$&130&X$(0,\frac{1}{2},0)$&X1,X2&F$(u,\frac{1}{2},w)$\bigstrut\\
\hline
135&R$(0,\frac{1}{2},\frac{1}{2})$&R1,R2&F$(u,\frac{1}{2},w)$&135&M$(\frac{1}{2},\frac{1}{2},0)$&M1+,4+,M1-,4-,M2+,3+,M2-,3-,M5+,M5-&F$(u,\frac{1}{2},w)$\bigstrut\\
\hline
135&X$(0,\frac{1}{2},0)$&X1,X2&F$(u,\frac{1}{2},w)$&136&A$(\frac{1}{2},\frac{1}{2},\frac{1}{2})$&A1,A2,A3,A4&F$(u,\frac{1}{2},w)$\bigstrut\\
\hline
136&R$(0,\frac{1}{2},\frac{1}{2})$&R1+,R1-&F$(u,\frac{1}{2},w)$&136&M$(\frac{1}{2},\frac{1}{2},0)$&M1+,4+,M1-,4-,M2+,3+,M2-,3-,M5+,M5-&F$(u,\frac{1}{2},w)$\bigstrut\\
\hline
136&X$(0,\frac{1}{2},0)$&X1,X2&F$(u,\frac{1}{2},w)$&137&A$(\frac{1}{2},\frac{1}{2},\frac{1}{2})$&A1,A2&F$(u,\frac{1}{2},w)$\bigstrut\\
\hline
137&M$(\frac{1}{2},\frac{1}{2},0)$&M1,M2,M3,M4&F$(u,\frac{1}{2},w)$&137&R$(0,\frac{1}{2},\frac{1}{2})$&R1,R2&F$(u,\frac{1}{2},w)$\bigstrut\\
\hline
137&X$(0,\frac{1}{2},0)$&X1,X2&F$(u,\frac{1}{2},w)$&138&A$(\frac{1}{2},\frac{1}{2},\frac{1}{2})$&A1+,4+,A1-,4-,A2+,3+,A2-,3-,A5+,A5-&F$(u,\frac{1}{2},w)$\bigstrut\\
\hline
138&M$(\frac{1}{2},\frac{1}{2},0)$&M1,M2,M3,M4&F$(u,\frac{1}{2},w)$&138&X$(0,\frac{1}{2},0)$&X1,X2&F$(u,\frac{1}{2},w)$\bigstrut\\
\hline
169&A$(0,0,\frac{1}{2})$&A1,6,A2,5,A3,4&E$(u,v,\frac{1}{2})$&169&H$(\frac{1}{3},\frac{1}{3},\frac{1}{2})$&H1,3,H2,2&E$(u,v,\frac{1}{2})$\bigstrut\\
\hline
169&L$(\frac{1}{2},0,\frac{1}{2})$&L1,2&E$(u,v,\frac{1}{2})$&170&A$(0,0,\frac{1}{2})$&A1,2,A3,6,A4,5&E$(u,v,\frac{1}{2})$\bigstrut\\
\hline
170&H$(\frac{1}{3},\frac{1}{3},\frac{1}{2})$&H1,1,H2,3&E$(u,v,\frac{1}{2})$&170&L$(\frac{1}{2},0,\frac{1}{2})$&L1,2&E$(u,v,\frac{1}{2})$\bigstrut\\
\hline
173&A$(0,0,\frac{1}{2})$&A1,2,A3,6,A4,5&E$(u,v,\frac{1}{2})$&173&H$(\frac{1}{3},\frac{1}{3},\frac{1}{2})$&H1,1,H2,3&E$(u,v,\frac{1}{2})$\bigstrut\\
\hline
173&L$(\frac{1}{2},0,\frac{1}{2})$&L1,2&E$(u,v,\frac{1}{2})$&176&A$(0,0,\frac{1}{2})$&A1&E$(u,v,\frac{1}{2})$\bigstrut\\
\hline
176&H$(\frac{1}{3},\frac{1}{3},\frac{1}{2})$&H1,2,H3,6,H4,5&E$(u,v,\frac{1}{2})$&176&L$(\frac{1}{2},0,\frac{1}{2})$&L1&E$(u,v,\frac{1}{2})$\bigstrut\\
\hline
178&A$(0,0,\frac{1}{2})$&A1,A2,A3&E$(u,v,\frac{1}{2})$&178&H$(\frac{1}{3},\frac{1}{3},\frac{1}{2})$&H1,2,H3&E$(u,v,\frac{1}{2})$\bigstrut\\
\hline
178&L$(\frac{1}{2},0,\frac{1}{2})$&L1&E$(u,v,\frac{1}{2})$&179&A$(0,0,\frac{1}{2})$&A1,A2,A3&E$(u,v,\frac{1}{2})$\bigstrut\\
\hline
179&H$(\frac{1}{3},\frac{1}{3},\frac{1}{2})$&H1,2,H3&E$(u,v,\frac{1}{2})$&179&L$(\frac{1}{2},0,\frac{1}{2})$&L1&E$(u,v,\frac{1}{2})$\bigstrut\\
\hline
182&A$(0,0,\frac{1}{2})$&A1,A2,A3&E$(u,v,\frac{1}{2})$&182&H$(\frac{1}{3},\frac{1}{3},\frac{1}{2})$&H1,2,H3&E$(u,v,\frac{1}{2})$\bigstrut\\
\hline
182&L$(\frac{1}{2},0,\frac{1}{2})$&L1&E$(u,v,\frac{1}{2})$&185&A$(0,0,\frac{1}{2})$&A1,3,A2,4&E$(u,v,\frac{1}{2})$\bigstrut\\
\hline
185&H$(\frac{1}{3},\frac{1}{3},\frac{1}{2})$&H1,1,H2,2&E$(u,v,\frac{1}{2})$&185&L$(\frac{1}{2},0,\frac{1}{2})$&L1,3,L2,4&E$(u,v,\frac{1}{2})$\bigstrut\\
\hline
186&A$(0,0,\frac{1}{2})$&A1,4,A2,3&E$(u,v,\frac{1}{2})$&186&H$(\frac{1}{3},\frac{1}{3},\frac{1}{2})$&H1,2,H3&E$(u,v,\frac{1}{2})$\bigstrut\\
\hline
186&L$(\frac{1}{2},0,\frac{1}{2})$&L1,4,L2,3&E$(u,v,\frac{1}{2})$&193&A$(0,0,\frac{1}{2})$&A1,A2&E$(u,v,\frac{1}{2})$\bigstrut\\
\hline
193&H$(\frac{1}{3},\frac{1}{3},\frac{1}{2})$&H1,3,H2,4&E$(u,v,\frac{1}{2})$&193&L$(\frac{1}{2},0,\frac{1}{2})$&L1,L2&E$(u,v,\frac{1}{2})$\bigstrut\\
\hline
194&A$(0,0,\frac{1}{2})$&A1,A2&E$(u,v,\frac{1}{2})$&194&H$(\frac{1}{3},\frac{1}{3},\frac{1}{2})$&H1,H2,H3&E$(u,v,\frac{1}{2})$\bigstrut\\
\hline
194&L$(\frac{1}{2},0,\frac{1}{2})$&L1,L2&E$(u,v,\frac{1}{2})$&198&M$(\frac{1}{2},\frac{1}{2},0)$&M1,2,M3,4&B$(u,\frac{1}{2},w)$\bigstrut\\
\hline
198&X$(0,\frac{1}{2},0)$&X1&B$(u,\frac{1}{2},w)$&205&X$(0,\frac{1}{2},0)$&X1,X2&B$(u,\frac{1}{2},w)$\bigstrut\\
\hline
212&M$(\frac{1}{2},\frac{1}{2},0)$&M1,4,M2,3,M5&B$(u,\frac{1}{2},w)$&212&X$(0,\frac{1}{2},0)$&X1,X2&B$(u,\frac{1}{2},w)$\bigstrut\\
\hline
213&M$(\frac{1}{2},\frac{1}{2},0)$&M1,4,M2,3,M5&B$(u,\frac{1}{2},w)$&213&X$(0,\frac{1}{2},0)$&X1,X2&B$(u,\frac{1}{2},w)$\bigstrut\\
\hline
\hline
\end{tabular}
\end{table*}\end{center}

\section{High-symmetry line}\label{hsl}
\subsection{Nodal line}\label{hsl1}
In this section, we list all the positions of HSLs with straight nodal line structure for four settings. Take G in SG 25 in the setting with TRS and significant SOC as the example, namely, the first item in the table of Sec.~\ref{hsl1trssoc}. G is an HSL in SG 25 connecting two HSPs U and X, and its coordinate in the conventional basis is $(\frac{1}{2}, 0, w)$ and the coordinates of two HSPs U and X are $(\frac{1}{2}, 0, \frac{1}{2})$ and $(\frac{1}{2}, 0, 0)$. The HSL G could be a nodal line  which corresponds the 2D irrep G5. We further find that such a nodal line is essential so G$(\frac{1}{2}, 0, w)$ is printed in red. For the other three settings, the corresponding tables can also be read following the same way as described here. All positions of HSLs being essentially straight nodal lines are listed in Sec. \ref{esshslnl}, where the first column contains labels for the HSLs and in the second column, it contains SG numbers for the corresponding HSLs.
\subsubsection{Results for materials with TRS and significant SOC}\label{hsl1trssoc}
\begin{center}\begin{table*}[!h]

\end{table*}\end{center}

\newpage
\begin{center}\begin{table*}[!h]
(Continued from previous table: Time-reversal symmetric, considering SOC)\\
%
\end{table*}\end{center}

\newpage
\begin{center}\begin{table*}[!h]
(Continued from previous table: Time-reversal symmetric, considering SOC)\\
\resizebox{\textwidth}{!}{%
}
\end{table*}\end{center}

\newpage
\begin{center}\begin{table*}[!h]
(Continued from previous table: Time-reversal symmetric, considering SOC)\\
%
\end{table*}\end{center}

\subsubsection{Results for materials with TRS and negligible SOC}
\newpage
\begin{center}\begin{table*}[!h]
(Continued from previous table: Time-reversal symmetric, neglecting SOC)\\
%
\end{table*}\end{center}

\newpage
\begin{center}\begin{table*}[!h]
(Continued from previous table: Time-reversal symmetric, neglecting SOC)\\
%
\end{table*}\end{center}

\newpage
\begin{center}\begin{table*}[!h]
(Continued from previous table: Time-reversal symmetric, neglecting SOC)\\
%
\end{table*}\end{center}

\newpage
\begin{center}\begin{table*}[!h]
(Continued from previous table: Time-reversal symmetric, neglecting SOC)\\
\resizebox{\textwidth}{!}{%
}
\end{table*}\end{center}

\newpage
\begin{center}\begin{table*}[!h]
(Continued from previous table: Time-reversal symmetric, neglecting SOC)\\
%
\end{table*}\end{center}

\subsubsection{Results for materials without TRS and significant SOC}
\newpage
\begin{center}\begin{table*}[!h]
(Continued from previous table: Time-reversal symmetry broken, considering SOC)\\
%
\end{table*}\end{center}

\newpage
\begin{center}\begin{table*}[!h]
(Continued from previous table: Time-reversal symmetry broken, considering SOC)\\
%
\end{table*}\end{center}

\newpage
\begin{center}\begin{table*}[!h]
(Continued from previous table: Time-reversal symmetry broken, considering SOC)\\
%
\end{table*}\end{center}

\newpage
\begin{center}\begin{table*}[!h]
(Continued from previous table: Time-reversal symmetry broken, considering SOC)\\
%
\end{table*}\end{center}

\newpage
\begin{center}\begin{table*}[!h]
(Continued from previous table: Time-reversal symmetry broken, considering SOC)\\
\resizebox{\textwidth}{!}{%
}
\end{table*}\end{center}

\newpage
\begin{center}\begin{table*}[!h]
(Continued from previous table: Time-reversal symmetry broken, considering SOC)\\
%
\end{table*}\end{center}

\subsubsection{Results for materials without TRS and negligible SOC}
\newpage
\begin{center}\begin{table*}[!h]
(Continued from previous table: Time-reversal symmetry broken, neglecting SOC)\\
%
\end{table*}\end{center}

\newpage
\begin{center}\begin{table*}[!h]
(Continued from previous table: Time-reversal symmetry broken, neglecting SOC)\\
%
\end{table*}\end{center}

\newpage
\begin{center}\begin{table*}[!h]
(Continued from previous table: Time-reversal symmetry broken, neglecting SOC)\\
%
\end{table*}\end{center}

\newpage
\begin{center}\begin{table*}[!h]
(Continued from previous table: Time-reversal symmetry broken, neglecting SOC)\\
%
\end{table*}\end{center}

\subsubsection{Results for HSLs with essential straight nodal line structures}\label{esshslnl}
% Table generated by Excel2LaTeX from sheet 'ess_HSL_degenerate_2021'
\newpage
\begin{table}[!h]
  \centering
\resizebox{\textwidth}{!}{
    \begin{tabular}{cl|cl}
    \hline
    \hline
    HSL    & \multicolumn{1}{c|}{SG} & HSL    & \multicolumn{1}{c}{SG} \bigstrut\\
    \hline
    \multicolumn{2}{c|}{Time-reversal symmetry broken,  neglecting SOC} & \multicolumn{2}{c}{Time-reversal symmetric, neglecting SOC} \bigstrut\\
    \hline
    G      & 28,29,31-34,48,50,52,53,55,58,230 & A      & \multicolumn{1}{l}{27,30,34,37,40,41,43,48,49,52,54,56,58,66,68,70} \bigstrut[t]\\
    Q      & 28-31,53,57,60-62,190,192,194 & B      & \multicolumn{1}{l}{27,30,34,37,40,41,43,48,49,52,54,56,58,66,68,70} \\
    H      & 30,32-34,43,48,50,52,55,57,58,60-62,70 & E      & \multicolumn{1}{l}{27,32,49,50,54,57,61} \\
    A      & 40,41,48,49,52-54,56-58,60,61,63,64,66,68,70 & P      & \multicolumn{1}{l}{27,28,30,32,49,50,60,61,147,157,159,162-165,175,176,183-186,189-194} \\
    B      & 48,49,52,54,56,58,62,66,68,70 & D      & \multicolumn{1}{l}{28,29,31-34,48,50,52,53,206,220,230} \\
    C      & 48,50,56,59,60 & G      & \multicolumn{1}{l}{28,29,31-34,48,50,52,53,220,230} \\
\hline
    \hline
    \end{tabular}}
  \label{ess-straightnl}%
\end{table}%

\subsection{Nodal surface}\label{hsl2}
Similar to Sec.~\ref{hsp3}, in this section we list all the positions of HSLs which could host degenerate energy levels implying nodal surfaces on neighboring  HSPLs.

%The first column contain SG numbers. labels for the HSLs. In the second column, it contains the HSL with 2d double-valued irrep(s) which are listed in the third column. The degenerate irrep would not decompose and form the nodal surface structure on the neighboring HSPL, which is listed in the fourth column.
%The results for all result are shown in the table below.
\subsubsection{Results for materials with TRS and significant SOC}\label{hsl2trssoc}

\begin{center}\begin{table*}[!h]
\begin{tabular}{c|c|c|c||c|c|c|c||c|c|c|c}\hline\hline
\multicolumn{1}{l|}{SG} & HSL & Irrep & HSPL & \multicolumn{1}{l|}{SG} & HSL & Irrep & HSPL & \multicolumn{1}{l|}{SG} & HSL & Irrep & HSPL \bigstrut\\
\hline
17&A$(u,0,\frac{1}{2})$&A3,4&W$(u,v,\frac{1}{2})$&17&B$(0,v,\frac{1}{2})$&B3,4&W$(u,v,\frac{1}{2})$&17&E$(u,\frac{1}{2},\frac{1}{2})$&E3,4&W$(u,v,\frac{1}{2})$\bigstrut\\
\hline
17&P$(\frac{1}{2},v,\frac{1}{2})$&P3,4&W$(u,v,\frac{1}{2})$&18&C$(u,\frac{1}{2},0)$&C3,4&N$(u,\frac{1}{2},w)$&18&D$(\frac{1}{2},v,0)$&D3,4&L$(\frac{1}{2},v,w)$\bigstrut\\
\hline
18&E$(u,\frac{1}{2},\frac{1}{2})$&E3,4&N$(u,\frac{1}{2},w)$&18&G$(\frac{1}{2},0,w)$&G3,4&L$(\frac{1}{2},v,w)$&18&H$(0,\frac{1}{2},w)$&H3,4&N$(u,\frac{1}{2},w)$\bigstrut\\
\hline
18&P$(\frac{1}{2},v,\frac{1}{2})$&P3,4&L$(\frac{1}{2},v,w)$&18&Q$(\frac{1}{2},\frac{1}{2},w)$&Q3,3,Q4,4&L$(\frac{1}{2},v,w)$&18&Q$(\frac{1}{2},\frac{1}{2},w)$&Q3,3,Q4,4&N$(u,\frac{1}{2},w)$\bigstrut\\
\hline
19&A$(u,0,\frac{1}{2})$&A3,4&W$(u,v,\frac{1}{2})$&19&B$(0,v,\frac{1}{2})$&B3,4&W$(u,v,\frac{1}{2})$&19&C$(u,\frac{1}{2},0)$&C3,4&N$(u,\frac{1}{2},w)$\bigstrut\\
\hline
19&D$(\frac{1}{2},v,0)$&D3,4&L$(\frac{1}{2},v,w)$&19&E$(u,\frac{1}{2},\frac{1}{2})$&E3,3,E4,4&N$(u,\frac{1}{2},w)$&19&E$(u,\frac{1}{2},\frac{1}{2})$&E3,3,E4,4&W$(u,v,\frac{1}{2})$\bigstrut\\
\hline
19&G$(\frac{1}{2},0,w)$&G3,4&L$(\frac{1}{2},v,w)$&19&H$(0,\frac{1}{2},w)$&H3,4&N$(u,\frac{1}{2},w)$&19&P$(\frac{1}{2},v,\frac{1}{2})$&P3,3,P4,4&L$(\frac{1}{2},v,w)$\bigstrut\\
\hline
19&P$(\frac{1}{2},v,\frac{1}{2})$&P3,3,P4,4&W$(u,v,\frac{1}{2})$&19&Q$(\frac{1}{2},\frac{1}{2},w)$&Q3,3,Q4,4&L$(\frac{1}{2},v,w)$&19&Q$(\frac{1}{2},\frac{1}{2},w)$&Q3,3,Q4,4&N$(u,\frac{1}{2},w)$\bigstrut\\
\hline
20&A$(u,0,\frac{1}{2})$&A3,4&Q$(u,v,\frac{1}{2})$&20&B$(0,v,\frac{1}{2})$&B3,4&Q$(u,v,\frac{1}{2})$&26&A$(u,0,\frac{1}{2})$&A3,4&W$(u,v,\frac{1}{2})$\bigstrut\\
\hline
26&B$(0,v,\frac{1}{2})$&B3,3,B4,4&W$(u,v,\frac{1}{2})$&26&E$(u,\frac{1}{2},\frac{1}{2})$&E3,4&W$(u,v,\frac{1}{2})$&26&P$(\frac{1}{2},v,\frac{1}{2})$&P3,3,P4,4&W$(u,v,\frac{1}{2})$\bigstrut\\
\hline
29&A$(u,0,\frac{1}{2})$&A3,3,A4,4&W$(u,v,\frac{1}{2})$&29&B$(0,v,\frac{1}{2})$&B3,4&W$(u,v,\frac{1}{2})$&29&E$(u,\frac{1}{2},\frac{1}{2})$&E3,3,E4,4&W$(u,v,\frac{1}{2})$\bigstrut\\
\hline
29&P$(\frac{1}{2},v,\frac{1}{2})$&P3,3,P4,4&W$(u,v,\frac{1}{2})$&31&A$(u,0,\frac{1}{2})$&A3,4&W$(u,v,\frac{1}{2})$&31&B$(0,v,\frac{1}{2})$&B3,3,B4,4&W$(u,v,\frac{1}{2})$\bigstrut\\
\hline
31&E$(u,\frac{1}{2},\frac{1}{2})$&E3,4&W$(u,v,\frac{1}{2})$&31&P$(\frac{1}{2},v,\frac{1}{2})$&P3,4&W$(u,v,\frac{1}{2})$&33&A$(u,0,\frac{1}{2})$&A3,3,A4,4&W$(u,v,\frac{1}{2})$\bigstrut\\
\hline
33&B$(0,v,\frac{1}{2})$&B3,4&W$(u,v,\frac{1}{2})$&33&E$(u,\frac{1}{2},\frac{1}{2})$&E3,4&W$(u,v,\frac{1}{2})$&33&P$(\frac{1}{2},v,\frac{1}{2})$&P3,3,P4,4&W$(u,v,\frac{1}{2})$\bigstrut\\
\hline
36&A$(u,0,\frac{1}{2})$&A3,4&Q$(u,v,\frac{1}{2})$&36&B$(0,v,\frac{1}{2})$&B3,3,B4,4&Q$(u,v,\frac{1}{2})$&76&S$(u,u,\frac{1}{2})$&S2,2&E$(u,v,\frac{1}{2})$\bigstrut\\
\hline
76&T$(u,\frac{1}{2},\frac{1}{2})$&T2,2&E$(u,v,\frac{1}{2})$&76&U$(0,v,\frac{1}{2})$&U2,2&E$(u,v,\frac{1}{2})$&78&S$(u,u,\frac{1}{2})$&S2,2&E$(u,v,\frac{1}{2})$\bigstrut\\
\hline
78&T$(u,\frac{1}{2},\frac{1}{2})$&T2,2&E$(u,v,\frac{1}{2})$&78&U$(0,v,\frac{1}{2})$&U2,2&E$(u,v,\frac{1}{2})$&90&T$(u,\frac{1}{2},\frac{1}{2})$&T3,4&F$(u,\frac{1}{2},w)$\bigstrut\\
\hline
90&V$(\frac{1}{2},\frac{1}{2},w)$&V5,6,V7,8&F$(u,\frac{1}{2},w)$&90&W$(0,\frac{1}{2},w)$&W3,4&F$(u,\frac{1}{2},w)$&90&Y$(u,\frac{1}{2},0)$&Y3,4&F$(u,\frac{1}{2},w)$\bigstrut\\
\hline
91&S$(u,u,\frac{1}{2})$&S3,4&E$(u,v,\frac{1}{2})$&91&T$(u,\frac{1}{2},\frac{1}{2})$&T3,4&E$(u,v,\frac{1}{2})$&91&U$(0,v,\frac{1}{2})$&U3,4&E$(u,v,\frac{1}{2})$\bigstrut\\
\hline
92&S$(u,u,\frac{1}{2})$&S3,4&E$(u,v,\frac{1}{2})$&92&T$(u,\frac{1}{2},\frac{1}{2})$&T3,3,T4,4&E$(u,v,\frac{1}{2})$&92&T$(u,\frac{1}{2},\frac{1}{2})$&T3,3,T4,4&F$(u,\frac{1}{2},w)$\bigstrut\\
\hline
92&U$(0,v,\frac{1}{2})$&U3,4&E$(u,v,\frac{1}{2})$&92&V$(\frac{1}{2},\frac{1}{2},w)$&V5,6,V7,8&F$(u,\frac{1}{2},w)$&92&W$(0,\frac{1}{2},w)$&W3,4&F$(u,\frac{1}{2},w)$\bigstrut\\
\hline
92&Y$(u,\frac{1}{2},0)$&Y3,4&F$(u,\frac{1}{2},w)$&94&T$(u,\frac{1}{2},\frac{1}{2})$&T3,4&F$(u,\frac{1}{2},w)$&94&V$(\frac{1}{2},\frac{1}{2},w)$&V5,6,V7,8&F$(u,\frac{1}{2},w)$\bigstrut\\
\hline
94&W$(0,\frac{1}{2},w)$&W3,4&F$(u,\frac{1}{2},w)$&94&Y$(u,\frac{1}{2},0)$&Y3,4&F$(u,\frac{1}{2},w)$&95&S$(u,u,\frac{1}{2})$&S3,4&E$(u,v,\frac{1}{2})$\bigstrut\\
\hline
95&T$(u,\frac{1}{2},\frac{1}{2})$&T3,4&E$(u,v,\frac{1}{2})$&95&U$(0,v,\frac{1}{2})$&U3,4&E$(u,v,\frac{1}{2})$&96&S$(u,u,\frac{1}{2})$&S3,4&E$(u,v,\frac{1}{2})$\bigstrut\\
\hline
96&T$(u,\frac{1}{2},\frac{1}{2})$&T3,3,T4,4&E$(u,v,\frac{1}{2})$&96&T$(u,\frac{1}{2},\frac{1}{2})$&T3,3,T4,4&F$(u,\frac{1}{2},w)$&96&U$(0,v,\frac{1}{2})$&U3,4&E$(u,v,\frac{1}{2})$\bigstrut\\
\hline
96&V$(\frac{1}{2},\frac{1}{2},w)$&V5,6,V7,8&F$(u,\frac{1}{2},w)$&96&W$(0,\frac{1}{2},w)$&W3,4&F$(u,\frac{1}{2},w)$&96&Y$(u,\frac{1}{2},0)$&Y3,4&F$(u,\frac{1}{2},w)$\bigstrut\\
\hline
113&T$(u,\frac{1}{2},\frac{1}{2})$&T3,4&F$(u,\frac{1}{2},w)$&113&W$(0,\frac{1}{2},w)$&W3,4&F$(u,\frac{1}{2},w)$&113&Y$(u,\frac{1}{2},0)$&Y3,4&F$(u,\frac{1}{2},w)$\bigstrut\\
\hline
114&T$(u,\frac{1}{2},\frac{1}{2})$&T3,4&F$(u,\frac{1}{2},w)$&114&W$(0,\frac{1}{2},w)$&W3,4&F$(u,\frac{1}{2},w)$&114&Y$(u,\frac{1}{2},0)$&Y3,4&F$(u,\frac{1}{2},w)$\bigstrut\\
\hline
169&Q$(u,u,\frac{1}{2})$&Q2,2&E$(u,v,\frac{1}{2})$&169&R$(u,0,\frac{1}{2})$&R2,2&E$(u,v,\frac{1}{2})$&170&Q$(u,u,\frac{1}{2})$&Q2,2&E$(u,v,\frac{1}{2})$\bigstrut\\
\hline
170&R$(u,0,\frac{1}{2})$&R2,2&E$(u,v,\frac{1}{2})$&173&Q$(u,u,\frac{1}{2})$&Q2,2&E$(u,v,\frac{1}{2})$&173&R$(u,0,\frac{1}{2})$&R2,2&E$(u,v,\frac{1}{2})$\bigstrut\\
\hline
178&Q$(u,u,\frac{1}{2})$&Q3,4&E$(u,v,\frac{1}{2})$&178&R$(u,0,\frac{1}{2})$&R3,4&E$(u,v,\frac{1}{2})$&179&Q$(u,u,\frac{1}{2})$&Q3,4&E$(u,v,\frac{1}{2})$\bigstrut\\
\hline
179&R$(u,0,\frac{1}{2})$&R3,4&E$(u,v,\frac{1}{2})$&182&Q$(u,u,\frac{1}{2})$&Q3,4&E$(u,v,\frac{1}{2})$&182&R$(u,0,\frac{1}{2})$&R3,4&E$(u,v,\frac{1}{2})$\bigstrut\\
\hline
185&Q$(u,u,\frac{1}{2})$&Q3,3,Q4,4&E$(u,v,\frac{1}{2})$&185&R$(u,0,\frac{1}{2})$&R3,4&E$(u,v,\frac{1}{2})$&186&Q$(u,u,\frac{1}{2})$&Q3,4&E$(u,v,\frac{1}{2})$\bigstrut\\
\hline
186&R$(u,0,\frac{1}{2})$&R3,3,R4,4&E$(u,v,\frac{1}{2})$&198&S$(u,\frac{1}{2},u)$&S2,2&B$(u,\frac{1}{2},w)$&198&T$(\frac{1}{2},\frac{1}{2},w)$&T3,3,T4,4&B$(u,\frac{1}{2},w)$\bigstrut\\
\hline
198&Z$(u,\frac{1}{2},0)$&Z3,4&B$(u,\frac{1}{2},w)$&198&ZA$(\frac{1}{2},u,0)$&ZA3,4&B$(u,\frac{1}{2},w)$&212&S$(u,\frac{1}{2},u)$&S3,4&B$(u,\frac{1}{2},w)$\bigstrut\\
\hline
212&T$(\frac{1}{2},\frac{1}{2},w)$&T5,6,T7,8&B$(u,\frac{1}{2},w)$&212&Z$(u,\frac{1}{2},0)$&Z3,4&B$(u,\frac{1}{2},w)$&213&S$(u,\frac{1}{2},u)$&S3,4&B$(u,\frac{1}{2},w)$\bigstrut\\
\hline
213&Z$(u,\frac{1}{2},0)$&Z3,4&B$(u,\frac{1}{2},w)$& & & & \bigstrut\\
\hline\hline
\end{tabular}
\end{table*}\end{center}

\clearpage
\subsubsection{Results for materials with TRS and negligible SOC}\label{hsl2trsnosoc}

\begin{center}\begin{table*}[!h]

\begin{tabular}{c|c|c|c||c|c|c|c||c|c|c|c}\hline\hline
\multicolumn{1}{l|}{SG} & HSL & Irrep & HSPL & \multicolumn{1}{l|}{SG} & HSL & Irrep & HSPL & \multicolumn{1}{l|}{SG} & HSL & Irrep & HSPL \bigstrut\\
\hline
17&A$(u,0,\frac{1}{2})$&A1,2&W$(u,v,\frac{1}{2})$&17&B$(0,v,\frac{1}{2})$&B1,2&W$(u,v,\frac{1}{2})$&17&E$(u,\frac{1}{2},\frac{1}{2})$&E1,2&W$(u,v,\frac{1}{2})$\bigstrut\\
\hline
17&P$(\frac{1}{2},v,\frac{1}{2})$&P1,2&W$(u,v,\frac{1}{2})$&18&C$(u,\frac{1}{2},0)$&C1,2&N$(u,\frac{1}{2},w)$&18&D$(\frac{1}{2},v,0)$&D1,2&L$(\frac{1}{2},v,w)$\bigstrut\\
\hline
18&E$(u,\frac{1}{2},\frac{1}{2})$&E1,2&N$(u,\frac{1}{2},w)$&18&G$(\frac{1}{2},0,w)$&G1,2&L$(\frac{1}{2},v,w)$&18&H$(0,\frac{1}{2},w)$&H1,2&N$(u,\frac{1}{2},w)$\bigstrut\\
\hline
18&P$(\frac{1}{2},v,\frac{1}{2})$&P1,2&L$(\frac{1}{2},v,w)$&18&Q$(\frac{1}{2},\frac{1}{2},w)$&Q1,1,Q2,2&L$(\frac{1}{2},v,w)$&18&Q$(\frac{1}{2},\frac{1}{2},w)$&Q1,1,Q2,2&N$(u,\frac{1}{2},w)$\bigstrut\\
\hline
19&A$(u,0,\frac{1}{2})$&A1,2&W$(u,v,\frac{1}{2})$&19&B$(0,v,\frac{1}{2})$&B1,2&W$(u,v,\frac{1}{2})$&19&C$(u,\frac{1}{2},0)$&C1,2&N$(u,\frac{1}{2},w)$\bigstrut\\
\hline
19&D$(\frac{1}{2},v,0)$&D1,2&L$(\frac{1}{2},v,w)$&19&E$(u,\frac{1}{2},\frac{1}{2})$&E1,1,E2,2&N$(u,\frac{1}{2},w)$&19&E$(u,\frac{1}{2},\frac{1}{2})$&E1,1,E2,2&W$(u,v,\frac{1}{2})$\bigstrut\\
\hline
19&G$(\frac{1}{2},0,w)$&G1,2&L$(\frac{1}{2},v,w)$&19&H$(0,\frac{1}{2},w)$&H1,2&N$(u,\frac{1}{2},w)$&19&P$(\frac{1}{2},v,\frac{1}{2})$&P1,1,P2,2&L$(\frac{1}{2},v,w)$\bigstrut\\
\hline
19&P$(\frac{1}{2},v,\frac{1}{2})$&P1,1,P2,2&W$(u,v,\frac{1}{2})$&19&Q$(\frac{1}{2},\frac{1}{2},w)$&Q1,1,Q2,2&L$(\frac{1}{2},v,w)$&19&Q$(\frac{1}{2},\frac{1}{2},w)$&Q1,1,Q2,2&N$(u,\frac{1}{2},w)$\bigstrut\\
\hline
20&A$(u,0,\frac{1}{2})$&A1,2&Q$(u,v,\frac{1}{2})$&20&B$(0,v,\frac{1}{2})$&B1,2&Q$(u,v,\frac{1}{2})$&26&A$(u,0,\frac{1}{2})$&A1,2&W$(u,v,\frac{1}{2})$\bigstrut\\
\hline
26&B$(0,v,\frac{1}{2})$&B1,1,B2,2&W$(u,v,\frac{1}{2})$&26&E$(u,\frac{1}{2},\frac{1}{2})$&E1,2&W$(u,v,\frac{1}{2})$&26&P$(\frac{1}{2},v,\frac{1}{2})$&P1,1,P2,2&W$(u,v,\frac{1}{2})$\bigstrut\\
\hline
29&A$(u,0,\frac{1}{2})$&A1,1,A2,2&W$(u,v,\frac{1}{2})$&29&B$(0,v,\frac{1}{2})$&B1,2&W$(u,v,\frac{1}{2})$&29&E$(u,\frac{1}{2},\frac{1}{2})$&E1,1,E2,2&W$(u,v,\frac{1}{2})$\bigstrut\\
\hline
29&P$(\frac{1}{2},v,\frac{1}{2})$&P1,1,P2,2&W$(u,v,\frac{1}{2})$&31&A$(u,0,\frac{1}{2})$&A1,2&W$(u,v,\frac{1}{2})$&31&B$(0,v,\frac{1}{2})$&B1,1,B2,2&W$(u,v,\frac{1}{2})$\bigstrut\\
\hline
31&E$(u,\frac{1}{2},\frac{1}{2})$&E1,2&W$(u,v,\frac{1}{2})$&31&P$(\frac{1}{2},v,\frac{1}{2})$&P1,2&W$(u,v,\frac{1}{2})$&33&A$(u,0,\frac{1}{2})$&A1,1,A2,2&W$(u,v,\frac{1}{2})$\bigstrut\\
\hline
33&B$(0,v,\frac{1}{2})$&B1,2&W$(u,v,\frac{1}{2})$&33&E$(u,\frac{1}{2},\frac{1}{2})$&E1,2&W$(u,v,\frac{1}{2})$&33&P$(\frac{1}{2},v,\frac{1}{2})$&P1,1,P2,2&W$(u,v,\frac{1}{2})$\bigstrut\\
\hline
36&A$(u,0,\frac{1}{2})$&A1,2&Q$(u,v,\frac{1}{2})$&36&B$(0,v,\frac{1}{2})$&B1,1,B2,2&Q$(u,v,\frac{1}{2})$&51&D$(\frac{1}{2},v,0)$&D1&L$(\frac{1}{2},v,w)$\bigstrut\\
\hline
51&G$(\frac{1}{2},0,w)$&G1,4,G2,3&L$(\frac{1}{2},v,w)$&51&P$(\frac{1}{2},v,\frac{1}{2})$&P1&L$(\frac{1}{2},v,w)$&51&Q$(\frac{1}{2},\frac{1}{2},w)$&Q1,4,Q2,3&L$(\frac{1}{2},v,w)$\bigstrut\\
\hline
52&C$(u,\frac{1}{2},0)$&C1,4,C2,3&N$(u,\frac{1}{2},w)$&52&E$(u,\frac{1}{2},\frac{1}{2})$&E1&N$(u,\frac{1}{2},w)$&52&H$(0,\frac{1}{2},w)$&H1&N$(u,\frac{1}{2},w)$\bigstrut\\
\hline
52&Q$(\frac{1}{2},\frac{1}{2},w)$&Q1,3,Q2,4&N$(u,\frac{1}{2},w)$&53&A$(u,0,\frac{1}{2})$&A1&W$(u,v,\frac{1}{2})$&53&B$(0,v,\frac{1}{2})$&B1,3,B2,4&W$(u,v,\frac{1}{2})$\bigstrut\\
\hline
53&E$(u,\frac{1}{2},\frac{1}{2})$&E1&W$(u,v,\frac{1}{2})$&53&P$(\frac{1}{2},v,\frac{1}{2})$&P1&W$(u,v,\frac{1}{2})$&54&D$(\frac{1}{2},v,0)$&D1&L$(\frac{1}{2},v,w)$\bigstrut\\
\hline
54&G$(\frac{1}{2},0,w)$&G1,4,G2,3&L$(\frac{1}{2},v,w)$&54&P$(\frac{1}{2},v,\frac{1}{2})$&P1,4,P2,3&L$(\frac{1}{2},v,w)$&54&Q$(\frac{1}{2},\frac{1}{2},w)$&Q1,4,Q2,3&L$(\frac{1}{2},v,w)$\bigstrut\\
\hline
55&C$(u,\frac{1}{2},0)$&C1,4,C2,3&N$(u,\frac{1}{2},w)$&55&D$(\frac{1}{2},v,0)$&D1,4,D2,3&L$(\frac{1}{2},v,w)$&55&E$(u,\frac{1}{2},\frac{1}{2})$&E1,4,E2,3&N$(u,\frac{1}{2},w)$\bigstrut\\
\hline
55&G$(\frac{1}{2},0,w)$&G1&L$(\frac{1}{2},v,w)$&55&H$(0,\frac{1}{2},w)$&H1&N$(u,\frac{1}{2},w)$&55&P$(\frac{1}{2},v,\frac{1}{2})$&P1,4,P2,3&L$(\frac{1}{2},v,w)$\bigstrut\\
\hline
55&Q$(\frac{1}{2},\frac{1}{2},w)$&Q1,2,Q3,4&L$(\frac{1}{2},v,w)$&55&Q$(\frac{1}{2},\frac{1}{2},w)$&Q1,2,Q3,4&N$(u,\frac{1}{2},w)$&56&C$(u,\frac{1}{2},0)$&C1&N$(u,\frac{1}{2},w)$\bigstrut\\
\hline
56&D$(\frac{1}{2},v,0)$&D1&L$(\frac{1}{2},v,w)$&56&E$(u,\frac{1}{2},\frac{1}{2})$&E1,4,E2,3&N$(u,\frac{1}{2},w)$&56&G$(\frac{1}{2},0,w)$&G1,4,G2,3&L$(\frac{1}{2},v,w)$\bigstrut\\
\hline
56&H$(0,\frac{1}{2},w)$&H1,3,H2,4&N$(u,\frac{1}{2},w)$&56&P$(\frac{1}{2},v,\frac{1}{2})$&P1,4,P2,3&L$(\frac{1}{2},v,w)$&56&Q$(\frac{1}{2},\frac{1}{2},w)$&Q1,2,Q3,4&L$(\frac{1}{2},v,w)$\bigstrut\\
\hline
56&Q$(\frac{1}{2},\frac{1}{2},w)$&Q1,2,Q3,4&N$(u,\frac{1}{2},w)$&57&A$(u,0,\frac{1}{2})$&A1&W$(u,v,\frac{1}{2})$&57&B$(0,v,\frac{1}{2})$&B1,3,B2,4&W$(u,v,\frac{1}{2})$\bigstrut\\
\hline
57&C$(u,\frac{1}{2},0)$&C1,4,C2,3&N$(u,\frac{1}{2},w)$&57&H$(0,\frac{1}{2},w)$&H1&N$(u,\frac{1}{2},w)$&57&P$(\frac{1}{2},v,\frac{1}{2})$&P1,3,P2,4&W$(u,v,\frac{1}{2})$\bigstrut\\
\hline
57&Q$(\frac{1}{2},\frac{1}{2},w)$&Q1&N$(u,\frac{1}{2},w)$&58&C$(u,\frac{1}{2},0)$&C1,4,C2,3&N$(u,\frac{1}{2},w)$&58&D$(\frac{1}{2},v,0)$&D1,4,D2,3&L$(\frac{1}{2},v,w)$\bigstrut\\
\hline
58&E$(u,\frac{1}{2},\frac{1}{2})$&E1&N$(u,\frac{1}{2},w)$&58&G$(\frac{1}{2},0,w)$&G1&L$(\frac{1}{2},v,w)$&58&H$(0,\frac{1}{2},w)$&H1&N$(u,\frac{1}{2},w)$\bigstrut\\
\hline
58&P$(\frac{1}{2},v,\frac{1}{2})$&P1&L$(\frac{1}{2},v,w)$&58&Q$(\frac{1}{2},\frac{1}{2},w)$&Q1,2,Q3,4&L$(\frac{1}{2},v,w)$&58&Q$(\frac{1}{2},\frac{1}{2},w)$&Q1,2,Q3,4&N$(u,\frac{1}{2},w)$\bigstrut\\
\hline
59&C$(u,\frac{1}{2},0)$&C1&N$(u,\frac{1}{2},w)$&59&D$(\frac{1}{2},v,0)$&D1&L$(\frac{1}{2},v,w)$&59&E$(u,\frac{1}{2},\frac{1}{2})$&E1&N$(u,\frac{1}{2},w)$\bigstrut\\
\hline
59&G$(\frac{1}{2},0,w)$&G1,4,G2,3&L$(\frac{1}{2},v,w)$&59&H$(0,\frac{1}{2},w)$&H1,3,H2,4&N$(u,\frac{1}{2},w)$&59&P$(\frac{1}{2},v,\frac{1}{2})$&P1&L$(\frac{1}{2},v,w)$\bigstrut\\
\hline
59&Q$(\frac{1}{2},\frac{1}{2},w)$&Q1,2,Q3,4&L$(\frac{1}{2},v,w)$&59&Q$(\frac{1}{2},\frac{1}{2},w)$&Q1,2,Q3,4&N$(u,\frac{1}{2},w)$&60&A$(u,0,\frac{1}{2})$&A1&W$(u,v,\frac{1}{2})$\bigstrut\\
\hline
60&B$(0,v,\frac{1}{2})$&B1,3,B2,4&W$(u,v,\frac{1}{2})$&60&D$(\frac{1}{2},v,0)$&D1&L$(\frac{1}{2},v,w)$&60&E$(u,\frac{1}{2},\frac{1}{2})$&E1,3,E2,4&W$(u,v,\frac{1}{2})$\bigstrut\\
\hline
60&G$(\frac{1}{2},0,w)$&G1,4,G2,3&L$(\frac{1}{2},v,w)$&60&Q$(\frac{1}{2},\frac{1}{2},w)$&Q1&L$(\frac{1}{2},v,w)$&61&A$(u,0,\frac{1}{2})$&A1&W$(u,v,\frac{1}{2})$\bigstrut\\
\hline
61&B$(0,v,\frac{1}{2})$&B1,3,B2,4&W$(u,v,\frac{1}{2})$&61&C$(u,\frac{1}{2},0)$&C1,4,C2,3&N$(u,\frac{1}{2},w)$&61&D$(\frac{1}{2},v,0)$&D1&L$(\frac{1}{2},v,w)$\bigstrut\\
\hline
61&G$(\frac{1}{2},0,w)$&G1,4,G2,3&L$(\frac{1}{2},v,w)$&61&H$(0,\frac{1}{2},w)$&H1&N$(u,\frac{1}{2},w)$&62&A$(u,0,\frac{1}{2})$&A1,3,A2,4&W$(u,v,\frac{1}{2})$\bigstrut\\
\hline
62&B$(0,v,\frac{1}{2})$&B1&W$(u,v,\frac{1}{2})$&62&C$(u,\frac{1}{2},0)$&C1,4,C2,3&N$(u,\frac{1}{2},w)$&62&D$(\frac{1}{2},v,0)$&D1&L$(\frac{1}{2},v,w)$\bigstrut\\
\hline
62&E$(u,\frac{1}{2},\frac{1}{2})$&E1,2,E3,4&N$(u,\frac{1}{2},w)$&62&E$(u,\frac{1}{2},\frac{1}{2})$&E1,2,E3,4&W$(u,v,\frac{1}{2})$&62&G$(\frac{1}{2},0,w)$&G1,4,G2,3&L$(\frac{1}{2},v,w)$\bigstrut\\
\hline
62&H$(0,\frac{1}{2},w)$&H1&N$(u,\frac{1}{2},w)$&62&P$(\frac{1}{2},v,\frac{1}{2})$&P1,2,P3,4&L$(\frac{1}{2},v,w)$&62&P$(\frac{1}{2},v,\frac{1}{2})$&P1,2,P3,4&W$(u,v,\frac{1}{2})$\bigstrut\\
\hline
\hline
\end{tabular}
\end{table*}\end{center}

\newpage
\begin{center}\begin{table*}[!h]
(Continued from previous table: Time-reversal symmetric, neglecting SOC)\\
\begin{tabular}{c|c|c|c||c|c|c|c||c|c|c|c}\hline\hline
\multicolumn{1}{l|}{SG} & HSL & Irrep & HSPL & \multicolumn{1}{l|}{SG} & HSL & Irrep & HSPL & \multicolumn{1}{l|}{SG} & HSL & Irrep & HSPL \bigstrut\\
\hline
63&A$(u,0,\frac{1}{2})$&A1&Q$(u,v,\frac{1}{2})$&63&B$(0,v,\frac{1}{2})$&B1,3,B2,4&Q$(u,v,\frac{1}{2})$&64&A$(u,0,\frac{1}{2})$&A1&Q$(u,v,\frac{1}{2})$\bigstrut\\
\hline
64&B$(0,v,\frac{1}{2})$&B1,3,B2,4&Q$(u,v,\frac{1}{2})$&76&S$(u,u,\frac{1}{2})$&S1,1&E$(u,v,\frac{1}{2})$&76&T$(u,\frac{1}{2},\frac{1}{2})$&T1,1&E$(u,v,\frac{1}{2})$\bigstrut\\
\hline
76&U$(0,v,\frac{1}{2})$&U1,1&E$(u,v,\frac{1}{2})$&78&S$(u,u,\frac{1}{2})$&S1,1&E$(u,v,\frac{1}{2})$&78&T$(u,\frac{1}{2},\frac{1}{2})$&T1,1&E$(u,v,\frac{1}{2})$\bigstrut\\
\hline
78&U$(0,v,\frac{1}{2})$&U1,1&E$(u,v,\frac{1}{2})$&90&T$(u,\frac{1}{2},\frac{1}{2})$&T1,2&F$(u,\frac{1}{2},w)$&90&V$(\frac{1}{2},\frac{1}{2},w)$&V1,2,V3,4&F$(u,\frac{1}{2},w)$\bigstrut\\
\hline
90&W$(0,\frac{1}{2},w)$&W1,2&F$(u,\frac{1}{2},w)$&90&Y$(u,\frac{1}{2},0)$&Y1,2&F$(u,\frac{1}{2},w)$&91&S$(u,u,\frac{1}{2})$&S1,2&E$(u,v,\frac{1}{2})$\bigstrut\\
\hline
91&T$(u,\frac{1}{2},\frac{1}{2})$&T1,2&E$(u,v,\frac{1}{2})$&91&U$(0,v,\frac{1}{2})$&U1,2&E$(u,v,\frac{1}{2})$&92&S$(u,u,\frac{1}{2})$&S1,2&E$(u,v,\frac{1}{2})$\bigstrut\\
\hline
92&T$(u,\frac{1}{2},\frac{1}{2})$&T1,1,T2,2&E$(u,v,\frac{1}{2})$&92&T$(u,\frac{1}{2},\frac{1}{2})$&T1,1,T2,2&F$(u,\frac{1}{2},w)$&92&U$(0,v,\frac{1}{2})$&U1,2&E$(u,v,\frac{1}{2})$\bigstrut\\
\hline
92&V$(\frac{1}{2},\frac{1}{2},w)$&V1,2,V3,4&F$(u,\frac{1}{2},w)$&92&W$(0,\frac{1}{2},w)$&W1,2&F$(u,\frac{1}{2},w)$&92&Y$(u,\frac{1}{2},0)$&Y1,2&F$(u,\frac{1}{2},w)$\bigstrut\\
\hline
94&T$(u,\frac{1}{2},\frac{1}{2})$&T1,2&F$(u,\frac{1}{2},w)$&94&V$(\frac{1}{2},\frac{1}{2},w)$&V1,2,V3,4&F$(u,\frac{1}{2},w)$&94&W$(0,\frac{1}{2},w)$&W1,2&F$(u,\frac{1}{2},w)$\bigstrut\\
\hline
94&Y$(u,\frac{1}{2},0)$&Y1,2&F$(u,\frac{1}{2},w)$&95&S$(u,u,\frac{1}{2})$&S1,2&E$(u,v,\frac{1}{2})$&95&T$(u,\frac{1}{2},\frac{1}{2})$&T1,2&E$(u,v,\frac{1}{2})$\bigstrut\\
\hline
95&U$(0,v,\frac{1}{2})$&U1,2&E$(u,v,\frac{1}{2})$&96&S$(u,u,\frac{1}{2})$&S1,2&E$(u,v,\frac{1}{2})$&96&T$(u,\frac{1}{2},\frac{1}{2})$&T1,1,T2,2&E$(u,v,\frac{1}{2})$\bigstrut\\
\hline
96&T$(u,\frac{1}{2},\frac{1}{2})$&T1,1,T2,2&F$(u,\frac{1}{2},w)$&96&U$(0,v,\frac{1}{2})$&U1,2&E$(u,v,\frac{1}{2})$&96&V$(\frac{1}{2},\frac{1}{2},w)$&V1,2,V3,4&F$(u,\frac{1}{2},w)$\bigstrut\\
\hline
96&W$(0,\frac{1}{2},w)$&W1,2&F$(u,\frac{1}{2},w)$&96&Y$(u,\frac{1}{2},0)$&Y1,2&F$(u,\frac{1}{2},w)$&113&T$(u,\frac{1}{2},\frac{1}{2})$&T1,2&F$(u,\frac{1}{2},w)$\bigstrut\\
\hline
113&V$(\frac{1}{2},\frac{1}{2},w)$&V1,2,V3,3,V4,4&F$(u,\frac{1}{2},w)$&113&W$(0,\frac{1}{2},w)$&W1,2&F$(u,\frac{1}{2},w)$&113&Y$(u,\frac{1}{2},0)$&Y1,2&F$(u,\frac{1}{2},w)$\bigstrut\\
\hline
114&T$(u,\frac{1}{2},\frac{1}{2})$&T1,2&F$(u,\frac{1}{2},w)$&114&V$(\frac{1}{2},\frac{1}{2},w)$&V1,2,V3,3,V4,4&F$(u,\frac{1}{2},w)$&114&W$(0,\frac{1}{2},w)$&W1,2&F$(u,\frac{1}{2},w)$\bigstrut\\
\hline
114&Y$(u,\frac{1}{2},0)$&Y1,2&F$(u,\frac{1}{2},w)$&127&T$(u,\frac{1}{2},\frac{1}{2})$&T1,4,T2,3&F$(u,\frac{1}{2},w)$&127&V$(\frac{1}{2},\frac{1}{2},w)$&V1,3,V2,4,V5&F$(u,\frac{1}{2},w)$\bigstrut\\
\hline
127&W$(0,\frac{1}{2},w)$&W1&F$(u,\frac{1}{2},w)$&127&Y$(u,\frac{1}{2},0)$&Y1,4,Y2,3&F$(u,\frac{1}{2},w)$&128&T$(u,\frac{1}{2},\frac{1}{2})$&T1&F$(u,\frac{1}{2},w)$\bigstrut\\
\hline
128&V$(\frac{1}{2},\frac{1}{2},w)$&V1,3,V2,4,V5&F$(u,\frac{1}{2},w)$&128&W$(0,\frac{1}{2},w)$&W1&F$(u,\frac{1}{2},w)$&128&Y$(u,\frac{1}{2},0)$&Y1,4,Y2,3&F$(u,\frac{1}{2},w)$\bigstrut\\
\hline
129&T$(u,\frac{1}{2},\frac{1}{2})$&T1&F$(u,\frac{1}{2},w)$&129&V$(\frac{1}{2},\frac{1}{2},w)$&V1,3,V2,4,V5&F$(u,\frac{1}{2},w)$&129&W$(0,\frac{1}{2},w)$&W1,3,W2,4&F$(u,\frac{1}{2},w)$\bigstrut\\
\hline
129&Y$(u,\frac{1}{2},0)$&Y1&F$(u,\frac{1}{2},w)$&130&T$(u,\frac{1}{2},\frac{1}{2})$&T1,4,T2,3&F$(u,\frac{1}{2},w)$&130&V$(\frac{1}{2},\frac{1}{2},w)$&V1,3,V2,4,V5&F$(u,\frac{1}{2},w)$\bigstrut\\
\hline
130&W$(0,\frac{1}{2},w)$&W1,3,W2,4&F$(u,\frac{1}{2},w)$&130&Y$(u,\frac{1}{2},0)$&Y1&F$(u,\frac{1}{2},w)$&135&T$(u,\frac{1}{2},\frac{1}{2})$&T1,4,T2,3&F$(u,\frac{1}{2},w)$\bigstrut\\
\hline
135&V$(\frac{1}{2},\frac{1}{2},w)$&V1,3,V2,4,V5&F$(u,\frac{1}{2},w)$&135&W$(0,\frac{1}{2},w)$&W1&F$(u,\frac{1}{2},w)$&135&Y$(u,\frac{1}{2},0)$&Y1,4,Y2,3&F$(u,\frac{1}{2},w)$\bigstrut\\
\hline
136&T$(u,\frac{1}{2},\frac{1}{2})$&T1&F$(u,\frac{1}{2},w)$&136&V$(\frac{1}{2},\frac{1}{2},w)$&V1,3,V2,4,V5&F$(u,\frac{1}{2},w)$&136&W$(0,\frac{1}{2},w)$&W1&F$(u,\frac{1}{2},w)$\bigstrut\\
\hline
136&Y$(u,\frac{1}{2},0)$&Y1,4,Y2,3&F$(u,\frac{1}{2},w)$&137&T$(u,\frac{1}{2},\frac{1}{2})$&T1&F$(u,\frac{1}{2},w)$&137&V$(\frac{1}{2},\frac{1}{2},w)$&V1,3,V2,4,V5&F$(u,\frac{1}{2},w)$\bigstrut\\
\hline
137&W$(0,\frac{1}{2},w)$&W1,3,W2,4&F$(u,\frac{1}{2},w)$&137&Y$(u,\frac{1}{2},0)$&Y1&F$(u,\frac{1}{2},w)$&138&T$(u,\frac{1}{2},\frac{1}{2})$&T1,4,T2,3&F$(u,\frac{1}{2},w)$\bigstrut\\
\hline
138&V$(\frac{1}{2},\frac{1}{2},w)$&V1,3,V2,4,V5&F$(u,\frac{1}{2},w)$&138&W$(0,\frac{1}{2},w)$&W1,3,W2,4&F$(u,\frac{1}{2},w)$&138&Y$(u,\frac{1}{2},0)$&Y1&F$(u,\frac{1}{2},w)$\bigstrut\\
\hline
169&Q$(u,u,\frac{1}{2})$&Q1,1&E$(u,v,\frac{1}{2})$&169&R$(u,0,\frac{1}{2})$&R1,1&E$(u,v,\frac{1}{2})$&170&Q$(u,u,\frac{1}{2})$&Q1,1&E$(u,v,\frac{1}{2})$\bigstrut\\
\hline
170&R$(u,0,\frac{1}{2})$&R1,1&E$(u,v,\frac{1}{2})$&173&Q$(u,u,\frac{1}{2})$&Q1,1&E$(u,v,\frac{1}{2})$&173&R$(u,0,\frac{1}{2})$&R1,1&E$(u,v,\frac{1}{2})$\bigstrut\\
\hline
176&Q$(u,u,\frac{1}{2})$&Q1,2&E$(u,v,\frac{1}{2})$&176&R$(u,0,\frac{1}{2})$&R1,2&E$(u,v,\frac{1}{2})$&178&Q$(u,u,\frac{1}{2})$&Q1,2&E$(u,v,\frac{1}{2})$\bigstrut\\
\hline
178&R$(u,0,\frac{1}{2})$&R1,2&E$(u,v,\frac{1}{2})$&179&Q$(u,u,\frac{1}{2})$&Q1,2&E$(u,v,\frac{1}{2})$&179&R$(u,0,\frac{1}{2})$&R1,2&E$(u,v,\frac{1}{2})$\bigstrut\\
\hline
182&Q$(u,u,\frac{1}{2})$&Q1,2&E$(u,v,\frac{1}{2})$&182&R$(u,0,\frac{1}{2})$&R1,2&E$(u,v,\frac{1}{2})$&185&Q$(u,u,\frac{1}{2})$&Q1,1,Q2,2&E$(u,v,\frac{1}{2})$\bigstrut\\
\hline
185&R$(u,0,\frac{1}{2})$&R1,2&E$(u,v,\frac{1}{2})$&186&Q$(u,u,\frac{1}{2})$&Q1,2&E$(u,v,\frac{1}{2})$&186&R$(u,0,\frac{1}{2})$&R1,1,R2,2&E$(u,v,\frac{1}{2})$\bigstrut\\
\hline
193&Q$(u,u,\frac{1}{2})$&Q1,3,Q2,4&E$(u,v,\frac{1}{2})$&193&R$(u,0,\frac{1}{2})$&R1&E$(u,v,\frac{1}{2})$&194&Q$(u,u,\frac{1}{2})$&Q1&E$(u,v,\frac{1}{2})$\bigstrut\\
\hline
194&R$(u,0,\frac{1}{2})$&R1,3,R2,4&E$(u,v,\frac{1}{2})$&198&S$(u,\frac{1}{2},u)$&S1,1&B$(u,\frac{1}{2},w)$&198&T$(\frac{1}{2},\frac{1}{2},w)$&T1,1,T2,2&B$(u,\frac{1}{2},w)$\bigstrut\\
\hline
198&Z$(u,\frac{1}{2},0)$&Z1,2&B$(u,\frac{1}{2},w)$&198&ZA$(\frac{1}{2},u,0)$&ZA1,2&B$(u,\frac{1}{2},w)$&205&S$(u,\frac{1}{2},u)$&S1,2&B$(u,\frac{1}{2},w)$\bigstrut\\
\hline
205&Z$(u,\frac{1}{2},0)$&Z1,4,Z2,3&B$(u,\frac{1}{2},w)$&205&ZA$(\frac{1}{2},u,0)$&ZA1&B$(u,\frac{1}{2},w)$&212&S$(u,\frac{1}{2},u)$&S1,2&B$(u,\frac{1}{2},w)$\bigstrut\\
\hline
212&T$(\frac{1}{2},\frac{1}{2},w)$&T1,2,T3,4&B$(u,\frac{1}{2},w)$&212&Z$(u,\frac{1}{2},0)$&Z1,2&B$(u,\frac{1}{2},w)$&213&S$(u,\frac{1}{2},u)$&S1,2&B$(u,\frac{1}{2},w)$\bigstrut\\
\hline
213&Z$(u,\frac{1}{2},0)$&Z1,2&B$(u,\frac{1}{2},w)$& & & & \bigstrut\\
\hline\hline
\end{tabular}
\end{table*}\end{center}

\subsection{Nodal point}\label{hsl3}
In this section, we list all positions of HSLs which can host nodal point(s).
%We first briefly introduce how to obtain the results of all HSLs.
%After filtering all k-vectors with non-trivial little groups to obtain all HSLs, we traverse all irreps in the little group of HSLs to obtain HSLs with two or more different irreps which could form possible BCs. When we continuously move from a point on HSL to a point lying in HSPL or a general point, the BC must decompose, and a possible nodal point could exist on HSL that satisfy the above condition.
%The results for all HSLs with isolated nodal points are shown in the table below. The first column contains labels for the HSLs and the second column contains SG numbers for the corresponding HSLs. SGs hosting four-fold degenerate nodal loops are printed in bold style (which occurs only when TRS exist) while for others the degeneracy is simply two.
\clearpage
\begin{table}[!bhtp]
  \centering
\resizebox{\textwidth}{!}{
    \begin{tabular}{ll||ll}
    \hline
    \hline
    \multicolumn{1}{c}{HSL} & \multicolumn{1}{c||}{SG} & \multicolumn{1}{c}{HSL} & \multicolumn{1}{c}{SG} \\
    \hline
    \multicolumn{2}{c||}{Time-reversal symmetry broken,  neglecting SOC} & \multicolumn{2}{c}{Time-reversal symmetry broken,  considering SOC} \\
    \hline
    A      & 16-22  & A      & 16-22 \\
    \hline
    B      & 16-22  & B      & 16-22 \\
    \hline
    C      & 16-19  & C      & 16-19 \\
    \hline
    D      & 16-19,197,199,204,206,211,214,217,229 & D      & 16-19,197,199,204,206,211,214,220,230 \\
    \hline
    DT     & 16-24,89-98,111-114,119,120,143-145,147,149-154,168-182,195-220 & \multirow{2}[4]{*}{DT} & 16-24,89-98,111-114,119,120,143-145,147,149-154,168-182, \\
\cline{1-2}    E      & 16-19  &        & 195-199,207-214 \\
    \hline
    G      & 16-19,211,214 & E      & 16-19 \\
    \hline
    H      & 16-22  & G      & 16-19,211,214 \\
    \hline
    \multirow{2}[4]{*}{LD} & 16-24,75-98,111-122,146,148,150,152,154,155,164,165,177-182, & H      & 16-22 \\
\cline{3-4}           & 195-214 & LD     & 16-24,75-98,146,148,150,152,154,155,164,165,177-182,195-214 \\
    \hline
    P      & 16-19,143-145,147,149-154,156,158,164,165,168-182,187,188 & P      & 16-19,143-145,147,149-154,156,158,164,165,168-182,187,188 \\
    \hline
    Q      & 16-19,150,152,154,164,165,177-182 & Q      & 16-19,150,152,154,164,165,177-182 \\
    \hline
    \multirow{2}[2]{*}{SM} & 16-24,89-98,115-118,121,122,149,151,153,162,163,177-182, & \multirow{2}[2]{*}{SM} & 16-24,89-98,115-118,121,122,149,151,153,162,163,177-182, \\
           & 207-214 &        & 207-214 \\
    \hline
    V      & 75-78,81,83-86,89-96,111-118,196,202,209,210,216,219,225,226 & V      & 75-78,81,83-86,89-96,196,203,209,210,216,219,227,228 \\
    \hline
    W      & 75-98,107,108,111-114,119-121,139,140 & W      & 75-98,109-114,119,120,122,141,142 \\
    \hline
    S      & 89-96,115-118,207,208,212,213 & S      & 89-96,115-118,207,208,212,213 \\
    \hline
    T      & 89-96,111-114,195,198,200,201,207,208,212,213,215,218 & T      & 89-96,111-114,195,198,205,207,208,212,213 \\
    \hline
    U      & 89-96,111-114,168-173,175-182 & U      & 89-96,111-114,168-173,175-182 \\
    \hline
    Y      & 89-98,111-114,119,120 & Y      & 89-98,111-114,119,120 \\
    \hline
    PA     & 143-145,150,152,154,174 & PA     & 143-145,150,152,154,174 \\
    \hline
    R      & 149,151,153,162,163,177-182 & R      & 149,151,153,162,163,177-182 \\
    \hline
    Z      & 195,198,207,208,212,213,215,218 & Z      & 195,198,207,208,212,213,215,218 \\
    \hline
    ZA     & 195,198 & ZA     & 195,198 \\
    \hline
    \multicolumn{2}{c||}{Time-reversal symmetric, neglecting SOC} & \multicolumn{2}{c}{Time-reversal symmetric, considering SOC} \\
    \hline
    A      & 16,18,21,22 & C      & 16,17,\textbf{48},\textbf{50},\textbf{60} \\
    \hline
    B      & 16,18,21,22 & D      & 16,17,\textbf{48},\textbf{50},\textbf{52},\textbf{53},197,199,\textbf{206},211,214,220,\textbf{230} \\
    \hline
    C      & 16,17  & \multirow{2}[4]{*}{DT} & 16-24,89-98,111-114,119,120,143-145,\textbf{147},149-154,\textbf{162-165}, \\
\cline{1-2}    D      & 16,17,197,199,204,211,214,217,229 &        & 168-173,\textbf{175},\textbf{176},177-182,\textbf{191-194},195-199,207-214,\textbf{221-230} \\
    \hline
    \multirow{2}[4]{*}{DT} & 16-24,89-98,111-114,119,120,143-145,149-154,168-173,175-182, & E      & 16,\textbf{19},\textbf{49},\textbf{50},\textbf{54} \\
\cline{3-4}           & 195-214 & G      & 16,17,\textbf{48},\textbf{50},\textbf{52},\textbf{53},211,214,\textbf{230} \\
    \hline
    E      & 16,\textbf{19},\textbf{62} & H      & 16,17,20,21,22,\textbf{48},\textbf{50},\textbf{60},\textbf{70} \\
    \hline
    G      & 16,17,211,214 & \multirow{2}[4]{*}{LD} & 16-24,75-80,\textbf{83-88},89-98,\textbf{123-142},146,\textbf{148},150,152,154,155, \\
\cline{1-2}    H      & 16,17,20-22 &        & \textbf{166},\textbf{167},177-182,195-199,\textbf{200-206},207-214,\textbf{221-230} \\
    \hline
    \multirow{2}[4]{*}{LD} & 16-24,75-80,83-98,146,150,152,154,155,164,165,177-182,195-199, & \multirow{2}[4]{*}{P} & 16,\textbf{19},\textbf{49},\textbf{50},143-145,\textbf{147},149-154,156,158,\textbf{162-165},168-174, \\
\cline{2-2}           & 207-214 &        & \textbf{175},\textbf{176},177-182,187,188,\textbf{191-194} \\
    \hline
    P      & 16,\textbf{19},\textbf{62},143-145,149-154,156,158,168-174,177-182,187,188 & Q      & 16,17,\textbf{18},\textbf{19},\textbf{53},150,152,154,\textbf{165},177,180,181,\textbf{192} \\
    \hline
    Q      & 16,17,\textbf{18},\textbf{19},\textbf{55},\textbf{56},\textbf{58},\textbf{59},150,152,154,164,177,180,181 & SM     & 16-24,89-98,115-118,121,122,149,151,153,177-182,207-214 \\
    \hline
    \multirow{2}[2]{*}{SM} & 16-24,89-98,115-118,121,122,149,151,153,162,163,177-182, & \multirow{2}[2]{*}{V} & 75-78,\textbf{83-86},89,91,93,95,\textbf{123-126},\textbf{131-134},196,\textbf{203},209,210, \\
           & 207-214 &        & 216,219,\textbf{227},\textbf{228} \\
    \hline
    V      & 75-78,83-86,89,91,93,95,196,202,209,210,216,219,225,226 & \multirow{2}[4]{*}{W} & 75-82,\textbf{85},\textbf{86},\textbf{88},89,91,93,95,97,98,109-112,119,120,122,\textbf{125}, \\
\cline{1-2}    W      & 75-84,87,89,91,93,95,97,98,107,108,111,112,119-121,139,140 &        & \textbf{126},\textbf{133},\textbf{134},\textbf{141},\textbf{142} \\
    \hline
    S      & 89,90,93,94,115-118,207,208 & \multirow{2}[4]{*}{S} & 89,90,93,94,115-118,\textbf{124},\textbf{126},\textbf{128},\textbf{130},\textbf{131},\textbf{133},\textbf{135},\textbf{137}, \\
\cline{1-2}    T      & 89,\textbf{92},93,\textbf{96},111,112,195,200,201,207,208 &        & 207,208,\textbf{222},\textbf{223} \\
    \hline
    U      & 89,90,93,94,111-114,168-173,175-182 & \multirow{2}[4]{*}{T} & 89,\textbf{92},93,\textbf{96},111,112,\textbf{124},\textbf{125},\textbf{132},\textbf{133},195,\textbf{205},207,208, \\
\cline{1-2}    Y      & 89,91,93,95,97,98,111,112,119,120 &        & \textbf{221-224} \\
    \hline
    PA     & 143-145,150,152,154,174 & \multirow{2}[4]{*}{U} & 89,90,93,94,111-114,\textbf{124},\textbf{126},\textbf{128},\textbf{130},\textbf{132},\textbf{134},\textbf{136},\textbf{138}, \\
\cline{1-2}    R      & 149,151,153,162,177,180,181 &        & 168-173,177-182 \\
    \hline
    Z      & 195,207,208,215,218 & Y      & 89,91,93,95,97,98,111,112,119,120,\textbf{125},\textbf{126},\textbf{133},\textbf{134},\textbf{141},\textbf{142} \\
    \hline
    ZA     & 195    & PA     & 143-145,150,152,154,174 \\
    \hline
  \multicolumn{2}{c||}{Time-reversal symmetric, considering SOC} & R      & 149,151,153,\textbf{163},177,180,181,\textbf{192} \\
    \hline
    A      & 16,18,21,22,\textbf{48},\textbf{49},\textbf{52},\textbf{54},\textbf{56},\textbf{58},\textbf{66},\textbf{68},\textbf{70} & Z      & 195,\textbf{201},207,208,215,218,\textbf{222},\textbf{224} \\
    \hline
    B      & 16,18,21,22,\textbf{48},\textbf{49},\textbf{52},\textbf{54},\textbf{56},\textbf{58},\textbf{66},\textbf{68},\textbf{70} & ZA     & 195,\textbf{201} \\
    \hline
    \hline
    \end{tabular}}
\end{table}%

\subsection{Nodal loop}\label{hsl4}
Here we list all the positions of HSLs which host BCs implying nodal loops within  neighboring HSPLs for four settings. % Take A in SG 28 in the setting with TRS and significant SOC as the example. A is a HSL in SG 28 and its coordinate in the conventional basis is $(u, 0, \frac{1}{2})$, could host a BC and the nodal line structure is occured on the corresponding HSPL M with coordinate $(u, 0, w)$. Two crossing bands could be labeled by two 1-dimendion double-valued irreps A3 and A4 and the crossing point is double degeneracy, which means the nodal line lying in M is double degeneracy. We further find that such a nodal line is essential so A$(u, 0, \frac{1}{2})$ is printed in red. For other three settings, the corresponding tables can alo be read following the same way as described here.
\subsubsection{Results for materials with TRS and significant SOC}
\begin{center}\begin{table*}[!h]

\end{table*}\end{center}

\newpage
\begin{center}\begin{table*}[!h]
(Continued from previous table: Time-reversal symmetric, considering SOC)\\
%
\end{table*}\end{center}

\newpage
\begin{center}\begin{table*}[!h]
(Continued from previous table: Time-reversal symmetric, considering SOC)\\
%
\end{table*}\end{center}

\newpage
\begin{center}\begin{table*}[!h]
(Continued from previous table: Time-reversal symmetric, considering SOC)\\
%
\end{table*}\end{center}

\subsubsection{Results for materials with TRS and negligible SOC}
\newpage
\begin{center}\begin{table*}[!h]
(Continued from previous table: Time-reversal symmetric, neglecting SOC)\\
%
\end{table*}\end{center}

\newpage
\begin{center}\begin{table*}[!h]
(Continued from previous table: Time-reversal symmetric, neglecting SOC)\\
%
\end{table*}\end{center}

\newpage
\begin{center}\begin{table*}[!h]
(Continued from previous table: Time-reversal symmetric, neglecting SOC)\\
%
\end{table*}\end{center}

\newpage
\begin{center}\begin{table*}[!h]
(Continued from previous table: Time-reversal symmetric, neglecting SOC)\\
%
\end{table*}\end{center}

\subsubsection{Results for materials without TRS and with significant SOC}
\newpage
\begin{center}\begin{table*}[!h]
(Continued from previous table: Time-reversal symmetry broken, considering SOC)\\
%
\end{table*}\end{center}

\newpage
\begin{center}\begin{table*}[!h]
(Continued from previous table: Time-reversal symmetry broken, considering SOC)\\
%
\end{table*}\end{center}

\newpage
\begin{center}\begin{table*}[!h]
(Continued from previous table: Time-reversal symmetry broken, considering SOC)\\
%
\end{table*}\end{center}

\newpage
\begin{center}\begin{table*}[!h]
(Continued from previous table: Time-reversal symmetry broken, considering SOC)\\
%
\end{table*}\end{center}

\subsubsection{Results for materials without TRS and with negligible SOC}
\newpage
\begin{center}\begin{table*}[!h]
(Continued from previous table: Time-reversal symmetry broken, neglecting SOC)\\
%
\end{table*}\end{center}

\newpage
\begin{center}\begin{table*}[!h]
(Continued from previous table: Time-reversal symmetry broken, neglecting SOC)\\
%
\end{table*}\end{center}

\newpage
\begin{center}\begin{table*}[!h]
(Continued from previous table: Time-reversal symmetry broken, neglecting SOC)\\
%
\end{table*}\end{center}

\newpage
\begin{center}\begin{table*}[!h]
(Continued from previous table: Time-reversal symmetry broken, neglecting SOC)\\
%
\end{table*}\end{center}

\section{High-symmetry plane}\label{hspl}
\subsection{Nodal surface}\label{hspl1}
%Here we briefly introduce how to obtain the results of all nodal surfaces.
%First we filter all k-vectors with non-trivial little groups to obtain all HSPLs, then we traverse all irreps in the little group of HSPLs to obtain HSPLs with degenerate irreps. It should be noted that when SOC is not negligible, Kramers degeneracy would occur in the centrosymmetric SGs, and the degenerate energy band on the HSPL does not decompose as we continuously move from a point on HSPL to a general point. It is necessary to further determine whether there is inversion symmetry in SG when SOC is included.
The results for all HSPLs being nodal surfaces are shown in the table below. %The first column contains labels for the HSPLs and the second column contains SG numbers for the corresponding HSPLs.
\begin{table}[!bhtp]
	\centering
%
	\label{table-surface}%
\end{table}%

\subsection{Nodal loop}\label{hspl2}

The results for all HSPLs with possible nodal loops are shown in Sec.~\ref{hspl2-pos}.
The results for all HSPLs with essential nodal loops are shown in Sec.~\ref{hspl2-ess}.
\clearpage
\subsubsection{Results for HSPLs with possible nodal line structures}\label{hspl2-pos}
\begin{table}[!bhtp]
	\centering
	%
%
	\label{possiblenl-hspl}%
\end{table}%
\clearpage
\subsubsection{Results for HSPLs with essential nodal line structures}\label{hspl2-ess}
\begin{table}[!bhtp]
	\centering
	%
\label{essential-NL}
\end{table}

\section{Type-I Hopf-link}\label{hopf1}
In this section, we list all the SGs allowing Type-I Hopf-link structure consisting of one nodal line threading a nodal loop.

\subsection{Results for materials with TRS and significant SOC}\label{hl1line1quantrssoc}
\begin{center}\begin{table*}[!h]
%
\end{table*}\end{center}

\newpage
\begin{center}\begin{table*}[!h]
(Continued from previous table: Time-reversal symmetric, considering SOC)\\
%
\end{table*}\end{center}

\subsection{Results for materials with TRS and negligible SOC}
\newpage
\begin{center}\begin{table*}[!h]
(Continued from previous table: Time-reversal symmetric, neglecting SOC)\\
%
\end{table*}\end{center}

\newpage
\begin{center}\begin{table*}[!h]
(Continued from previous table: Time-reversal symmetric, neglecting SOC)\\
%
\end{table*}\end{center}

\newpage
\begin{center}\begin{table*}[!h]
(Continued from previous table: Time-reversal symmetric, neglecting SOC)\\
%
\end{table*}\end{center}

\newpage
\begin{center}\begin{table*}[!h]
(Continued from previous table: Time-reversal symmetric, neglecting SOC)\\
%
\end{table*}\end{center}

\subsection{Results for materials without TRS and with significant SOC}
\newpage
\begin{center}\begin{table*}[!h]
(Continued from previous table: Time-reversal symmetry broken, considering SOC)\\
%
\end{table*}\end{center}

\newpage
\begin{center}\begin{table*}[!h]
(Continued from previous table: Time-reversal symmetry broken, considering SOC)\\
%
\end{table*}\end{center}

\newpage
\begin{center}\begin{table*}[!h]
(Continued from previous table: Time-reversal symmetry broken, considering SOC)\\
%
\end{table*}\end{center}

\newpage
\begin{center}\begin{table*}[!h]
(Continued from previous table: Time-reversal symmetry broken, considering SOC)\\
%
\end{table*}\end{center}

\subsection{Results for materials without TRS and with negligible SOC}

\begin{center}\begin{table*}[!h]
%
\end{table*}\end{center}

\newpage
\begin{center}\begin{table*}[!h]
(Continued from previous table: Time-reversal symmetry broken, neglecting SOC)\\
%
\end{table*}\end{center}

\newpage
\begin{center}\begin{table*}[!h]
(Continued from previous table: Time-reversal symmetry broken, neglecting SOC)\\
%
\end{table*}\end{center}

\newpage
\begin{center}\begin{table*}[!h]
(Continued from previous table: Time-reversal symmetry broken, neglecting SOC)\\
%
\end{table*}\end{center}

\section{Two nonequivalent HSPLs}\label{twohspls}
In this section, we list all irrep doublets in each HSL which could imply possible nodal loop  on corresponding HSPLs for four settings.
The results of all essential nodal chain structures are given in Sec.~\ref{chain}.
The results of possible hopf-link structures with two nodal loops are listed in Sec.~\ref{hopf2}.
\subsection{Results for materials with TRS and significant SOC}\label{allcoupletrssoc}
\begin{center}\begin{table*}[!h]
%
\end{table*}\end{center}

\newpage
\begin{center}\begin{table*}[!h]
(Continued from previous table: Time-reversal symmetric, considering SOC)\\
%
\end{table*}\end{center}

\clearpage
\subsection{Results for materials with TRS and negligible SOC}\label{allcoupletrsnosoc}

\begin{center}\begin{table*}[!h]
%
\end{table*}\end{center}

\clearpage
\subsection{Results for materials without TRS and with significant SOC}\label{allcouplenotrssoc}
\begin{center}\begin{table*}[!h]
%
\end{table*}\end{center}

\subsection{Results for materials without TRS and with negligible SOC}\label{allcouplenotrsnosoc}
\newpage
\begin{center}\begin{table*}[!h]
(Continued from previous table: Time-reversal symmetry broken, neglecting SOC)\\
%
\end{table*}\end{center}

\subsection{Results for all the positions of essential nodal chain structure}\label{chain}
\begin{table}[!h]
  \centering
   %
%
\end{table}%

\newpage
\begin{table}[!h]
  \centering
(Continued from previous table: Results for all the positions of essential nodal chain structure)\\
  %
%
\end{table}%

\section{Type-II Hopf-link}\label{hopf2}

\clearpage
\subsection{Results for materials with TRS and significant SOC}\label{hl2looptrssoc}
\begin{center}\begin{table*}[!h]
%
\end{table*}\end{center}

\clearpage
\subsection{Results for materials with TRS and negligible SOC}
\begin{center}\begin{table*}[!h]
%
\end{table*}\end{center}

\clearpage
\subsection{Results for materials without TRS and with significant SOC}

\begin{center}\begin{table*}[!h]
%(Continued from previous table: Time-reversal symmetry broken, considering SOC)\\
%
\end{table*}\end{center}

\clearpage
\subsection{Results for materials without TRS and with negligible SOC}

\begin{center}\begin{table*}[!h]
%(Continued from previous table: Time-reversal symmetry broken, neglecting SOC)\\
%
\end{table*}\end{center}

\clearpage
\section{Nodal line/loop materials}\label{materials}
Based on the Sec. \ref{hsl1} and the Sec. \ref{hspl2}, in the SM, we enumerate some example materials which host essential nodal lines/loops, as shown in the following table. We take Y$_3$Pt, Bi$_2$PdO$_4$, CuBi$_2$O$_4$, HfCoSn and IrTe$_2$ as materials examples and show their band structure in detail in Sec. \ref{materials-demon}.
\begin{table}[htbp]
	\centering
	\caption{Example materials with essential nodal lines. The first column lists the corresponding SGs. In the second and third column, it contains the positions of HSLs and HSPLs where the essential nodal lines occur. The dimensions of nodal lines are given in the fourth column. And the example materials are given in the last column. }
	\begin{tabular}{c|p{15em}|p{11em}|c|p{20em}}
		\hline
		\hline
		SG     & {\makecell [c]{HSL} }    & {\makecell [c]{HSPL} }   & Dimension & {\makecell [c]{Materials} } \bigstrut\\
		\hline
		7      &  & F$(u, 0, w)$, G$(u, \frac{1}{2}, w)$ & 2      & Ga$_9$Rh$_2$, Ga$_9$Ir$_2$, Te$_2$Au, BaAs$_2$ \bigstrut\\
		\hline
		9      &  & B$(u, 0, w)$ & 2      & Ga$_2$X$_3$(X=S, Se, Te), Al$_2$Se$_3$, LiAsS$_2$, Cu$_2$SiSe$_3$ \bigstrut\\
		\hline
		26     & G$( \frac{1}{2}, 0, w)$, H$(0,\frac{1}{2},w)$, LD$(0,0,w)$,  Q$(\frac{1}{2},\frac{1}{2},w)$ & K$(0, v, w)$, L$(\frac{1}{2}, v, w)$ & 2      & SrCa(CO$_3$)$_2$, BaHgS$_2$, CaPPt \bigstrut\\
		\hline
		28     & D$(\frac{1}{2},v,0)$, H$(0,\frac{1}{2},w)$, LD$(0,0,w)$, P$(\frac{1}{2},v,\frac{1}{2})$ & M$(u, 0, w)$, N$(u, \frac{1}{2}, w)$ & 2      & Hg$_2$TeO$_3$ \bigstrut\\
		\hline
		29     & D$(\frac{1}{2},v,0)$, H$(0,\frac{1}{2},w)$, LD$(0,0,w)$ & L$(\frac{1}{2}, v, w)$, M$(u, 0, w)$, N$(u, \frac{1}{2}, w)$ & 2      & SbIrS, KGeNO, CoAsS, GePtSe, GePtS, K$_2$I$_2$ClO$_6$,Li$_2$AgN, LiSbS, Mg(ScSe$_2$)$_2$, Ni$_6$Se$_5$, SiPtSe, Te$_2$PO$_7$, Y$_2$MgSe$_4$, AsS, Ba$_2$Tl(NO$_2$)$_5$\newline{} \bigstrut\\
		\hline
		30     & A$(u,0,\frac{1}{2})$, B$(0,v,\frac{1}{2})$, C$(u,\frac{1}{2},0)$, G$(\frac{1}{2},0,w)$, LD$(0,0,w)$, P$(\frac{1}{2},v,\frac{1}{2})$ & K$(0, v, w)$, L$(\frac{1}{2}, v, w)$, N$(u, \frac{1}{2}, w)$ & 2      & Nb$_3$(BiO$_3$)$_5$ \bigstrut\\
		\hline
		31     & D$(\frac{1}{2},v,0)$, H$(0,\frac{1}{2},w)$, LD$(0,0,w)$ & K$(0, v, w)$, M$(u, 0, w)$, N$(u, \frac{1}{2}, w)$   & 2      & Te$_2$W, Ga$_3$Te$_3$As, ZnCu$_2$SnS$_4$ \bigstrut\\
		\hline
		33     & C$(u,\frac{1}{2},0)$, D$(\frac{1}{2},v,0)$, LD$(0,0,w)$, Q$(\frac{1}{2},\frac{1}{2},w)$ & K$(0, v, w)$, L$(\frac{1}{2}, v, w)$, M$(u, 0, w)$ & 2      & TiZnN$_2$, ZnSnN$_2$, CaSnN$_2$, CdP$_2$, TlSbAu, CaCuGe, CuHgSI, Hg$_2$TeO$_5$, LiP$_5$, MgSnN$_2$ \bigstrut\\
		\hline
		34     & A$(u,0,\frac{1}{2})$, B$(0,v,\frac{1}{2})$, C$(u,\frac{1}{2},0)$, D$(\frac{1}{2},v,0)$, LD$(0,0,w)$, Q$(\frac{1}{2},\frac{1}{2},w)$ & K$(0, v, w)$, L$(\frac{1}{2}, v, w)$, M$(u, 0, w)$,  N$(u, \frac{1}{2}, w)$ & 2      & Li$_2$SnTeO$_6$, Li$_2$TiTeO$_6$, K$_2$AgSbS$_4$,Ag$_2$BiO$_3$, Ag$_2$WO$_4$, B$_5$Pb$_2$IO$_9$, BaSnHgS$_4$, K$_2$Na(SbO$_3$)$_3$, LiTeO$_3$, Rb$_2$Be$_2$Si$_2$O$_7$, Sb$_5$IO$_7$, Sr(ReO$_3$)$_3$, ZnC$_4$(NO$_3$)$_2$ \bigstrut\\
		\hline
		36     & H$(1,0,w)$, LD$(0,0,w)$ & K$(0, v, w)$ & 2      & AsPd$_2$, P$_2$W, Na$_2$PdS$_2$ \bigstrut\\
		\hline
		39     & A$(\frac{1}{2},0,w)$, SM$(0,0,w)$ & P$(0, v, w)$, Q$(\frac{1}{2}, v, w)$ & 2      & Bi$_2$TeO$_5$, Bi$_2$SeO$_5$ \bigstrut\\
		\hline
		40     &  B$(\frac{1}{2},v,0)$, SM$(0,0,w)$ & M$(u, 0, w)$ & 2      & K$_2$CdPb \bigstrut\\
		\hline
		41     &  B$(\frac{1}{2},v,0)$, SM$(0,0,w)$ & M$(u, 0, w)$, P$(0, v, w)$, Q$(\frac{1}{2}, v, w)$ & 2      & SeBr, SrS$_3$ \bigstrut\\
		\hline
		43     & A$(u,0,1)$, B$(0,v,1)$, LD$(0,0,w)$ & E$(0, v, w)$, J$(u, 0, w)$ & 2      & X$_2$Y$_3$(X=Hf, Zr; Y=Al, Ga), Y$_3$Ge$_5$, Ba$_5$P$_9$ \bigstrut\\
		\hline
		46     & LD$(0,0,w)$ & B$(0, v, w)$ & 2      & TiSiRu \bigstrut\\
		\hline
		57     & B$(0,v,\frac{1}{2})$, C$(u,\frac{1}{2},0)$, P$(\frac{1}{2},v,\frac{1}{2})$ & N$(u, \frac{1}{2}, w)$ & 4      & NdAl, HfGa, CaAlPd, Ca$_3$(GaPt)$_2$, Ca$_3$(GaPd)$_2$ \bigstrut\\
		\hline
		60     & B$(0,v,\frac{1}{2})$, E$(u,\frac{1}{2},\frac{1}{2})$, G$(\frac{1}{2},0,w)$ & L$(\frac{1}{2}, v, w)$, W$(u, v, \frac{1}{2} )$ & 4      & NiS$_2$, AlTeO$_4$, Sn$_2$(SO$_4$)$_3$ \bigstrut\\
		\hline
		61     & B$(0,v,\frac{1}{2})$, C$(u,\frac{1}{2},0)$, G$(\frac{1}{2},0,w)$ & L$(\frac{1}{2}, v, w)$, W$(u, v, \frac{1}{2} )$, N$(u, \frac{1}{2}, w)$ & 4      & NiAs$_2$, NiP, Bi$_2$Pt, SiNi$_2$, SiNi$_2$As, SnTePd \bigstrut\\
		\hline
		62     & A$(u,0,\frac{1}{2})$, C$(u,\frac{1}{2},0)$, E$(u,\frac{1}{2},\frac{1}{2})$, G$(\frac{1}{2},0,w)$, P$(\frac{1}{2},v,\frac{1}{2})$ & L$(\frac{1}{2}, v, w)$ & 4      & SnRh$_2$, GePt, SiPt, TiSiRh, HfSiPt, TaNiGe, NbSiPd, SrAl$_2$Au$_3$, Y$_3$Pt \bigstrut\\
		\hline
		100    & LD$(0,0,w)$, T$(u,\frac{1}{2},\frac{1}{2})$, V$(\frac{1}{2},\frac{1}{2},w)$, Y$(u,\frac{1}{2},0)$ & B$(0, v, w)$ & 2      & Ba$_2$Ti(XO$_4$)$_2$(X=Si, Ge) \bigstrut\\
		\hline
		\hline
	\end{tabular}%
	\label{materials1}%
\end{table}%

\clearpage
% Table generated by Excel2LaTeX from sheet 'Sheet2'
\begin{table}[htbp]
	\centering
	(Continued from previous table: example materials)
	\begin{tabular}{c|p{15em}|p{11em}|c|p{20em}}
		\hline
		\hline
		SG     & {\makecell [c]{HSL} }    & {\makecell [c]{HSPL} }   & Dimension & {\makecell [c]{Materials} } \bigstrut\\
		\hline
		102    & LD$(0,0,w)$, U$(0,v,\frac{1}{2})$, V$(\frac{1}{2},\frac{1}{2},w)$, Y$(u,\frac{1}{2},0)$ & B$(0, v, w)$, F$(u, \frac{1}{2}, w)$ & 2      & Be(CN)$_2$, Cd(CN)$_2$, Cs$_2$Hg$_6$S$_7$, K$_2$NaMo(OF)$_3$, K$_2$Zn$_6$O$_7$, Li$_6$Ti$_2$O$_7$, LiVF$_6$, Mg(CN)$_2$, S$_2$N, TaTiGaO$_6$, Zn(CN)$_2$ \bigstrut\\
		\hline
		104    & LD$(0,0,w)$, S$(u,u,\frac{1}{2})$, U$(0,v,\frac{1}{2})$, V$(\frac{1}{2},\frac{1}{2},w)$, Y$(u,\frac{1}{2},0)$ & B$(0, v, w)$, F$(u, \frac{1}{2}, w)$ & 2      & Ba$_5$In$_4$Bi$_5$ \bigstrut\\
		\hline
		108    & LD$(0,0,w)$, W$(\frac{1}{2},\frac{1}{2},w)$ & B$(u, 0, w)$ & 2      & Cu(BiO$_2$)$_2$, Bi$_2$PdO$_4$, NClOF$_4$, K$_3$NO$_3$ \bigstrut\\
		\hline
		109    & LD$(0,0,w)$, Y$(u,1-u,0)$ & A$(u, u, w)$ & 2      & Ga$_2$TeX$_2$(X=S, Se), AgPb$_2$Cl$_3$OF, AsRhO$_4$, Ba$_2$S$_3$, Li$_8$TeN$_2$, LiInO$_2$, NbAs, NbP, Re$_2$O$_7$, TaAlO$_4$, TaAs, TaP \bigstrut\\
		\hline
		110    & LD$(0,0,w)$, Y$(u,1-u,0)$ & B$(u, 0, w)$, A$(u, u, w)$ & 2      & BrN$_3$, CuReO$_4$, Li$_2$B$_4$O$_7$, Li$_2$Si$_4$O$_7$, V$_2$Ni$_2$PbO$_8$, Zn$_3$As$_2$, ZrMo$_2$O$_{11}$ \bigstrut\\
		\hline
		118    & LD$(0,0,w)$, U$(0,v,\frac{1}{2})$, V$(\frac{1}{2},\frac{1}{2},w)$, Y$(u,\frac{1}{2},0)$ & B$(0, v, w)$, F$(u, \frac{1}{2}, w)$ & 2      & CeSe$_2$, Ga$_5$Ir$_3$, In$_3$Ru, Nb(PS$_4$)$_2$, Zn$_3$(PS$_4$)$_2$, ZnSb$_2$ \bigstrut\\
		\hline
		120    & LD$(0,0,w)$ & B$(u, 0, w)$ & 2      & K(SnAu$_2$)$_2$ \bigstrut\\
		\hline
		122    & LD$(0,0,w)$, Y$(u,1-u,0)$ & A$(u, u, w)$ & 2      & CdXY$_2$(X=Sn, Ge; Y=P, As), ZnSnX$_2$(X=P, Sb, As), ZnSiAs$_2$, MgSiX$_2$(S=P, Sb), MgXAs$_2$(X=Sn, Ge), GaAgX$_2$(X=S, Te), LiXN$_2$(X=Os, Re), GeSe$_2$, AlCuSe$_2$, TlCuS$_2$, B$_2$PdO$_4$, Cu$_2$S, MoS$_2$, SbS$_2$,  \bigstrut\\
		\hline
		159    & R$(u,0,\frac{1}{2})$  & C$(u, u, w)$ & 2      & RbLiCrO$_4$, KAlSiO$_4$, Ag$_2$CO$_3$, Si$_3$N$_4$, Ge$_3$N$_4$ \bigstrut\\
		\hline
		161    &  & C$(u, -u, w)$ & 2      & AgNO$_3$, ZnSnO$_3$, ZnGeO$_3$ \bigstrut\\
		\hline
		185    & DT$(0,0,w)$, U$(\frac{1}{2},0,w)$ & C$(u, u, w)$ & 2      & Na$_3$As \bigstrut\\
		\hline
		186    & DT$(0,0,w)$, U$(\frac{1}{2},0,w)$ & D$(u, 0, w)$ & 2      & YSiAu, YGeAu, YCuPb, TiCuSn \bigstrut\\
		\hline
		188    & Q$(u,u,\frac{1}{2})$, SM$(u,0,0)$ & D$(u, 0, w)$ & 2      & BaX(SiO$_3$)$_3$(X=Sn, Ti), XTa(GeO$_3$)$_3$(X=K, Rb, Tl), BaSn(GeO$_3$)$_3$, RbNb(GeO$_3$)$_3$, Sr$_2$Be$_2$B$_2$O$_7$, LiScI$_3$ \bigstrut\\
		\hline
		190    & LD$(u,u,0)$, R$(u,0,\frac{1}{2})$ & C$(u, u, w)$ & 2      & XBe$_3$O$_4$(X=Ca, Sr), Tl$_3$AgI$_5$, Li$_2$Sb, HfCoSn \bigstrut\\
		\hline
		205    & Z$(u,\frac{1}{2},0)$ & B$(u, \frac{1}{2}, w)$ & 4      & Sb$_2$X(X=Pd, Au),  As$_2$X(X=Pd, Pt), Te$_2$X(X=Rh, Ir, Os), OsS$_2$, RhS$_2$, Bi$_2$Pt \bigstrut\\
		\hline
		220    & DT$(0,v,0)$, G$(1+u,1-u,1)$ & C$(u, u, w)$ & 2      & X$_3$N$_4$(X=Zr, Hf, Ge), Th$_3$X$_4$(X=P, Bi, As, Sb), Y$_2$C$_3$, Sr$_4$Bi$_3$, Sc$_4$C$_3$, Zr$_3$Ni$_3$Sb$_4$, Y$_3$Sb$_4$Au$_3$ \bigstrut\\
		\hline
		\hline
	\end{tabular}%
	\label{materials2}%
\end{table}%

\section{Methods}
Our ab initio calculations are based on WIEN2K \cite{WIEN2k} using the linearized augmented plane wave (LAPW) method. We consider spin-orbital coupling in all of our calculations. The local-density approximation(LDA) was adopted for the exchange-correlation functional.
\section{Materials demonstration}\label{materials-demon}

\begin{figure*}[!hbtp]
	% Requires \usepackage{graphicx}
	\includegraphics[width=1 \textwidth]{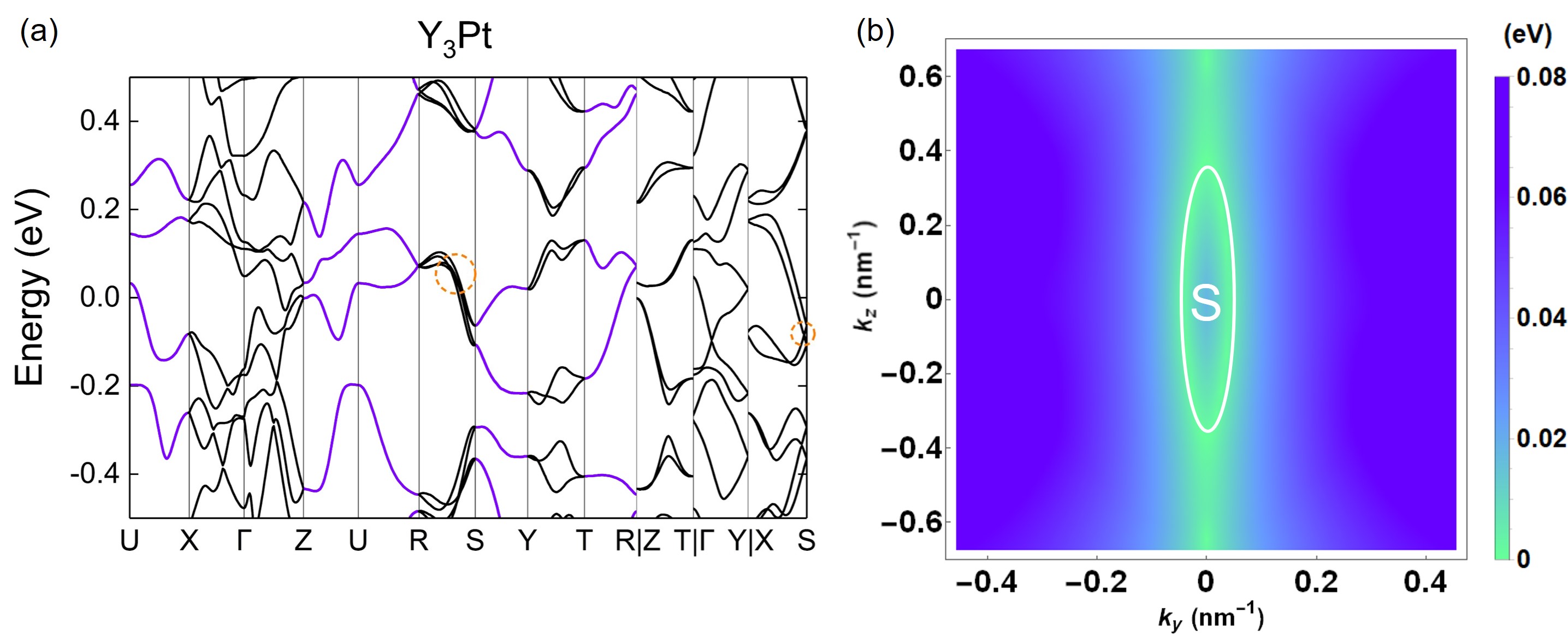}\\
	\caption{ (a)Electronic band structure of Y$_3$Pt (SG 62)  by first principle calculations. The energy bands forming essential nodal lines coinciding  with HSLs are printed in purple. We highlight two BCs lying in the HSLs S-D-X and S-Q-R by dashed curve circle. (b) The essential nodal loop around S for  Y$_3$Pt in HSPL L is shown by first-principles calculations. }
	\label{y3pt}
\end{figure*}

\begin{figure*}[!hbtp]
	% Requires \usepackage{graphicx}
	\includegraphics[width=1 \textwidth]{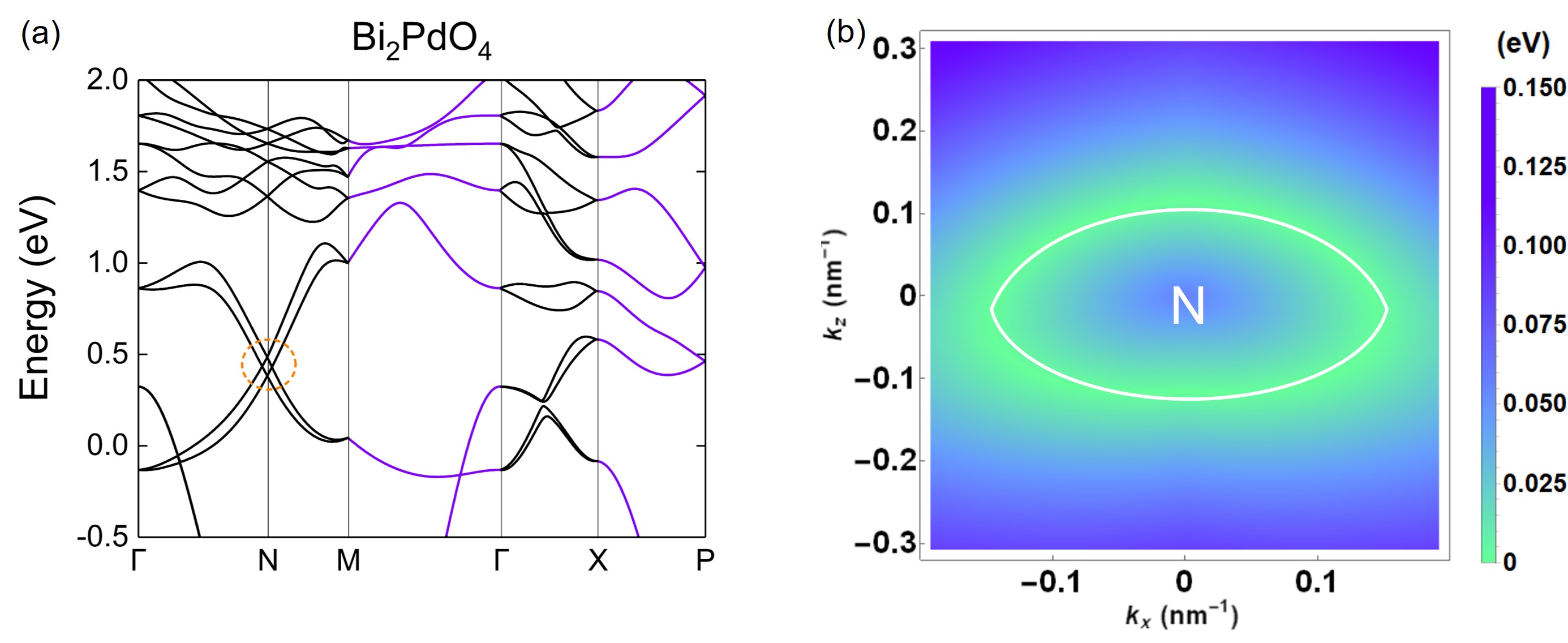}\\
	\caption{ (a)First principles calculated electronic band structure  Bi$_2$PdO$_4$ ( SG 108) \cite{bi2pdo4}. HSLs M-LD-GM and X-W-P can host essential nodal lines which are printed in purple, and the dashed circle indicates two hourglass BCs which actually lie in the nodal loop in B$(u,0,w)$($k_zk_x$ plane). The nodal loop is around N point and is shown in (b). }
	\label{bi2pdo4}
\end{figure*}

\begin{figure*}[!hbtp]
	% Requires \usepackage{graphicx}
	\includegraphics[width=1 \textwidth]{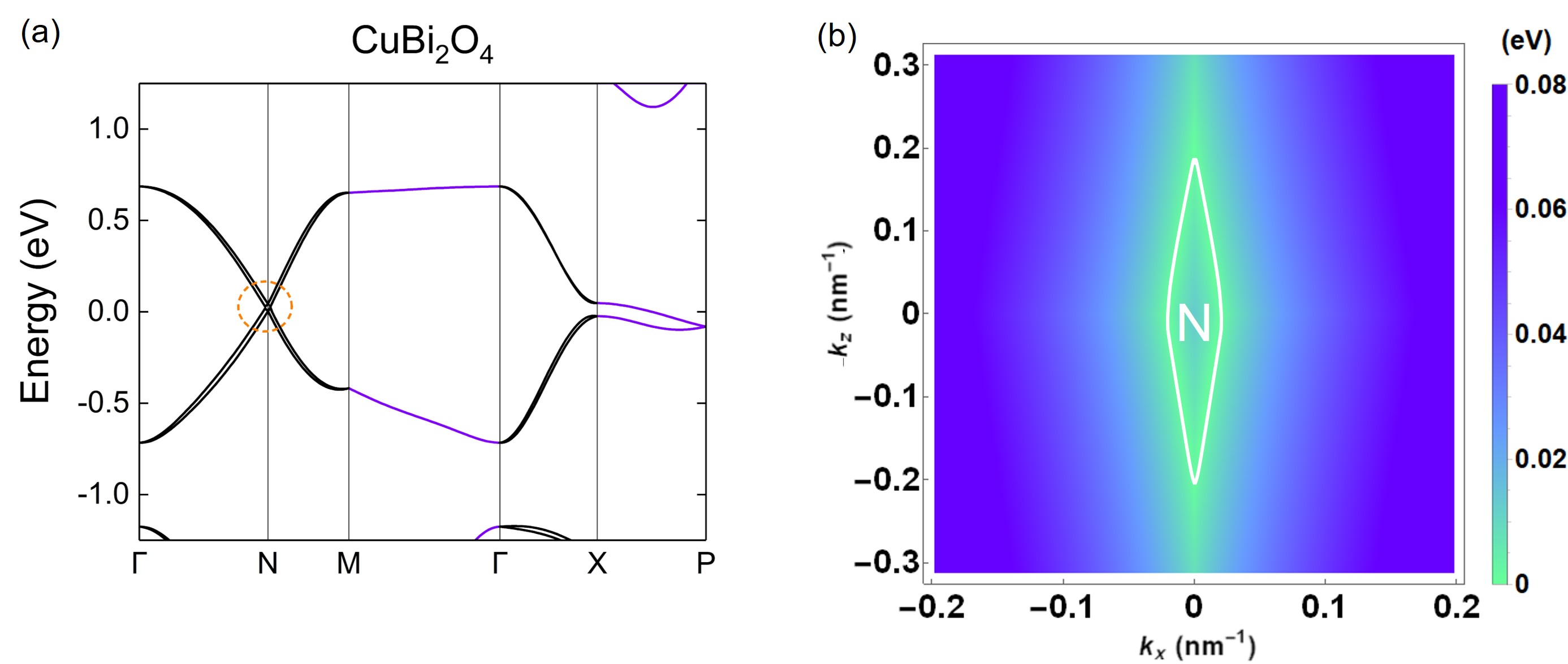}\\
	\caption{ (a)First principles calculated electronic band structure of  CuBi$_2$O$_4$ (SG 108) \cite{cubi2o4}. HSLs M-LD-GM and X-W-P can host  essential nodal lines which are printed in purple, and the dashed circle indicates two hourglass BCs. The BCs lie in the nodal loop in B which is around N point and is shown in (b). }
	\label{cubi2o4}
\end{figure*}

\begin{figure*}[!hbtp]
	% Requires \usepackage{graphicx}
	\includegraphics[width=1 \textwidth]{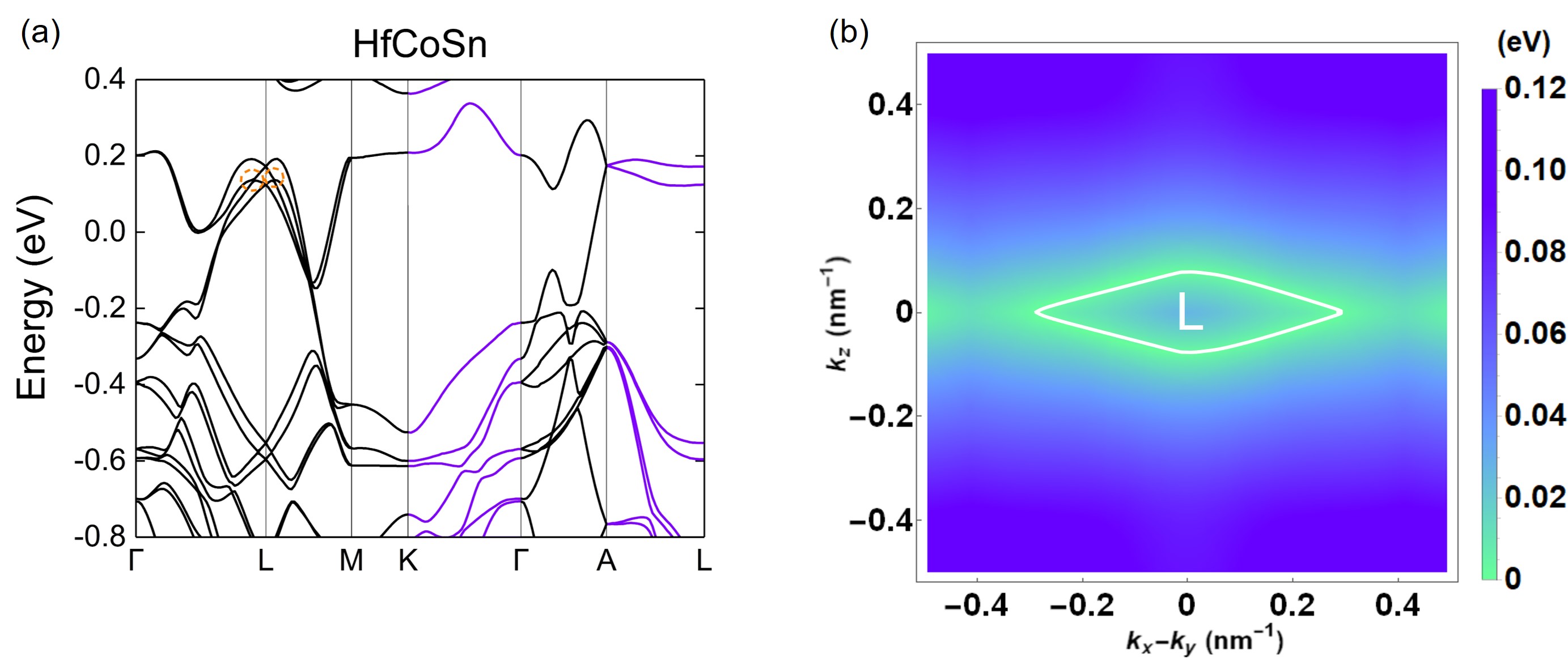}\\
	\caption{ (a)First principles calculated electronic band structure of HfCoSn (SG 190) \cite{hfcosn}. HSLs K-LD-GM and A-R-L can host  essential nodal lines which are printed in purple, and the dashed circle indicates two hourglass BCs. The BCs lie in the nodal loop in C$(u,u,w)$ which is around L$(\frac{1}{2},\frac{1}{2},\frac{1}{2})$ point and is shown in (b). }
	\label{hfcosn}
\end{figure*}

\begin{figure*}[!hbtp]
	% Requires \usepackage{graphicx}
	\includegraphics[width=1 \textwidth]{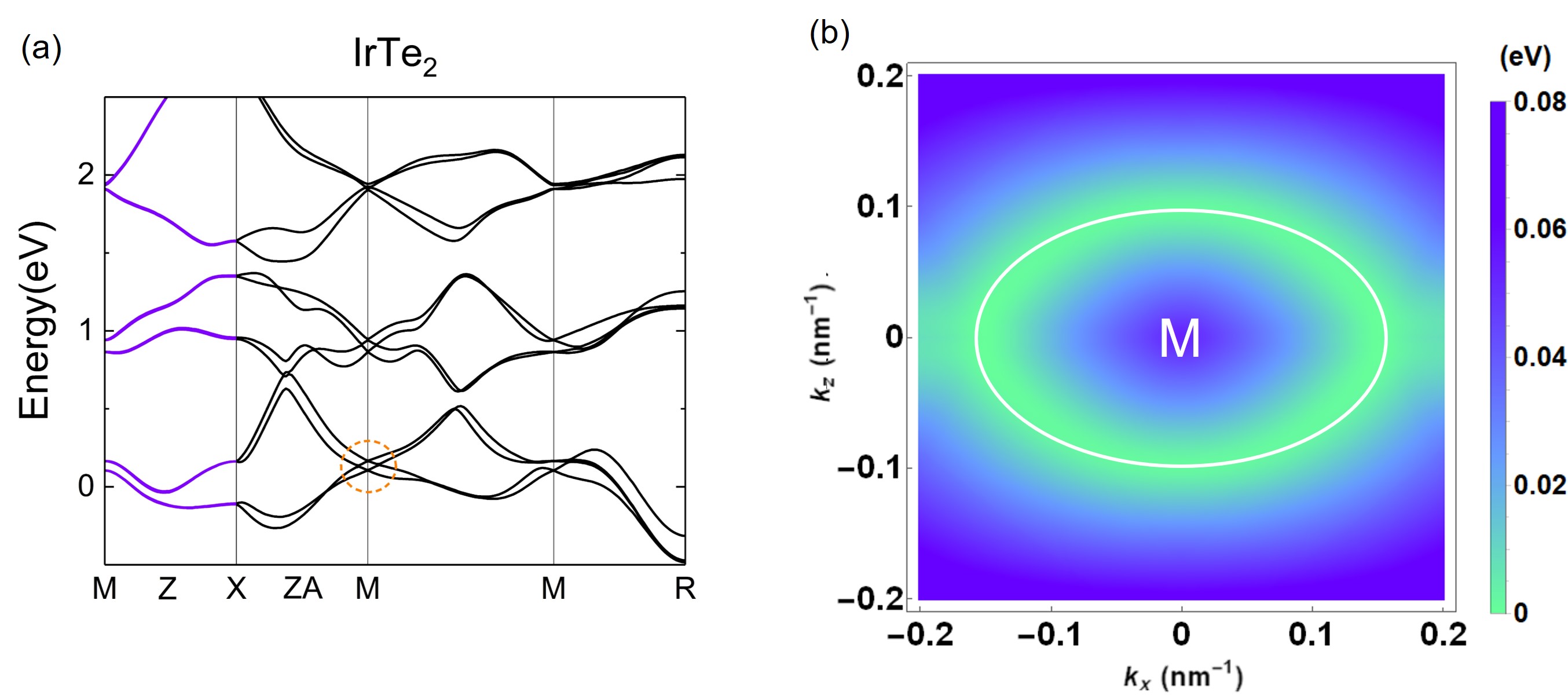}\\
	\caption{(a) Electronic band structure of IrTe$_2$ (SG 205) \cite{irte2} where the line M-ZA-X hosts an essential nodal line which is printed in purple, and two hourglass BCs around M point are indicated by a dashed circle. (b) The essential nodal loop around M for IrTe$_2$ in HSPL B $(u,\frac{1}{2},w)$ is shown by first-principles calculations. }
	\label{irte2}
\end{figure*}

\end{document}